\documentclass[12pt]{article}

\usepackage[T1]{fontenc}

\usepackage{amsmath}
\usepackage{amssymb}
\usepackage{times}
\usepackage{txfonts}
\usepackage[mathlines,displaymath]{lineno}
\DeclareMathAlphabet{\mathbold}{OML}{txr}{b}{it}

\usepackage{multirow}
\usepackage{hhline}
\usepackage{units}
\usepackage{epsfig}
\usepackage{longtable}
\usepackage{xspace}
\usepackage{bm} 
\usepackage{acronym}

\usepackage{graphicx}
\usepackage{rotating}

\usepackage{hyperref} 
\hypersetup{colorlinks=true, urlcolor=blue}
\usepackage{cite} 
\hypersetup{
  colorlinks,
  citecolor=blue,
  linkcolor=red,
  urlcolor=blue
  }
  
\renewcommand{\arraystretch}{1.4} 
\newlength{\dinwidth}
\newlength{\dinmargin}
\usepackage{parskip}

\begin{document}  




%
\newcommand{\TODO}{{\color{red}TODO}\xspace}
\newcommand{\kd}{\textcolor{black}} 

\newcommand{\muf}{\ensuremath{\mu_{f}}\xspace}
\newcommand{\mur}{\ensuremath{\mu_{r}}\xspace}
\newcommand{\as}{\ensuremath{\alpha_s}\xspace}
\newcommand{\asmz}{\ensuremath{\alpha_s(M_Z)}\xspace}
\newcommand{\asmur}{\ensuremath{\alpha_s(\mur)}\xspace}
\newcommand{\aem}{\ensuremath{\alpha_{\mathrm{em}}}\xspace}
\newcommand{\Lumi}{\ensuremath{\mathcal{L}}}
\newcommand{\pb}{\rm pb}
\newcommand{\invpb}{\ensuremath{\rm{pb}^{-1}}}
\newcommand{\PDF}{\ensuremath{{\rm PDF}}\xspace}
\renewcommand{\deg}{\ensuremath{^\circ}\xspace}
\newcommand{\unitmatrix}{1\!\!1}
\newcommand{\fC}{\ensuremath{f^{\rm C}}\xspace}
\newcommand{\fU}{\ensuremath{f^{\rm U}}\xspace}
\newcommand{\bas}{\ensuremath{\boldsymbol{{\alpha_s}}}\xspace} 
\newcommand{\basmz}{\ensuremath{\boldsymbol{\alpha_s(M_Z)}}\xspace} 
\newcommand{\bmur}{\ensuremath{\boldsymbol{\mu_{r}}}\xspace}
\newcommand{\basmur}{\ensuremath{\boldsymbol{\alpha_s(\mur)}}\xspace} 

\newcommand{\chisq}{\ensuremath{\chi^{2}}}
\newcommand{\chisqA}{\ensuremath{\chi_{\rm A}^{2}}}
\newcommand{\chisqL}{\ensuremath{\chi_{\rm L}^{2}}}
\newcommand{\ndf}{\ensuremath{n_{\rm dof}}}
\newcommand{\A}{\ensuremath{\bm{A}}}
\newcommand{\M}{\ensuremath{\bm{M}}}
\newcommand{\V}{\ensuremath{\bm{V}}}
\newcommand{\B}{\ensuremath{\bm{B}}}
\newcommand{\J}{\ensuremath{\bm{J}}}
\newcommand{\N}{\ensuremath{\bm{N}}}
\newcommand{\LL}{\ensuremath{\bm{L}}}

\newcommand{\femjet}{\ensuremath{f_{\mathrm{em,jet}}}\xspace}
\newcommand{\femjetgen}{\ensuremath{f_{\mathrm{em,jet}}^{\mathrm{gen}}}\xspace}
\newcommand{\femjetrec}{\ensuremath{f_{\mathrm{em,jet}}^{\mathrm{rec}}}\xspace}
\newcommand{\Pem}{\ensuremath{P_{\mathrm{em}}}\xspace}
\newcommand{\Eem}{\ensuremath{E_{\mathrm{em}}}\xspace}
\newcommand{\Ehad}{\ensuremath{E_{\mathrm{had}}}\xspace}
\newcommand{\Ptbal}{\ensuremath{P_{\mathrm{T}}}--balance\ }
\newcommand{\PTbal}{\ensuremath{P_{\mathrm{T,bal}}}\xspace}
\newcommand{\Ptgen}{\ensuremath{P_{\mathrm{T}}^\mathrm{gen}}\xspace}
\newcommand{\Ptda}{\ensuremath{P_{\mathrm{T}}^{\mathrm{da}}}\xspace}
\newcommand{\thetajet}{\ensuremath{\theta_{\mathrm{jet}}}\xspace}
\newcommand{\etajet}{\ensuremath{\eta_{\mathrm{jet}}}\xspace}
\newcommand{\Ejet}{\ensuremath{E_{\mathrm{jet}}}\xspace}
\newcommand{\Ejetda}{\ensuremath{E^{\mathrm{da}}_{\mathrm{jet}}}\xspace}
\newcommand{\relres}{\ensuremath{\sigma(\Ejet)/\Ejet}\xspace}

\newcommand{\Empz}{\ensuremath{E-p_z}\xspace}
\newcommand{\gammah}{\ensuremath{\gamma_{\mathrm{h}}}\xspace}
\newcommand{\Pth}{\ensuremath{P_{\mathrm{T}}^{\mathrm{h}}}\xspace}
\newcommand{\Pzh}{\ensuremath{p_z^{\mathrm{h}}}\xspace}
\newcommand{\h}{\ensuremath{\mathrm{h}}}

\newcommand{\e}{\ensuremath{\mathrm{e}}}
\newcommand{\Ee}{\ensuremath{E_\mathrm{e}^\prime}\xspace}
\newcommand{\Pte}{\ensuremath{P_{\mathrm{T}}^{\mathrm{e}}}\xspace}
\newcommand{\thetae}{\ensuremath{\theta_\mathrm{e}}\xspace}
\newcommand{\phie}{\ensuremath{\phi_\mathrm{e}}\xspace}
\newcommand{\Eda}{\ensuremath{E^{\mathrm{da}}}\xspace}

\newcommand{\Ebeam}{\ensuremath{E_{\mathrm{e}}}\xspace}
\newcommand{\Pbeam}{\ensuremath{E_{\mathrm{p}}}\xspace}

\newcommand{\xbj}{\ensuremath{x}\xspace}
\newcommand{\Qsq}{\ensuremath{Q^2}\xspace}

\newcommand{\ptjet}{\ensuremath{P_{\rm T}^{\rm jet}}\xspace}
\newcommand{\meanpt}{\ensuremath{\langle P_{\rm T} \rangle}}
\newcommand{\meanptdi}{\ensuremath{\langle P_{\mathrm{T}} \rangle_{2}}\xspace}
\newcommand{\meanpttri}{\ensuremath{\langle P_{\mathrm{T}} \rangle_{3}}\xspace}
\newcommand{\pt}{\ensuremath{P_{\rm T}}\xspace}
\newcommand{\Pt}{\ensuremath{P_{\mathrm{T}}}\xspace}
\newcommand{\Ptone}{\ensuremath{P_{\mathrm{T,1}}}\xspace}
\newcommand{\Pttwo}{\ensuremath{P_{\mathrm{T,2}}}\xspace}

\newcommand{\kt}{\ensuremath{{k_{\mathrm{T}}}}\xspace}
\newcommand{\antikt}{\ensuremath{{\mathrm{anti-}k_{\mathrm{T}}}}\xspace}
\newcommand{\bkt}{\ensuremath{\bm{k}_{\boldsymbol{\mathrm{T}}} }\xspace}
\newcommand{\bantikt}{\ensuremath{\boldsymbol{\mathrm{anti-}}\bm{k}_{\boldsymbol{\mathrm{T}}}}\xspace}

\newcommand{\Et}{\ensuremath{E_\mathrm{T}}\xspace}
\newcommand{\etalab}{\ensuremath{\eta^{\mathrm{jet}}_{\mathrm{lab}}}\xspace}
\newcommand{\ptlab}{\ensuremath{P^{\mathrm{jet}}_{\mathrm{T,lab}}}\xspace}
\newcommand{\Mjj}{\ensuremath{M_{\mathrm{12}}}\xspace}
\newcommand{\Mjjj}{\ensuremath{M_{\rm 123}}}
\newcommand{\xij}{\ensuremath{\xi}\xspace}
\newcommand{\xidi}{\ensuremath{\xi_2}\xspace}
\newcommand{\xitri}{\ensuremath{\xi_3}\xspace}

\newcommand{\ud}{\ensuremath{\mathrm{d}}\xspace}
\newcommand{\LO}{\ensuremath{\mathcal{O}(\alpha_s^0)}\xspave}
\newcommand{\Oa}{\ensuremath{\mathcal{O}(\alpha_s)}\xspace}
\newcommand{\Oaa}{\ensuremath{\mathcal{O}(\alpha_s^2)}\xspace}
\newcommand{\Oaaa}{\ensuremath{\mathcal{O}(\alpha_s^3)}\xspace}

\newcommand{\eq}{equation}
\newcommand{\fig}{figure}
\newcommand{\tab}{table}

\newcommand{\GeV}{\ensuremath{\mathrm{GeV}}\xspace}
\newcommand{\GeVsq}{\ensuremath{\mathrm{GeV}^2}\xspace}

\newcommand{\fifteen}{\ensuremath{15\,000}}
\newcommand{\dd}{\mathrm{d}}
\newcommand{\Ord}{\ensuremath{\mathcal{O}}}

\newcommand{\delrel}{\ensuremath{\delta}}
\newcommand{\delabs}{\ensuremath{\Delta}}
\newcommand{\dabs}[2][1]{\ensuremath{\delabs^{{\rm{#1}}}_{{#2}}}}
\newcommand{\drel}[2][1]{\ensuremath{\delrel^{{\rm{#1}}}_{{#2}}}}
\newcommand{\Nsys}{{N_{\rm sys}}}
\newcommand{\stat}{{\rm stat}}
\newcommand{\sys}{{\rm sys}}

\newcommand{\dHFS}[2][] {\dabs[RCES]{#2}\xspace}
\newcommand{\dJES}[2][] {\dabs[JES]{#2}\xspace}
\newcommand{\dLAr}[2][] {\dabs[LAr Noise]{#2}\xspace}
\newcommand{\dEe}[2][]  {\dabs[E_{e}]{#2}\xspace}
\newcommand{\dThe}[2][] {\dabs[\theta_{e}]{#2}\xspace}
\newcommand{\dID}[2][]  {\dabs[ID(e)]{#2}\xspace}
\newcommand{\dMod}[2][] {\dabs[Model]{#2}\xspace}
\newcommand{\dLumi}[2][]{\dabs[Lumi]{#2}\xspace}
\newcommand{\dTrig}[2][]{\dabs[Trig]{#2}\xspace}
\newcommand{\dTCl}[2][] {\dabs[TrkCl]{#2}\xspace}
\newcommand{\dNorm}[2][]{\dabs[Norm]{#2}\xspace}

\newcommand{\csdsub}{\ensuremath{}}
\newcommand{\csdsubn}{\ensuremath{}}
\newcommand{\CS}{\ensuremath{\sigma}}
\newcommand{\CSN}{\ensuremath{\sigma/\sigma_{\rm NC}}}
\newcommand{\trenn}{\cdot}
\newcommand{\bQsq}{\ensuremath{\bm{Q^2}}}
\newcommand{\bxidi}{\ensuremath{\bm{\xi_2}}}
\newcommand{\bxitri}{\ensuremath{\bm{\xi_3}}}
\newcommand{\bmeanptdi}{\ensuremath{\bm{\langle P_{\mathrm{T}} \rangle_{2}}}}
\newcommand{\bmeanpttri}{\ensuremath{\bm{\langle P_{\mathrm{T}} \rangle_{3}}}}
\newcommand{\bptjet}{\ensuremath{\bm{P_{\rm T}^{\rm jet}}}}
\newcommand{\bpt}{\ensuremath{\bm{P_{\rm T}}}}

\newcommand{\DStat}[1]{\drel[\stat]{#1}\xspace}
\newcommand{\DSys}[1] {\drel[\sys]{#1}\xspace}
\newcommand{\DHFS}[1] {\drel[RCES]{#1}\xspace}
\newcommand{\DJES}[1] {\drel[JES]{#1}\xspace}
\newcommand{\DLAr}[1] {\drel[LAr Noise]{#1}\xspace}
\newcommand{\DEe}[1]  {\drel[E_{e}^\prime]{#1}\xspace}
\newcommand{\DThe}[1] {\drel[\theta_{e}]{#1}\xspace}
\newcommand{\DID}[1]  {\drel[ID(e)]{#1}\xspace}
\newcommand{\DLumi}[1]{\drel[Lumi]{#1}\xspace}
\newcommand{\DTrig}[1]{\drel[Trig]{#1}\xspace}
\newcommand{\DTCL}[1] {\drel[TrkCl]{#1}\xspace}
\newcommand{\DMod}[1] {\drel[Model]{#1}\xspace}
\newcommand{\DNorm}[1]{\drel[Norm]{#1}\xspace}

\newcommand{\cHad} {\ensuremath{c^{\rm had}}    }
\newcommand{\cEW}  {\ensuremath{c^{\rm ew}}}
\newcommand{\cHadi} {\ensuremath{c_i^{\rm had}}}
\newcommand{\cEWi}  {\ensuremath{c_i^{\rm ew}}}
\newcommand{\DHad} {\drel[had]{}\xspace}

\newcommand{\sI} {\ensuremath{\sigma_{\rm jet}}\xspace}
\newcommand{\sD} {\ensuremath{\sigma_{\rm dijet}}\xspace}
\newcommand{\sT} {\ensuremath{\sigma_{\rm trijet}}\xspace}
\newcommand{\sNC} {\ensuremath{\sigma_{\rm NC}}\xspace}
\newcommand{\sIN} {\ensuremath{\displaystyle\frac{\sI}{\sNC}}\xspace}
\newcommand{\sDN} {\ensuremath{\displaystyle\frac{\sD}{\sNC}}\xspace}
\newcommand{\sTN} {\ensuremath{\displaystyle\frac{\sT}{\sNC}}\xspace}

\newcommand{\NLOJet}{NLOJet++}
\newcommand{\Rapgap}{RAPGAP\xspace}
\newcommand{\Heracles}{HERACLES\xspace}
\newcommand{\Lepto}{LEPTO\xspace}
\newcommand{\Django}{DJANGO\xspace}
\newcommand{\Sherpa}{SHERPA\xspace}

\newcommand{\HERAI}{HERA-1\xspace}
\newcommand{\HERAII}{HERA-2\xspace}

\def\Journal#1#2#3#4{{#1} {\bf #2} (#3) #4}
\def\NCA{\em Nuovo Cimento}
\def\NIM{\em Nucl. Instrum. Methods}
\def\NIMA{{\em Nucl. Instrum. Methods} {\bf A}}
\def\NPB{{\em Nucl. Phys.}   {\bf B}}
\def\PLB{{\em Phys. Lett.}   {\bf B}}
\def\PRL{\em Phys. Rev. Lett.}
\def\PRD{{\em Phys. Rev.}    {\bf D}}
\def\ZPC{{\em Z. Phys.}      {\bf C}}
\def\EJC{{\em Eur. Phys. J.} {\bf C}}
\def\CPC{\em Comp. Phys. Commun.}

 

\acrodef{BBE}{Backward Barrel\acroextra{, Electromagnetic (calorimeter wheel)}}
\acrodef{BCDMS}{Bologna-Cern-Dubna-Munich-Saclay}
\acrodef{BGF}{Boson Gluon Fusion}
\acrodef{BPC}{Backward Proportional Chamber}
\acrodef{BST}{Backward Silicon Tracker}
\acrodef{CC}{charged current}
\acrodef{CDM}{Colour Dipole Model}
\acrodef{CB}{Central Barrel\acroextra{ (calorimter wheel)}}
\acrodef{CIP}{Central Inner Proportional Chamber}
\acrodef{CJC}{Central Jet Chamber}
\acrodef{COP}{Central Outer Proportional Chamber}
\acrodef{COZ}{Central Outer z-Chamber}
\acrodef{CST}{Central Silicon Tracker}
\acrodef{CTD}{Central Track Detector}
\acrodef{DESY}{Deutsches Elektronen Synchrotron}
\acrodef{DIS}{deep-inelastic scattering}
\acrodef{DGLAP}{Dokshitzer, Gribov, Lipatov, Altarelli, Parisi}
\acrodef{DREAM}{Dual-Readout Module}
\acrodef{DST}{Data Summary Tape}
\acrodef{DVCS}{Deeply Virtual Compton Scattering}
\acrodef{EW}{electroweak}
\acrodef{EMC}{Electromagnetic Calorimeter}
\acrodef{ET}{Electron Tagger}
\acrodef{FB}{Forward Barrel\acroextra{ (calorimeter wheel)}}
\acrodef{FMD}{Forward Muon Detector}
\acrodef{FST}{Forward Silicon Tracker}
\acrodef{FTD}{Forward Track Detector}
\acrodef{FTT}{Fast Track Trigger}
\acrodef{H1OO}{H1 Object Oriented Analysis Software}
\acrodef{HAC}{Hadronic Calorimeter}
\acrodef{HAT}{H1 Event Tag}
\acrodef{HERA}{Hadron-Elektron-Ring-Anlage}
\acrodef{HFS}{hadronic final state}
\acrodef{IF}{Inner Forward\acroextra{ (calorimeter wheel)}}
\acrodef{IP}{Interaction Point}
\acrodef{IR}{infrared}
\acrodef{LAr}{liquid argon}
\acrodef{LEP}{Large Electron Positron Collider}
\acrodef{LHC}{Large Hadron Collider}
\acrodef{LO}{leading order}
\acrodef{MC}{Monte Carlo\acroextra{ (event generator)}}
\acrodef{MEPS}{matrix elements and parton shower}
\acrodef{NC}{neutral current}
\acrodef{NMC}{New Muon Collaboration}
\acrodef{NNLO}{next-to-next-to-leading order}
\acrodef{NLO}{next-to-leading order}
\acrodef{OF}{Outer Forward\acroextra{ (calorimeter wheel)}}
\acrodef{PETRA}{Positron-Elektron-Ring-Anlage}
\acrodef{PD}{Photon Detector}
\acrodef{PDF}{Parton Distribution Function}
\acrodef{POT}{Production Output Tape}
\acrodef{pQCD}{perturbative \ac{QCD}}
\acrodef{QCD}{Quantum Chromodynamics}
\acrodef{QCDC}{\ac{QCD} Compton}
\acrodef{QED}{Quantum Electrodynamics}
\acrodef{QEDC}{\ac{QED} Compton}
\acrodef{QPM}{Quark Parton Model}
\acrodef{RGE}{Renormalisation Group Equation}
\acrodef{SLAC}{Stanford Linear Accelerator Center}
\acrodef{SM}{Standard Model}
\acrodef{SpaCal}{Spaghetti Calorimeter}
\acrodef{TC}{Tail Catcher}
\acrodef{ToF}{Time-of-Flight}
\acrodef{UV}{ultraviolet}


\begin{titlepage}

\noindent
\begin{flushleft}
{\tt DESY 14-089    \hfill    ISSN 0418-9833} \\
{\tt June 2014}                  \\
\end{flushleft}

\noindent
\noindent

\vspace{2cm}
\begin{center}
\begin{Large}

{\bf Measurement of Multijet Production in $\bm{e\!\!\;p}$ Collisions at High $\bQsq$ and Determination of the Strong Coupling $\bas$}

\vspace{2cm}

H1 Collaboration

\end{Large}
\end{center}

\vspace{2cm}

\begin{abstract}
Inclusive jet, dijet and trijet differential cross sections are measured in neutral current deep-inelastic scattering for exchanged boson virtualities $150 < \Qsq < \unit[\fifteen]{\GeVsq}$ using the H1 detector at HERA.
The data were taken in the years 2003 to 2007 and correspond to
an integrated luminosity of $351~\mathrm{pb}^{-1}$. 
Double differential Jet cross sections are obtained using a regularised unfolding procedure. 
They are presented as a function of \Qsq and the transverse momentum of the jet, \ptjet, 
and as a function  of \Qsq and the proton's longitudinal momentum fraction, \xij, carried by the parton participating in the hard interaction. 
In addition normalised double differential jet cross sections are measured as the ratio of the jet cross sections to the inclusive neutral current cross sections in the respective \Qsq\ bins of the jet measurements.
Compared to earlier work, the measurements benefit from an improved reconstruction and calibration of the hadronic final state.
The cross sections are compared to perturbative QCD calculations in next-to-leading order and are used to determine the running coupling and the value of the strong coupling constant as $\asmz = 0.1165 \;\, (8)_{\rm exp} \;\, (38)_{\rm pdf,theo} \, $.
\end{abstract}

\vspace{1.5cm}

\begin{center}
To be submitted to \EJC
\end{center}

\end{titlepage}

%
%
%
\begin{flushleft}

V.~Andreev$^{21}$,             
A.~Baghdasaryan$^{33}$,        
K.~Begzsuren$^{30}$,           
A.~Belousov$^{21}$,            
P.~Belov$^{10}$,               
V.~Boudry$^{24}$,              
G.~Brandt$^{45}$,              
M.~Brinkmann$^{10}$,           
V.~Brisson$^{23}$,             
D.~Britzger$^{10}$
% ...


\bigskip{\it
% ...
$ ^{10}$ DESY, Hamburg, Germany \\
% ...
$ ^{21}$ Lebedev Physical Institute, Moscow, Russia \\
% ...
$ ^{23}$ LAL, Universit\'e Paris-Sud, CNRS/IN2P3, Orsay, France \\
 $ ^{24}$ LLR, Ecole Polytechnique, CNRS/IN2P3, Palaiseau, France \\
% ...
$ ^{30}$ Institute of Physics and Technology of the Mongolian
          Academy of Sciences, Ulaanbaatar, Mongolia \\
% ...
$ ^{33}$ Yerevan Physics Institute, Yerevan, Armenia \\
% ...
$ ^{45}$ Department of Physics, Oxford University,
          Oxford, UK$^{ b}$ \\

\bigskip
% ...
$ ^b$ Supported by the UK Science and Technology Facilities Council,
      and formerly by the UK Particle Physics and
      Astronomy Research Council \\
% ...
}
\end{flushleft}
%


\clearpage
\section{Introduction}

Jet production in neutral current (NC) deep-inelastic $ep$ scattering (DIS) at HERA is an important process to study the strong interaction and its theoretical description by Quantum Chromodynamics (QCD)~\cite{Fritzsch:72:135,Gross:73:1343,Politzer:73:1346,Ali:2010tw}. 
Due to the asymptotic freedom of QCD, quarks and gluons participate as quasi-free particles in short distance interactions. 
At larger distances they hadronise into collimated jets of hadrons, which provide momentum information of the underlying partons. Thus, the jets can be measured and compared to perturbative QCD (pQCD) predictions, corrected for hadronisation effects. 
This way the theory can be tested, and the value of the strong coupling, \asmz, as well as its running can be measured with high precision. \kd{A comprehensive review of jets in $ep$ scattering at HERA is given in~\cite{SchornerSadenius:2012de}.}

\begin{figure}[h]
\centering
 \includegraphics[height=3.5cm]{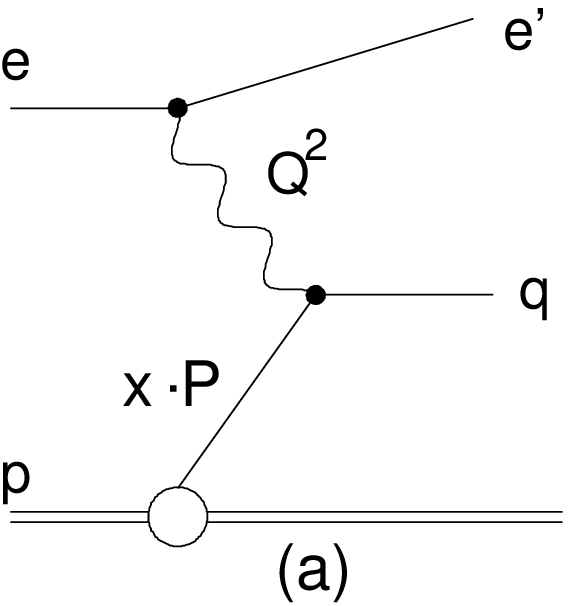}\hskip0.5cm
 \includegraphics[height=3.5cm]{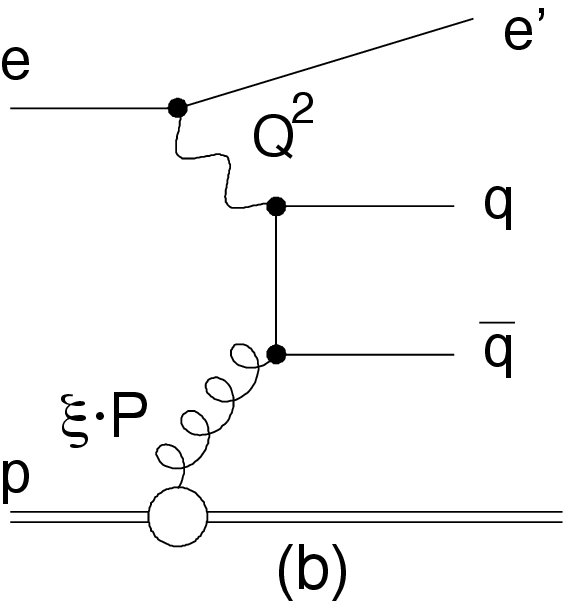}\hskip0.5cm
 \includegraphics[height=3.5cm]{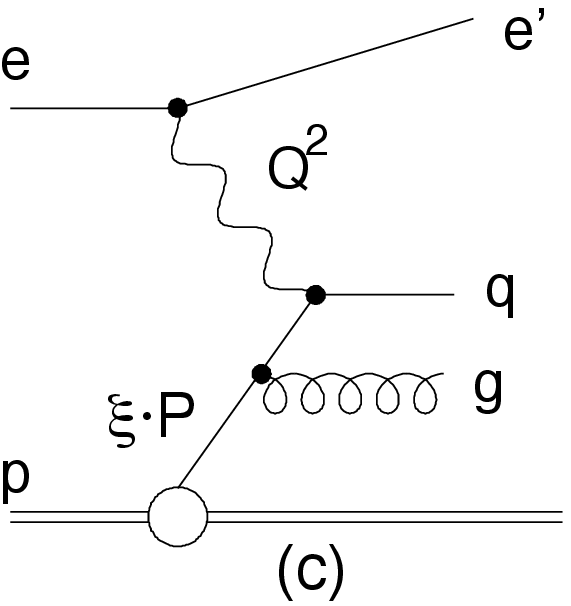}\hskip0.5cm
 \includegraphics[height=3.5cm]{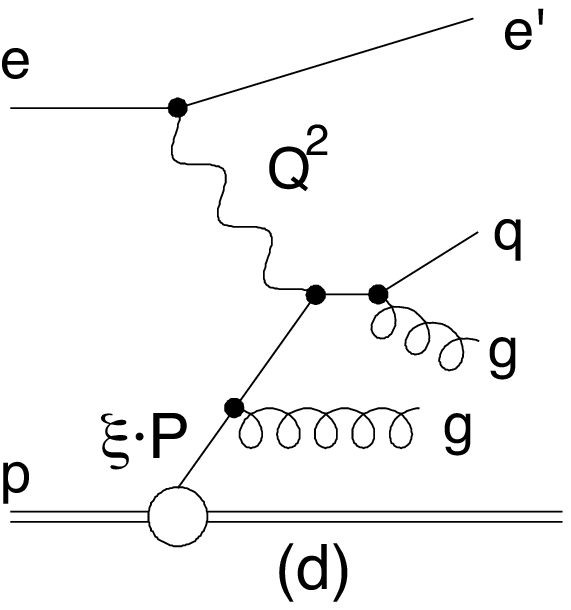}\hskip0.5cm 
\caption[Diagrams of different order in \as\ in deep-inelastic lepton-proton scattering]%
{
Deep-inelastic $ep$ scattering at different orders in \as: 
(a) Born contribution $\mathcal{O}(\aem^2)$, 
(b) example of boson-gluon fusion $\mathcal{O}(\aem^2 \as)$,
(c) example of QCD Compton scattering $\mathcal{O}(\aem^2 \as)$ and 
(d) example of a trijet process $\mathcal{O}(\aem^2 \as^2)$.
}
\label{fig:feynborn}
\end{figure}

In contrast to inclusive DIS, where the dominant effects of the strong interactions are the scaling violations of the proton structure functions, the production of jets allows for a direct measurement of the strong coupling \as. If the measurement is performed in the Breit frame of reference~\cite{Feynman:72,Streng:1979pv}, where the virtual boson collides head on with a parton from the proton, the Born level contribution to DIS (\fig~\ref{fig:feynborn}a) generates no transverse momentum. Significant transverse momentum \Pt in the Breit frame is produced at leading order (LO) in the strong coupling \as\ by boson-gluon fusion (\fig~\ref{fig:feynborn}b) and the QCD Compton (\fig~\ref{fig:feynborn}c) processes. 
In LO the proton's longitudinal momentum fraction carried by the parton participating in the hard interaction is given by $\xi =  x ( 1+ M_{12}^2 / \Qsq )$. 
The variables $x$, \Mjj and \Qsq denote the Bjorken scaling variable, the invariant mass of 
the two jets and the negative four-momentum transfer squared, respectively. 
In the kinematic regions of low \Qsq, low \Pt and low \xij, boson-gluon fusion dominates jet production and provides direct sensitivity to terms proportional to the product of \as and the gluon component of the proton structure.
At high \Qsq and high \Pt the QCD Compton processes are dominant, which are sensitive to the valence quark densities and \as. 
Calculations in pQCD in LO for inclusive jet and dijet production in the Breit frame are of $\mathcal{O}(\as)$ and for trijet production (\fig~\ref{fig:feynborn}d) of $\mathcal{O}(\as^2)$.

Recent publications by the ZEUS collaboration concerning jet production in DIS dealt with cross sections of dijet~\cite{Abramowicz:10:965} and inclusive jet production~\cite{Abramowicz:10:127}, whereas recent H1 publications dealt with multijet production and the determination of the strong coupling constant \asmz at low \Qsq~\cite{Aaron:10:1} and at high \Qsq~\cite{Aaron:10:363}. 
%
%
%
%
%

In this paper double-differential measurements are presented of absolute and normalised inclusive jet, dijet and trijet cross sections in the Breit frame. 
Two different jet algorithms, the \kt~\cite{Ellis:93:3160} and the \antikt~\cite{Cacciari:08:063} algorithm, are explored.
The cross sections are measured as a function of \Qsq\ and the transverse jet momentum \ptjet\ for the case of inclusive jets.
Dijet and trijet cross sections are measured as a function of \Qsq\ and the average jet transverse momentum.
In addition, dijet and trijet cross sections are measured as a function of \Qsq\ and the proton's longitudinal momentum fraction $\xi$.
The measurements of the ratios of the number of inclusive jets as well as dijet and trijet events to the number of inclusive NC DIS events in the respective bins of \Qsq, referred to as normalised multijet cross sections, are also reported. 
In comparison to absolute jet cross sections these measurements profit from a significant reduction of the systematic experimental uncertainties. 

The analysis reported here profits from improvements in the reconstruction of tracks and calorimetric energies, 
together with a new calibration of the hadronic energy. They lead to a reduction of the jet energy scale uncertainty to $\unit[1]{\%}$~\cite{Kogler:10} and allow an extension of the pseudorapidity\footnote{The pseudorapidity is related to the polar angle $\theta$, defined with respect to the proton beam direction, by $\eta=-\ln \tan(\theta/2)$.} range of the reconstructed jets in the laboratory rest frame from $2.0$ to $2.5$ in the proton direction and from $-0.8$ to $-1.0$ in the photon direction, compared to a previous analysis~\cite{Aaron:10:363}.
The increase in phase space allows the trijet cross section to be measured double-differentially for the first time at HERA.
The measurements presented in this paper supersede the previously published normalised multijet cross sections~\cite{Aaron:10:363}, 
which include in addition to the data used in the present analysis data from the HERA-I running period, yielding an increase in statistics of about $\unit[10]{\%}$.
However, the above mentioned improvements in the present analysis, which uses
only data from the HERA-II running period, outweigh the small benefit from the additional HERA-I data and yield an overall better precision of the results. 

In order to match the improved experimental precision, the results presented here are extracted using a regularised unfolding procedure which properly takes into account detector effects, like acceptance and migrations, as well as statistical correlations between the different observables.

The measurements are compared to perturbative QCD predictions at NLO corrected for hadronisation effects.
Next-to-next-to-leading order (NNLO) jet calculations in DIS or approximations beyond NLO are not available yet.
The strong coupling \as\ is extracted as a function of the hard scale chosen for jet production in DIS.

\section{Experimental Method}
The data sample was collected with the H1 detector at HERA in the years $2003$ to $2007$ 
when HERA collided electrons or positrons\footnote{
Unless otherwise stated, the term "electron" is used in the following to refer to both electron and positron.
} of energy $\Ebeam = 27.6$~GeV with protons of energy $\Pbeam = \unit[920]{GeV}$, 
providing a centre-of-mass energy of $\sqrt{s}=\unit[319]{GeV}$. The data sample used in this analysis 
corresponds to an integrated luminosity of \unit[351]{\invpb}, of which \unit[160]{\invpb} were recorded 
in $e^-p$ collisions and \unit[191]{\invpb} in $e^+p$ collisions.

\subsection{The H1 detector}
A detailed description of the H1 detector can be found elsewhere~\cite{Abt:97:310,Abt:97:348,Andrieu:93:460}. 
The right-handed coordinate system of H1 is defined such that the positive $z$-axis is in the direction of the proton beam (forward direction), and the nominal interaction point is located at $z = 0$. 
The polar angle $\theta$ and azimuthal angle $\phi$ are defined with respect to this axis. 

The essential detector components for this analysis are the Liquid Argon (LAr) calorimeter and the central tracking detector (CTD), which are both located inside a $1.16$~T solenoidal magnetic field.

Electromagnetic and hadronic energies are measured using the LAr calorimeter in the polar angular range $4^\circ <\theta <154^\circ$ and with full azimuthal coverage \cite{Andrieu:93:460}. 
The LAr calorimeter consists of an electromagnetic section made of lead absorbers between $20$ and $30$ radiation lengths and a hadronic section with steel absorbers. 
The total depth of the LAr calorimeter varies between $4.5$ and $8$ hadronic interaction lengths. 
The calorimeter is divided into eight wheels along the beam axis, each consisting of eight absorber stacks arranged in an octagonal formation around the beam axis. 
The electromagnetic and the hadronic sections are highly segmented in the transverse and the longitudinal directions with in total 45000 readout cells. 
The energy resolution is $\sigma_E/E = 11\,\% / \sqrt{E \,/\GeV}\oplus 1\%$ for electromagnetic energy deposits and 
$\sigma_{E}/E \simeq 50\,\%/\sqrt{E\,/\GeV}\oplus 3\,\%$ for pions, as obtained from electron and pion test beam 
measurements~\cite{Andrieu:94:57,Andrieu:93:499}. 
In the backward region ($153^\circ < \theta < 174^\circ$) energy deposits are measured by a
lead/scintillating fibre Spaghetti-type Calorimeter (SpaCal), composed of an electromagnetic and an hadronic section~\cite{Appuhn:97:397,Appuhn:96:395}. 

The CTD, covering $15^\circ < \theta < 165^\circ$, is located inside the LAr calorimeter and consists of drift and proportional chambers, complemented by a silicon vertex detector covering the range $30^\circ < \theta < 150^\circ$~\cite{Pitzl:00:334}. 
The trajectories of charged particles are measured with a transverse momentum resolution of $\sigma_{\pt}/\pt \simeq 0.2\,\% \; \pt / \GeV \oplus 1.5\,\%$.

The luminosity is determined from the rate of the elastic QED Compton process with the electron and the 
photon detected in the SpaCal calorimeter~\cite{Aaron:2012kn}. 


\subsection{Reconstruction and calibration of the hadronic final state}
\label{sect:calib_hfs}
In order to obtain a high experimental precision in the measurement of jet cross sections and the determination of \asmz, the hadronic jet energy scale uncertainty needs to be minimised. It has been so far the dominant experimental uncertainty in jet measurements.
Details on an improved procedure to achieve a jet energy scale uncertainty of $\unit[1]{\%}$ can be found elsewhere~\cite{Kogler:10} and are briefly summarised here.

After removal of the compact energy deposit (cluster) in the electromagnetic part of the LAr calorimeter 
and the track associated with the scattered electron, the remaining electromagnetic and hadronic clusters and charged tracks are attributed to the \ac{HFS}. 
It is reconstructed using an energy flow algorithm~\cite{Peez:03,Hellwig:04,Portheault:05}, combining information from tracking and calorimetric measurements, which avoids double counting of measured energies. This algorithm provides an improved jet resolution compared to a purely calorimetric jet measurement, due to the superior resolution of the tracking detectors for charged hadrons. 

For the final re-processing of the H1 data and subsequent analyses using these data, 
further improvements have been implemented.
The track and vertex reconstruction is performed
using a double-helix trajectory, 
thus taking multiple scatterings in the detector material better into account. 
The calorimetric measurement benefits from 
a separation of hadronic and electromagnetic showers based on shower shape estimators and neural networks~\cite{Feindt:2004wla, Feindt:06:190}  for determining the probability that the measured energy deposit of a cluster in the electromagnetic part of the LAr calorimeter is originating from an electromagnetic or hadronic shower. 
This improves the calorimetric measurement, since the non-compensating LAr calorimeter has a different response for incident particles leading to hadronic or electromagnetic showers.
The neural networks are trained~\cite{Kogler:10} for each calorimeter wheel separately, using a mixture of neutral pions,
photons and charged particles for the simulation of electromagnetic and hadronic showers.
The most important discriminants are the energy fractions in the calorimeter layers and the longitudinal first and second moments. 
Additional separation power is gained by the covariance between the longitudinal and radial shower extent and the longitudinal 
and radial kurtosis. The neural network approach was tested on data using identified electrons and jets and shows an 
improved efficiency for the identification of purely electromagnetic or hadronic clusters, compared to the previously used algorithm.


neutral pions and photons for the generation of electromagnetic showers and charged pions are used for simulating hadronic showers.

The overconstrained NC DIS kinematics allows for the in situ calibration of the energy scale of the \ac{HFS}
using a 
single-jet calibration event sample~\cite{Kogler:10}, 
%
employing the mean value of the \Ptbal distribution, defined as $\PTbal=\langle\Pth/\Ptda\rangle$.
The transverse momentum of the \ac{HFS}, \Pth, is calculated by summing the momentum components $P_{i,x}$ and $P_{i,y}$ of 
all \ac{HFS} objects $i$,
\begin{equation}
\Pth = \sqrt{ \left( \sum_{i\in\h} P_{i,x} \right)^2 + \left( \sum_{i\in\h} P_{i,y} \right)^2} \, .
\label{eq:Pth}
\end{equation}
The expected transverse momentum \Ptda is calculated using the double-angle method, which, to a good approximation, is insensitive to the absolute energy scale of the \ac{HFS} measurement.
It makes use of the angles of the scattered electron $\thetae$ and of the inclusive hadronic 
angle $\gammah$~\cite{Bentvelsen:1992fu, Hoeger:1992}, 
to define \Ptda as
\begin{equation}
\Ptda = \frac{2\Ebeam}{\tan\frac{\gammah}{2}+\tan\frac{\thetae}{2}} \, .
\label{eq:Ptda}
\end{equation}

Calibration functions for calorimeter clusters are derived, 
depending on their probability to originate from electromagnetically or hadronically induced showers.
They are chosen to be smooth functions depending on the cluster energy and polar angle.
The free parameters of the calibration functions are obtained in a global $\chi^2$ minimisation
procedure, where $\chi^2$ is calculated from the deviation of the value of \PTbal from unity 
in bins of several variables. Since no jets are required at this stage, all calorimeter clusters are calibrated. The uncertainty on the energy measurement of 
individual clusters is referred to as residual cluster energy scale (RCES). 
In addition, further calibration functions for clusters associated to jets measured in the 
laboratory frame are derived. This function depends on the jet pseudorapidity, 
\etalab, and transverse momentum, \ptlab.
It provides an improved calibration for those clusters which are detected in the dense environment of a jet.
The calibration procedure described above is applied both to data and to Monte Carlo (MC) event simulations. 
Track-based four-vectors of the \ac{HFS} are not affected by the new calibration procedure.

\begin{figure}
  \centering
    \includegraphics[width=0.48\textwidth]{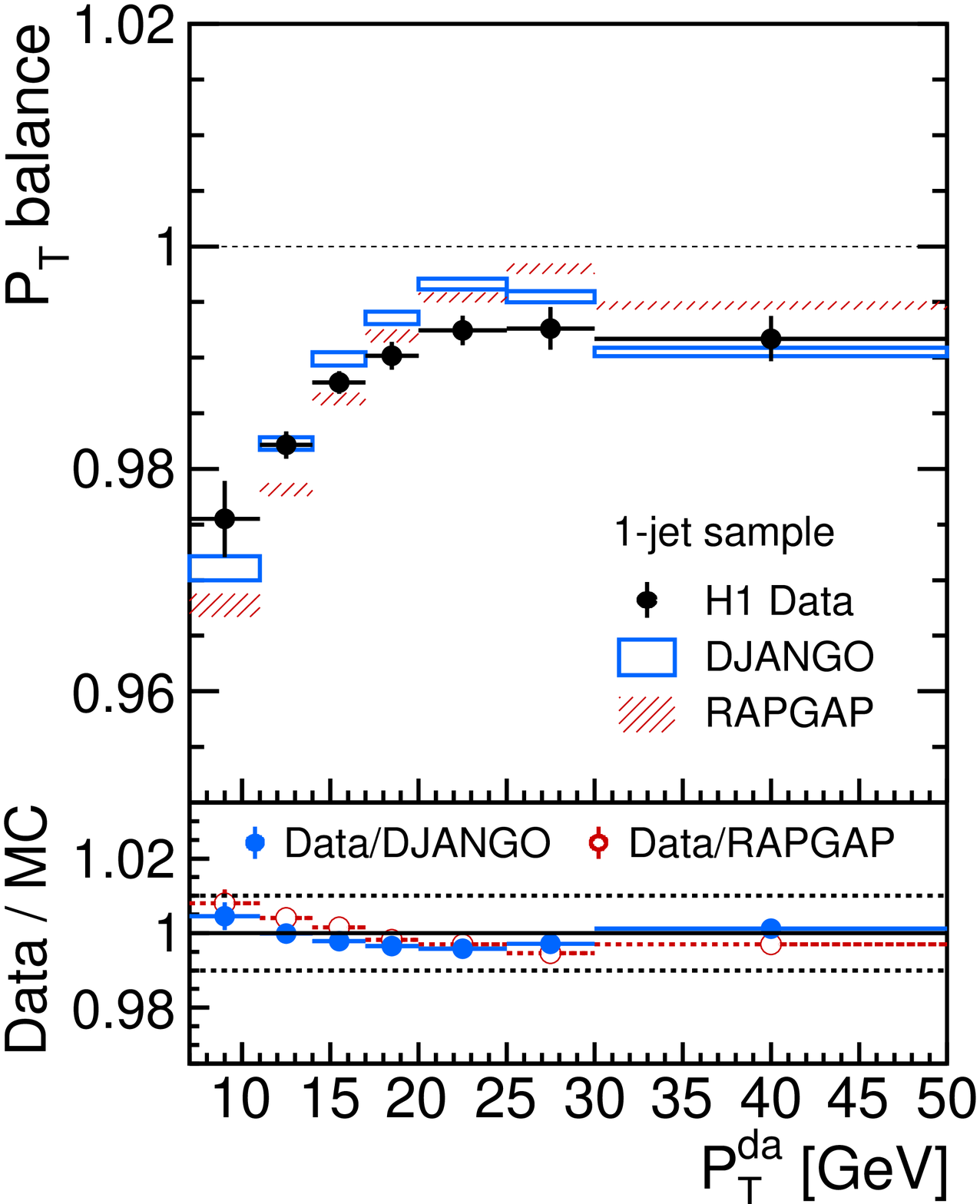}
    \includegraphics[width=0.48\textwidth]{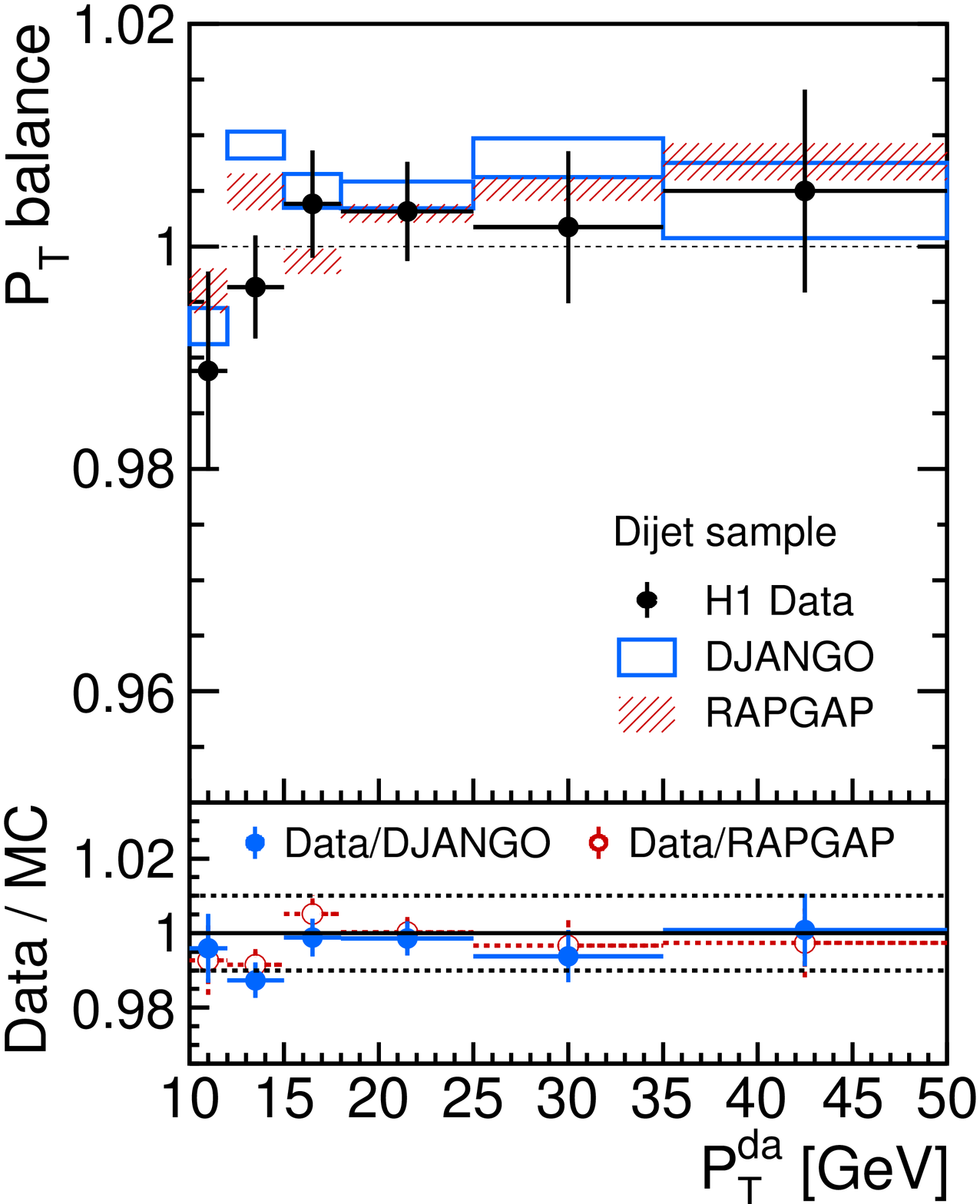}
    \caption{Mean values of the \PTbal-distributions and the double-ratio of data to MC simulations
      as function of \Ptda, as measured in the one-jet calibration sample and in an independent dijet sample.
      Results for data are compared to \Rapgap and \Django.
      The open boxes and the shaded areas illustrate the statistical uncertainties of the MC simulations.
      The dashed lines in the double-ratio figure indicate a $\pm\unit[1]{\%}$ deviation.
    \label{fig:ptbal_ptda_dr}
    }
\end{figure}

The double-ratio of the \PTbal-ratio of data to MC simulations, after the application of the new calibration constants, is shown for the one-jet calibration sample and for a statistically independent dijet sample in \fig~\ref{fig:ptbal_ptda_dr} as a function of \Ptda.
Good agreement between data and simulation is observed over the full detector acceptance. 
This corresponds to a precision of $\unit[1]{\%}$ on the jet energy scale in the kinematic domain of the measurements.

\subsection{Event selection}
\label{sect:event_sel_rec}
The NC DIS events are triggered and selected by requiring a 
cluster in the electromagnetic part of the LAr calorimeter.
The scattered electron is identified as the isolated cluster of highest transverse momentum, with a track associated to it.
Details of the isolation criteria and the electron finding algorithm can be found elsewhere~\cite{Adloff:2003uh}.
The electromagnetic energy calibration and the alignment of the H1 detector are performed following the procedure as in~\cite{Adloff:2003uh}. 
The reconstructed electron energy \Ee is required to exceed \unit[11]{\GeV}, for which the trigger efficiency is close to unity. 
Only those regions of the calorimeter where the trigger efficiency is greater than $\unit[98]{\%}$ are used for the detection of the scattered electron, which corresponds to about $\unit[90]{\%}$ of the $\eta$--$\phi$-region covered by the LAr calorimeter. 
These two requirements, on \Ee and $\eta$--$\phi$, ensure the overall trigger efficiency to be above $\unit[99.5]{\%}$~\cite{Aaron:2012qi}. In the central region, $30^\circ < \thetae < 152^\circ$, where \thetae denotes the polar angle of the reconstructed scattered electron, the cluster is required to be associated with a track measured in the CTD, matched to the primary event vertex. 
The requirement of an associated track reduces the amount of wrongly identified scattered leptons to below $\unit[0.3]{\%}$. The $z$-coordinate of the primary event vertex is required to be within $\pm\unit[35]{cm}$ of the nominal position of the interaction point.

The total longitudinal energy balance, calculated as the difference of the total energy $E$ and the longitudinal component of the total momentum $P_{\rm z}$, using all detected particles including the scattered electron, has little sensitivity to losses in the proton beam direction and is thus only weakly affected by the incomplete reconstruction of the proton remnant. Using energy-momentum conservation, the relation $E - P_{\rm z} \simeq 2\Ebeam = \unit[55.2]{GeV}$ holds for DIS events.
The requirement $45 < E - P_{\rm z} < 65$~GeV thus reduces the contribution of DIS events with hard initial state photon radiation. For the latter events, the undetected photons, propagating predominantly in the negative $z$-direction, lead to values of $E-P_{\rm z}$ significantly lower than the expected value of $\unit[55.2]{GeV}$. The $E - P_{\rm z}$ requirement together with the scattered electron selection also reduces background contributions from photoproduction, where no scattered electron is expected to be detected, to less than $\unit[0.2]{\%}$.
Cosmic muon and beam induced backgrounds are reduced to a negligible level 
after the application of a dedicated cosmic muon finder algorithm.
QED Compton processes are reduced to $\unit[1]{\%}$ by requiring the acoplanarity $A=\cos(|\pi-\Delta\phi|)$ to be smaller than $0.95$, with $\Delta\phi$ being the azimuthal angle between the scattered lepton and an identified photon with energy larger than \unit[4]{GeV}.
The background from lepton pair production processes 
is found to be negligible. 
Also backgrounds from charged current processes and deeply virtual Compton scattering are found to be negligible. 
The backgrounds originating from the sources discussed above are modelled using a variety of MC event generators as described in \cite{Kogler:10}.

The event selection of the analysis is based on an extended analysis phase space defined 
by $100 < \Qsq < \unit[40\,000]{\GeVsq}$ and $0.08 < y < 0.7$, where $y = \Qsq /(s\xbj)$ quantifies the inelasticity of the interaction. Jets are also selected within an extended range in \ptjet\ and \etalab\ as described in sect.~\ref{sect:recon_jets}. The extended analysis phase space and the measurement phase space are summarised in \tab~\ref{tab:ps}.

The variables \Qsq and $y$ are reconstructed from the four-momenta of the scattered electron and the hadronic final state particles using the electron-sigma method~\cite{Bassler:95:197,Bassler:99:583},
\begin{eqnarray} 
\Qsq  = 4 \Ebeam \Ee \cos^2\frac{\thetae}{2} \;\;\; {\rm and} \;\;\;
y = y_\Sigma \; \frac{2 \Ebeam}{\Sigma + \Ee (1-cos\thetae)}  
\\
{\rm with} \;\;\; y_\Sigma = \frac{\Sigma}{\Sigma + \Ee (1-\cos\thetae)} \;\;\; 
{\rm and} \;\;\; \Sigma = \sum_{i\in\h} (E_i-P_{i,z}) \;\; ,
\label{eq:ESigma}
\end{eqnarray}
where $\Sigma$ is calculated by summing over all hadronic final state particles $i$ with energy $E_i$ and longitudinal momentum $P_{i,z}$.


\subsection{Reconstruction of jet observables}
\label{sect:recon_jets}
The jet finding is performed in the Breit frame of reference, where the boost from the laboratory system is determined by \Qsq, $y$ and the azimuthal angle $\phie$ of the scattered electron~\cite{Wobisch:00}. Particles of the 
hadronic final state are clustered into jets using the inclusive \kt~\cite{Ellis:93:3160} or alternatively 
the \antikt~\cite{Cacciari:08:063} jet algorithm.
The jet finding is implemented in FastJet~\cite{Cacciari:06:57}, and the 
massless \pt recombination scheme and the distance parameter $R_0 = 1$ in the $\eta$--$\phi$ plane are used. 
\kd{MC studies of the reconstruction performance and comparisons between jets on detector, hadron and parton level indicate that $R_0 = 1$ is a good choice for the phase space of this analysis. This is also in agreement with the result reported in \cite{Chekanov:2006yc}.}
The transverse component of the jet four-vector with respect to the $z$-axis in the Breit frame is referred to as \ptjet.
The jets are required to have $\ptjet > 3$~GeV. 

The jet axis is transformed to the laboratory rest frame, and jets with a pseudorapidity in the laboratory 
frame of $-1.5 < \etalab < 2.75$ are selected. Furthermore, the transverse momentum of jets with respect to 
the beam-axis in the laboratory frame is restricted to $\ptlab > \unit[2.5]{\GeV}$. 
\kd{This requirement removes only a few very soft jets which are not well measured and is not part of the phase space definition.}

Inclusive jets are defined by counting all jets in a given event with $\ptjet > \unit[3]{\GeV}$. 
Dijet and trijet events are selected by requiring at least two or three jets with $3 < \ptjet < \unit[50]{\GeV}$, such that the trijet sample is a subset of the dijet sample.
The measurement is performed as a function of the average transverse momentum 
$\meanptdi = \tfrac{1}{2}(\pt^{\rm jet1}+\pt^{\rm jet2})$ and $\meanpttri = \tfrac{1}{3}(\pt^{\rm jet1}+\pt^{\rm jet2}+\pt^{\rm jet3})$ 
of the two or three leading jets for the dijet and trijet measurement, respectively. 
Furthermore, dijet and trijet cross sections are measured as a function of the observables
$\xidi=\xbj\left(1+\Mjj^2/\Qsq\right)$ and $\xitri=\xbj\left(1+\Mjjj^2/\Qsq\right)$, respectively, with $\Mjjj$ being the invariant mass of the three leading jets.
The observables \xidi and \xitri provide a good approximation of the proton's longitudinal momentum fraction \xij carried by the 
parton which participates in the hard interaction.

\subsection{Measurement phase space and extended analysis phase space}
The NC DIS and the jet phase space described above refers to an extended 
analysis phase space compared to the measurement phase space for which the results are quoted. 
Extending the event selection to a larger phase space helps to quantify 
migrations at the phase space boundaries, thereby improving the precision of the measurement.
The actual measurement is performed in the NC DIS phase space given by 
$150 < \Qsq < \unit[15\,000]{\GeVsq}$ and $0.2 < y < 0.7$. 
Jets are required to have $-1.0 < \etalab < 2.5$, which ensures that they are well contained 
within the acceptance of the LAr calorimeter and well calibrated. For the inclusive 
jet measurement, each jet has to fulfil the requirement $7<\ptjet<\unit[50]{GeV}$. For 
the dijet and trijet measurements jets are considered with $5<\ptjet<\unit[50]{GeV}$, and, 
in order to avoid regions of phase space where calculations in fixed order perturbation theory are not reliable~\cite{Frixione:97:315,Gouzevitch:08}, an additional requirement on the invariant mass of $\Mjj>\unit[16]{\GeV}$ is imposed. 
This ensures a better convergence of the perturbative series at NLO~, 
which is essential for the comparison of the NLO calculation with data and the extraction of \as.  
The extended analysis and the measurement phase space are summarised in \tab~\ref{tab:ps}.


\begin{table}[tbp]
\begin{center}
\footnotesize
\begin{tabular}{l|cc}
\hline
  & \textbf{Extended analysis phase space} & \textbf{Measurement phase space} \\[-0.2cm] 
  & \textbf{ } & \textbf{for jet cross sections} \\
\hline 
NC DIS phase space &
 $100 < \Qsq < \unit[40\,000]{\GeVsq}$ &  $150 < \Qsq < \unit[\fifteen]{\GeVsq}$   \\[-0.1cm]
 &  $0.08 < y < 0.7$                    &      $0.2 < y < 0.7$                \\
\hline
Jet polar angular range  &
 $-1.5 < \etalab < 2.75$              &    $-1.0 < \etalab < 2.5$          \\
\hline 
Inclusive jets &
 $\ptjet > \unit[3]{GeV}$         &    $7 < \ptjet < \unit[50]{GeV}$           \\
\hline
Dijets and trijets &
 $3 < \ptjet < \unit[50]{GeV}$             &     $5 < \ptjet < \unit[50]{GeV}$             \\[-0.1cm]
 &  &   $\Mjj > \unit[16]{GeV}$                                                       \\
\hline
\end{tabular}
\caption{Summary of the extended analysis phase space and the measurement phase space of the jet cross sections.}
\label{tab:ps}
\end{center}
\end{table}


\subsection{Monte Carlo simulations}
\label{sect:mc_simulation}
The migration matrices needed for the unfolding procedure (see section~\ref{sect:unfolding}) are determined using simulated NC DIS events. The generated events are passed through a detailed GEANT3~\cite{Brun:87} based simulation of the H1 detector and subjected to the same reconstruction and analysis chains as are used for the data. 
The following two Monte Carlo (MC) event generators are used for this purpose, both implementing LO matrix elements for NC DIS, boson-gluon fusion and QCD Compton events. 
The CTEQ6L~\cite{Pumplin:2002vw} parton density functions (PDFs) are used.
Higher order parton emissions are simulated in \Django~\cite{Charchula:94:381} according to the colour dipole model, as implemented in Ariadne~\cite{Lonnblad:92:15,Lonnblad:01}, and in \Rapgap~\cite{Jung:95:147,Jung:06} with parton showers in the leading-logarithmic approximation. In both MC programs hadronisation is modelled with the Lund string fragmentation~\cite{Andersson:83:31,Sjostrand:95iq} using the ALEPH tune~\cite{Schael:2004ux}. 
The effects of QED radiation and electroweak effects are simulated using the \Heracles~\cite{Kwiatkowski:92:155} program, which is interfaced to the \Rapgap, \Django and \Lepto~\cite{Ingelman:97:108} event generators. The latter one is used to correct the $e^+p$ and $e^-p$ data for their different electroweak effects (see section~\ref{sec:ewcorr}).

\section{Unfolding} 
\label{sect:unfolding}

The jet data are corrected for detector effects using a regularised unfolding method which 
is described in the following. 
The matrix based unfolding method as implemented in the 
TUnfold package~\cite{Schmitt:2012kp} is employed.
A detector response matrix is constructed for the unfolding of the neutral current DIS, the inclusive jet, the dijet and the trijet measurements simultaneously~\cite{Britzger:13}. 
The unfolding takes into account the statistical correlations between these measurements as well as the statistical correlations of several jets originating from a single event. 
The corrections for QED radiation are included in the unfolding procedure.
Jet cross sections and normalised jet cross sections at hadron level are determined using this method. 
%
The hadron level refers to all stable particles in an event 
\kd {with a proper lifetime larger than $c\tau>\unit[10]{\rm mm}$.}
It is obtained from MC event generators by selecting all particles after hadronisation and subsequent particle decays.


\subsection{Weighting of MC models to describe data}
\label{sect:Reweighting}
Both \Rapgap and \Django provide a fair description of the experimental data for the inclusive NC DIS events and the multijet samples. 
To further improve the agreement between reconstructed Monte Carlo events and the data, weights are applied to selected observables on hadron level.
The weights are obtained iteratively from the ratio of data to the reconstructed MC distributions and are applied to events on hadron level. 
The observables of the inclusive NC DIS events are in general well described and 
are not weighted. An exception is the inelasticity $y$. The slope of this distribution is not described satisfactorily,
where at low values of $y$ the disagreement amounts to about $5\,\%$ between the data and the LO MC prediction. 
Since this quantity is important, as it enters in the calculation of the boost to the Breit frame, it was weighted to provide a good description of the data. 

\begin{figure}
  \centering
  \includegraphics[width=0.44\textwidth]{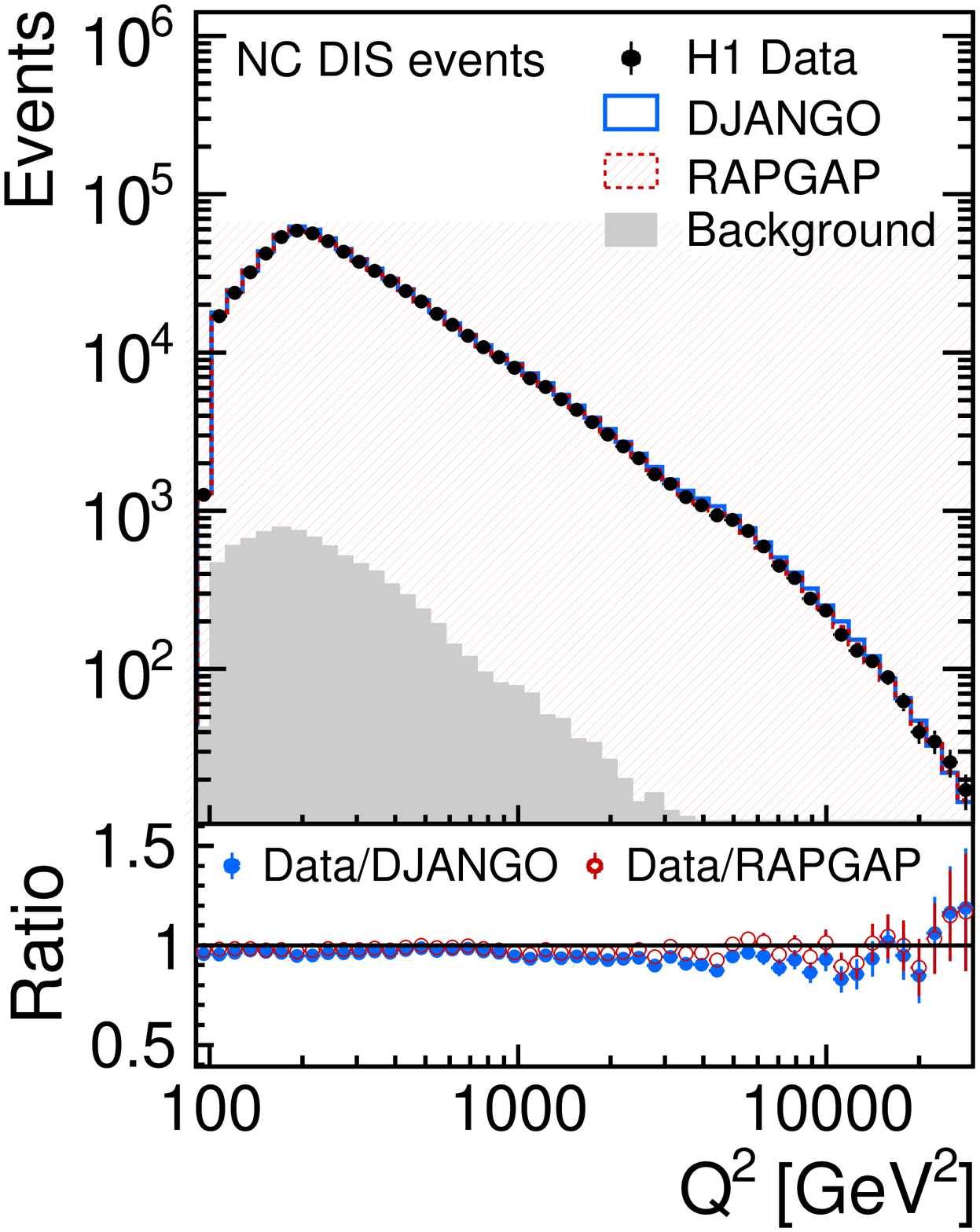}
  \includegraphics[width=0.44\textwidth]{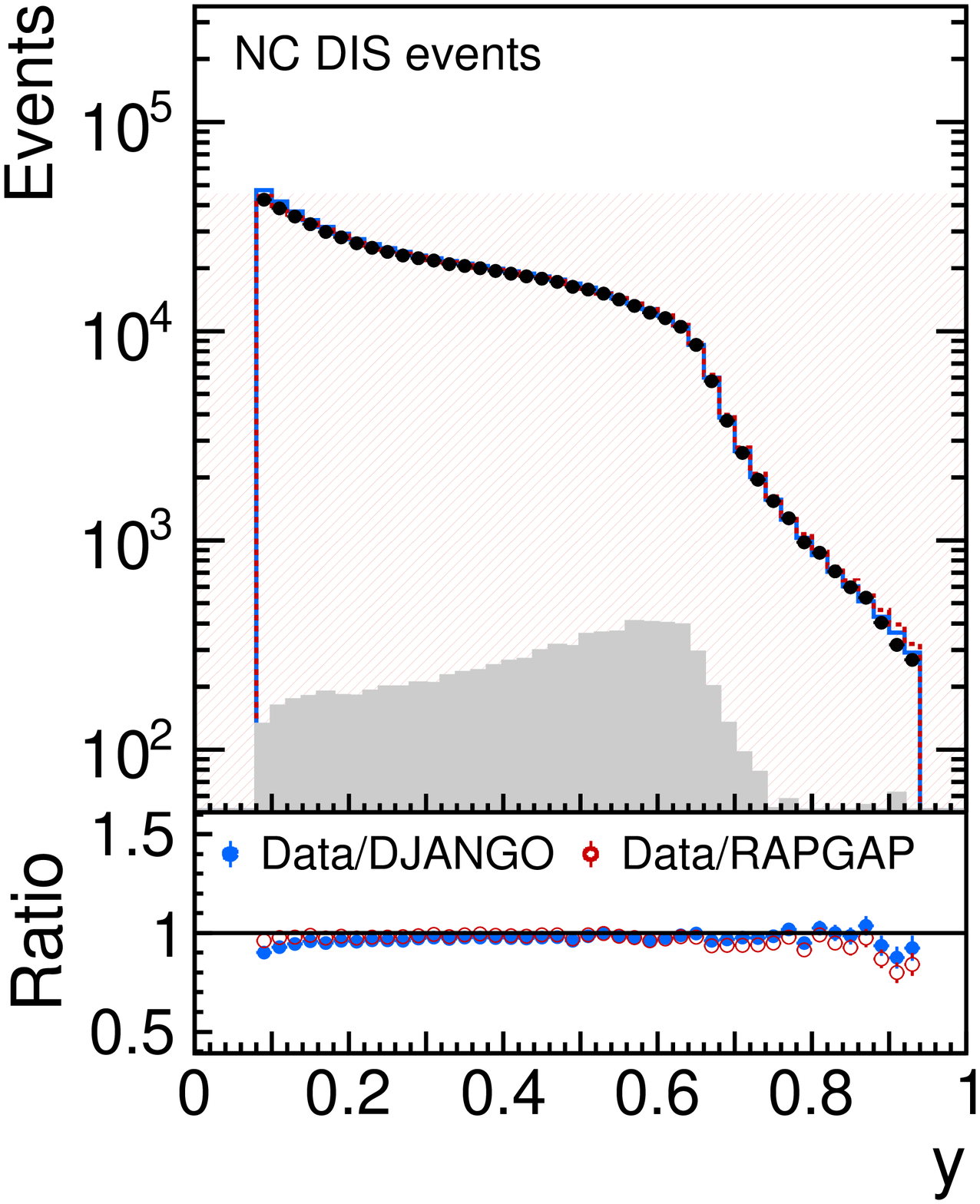}
  \caption{
  Distributions of \Qsq\ and $y$ for the selected NC DIS data on detector level in the extended analysis phase space.
  The data are corrected for the estimated background contributions, shown as gray area. The predictions from \Django and \Rapgap
  are weighted to achieve good agreement with the data.
  The ratio of data to prediction is shown at the bottom of each figure.
    \label{fig:CtrlPlotsNCDIS}
  }
\end{figure}

\begin{figure}
  \centering
  \includegraphics[width=0.44\textwidth]{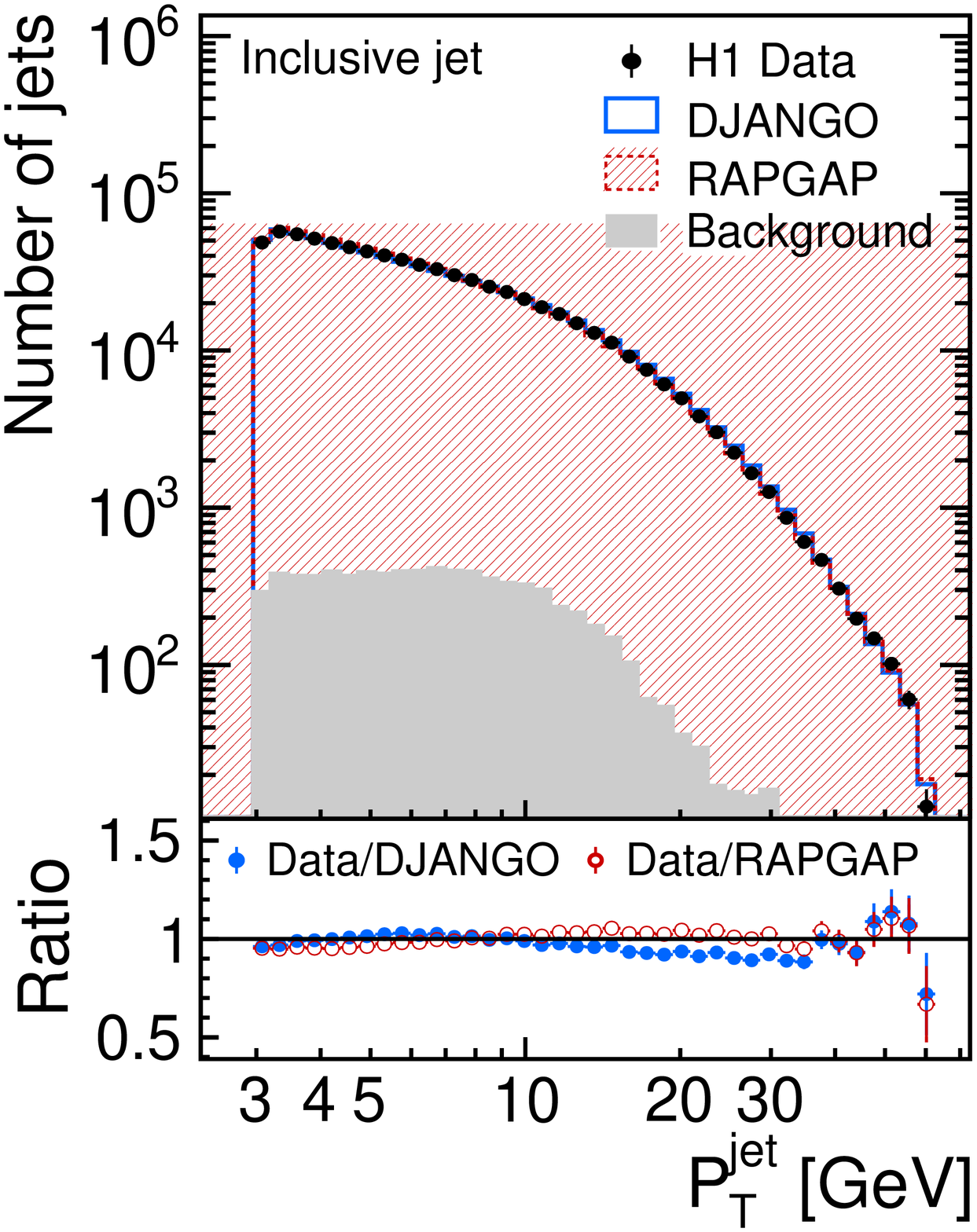}
  \includegraphics[width=0.44\textwidth]{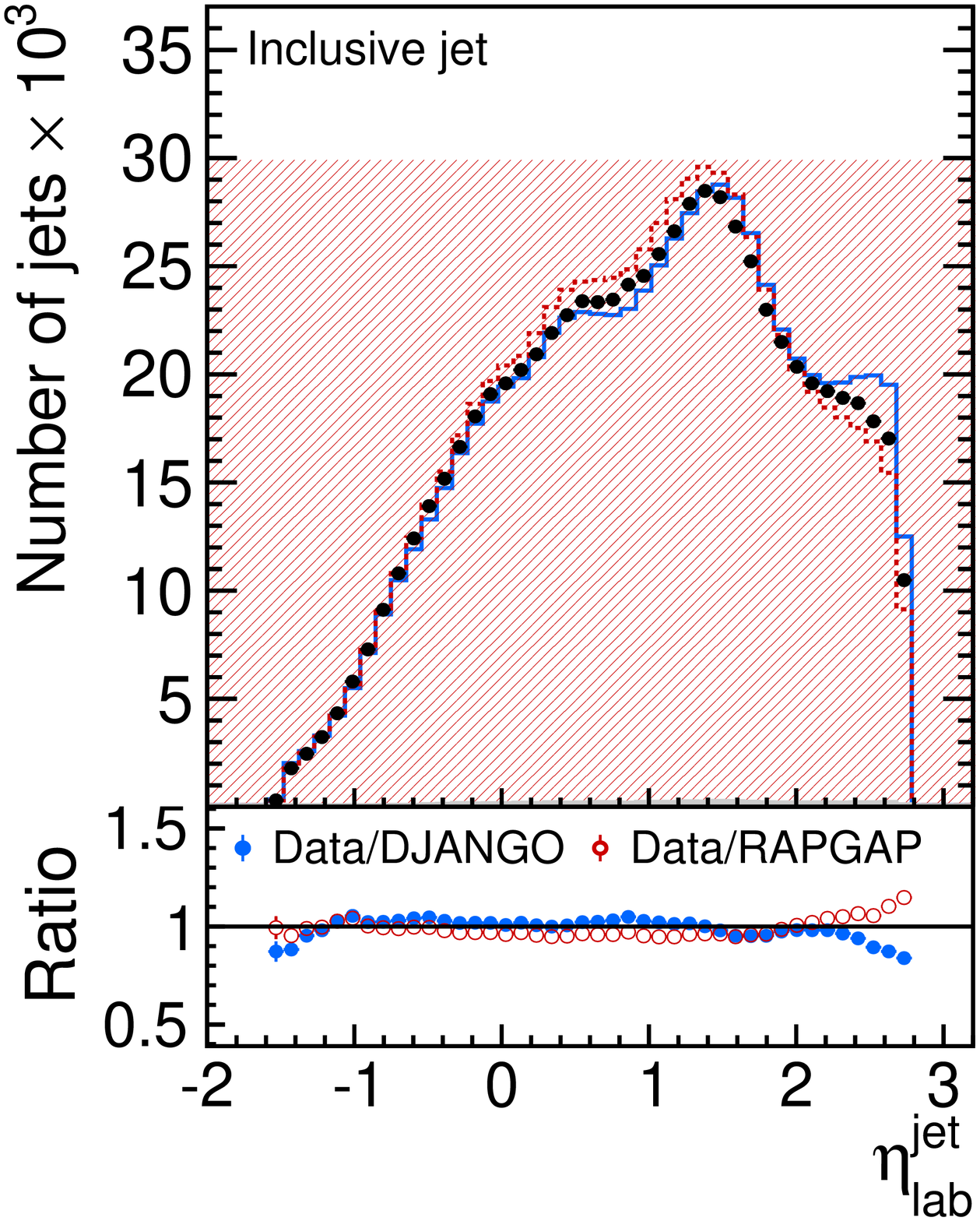}
  \caption{
   Distributions of \ptjet\ and $\etalab$ for the selected inclusive jet data on detector level in the extended analysis phase space.
   The are been corrected for the estimated background contributions, shown as gray area. The predictions from \Django and \Rapgap
   are weighted to achieve good agreement with the data.
   The ratio of data to prediction is shown at the bottom of each figure.
    \label{fig:CtrlPlotsIncJet}
  }
\end{figure}

\begin{figure}
  \centering
  \includegraphics[width=0.44\textwidth]{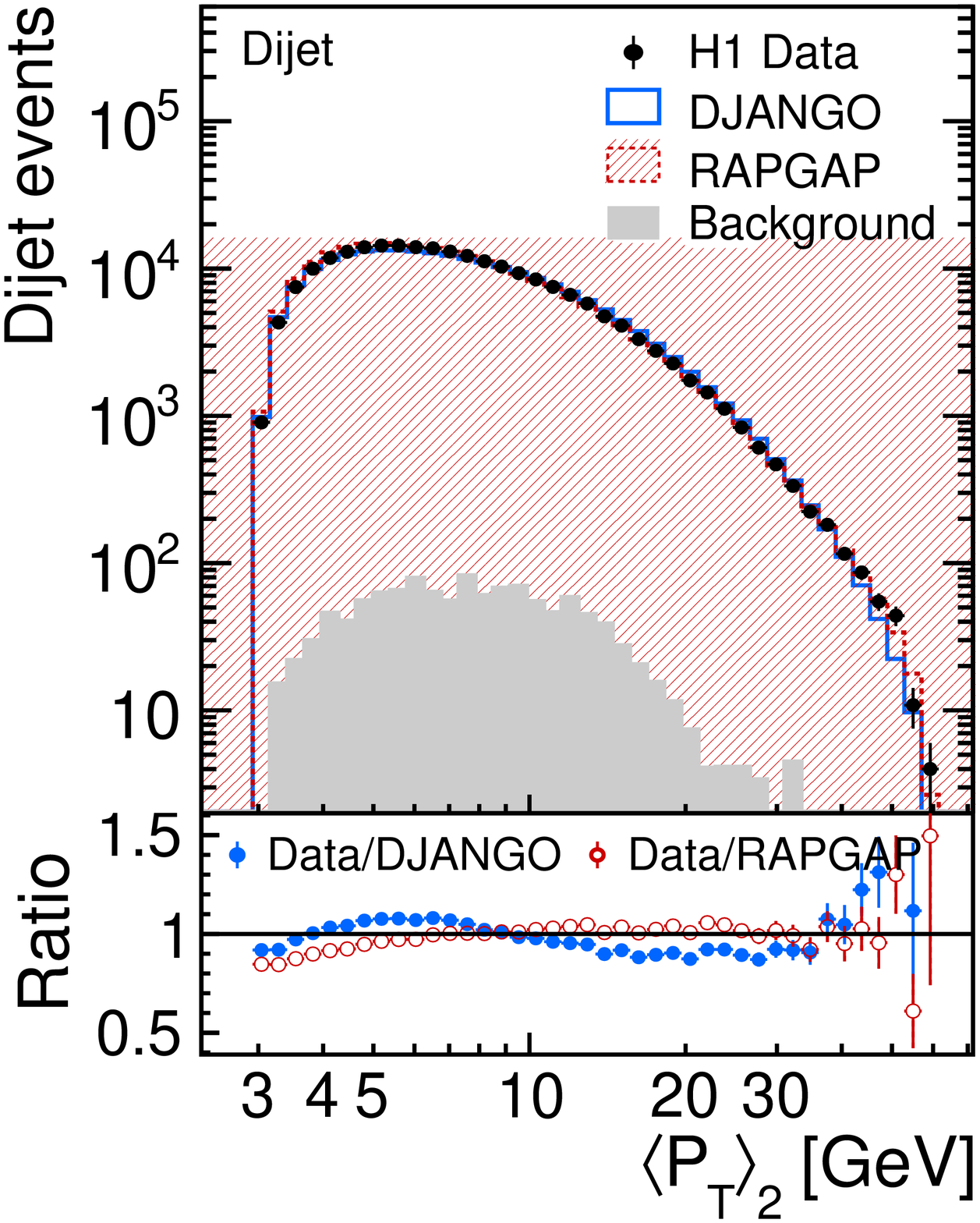}
  \includegraphics[width=0.44\textwidth]{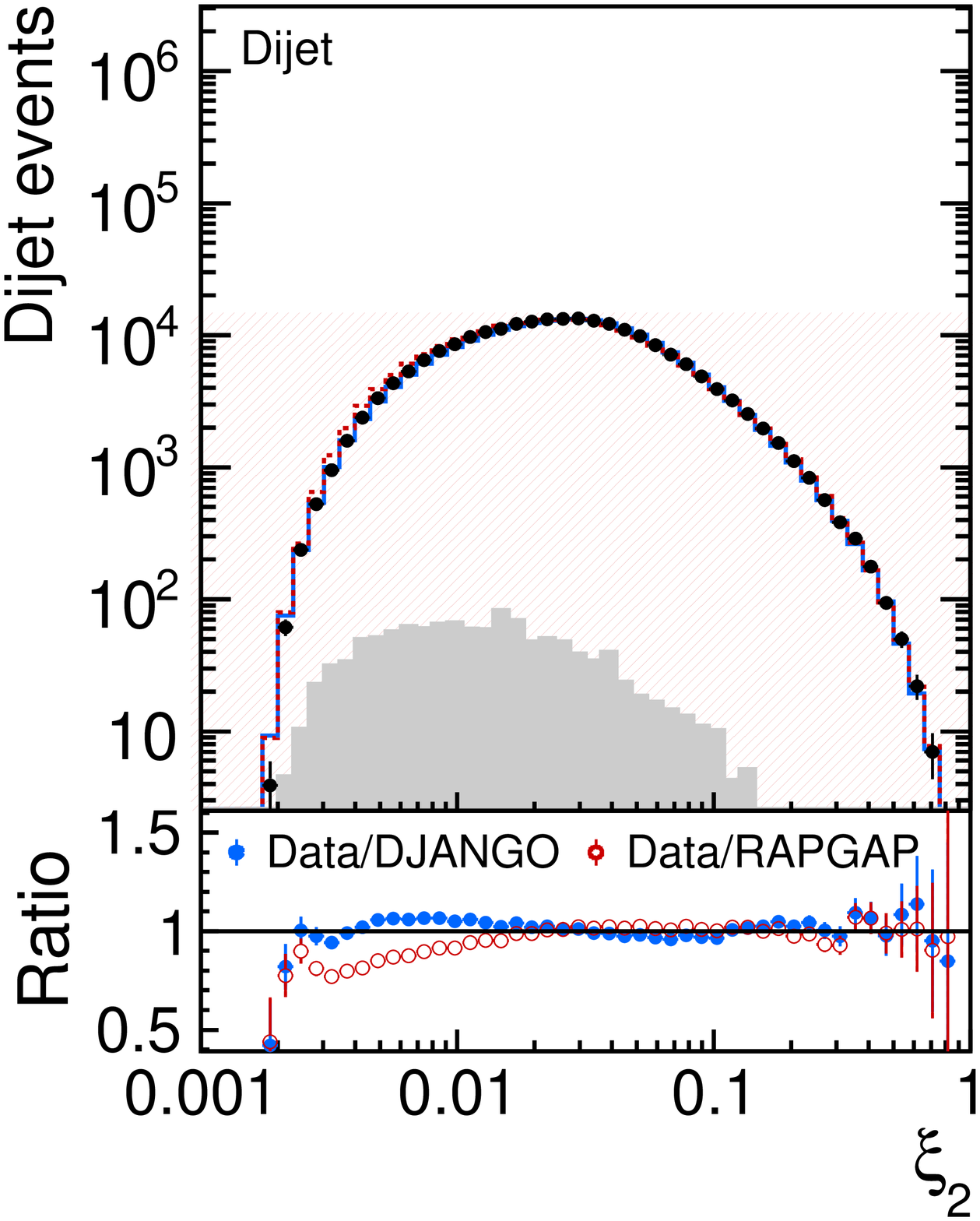}
  \caption{
  Distributions of \meanptdi\ and \xidi\ for the selected dijet data on detector level in the extended analysis phase space.
  The data are corrected for the estimated background contributions, shown as gray area. The predictions from \Django and \Rapgap
   are weighted to achieve good agreement with the data.
   The ratio of data to prediction is shown at the bottom of each figure.
    \label{fig:CtrlPlotsDijet}
  }
\end{figure}

\begin{figure}
  \centering
  \includegraphics[width=0.44\textwidth]{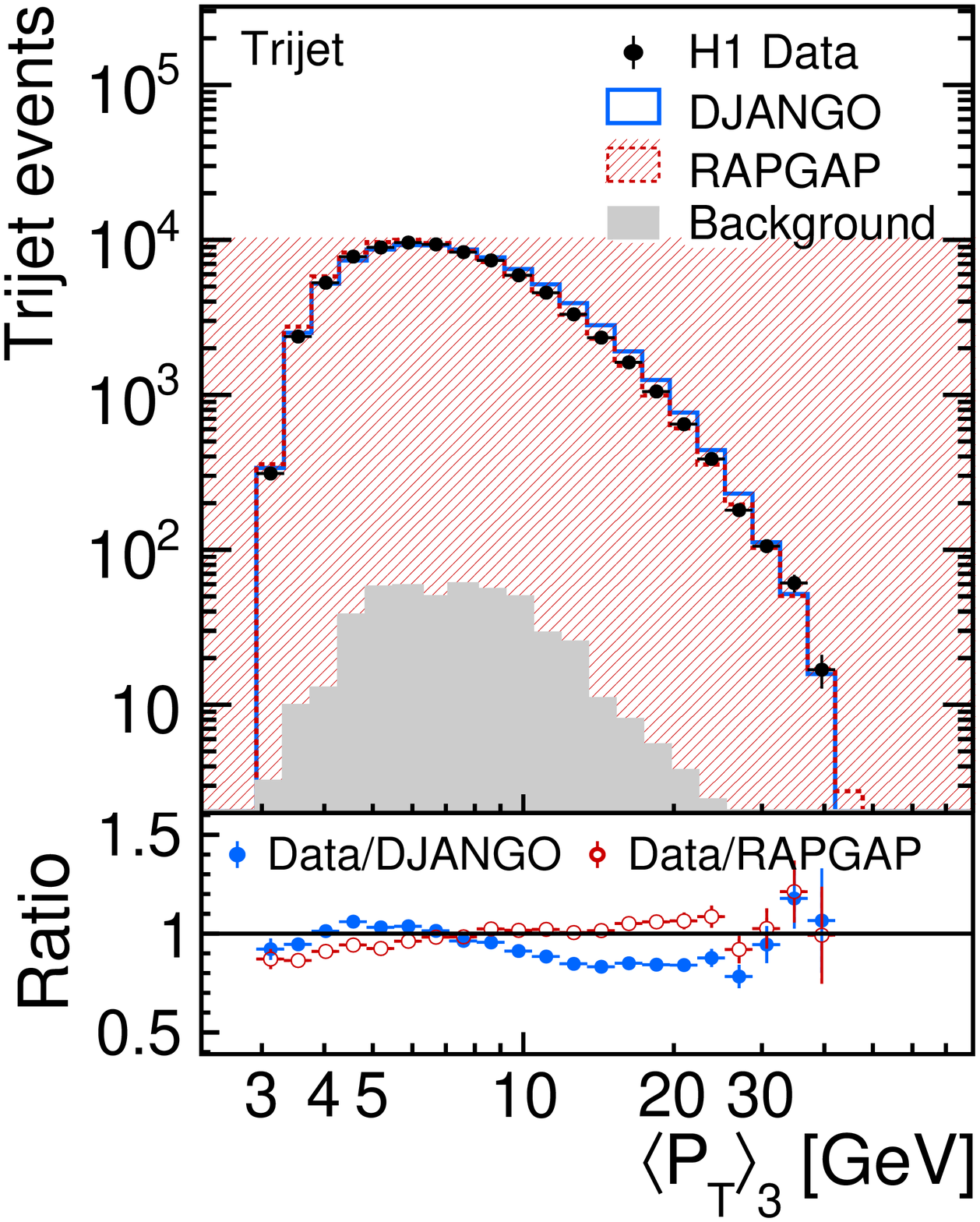}
  \includegraphics[width=0.44\textwidth]{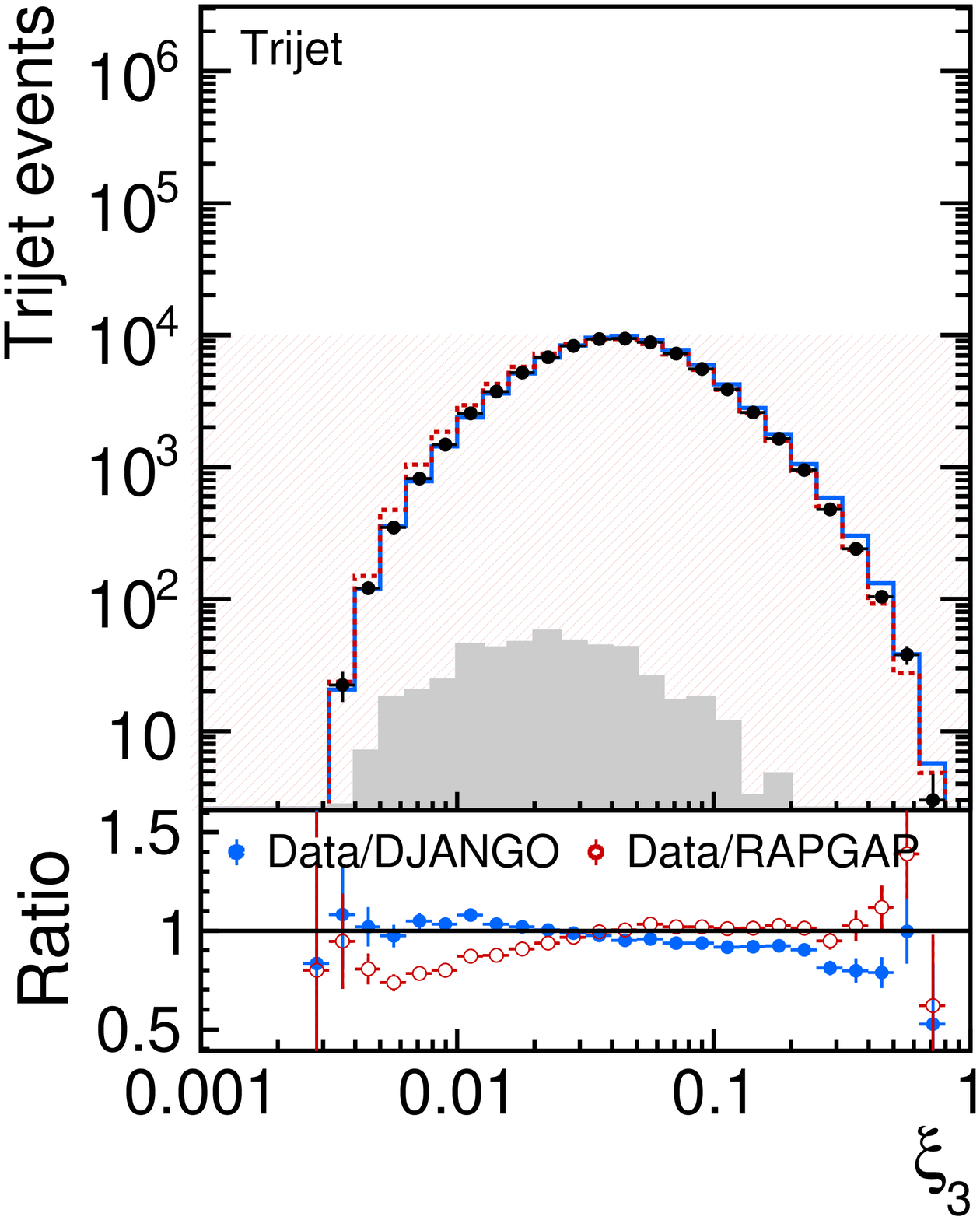}
  \caption{
  Distributions of \meanpttri\ and \xitri\ for the selected trijet data on detector level in the extended analysis phase space.
  The data are corrected for the estimated background contributions, shown as gray area. The predictions from \Django and \Rapgap
   are weighted to achieve good agreement with the data.
   The ratio of data to prediction is shown at the bottom of each figure.
    \label{fig:CtrlPlotsTrijet}
  }
\end{figure}

The MC models, simulating LO matrix elements and parton showers, do not 
provide a good description of higher jet multiplicities. 
Event weights are applied for the jet multiplicity as a function of \Qsq.
The MC models are also not able to reproduce well the observed \ptjet\ spectra at high \ptjet\ and the pseudorapidity distribution of the jets. 
Thus, weights are applied depending on the transverse momentum and pseudorapidity of the jet with the highest (most forward) pseudorapidity in the event as well as for the jet with the smallest (most backward) pseudorapidity in the event. Additional weights are applied for trijet events as a function of the sum of \ptjet\ of the three leading jets. The weights are typically determined as two-dimensional $2^{\rm nd}$ degree polynomials with either $P^{\rm jet}_{\rm T,fwd}$, $P^{\rm jet}_{\rm T,bwd}$ or \Qsq\ as the second observable to ensure that no discontinuities are introduced~\cite{Kogler:10}. 
These weights are derived and applied in the extended analysis phase space (see section~\ref{sect:event_sel_rec} and \tab~\ref{tab:ps}) in order to control migrations in the unfolding from outside into the measurement phase space. 
After application of the weights, the simulations provide a good description of the shapes of all data 
distributions, some of which are shown in \fig{}s~\ref{fig:CtrlPlotsNCDIS}, \ref{fig:CtrlPlotsIncJet}, \ref{fig:CtrlPlotsDijet} and \ref{fig:CtrlPlotsTrijet}.

\subsection{Regularised unfolding}
The events are counted in bins, where the bins on hadron level are arranged in a vector $\vec{x}$ with dimension $1370$, and the bins on detector level are arranged in a vector $\vec{y}$ with dimension $4562$.
The vectors $\vec{x}$ and $\vec{y}$ are connected by a folding equation $\vec{y}=\A \vec{x}$, where $\A$ is a matrix of probabilities, the detector response matrix.
It accounts for migration effects and efficiencies.
The element $A_{ij}$ of $\A$ quantifies the probability to detect an event in bin $i$ of $\vec{y}$, given that it was produced in bin $j$ of $\vec{x}$.
Given a vector of measurements $\vec{y}$, the unknown hadron level distribution $\vec{x}$ is estimated~\cite{Schmitt:2012kp}  in a linear fit, by determining the minimum of
\begin{equation}
\chisq = \chisqA + \chisqL := (\vec{y} - \A \vec{x})^{\rm T} \V_y^{-1} (\vec{y}-\A \vec{x}) +  \tau^2 (\vec{x}-\vec{x}_0)^{\rm T}({\bf L}^{\rm T}{\bf L})(\vec{x}-\vec{x}_0)~,
\label{eq:UnfoldingChi2}
\end{equation}
where $\V_y$ is the covariance matrix on detector level, and \chisqL\ is a regularisation term to suppress fluctuations of the result.
The regularisation parameter $\tau$ is a free parameter. The matrix $\bf{L}$ contains the 
regularisation condition and is set to unity. 
The bias vector $\vec{x}_0$ represents the hadron level distribution of the MC model.
The detector response matrix $\A$ is constructed from another matrix $\M$~\cite{Schmitt:2012kp}, called migration matrix throughout this paper.  
The migration matrix is obtained by counting MC jets or events in bins of $\vec{x}$ and $\vec{y}$. 
It is determined by averaging the matrices obtained from two independent samples of simulated events by the \Django and \Rapgap generators.
It also contains an extra row, $\vec{\varepsilon}$, to account for inefficiencies, i.e.\ for events which are not reconstructed in any bin of $\vec{y}$.

QED radiative corrections are included in the unfolding as efficiency corrections~\cite{Britzger:13}.
The running of the electromagnetic coupling $\aem(\mur)$ is not corrected for. 
The size of the radiative corrections is of order $\unit[10]{\%}$ for absolute jet cross sections and of order $\unit[5]{\%}$ for normalised jet cross sections. 

Prior to solving the folding equation, the remaining small backgrounds in the data from the QED Compton process and from photoproduction  after the event selection 
are subtracted from the input data~\cite{Schmitt:2012kp} using simulated MC jets or events. 
Also MC simulated DIS events with inelasticity $y>0.7$ on hadron level, and thus from outside the accepted phase space, are considered as background and are subtracted from data. 
These contributions cannot be determined reliably from data, since the cut on \Ee\ 
results in a low reconstruction efficiency for events with $y>0.7$ on detector level.
The contribution from such events is less than $\unit[1]{\%}$ in any bin of the cross section measurement.

A given event with jets may produce entries in several bins of $\vec{y}$.
This introduces correlations between bins of $\vec{y}$ which lead to off-diagonal entries in the covariance matrix $\V_y$.



\subsection{Definition of the migration matrix}
\begin{figure}[p]
\newpage
  \centering
  \includegraphics[width=0.69\textwidth]{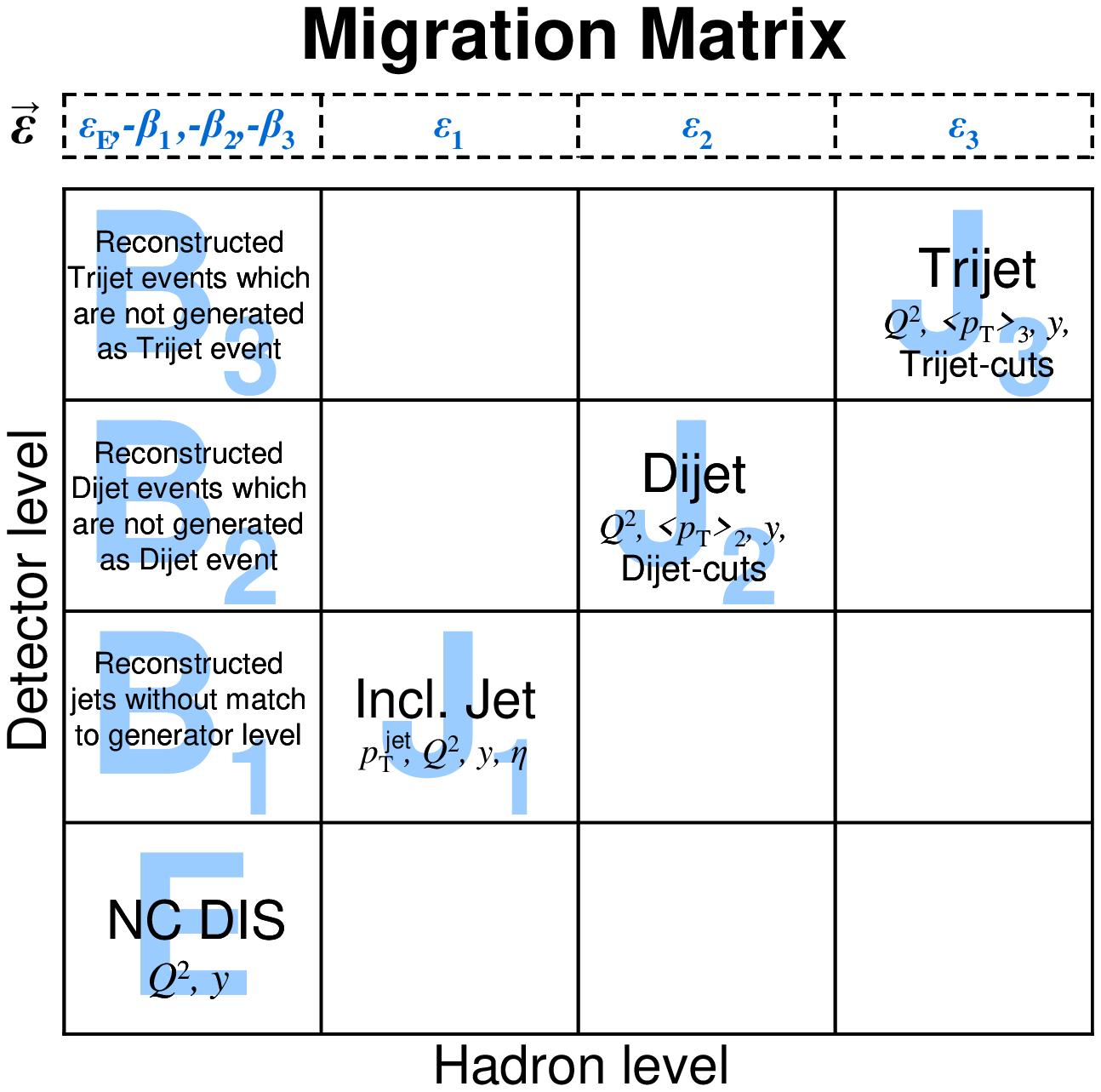}
  \caption{Schematic illustration of the migration matrix for the regularised unfolding, which includes the NC DIS (E), the inclusive jet $(\rm J_1)$, the dijet $(\rm J_2)$ and the trijet $(\rm J_3)$ MC events. The observables utilised for the description of migrations are given in the boxes referring to the respective submatrices. The submatrices which connect the hadron level NC DIS data with the detector level jet data ($(\rm B_1)$,$(\rm B_2)$, and $(\rm B_3)$)  help to control detector-level-only entries. An additional vector, $\vec{\varepsilon}$, is used for efficiency corrections and to preserve the normalisation.
  \label{fig:MigMaScheme}
  }
\end{figure}

The migration matrix is composed of a $4\times 4$ structure of submatrices representing the four different data samples (NC DIS, inclusive jet, dijet and trijet), thus enabling a simultaneous unfolding of NC DIS and jet cross sections. 
It is schematically illustrated in \fig~\ref{fig:MigMaScheme}.
The four submatrices $\bm E$, ${\bm J_1}$, ${\bm J_2}$ and  ${\bm J_3}$ represent the migration matrices for the NC DIS, the inclusive jet, the dijet and the trijet measurements, respectively. 
Hadron-level jets or events which do not fulfil the reconstruction cuts are filled into the additional vector $\vec{\varepsilon}$.
The three submatrices ${\bm B_1}$, ${\bm B_2}$ and ${\bm B_3}$ connect the jet measurements on 
detector level with the hadron level of the NC DIS measurement. 
They are introduced to account for cases where a jet or an event is reconstructed, although it is absent on hadron level.
Such detector-level-only contributions are present due to different jet multiplicities on detector and on hadron level, caused by limited detector resolution and by acceptance effects.
The unfolding procedure determines the normalisation of these detector-level-only contributions from data.
Each entry in one of the submatrices $\B_i$ is compensated by a negative entry in the efficiency bin 
(denoted as $\beta_i$ in \fig~\ref{fig:MigMaScheme}), in order to preserve the normalisation of the NC DIS measurement. 
The four submatrices, $\bm E$, ${\bm J_1}$, ${\bm J_2}$ and ${\bm J_3}$, are explained in the following.
More details can be found in~\cite{Britzger:13}. 

\begin{itemize}
\item {\bf NC DIS (${\bm E}$):}\
For the measurement of the NC DIS cross sections a two-dimensional unfolding considering migrations in \Qsq\ and $y$ is used. 
On detector level $14$ bins in \Qsq\ times $3$ bins in $y$ ($0.08<y<0.7$) are used to determine $8$ bins in \Qsq\ times $2$ bins in $y$ on hadron level. 
Out of these $16$ bins, only $6$ bins are used for the determination of the normalised cross sections.


\item {\bf Inclusive jets (${\bm J_1}$):}\
The unfolding of the inclusive jet measurement is performed as a four-dimensional unfolding, where migrations in the observables \Qsq, $y$, \ptjet and \etalab\ are considered.
To model the migrations, jets found on hadron level are matched to detector-level jets, 
employing a closest-pair algorithm with the distance parameter $R=\sqrt{\Delta\phi^2+\Delta\eta^2}$ and a requirement of $R<0.9$. 
Here $\Delta\phi$ and $\Delta\eta$ are the distances between detector level and 
hadron level jets in $\phi$ and $\eta$ in the laboratory rest frame, respectively.  
Detector-level-only jets which are not matched on hadron level are filled into the submatrix ${\bm{B}}_1$ and are therefore determined from data.
Hadron-level jets which are not matched on detector level are filled into the vector $\vec{\varepsilon_1}$.
The bin grid in \Qsq\ and $y$ is defined in the same way as for the NC DIS case.
Migrations in \ptjet are described using $16$ bins on detector level and $8$ bins on hadron level. 
Migrations in \etalab\ within $-1.0<\etalab <2.5$ are described by a $3$ times $2$ structure.
Additional bins (differential in \ptjet, \Qsq\ and $y$) are used to describe migrations of jets in \etalab\ with $\etalab < -1.0 $ or $\etalab > 2.5$. 
The results of the $7$ times $2$ bins within the measurement phase space in \ptjet and \etalab are finally combined to obtain the  $4$ bins for the cross section measurement for each \Qsq bin.

\item {\bf Dijet (${\bm J_2}$):}\
Dijet events are unfolded using a three-dimensional unfolding, where migrations in \Qsq, $y$ and \meanptdi\ are considered. Also taken into account are migrations at the phase space boundaries in \Mjj, $\pt^{\rm jet2}$ and \etalab.
The bin grid in \Qsq\ and $y$ is identical to the one used for the NC DIS unfolding.
Migrations in \meanptdi\ are described using $18$ bins on 
detector level and $11$ bins on hadron level, out of which $8$ bins are combined to obtain the $4$ data points of interest.
Migrations in \Mjj, $\pt^{\rm jet2}$ and \etalab\ are described by additional bins, which are each further binned in \meanptdi and $y$.

\item {\bf Trijet (${\bm J_3}$):}\
The unfolding of the trijet measurement is performed similarly to the dijet unfolding, using a 
three-dimensional submatrix in \Qsq, $y$ and \meanpttri. 
Migrations in $\Mjj$, $\pt^{\rm jet3}$ and \etalab\ are also considered.
Due to the limited number of trijet events, the number of bins is slightly reduced compared to the dijet measurement.

\end{itemize}

Unfolding in the extended analysis phase space increases the stability of the measurement in the measurement phase space to a large extent, in particular for the dijet and trijet data points with $\meanpt < \unit[11]{\GeV}$.
The resulting detector response matrix $\M$ has an overall size of $4562 \times 1370$ bins, of which about $\unit[3]{\%}$ have a non-zero content.
A finer bin grid than the actual measurement bin grid ensures a reduced model dependence in the unfolding procedure.
$148$ bins on hadron level, located in the measurement phase space, and additional adjacent bins, mostly at low transverse momenta, are combined to arrive at the final $64$ cross section bins~\cite{Britzger:13}.

For the dijet and trijet measurements as a function of $\xidi$ and $\xitri$ dedicated new submatrices ${\bm J_2}$ and ${\bm J_3}$ are set up.


\begin{itemize}
\item
The unfolding of the dijet measurement as a function of $\xidi$ is performed as a 
four-di\-men\-sio\-nal unfolding in the variables \Qsq, $y$, $\xidi$ and \Mjj. 
Including \Mjj\ in the unfolding reduces the model dependence considerably.
Additional bins are further used to account for migrations at the phase space boundaries in \Mjj, $\pt^{\rm jet2}$ and \etalab.

\item
A four-di\-men\-sio\-nal unfolding is employed in the variables \Qsq, $y$, $\xitri$ and $\Mjjj$. 
Additional bins are considered to describe migrations at the phase space boundaries in 
\Mjj, $\pt^{\rm jet3}$ and \etalab.
\end{itemize}

\subsection{Regularisation strength}
\kd{The regularisation parameter $\tau$ in equation~\ref{eq:UnfoldingChi2} is set to $\tau = 10^{-6}$.
In this region no dependence of the results on the value of $\tau$ is observed~\cite{Britzger:13}. 
When alternatively applying the method of the L-curve scan~\cite{Schmitt:2012kp} for the choice of $\tau$, a value of $\tau = 7.8 \cdot 10^{-5}$ is obtained with consistent results for the cross sections.}



\section{Jet cross section measurement}
\label{sec:def_of_CS}


\subsection{Observables and phase space}
The jet cross sections presented are hadron level cross sections.
For bin $i$, the cross section $\sigma_i$ is defined as
\begin{equation}
 \sigma_i = \frac{x_i^{\rm unfolded}}{\Lumi^+ + \Lumi^-}\, ,
 \label{eq:CSdefinition}
\end{equation}
where $x_i^{\rm unfolded}$ is the unfolded number of jets or events in bin $i$, including QED radiative corrections.
The integrated luminosities are $\Lumi^+=191$~pb$^{-1}$ and $\Lumi^-=160$~pb$^{-1}$for $e^+p$ and $e^-p$ scattering,respectively. 
The observed cross sections correspond to luminosity weighted averages of $e^+p$ and $e^-p$ processes (see section \ref{sec:ewcorr}).
Double-differential jet cross sections are presented for the measurement phase space given in table~\ref{tab:ps}. Inclusive jet, dijet and trijet cross sections are measured as a function of \Qsq\ and \ptjet or \meanptdi\ or \meanpttri. Dijets and trijets are also measured as a function of \Qsq\ and \xidi\ or \xitri. The phase space in \ptjet\ allows measuring the range $0.006 < \xidi < 0.316$ for dijets and $0.01 < \xitri < 0.50$ for trijets. The trijet phase space is a subset of the dijet phase space, but the observables \meanpttri\ and \xitri\ are calculated using the three leading jets. 
The phase space boundaries of the measurements are summarised in table \ref{tab:PhaseSpaceSummary}.
\begin{table}[htbp]
\footnotesize
\center
\begin{tabular}{l|c|c|c}
\hline
\textbf{Measurement}
        & \textbf{NC DIS phase space}
        & \multicolumn{2}{c}{\textbf{Phase space for jet cross sections}} \\
\hline

$\sI(\Qsq,\ptjet)$ &
        \multirow{9}{*}{\begin{minipage}{0.26\linewidth}
                        \begin{tabular}{r@{$\,<\,$}c@{$\,<\,$}l}
                                 $150$&$\Qsq$&$\unit[\fifteen]{\GeV^2}$\\[-0.1cm]
                                 $0.2$&$y$&$0.7$
                        \end{tabular}
                 \end{minipage}}
                 &
                \begin{minipage}{0.22\linewidth}
                        \begin{tabular}{r@{$\,<\,$}c@{$\,<\,$}l}
                                 $7$&$\ptjet$&$\unit[50]{\GeV}$ \\[-0.0cm]
                                 $-1.0$&$\etalab$&$2.5$
                        \end{tabular}
                 \end{minipage}
                 &
             \begin{minipage}{0.22\linewidth}
                        \begin{tabular}{c}
                                 \quad \quad \; $N_{\rm jet} \geq 1$
                        \end{tabular}
                 \end{minipage}
\\
\cline{1-1}\cline{3-4}
$\sD(\Qsq,\meanptdi)$ &
        &       \multirow{8}{*}{
                \begin{minipage}{0.23\linewidth}
                        \begin{tabular}{r@{$\,<\,$}c@{$\,<\,$}l}
                                 $5$&$\ptjet$&$\unit[50]{\GeV}$ \\[-0.0cm]
                                 $-1.0$&$\etalab$&$2.5$ \\[-0.0cm]
                                 \multicolumn{3}{c}{$\Mjj>\unit[16]{\GeV}$}\\
                        \end{tabular}
                 \end{minipage}
                 }
        &  
                \begin{minipage}{0.22\linewidth}
                        \begin{tabular}{c}
                                 $N_{\rm jet} \geq 2$\\[-0.1cm]
                                 $7<\meanptdi<\unit[50]{\GeV}$
                        \end{tabular}
                 \end{minipage}
\\
\cline{1-1}\cline{4-4}
$\sT(\Qsq,\meanpttri)$  &  & & 
                \begin{minipage}{0.22\linewidth}
                        \begin{tabular}{c}
                                 $N_{\rm jet} \geq 3$\\[-0.1cm]
                                 $7<\meanpttri<\unit[30]{\GeV}$
                        \end{tabular}
                 \end{minipage}
\\
\cline{1-1}\cline{4-4}
$\sD(\Qsq,\xidi)$ &  & & 
                \begin{minipage}{0.22\linewidth}
                        \begin{tabular}{c}
                                 $N_{\rm jet} \geq 2$\\[-0.1cm]
                                 $0.006<\xidi<0.316$
                        \end{tabular}
                 \end{minipage}
\\
\cline{1-1}\cline{4-4}
$\sT(\Qsq,\xitri)$ &  & & 
                \begin{minipage}{0.22\linewidth}
                        \begin{tabular}{c}
                                 $N_{\rm jet} \geq 3$\\[-0.1cm]
                                 $0.01<\xitri<0.50$
                        \end{tabular}
                 \end{minipage}
\\
\hline
\end{tabular}
\caption{Summary of the phase space boundaries of the measurements.}
\label{tab:PhaseSpaceSummary}
\end{table}

The simultaneous unfolding of the NC DIS and the jet measurements allows also the determination of jet cross sections normalised to the NC DIS cross sections. 
Normalised jet cross sections are defined as the ratio of the double-differential absolute jet 
cross sections to the NC DIS cross sections \sNC\ in the respective \Qsq-bin, where \sNC\ is 
calculated using equation~\ref{eq:CSdefinition}.
The phase space for the normalised inclusive jet $\sI / \sNC$, normalised dijet $\sD / \sNC$ and normalised trijet $\sT / \sNC$ cross sections is identical to the one of the corresponding absolute jet cross sections.
The covariance matrix of the statistical uncertainties is determined taking the statistical 
correlations between the NC DIS and the jet measurements into account.
The systematic experimental uncertainties are correlated between the NC DIS and the jet measurements.
Consequently, all normalisation uncertainties cancel, and many other systematic uncertainties are reduced signficantly.


\subsection{Experimental uncertainties}
\label{sec:exp_unc}
Statistical and other experimental uncertainties are propagated by analytical linear error propagation through the unfolding process~\cite{Schmitt:2012kp}.


Systematic uncertainties are estimated by varying the measurement of a given quantity within the experimental uncertainties in simulated events. 
For each `up' and `down' variation, for each source of uncertainty, a new migration matrix is obtained. 
The difference of these matrices with respect to the nominal unfolding matrix is 
propagated through the unfolding process~\cite{Schmitt:2012kp} to obtain the size of the 
uncertainty on the cross sections. 
To avoid fluctuations of the systematic uncertainties caused by limited number of data events, 
in most cases uncertainties are obtained by unfolding simulated data.

The following sources of systematic uncertainties are taken into account:
\begin{itemize}

\item 
The uncertainty of the energy scale of the HFS is subdivided into two components related to the two-stage calibration procedure described in section~\ref{sect:calib_hfs}.

The uncertainties on the cross sections due to the jet energy scale, \DJES{}, are 
determined by varying the energy of all \ac{HFS} objects clustered into jets 
with $\ptlab>\unit[7]{GeV}$ by $\pm\unit[1]{\%}$.
This results in \DJES{} ranging from $2$ to $\unit[6]{\%}$, with the larger values for high values of \ptjet.

The energy of HFS objects which are not part of a jet in the laboratory system 
with $\ptlab>\unit[7]{GeV}$ is varied separately by $\pm\unit[1]{\%}$. This uncertainty is 
determined using a dijet calibration sample, requiring jets with $\ptlab>\unit[3]{GeV}$. 
The resulting uncertainty on the jet cross section is referred to as remaining cluster energy scale uncertainty, \DHFS{}. 
The effect of this uncertainty plays a larger r\^ole at low transverse momenta, 
where jets in the Breit frame include a larger fraction of HFS objects which are not part of a calibrated jet in the laboratory rest frame. 
The resulting uncertainty on the jet cross sections is about $\unit[1]{\%}$ for the inclusive jet and the dijet cross sections, and up to $\unit[4]{\%}$ for the trijet cross sections at low transverse momenta.   
 
\item
The uncertainty \DLAr{}, due to subtraction of the electronic noise from the LAr electronics, is determined by adding randomly $\unit[20]{\%}$ of all rejected noise clusters to the signal. 
This increases the jet cross sections by $\unit[0.5]{\%}$ for the inclusive jet data, $\unit[0.6]{\%}$ for the dijet and $\unit[0.9]{\%}$ for the trijet data. 
  
\item 
The energy of the scattered lepton is measured with a precision of $\unit[0.5]{\%}$ in the central and backward region ($z_{\rm impact}<\unit[100]{\rm{cm}}$) and with $\unit[1]{\%}$ precision in the forward region of the detector, where $z_{\rm impact}$ is the $z$-coordinate of the electron's impact position at the LAr calorimeter.
The corresponding uncertainty on the jet cross sections, \DEe{}, lies between $0.5$ and $\unit[2]{\%}$, with the larger value at high \ptjet or high \Qsq. 
 
\item 
The position of the LAr calorimeter with respect to the CTD is aligned with a precision of $\unit[1]{\rm mrad}$~\cite{Aaron:2012qi}, resulting in a corresponding uncertainty of the electron polar angle measurement $\thetae$. The uncertainty on the jet cross sections, denoted as \DThe{}, is around $\unit[0.5]{\%}$. Only in the highest \Qsq bin it is up to $\unit[1.5]{\%}$.
 
\item 
The uncertainty on the electron identification is $\unit[0.5]{\%}$ in the central region
($z_{\rm impact}<\unit[100]{\rm{cm}}$) and $\unit[2]{\%}$ in the forward direction~\cite{Kogler:10} ($z_{\rm impact}>\unit[100]{\rm{cm}}$). 
This leads to a \Qsq\ dependent uncertainty on the jet cross sections, \DID{}, of around $\unit[0.5]{\%}$ for smaller values of \Qsq\ and up to $\unit[2]{\%}$ in the highest \Qsq\ bin.
 
\item 
The model uncertainty is estimated from the difference between the nominal result of the unfolding matrix and results obtained based on the migration matrices of either \Rapgap or \Django. 
These differences are calculated using data, denoted as $\DMod{\rm d, R}$ and $\DMod{\rm d, D}$, as well as using pseudodata, denoted as $\DMod{\rm p, R}$ and $\DMod{\rm p, D}$.
The model uncertainty on the cross sections is then calculated for each bin using
\begin{equation}
\DMod{} = \pm\sqrt{\frac{1}{2} \left( \max\left(\DMod{\rm d, R},\DMod{\rm p, R}\right)^2 
                + \max\left(\DMod{\rm d, D}, \DMod{\rm p, D}\right)^2 \right)} \; .
\end{equation}
The sign is given by the difference with the largest modulus.
The uncertainty due to the reweighting of the MC models is found to be negligible compared to the model uncertainty obtained in this way. 

\item
The uncertainty due to the requirement on the $z$-coordinate of the primary event vertex is found to be negligible. This is achieved by a detailed simulation of the time dependent longitudinal and lateral profiles of the HERA beams.

\item 
The uncertainty of the efficiency of the NC DIS trigger results in an overall uncertainty of the jet cross sections of $\DTrig{} = \unit[1.0]{\%}$.

\item 
The efficiency of the requirement of a link between the primary vertex, the electron track and the electron cluster in the LAr calorimeter is described by the simulation within $\unit[1]{\%}$, which is assigned as an overall track-cluster-link uncertainty, \DTCL{}, on the jet cross sections~\cite{Kogler:10}. 
 
\item 
The overall normalisation uncertainty due to the luminosity measurement is $\DLumi{} = \unit[2.5]{\%}$~\cite{Aaron:2012kn}. 

\end{itemize}

In case of the normalised jet cross sections all systematic uncertainties are varied simultaneously in the numerator and denominator.
Consequently, all normalisation uncertainties, \DLumi{}, \DTCL{} and \DTrig{}, cancel fully. 
Uncertainties due to the electron reconstruction, such as \DEe{}, \DID{} and \DThe{} cancel to a large extent, and uncertainties due to the reconstruction of the HFS cancel partially.

\begin{figure}[p]
\newpage
  \centering
  \includegraphics[width=0.89\textwidth]{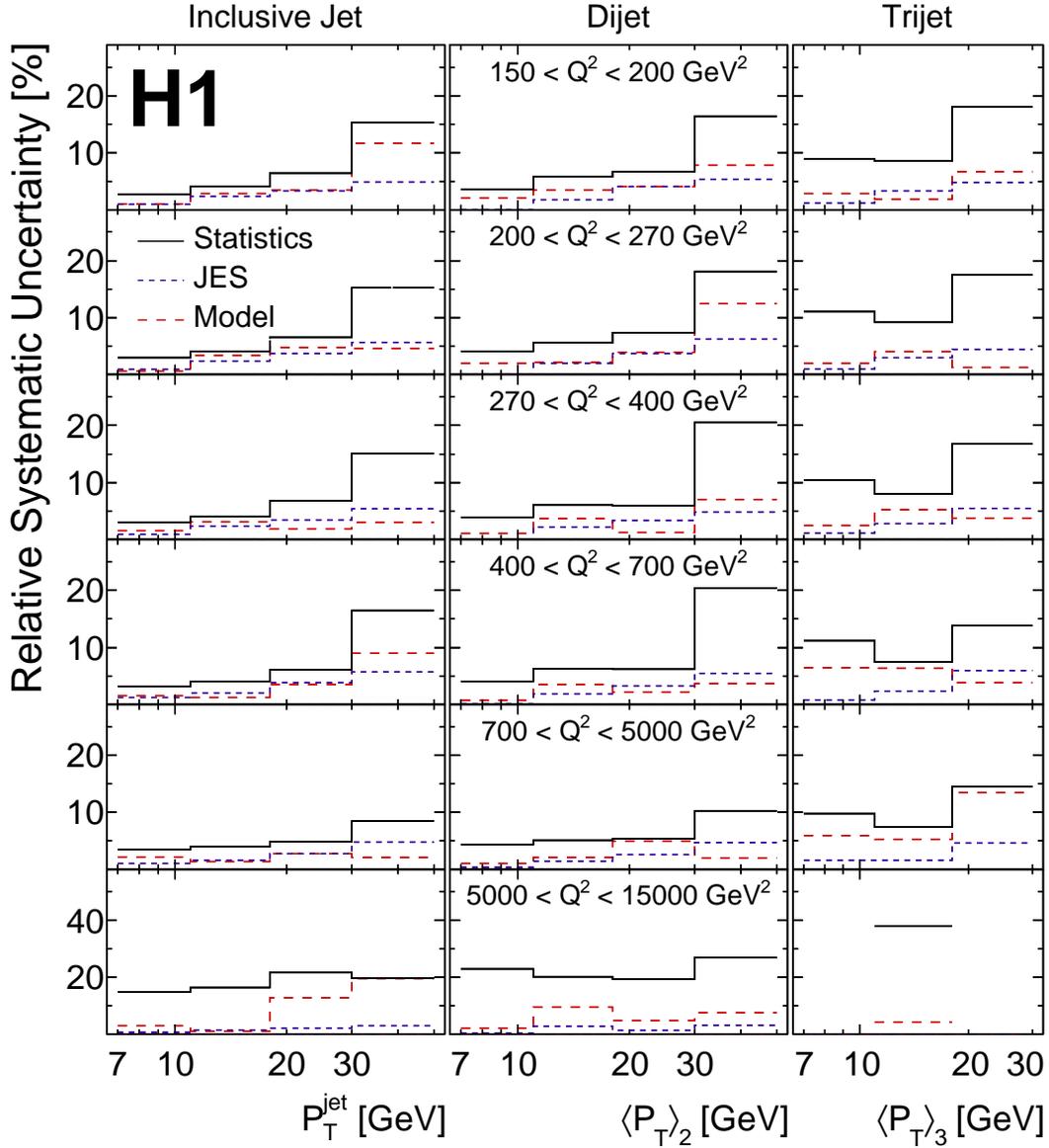}
  \caption{Illustration of the most prominent experimental uncertainties of the cross section measurement. Shown are the statistical uncertainties, the jet energy scale \DJES\ and the model uncertainty. Adjacent bins typically have negative correlation coefficients for the statistical uncertainty. 
The uncertainties shown are of comparable size for the corresponding normalised jet cross sections.
  \label{fig:RelevantUncertainties}
  }
\end{figure}

The relative size of the dominant experimental uncertainties \DStat{}, \DJES{} and 
\DMod{} are displayed in \fig~\ref{fig:RelevantUncertainties} for the absolute jet cross sections.
The jet energy scale \DJES{}\ becomes relevant for the high-\ptjet region, since these jets tend to go more in the direction of the incoming proton and are thus mostly made up from calorimetric information. 
The model uncertainty is sizeable mostly in the high-\ptjet region.

\section{Theoretical predictions}

Theoretical pQCD predictions in NLO accuracy are compared to the measured cross sections.
Hadronisation effects and effects of $Z$-exchange are not part of the pQCD predictions, 
and are therefore taken into account by correction factors.

\subsection{NLO calculations}
The parton level cross section $\sigma_i^{\rm parton}$ in each bin $i$ is predicted in pQCD as 
a power-series in \asmur, where $\mur$ is the renormalisation scale. 
The perturbative coefficients $c_{i,a,n}$ for a parton of flavour $a$ in order $n$ are convoluted 
in $x$ with the parton density functions $f_a$ of the proton,
\begin{equation}
 \sigma_i^{\rm parton} = \sum_{a,n} \as^n(\mur,\asmz)\,c_{i,a,n}\left(x,\mur,\muf\right)\otimes f_a(x,\muf) \, .
\end{equation}
The variable \muf denotes the factorisation scale, and \asmz is the value of the strong coupling constant
at the mass of the $Z$-boson. 
The first non-vanishing contribution to $\sigma_i^{\rm parton}$ is of order \as\
for inclusive jet and dijet cross sections and of order $\alpha_s^2$
for trijet cross sections.
The perturbative coefficients are currently known only to NLO.

The predictions $\sigma_i^{\rm parton}$ are obtained using the fastNLO framework~\cite{Kluge:2006xs,Britzger2012} with perturbative coefficients calculated by the \NLOJet\ program~\cite{Nagy:1998bb,Nagy:2001xb}.
The calculations are performed in NLO in the strong coupling and use the $\overline{\mathrm{MS}}$-scheme with five massless quark flavours.
The PDFs are accessed via the LHAPDF routines~\cite{Whalley:2005nh}. The MSTW2008 PDF 
set~\cite{Martin:09:189, Martin:2009bu} is used, determined with a value of the strong coupling constant of $\asmz=0.118$~\cite{Beringer:1900zz}.
The \as-evolution is performed using the evolution routines as provided together with the PDF sets in LHAPDF.
The running of the electromagnetic coupling $\aem(Q)$ is calculated using a recent determination of the hadronic contribution 
$\Delta \alpha_{\mathrm{had}}(M_Z^2) = 275.7(0.8)\times10^{-4}$~\cite{Bodenstein:2012pw}.
The renormalisation and factorisation scales are chosen to be
\begin{equation}
 \mur^2 = (\Qsq + \pt^2) / 2 \ \ \mathrm{and}\ \ \muf^2 = \Qsq~.
\end{equation}
The choice of \mur is motivated by the presence of two hard scales in the process, whereas \muf is chosen such that the same factorisation scale can be used in the calculation of jet and NC DIS cross sections.

The calculation of the NC DIS cross sections, $\sigma_{i}^{\rm NC}$, for the prediction of the normalised jet cross sections is performed using the QCDNUM program~\cite{Botje:11:490} in NLO in the zero mass variable flavour number scheme (ZM-VFNS). No contribution from $Z$-exchange is included, and both \muf and \mur are set to $Q$.

\subsection{Hadronisation corrections}
\label{sec:hadcorr}
The NLO calculations at parton level have to be corrected for non-perturbative hadronisation effects.
The hadronisation corrections \cHad\ account for long-range effects in the cross section calculation such as the fragmentation of partons into hadrons. 
It is given by the ratio of the jet cross section on hadron level to the jet cross section on parton level, i.e.\ for each bin $i$
$c^{\rm had}_i = \sigma_i^{\rm hadron} / \sigma_i^{\rm parton} $~.

The jet cross sections on parton and hadron level are calculated using \Django and 
\Rapgap. The parton level is obtained for MC event generators by selecting all partons before they are subjected to the fragmentation process.
Reweighting the MC distributions of jet observables on parton level to those obtained from the NLO calculation has negligible impact on the hadronisation corrections.
Hadronisation corrections are computed for both the \kt\ and the \antikt\ jet algorithm. 
They are very similar for inclusive jets and dijets, for trijets the corrections for \antikt\ tend to be somewhat smaller than for \kt.

The arithmetic average of \cHad is used, obtained from the weighted \Django and 
\Rapgap predictions (see section~\ref{sect:Reweighting}). 
Small differences of the correction factors between \Rapgap and \Django, which both use the Lund string fragmentation model, are 
observed, due to the different modelling of the partonic final state. 
The values of \cHad range from $0.8$ to $1$ and are given in the jet cross sections 
\tab{}s~\ref{tab:IncJet}--\ref{tab:NormTrijetXiAntiKt}.

\subsection{Electroweak corrections}
\label{sec:ewcorr}
Only virtual corrections for $\gamma$-exchange via the running of $\aem(\mur)$ are included in the pQCD calculations. 
The electroweak corrections \cEW\ account for the contributions from $\gamma Z$-interference and $Z$-exchange. 
They are estimated using the \Lepto event generator, where cross sections can be calculated including these effects ($\sigma^{\gamma,Z}$) and excluding them ($\sigma^{\gamma}$).
The electroweak correction factor \cEW\ is defined for each bin $i$ by 
$\cEWi=\sigma^{\gamma,Z}_i / \sigma^{\gamma}_i$.
It is close to unity at low \Qsq and becomes relevant for $\Qsq\rightarrow M_Z^2$, i.e.\ mainly in the highest \Qsq bin from $5000<\Qsq<\unit[\fifteen]{\GeV^2}$. In this bin 
the value of \cEW\ is around $1.1$ for the luminosity-weighted sum of $e^+p$ and $e^-p$ data corresponding to the full HERA-II dataset. 
The electroweak correction has some \pt-dependence, which, however, turns out to be negligible for the recorded mixture of $e^+p$ and $e^-p$ data.
In case of normalised jet cross sections, the electroweak corrections cancel almost completely 
such that they can be neglected.
The electroweak corrections are well-known compared to the statistical precision of those data 
points where the corrections deviate from unity, and therefore no uncertainty on \cEW\ is assigned. The values of \cEW\ are given in the jet cross sections 
\tab{}s~\ref{tab:IncJet}--\ref{tab:TrijetXiAntiKt}.

\subsection{QCD predictions on hadron level}
Given the parton level cross sections, $\sigma_i^{\rm parton}$, and the correction factors \cHadi\ and \cEWi\ in bin $i$, the hadron level jet cross sections are calculated as
\begin{equation}
 \sigma_i^{\rm hadron} =  \sigma_i^{\rm parton}\,\cHadi\,\cEWi \, ,
\end{equation}
while the predictions for the normalised jet cross sections are given by
\begin{equation}
 \left(\frac{\sigma}{\sNC}\right)^{\rm hadron}_{i} = \frac{\sigma_i^{\rm parton}\,\cHadi}{\sigma_{i}^{\rm NC}} \, .
\end{equation}

\subsection{Theoretical uncertainties}
\label{sect:xsec_theo_unc}
The following uncertainties on the NLO predictions are considered:

\begin{itemize}
\item   
The dominant theoretical uncertainty is attributed to the contribution from missing higher 
orders in the truncated perturbative expansion beyond NLO.
These contributions are estimated by a simultaneous variation of the chosen scales for \mur\ and \muf\ by the conventional factors of $0.5$ and $2$.
In case of normalised jet cross sections, the scales are varied simultaneously in the calculation of the numerator and denominator.
  
\item
The uncertainty on the hadronisation correction \DHad is estimated using the \Sherpa 
event generator~\cite{Gleisberg:2008ta}.
Processes including parton scattering of $2\rightarrow 5$ configurations are generated on tree level, providing a good description of jet production up to trijets. Also the parton level distributions are in reasonable agreement with the NLO calculation. The partons are hadronised once with the Lund string fragmentation model and once with the cluster fragmentation model~\cite{Webber:84:492}.
Half the difference between the two correction factors, derived from the two different fragmentation models, is taken as uncertainty on the hadronisation correction \DHad.
It is between $1$ to $\unit[2]{\%}$ for the inclusive jet and dijet measurements and between $0.5$ and $\unit[5]{\%}$ for the trijet measurements. These uncertainties are included in the cross section tables.
The absolute predictions from \Sherpa, however, are considered to be unreliable due to mismatches between the parton shower algorithm and the PDFs~\cite{Collins:2002ey}.
Therefore, only ratios of \Sherpa predictions are used for determining the uncertainty on the hadronisation corrections.
The uncertainties obtained in this way are typically between $30$ to $\unit[100]{\%}$ larger than
half the difference between the correction factors obtained using \Rapgap and \Django.

\item 
The uncertainty on the predictions due to the limited knowledge of the PDFs is determined 
at a confidence level of $\unit[68]{\%}$ from the MSTW2008 eigenvectors, following the formula 
for asymmetric PDF uncertainties~\cite{Campbell2007}.
The PDF uncertainty is found to be almost symmetric with a size of about $\unit[1]{\%}$ for all data points.
Predictions using other PDF sets do not deviate by more than two standard deviations of the PDF uncertainty.
\end{itemize}


\section{Experimental results}

\begin{figure}
  \centering
  \includegraphics[width=0.95\textwidth]{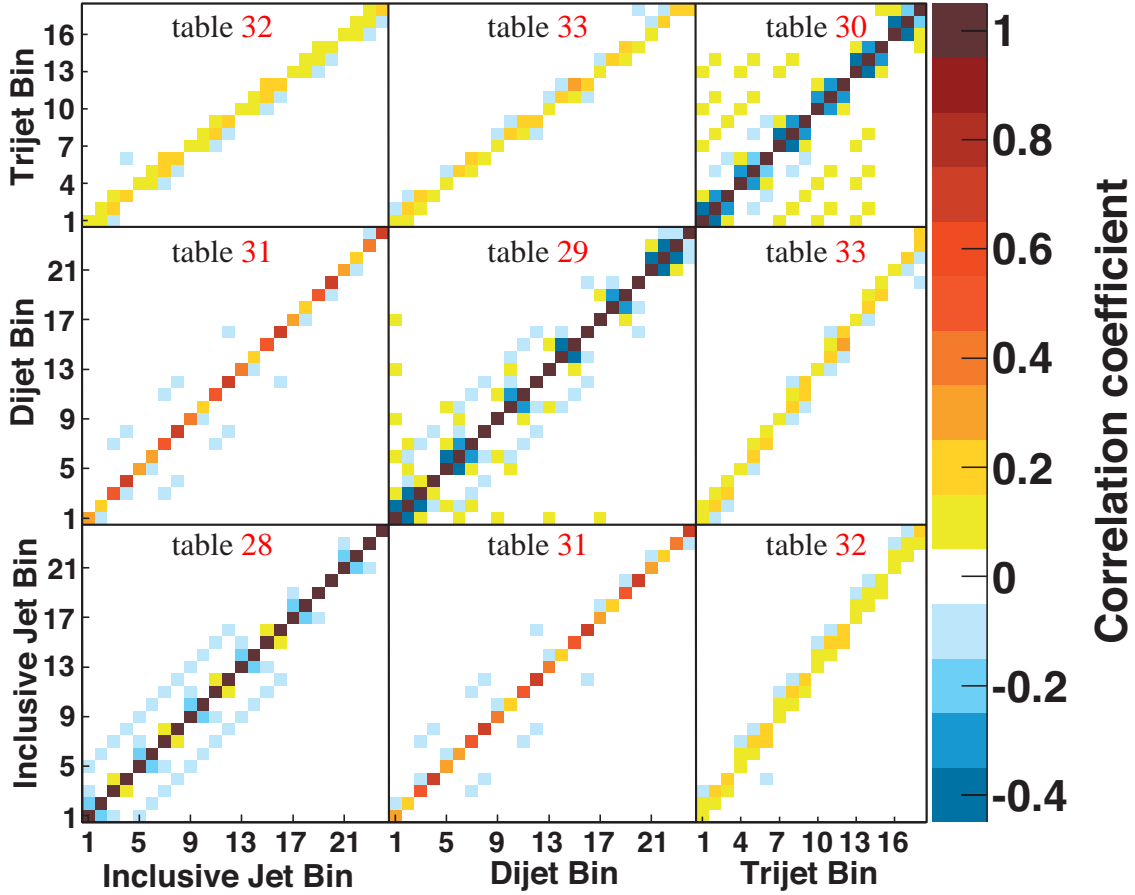}
  \put(-130, 50){\tab~\ref{tab:CorrInclIncl}}
  \put(-88, 50){\tab~\ref{tab:CorrInclDijet}}
  \put(-51, 50){\tab~\ref{tab:CorrInclTrijet}}
  \put(-130, 89){\tab~\ref{tab:CorrInclDijet}}
  \put(-88, 89){\tab~\ref{tab:CorrDijetDijet}}
  \put(-51, 89){\tab~\ref{tab:CorrDijetTrijet}}
  \put(-130, 119){\tab~\ref{tab:CorrInclTrijet}}
  \put(-88, 119){\tab~\ref{tab:CorrDijetTrijet}}
  \put(-51, 119){\tab~\ref{tab:CorrTrijetTrijet}}
  \caption{Correlation matrix of the three jet cross section measurements. The bin numbering is given by $b=(q-1)n_{\Pt} + p$,
  where $q$ stands for the bins in \Qsq and $p$ for the bins in \Pt (see table \tab~\ref{tab:BinNumbering}).
  For the inclusive jet and dijet measurements $n_{\Pt}=4$, and for the trijet measurement $n_{\Pt}=3$.
  The numerical values of the correlation coefficients are given in the tables indicated.
  }
  \label{fig:CorrMa}
\end{figure}

In the following the absolute and normalised double-differential jet cross sections are presented for inclusive jet, dijet, and trijet production using the \kt\ and the \antikt\ jet algorithms.
The labelling of the bins in the tables of cross sections is explained in \tab~\ref{tab:BinNumbering}.
\begin{table}
\footnotesize
\center
\begin{tabular}[h]{l|cccc}
  \hline
  Observable & \kt\ & \antikt\ & \kt\ (normalised) & \antikt\ (normalised) \\
  \hline 
   $\sI(\Qsq,\ptjet)$    & \tab~\ref{tab:IncJet}  & \tab~\ref{tab:IncJetAntiKt}  & \tab~\ref{tab:NormIncJet} & \tab~\ref{tab:NormIncJetAntiKt} \\
   $\sD(\Qsq,\meanptdi)$  & \tab~\ref{tab:Dijet}   & \tab~\ref{tab:DijetAntiKt}   & \tab~\ref{tab:NormDijet} & \tab~\ref{tab:NormDijetAntiKt} \\
   $\sD(\Qsq,\xidi)$     & \tab~\ref{tab:DijetXi} & \tab~\ref{tab:DijetXiAntiKt} & \tab~\ref{tab:NormDijetXi} & \tab~\ref{tab:NormDijetXiAntiKt} \\   
   $\sT(\Qsq,\meanpttri)$ & \tab~\ref{tab:Trijet}  & \tab~\ref{tab:TrijetAntiKt}  & \tab~\ref{tab:NormTrijet} & \tab~\ref{tab:NormTrijetAntiKt} \\
   $\sT(\Qsq,\xitri)$     & \tab~\ref{tab:TrijetXi}& \tab~\ref{tab:TrijetXiAntiKt}& \tab~\ref{tab:NormTrijetXi} & \tab~\ref{tab:NormTrijetXiAntiKt} \\
  \hline
\end{tabular}
\caption{Overview of the tables of cross sections.}
\label{tab:TableOfCrossSectionTables}
\end{table}
An overview of the tables of jet cross sections is summarised 
in~\tab~\ref{tab:TableOfCrossSectionTables} and of the tables of correlation coefficients, i.e. point-to-point statistical correlations, is provided 
in~\tab~\ref{tab:TableOfCorrelations}.
Figure~\ref{fig:CorrMa} shows the correlation matrix of the inclusive, dijet and trijet cross sections, corresponding to \tab{}s~\ref{tab:CorrInclIncl}--\ref{tab:CorrDijetTrijet}.
When looking at the inclusive jet, dijet or trijet cross sections alone, negative correlations down to $-0.5$ are observed between adjacent bins in \Pt, which reflects the moderate jet resolution in \Pt. 
In adjacent \Qsq bins, the negative correlations of about $-0.1$ are close to zero, due to the better resolution in \Qsq.
Sizeable positive correlations are observed between inclusive jet and dijet cross sections with the same \Qsq\ and similar \Pt. 
Positive correlations between the trijet and the inclusive jet and dijet measurements are 
smaller than those between the dijet and inclusive jet, because of the smaller statistical overlap. 
Within the accuracy of this measurement, the correlation coefficients are very similar no matter whether the \kt\ or \antikt\ jet algorithm are used. 
Similarly, the statistical correlations of the normalised and the absolute cross sections are almost identical.

\begin{table}
\footnotesize
\center
\begin{tabular}[h]{l|ccccc}
  \hline
  Observable & $\sI(\Qsq,\ptjet)$ & $\sD(\Qsq,\meanptdi)$ & $\sT(\Qsq,\meanpttri)$ & $\sD(\Qsq,\xidi)$ & $\sT(\Qsq,\xitri)$ \\
  \hline 
   $\sI(\Qsq,\ptjet)$    &  \tab~\ref{tab:CorrInclIncl} & \tab~\ref{tab:CorrInclDijet} & \tab~\ref{tab:CorrInclTrijet} & \tab~\ref{tab:CorrInclDijetXi} & \tab~\ref{tab:CorrInclTrijetXi} \\
   $\sD(\Qsq,\meanptdi)$ &  \tab~\ref{tab:CorrInclDijet} & \tab~\ref{tab:CorrDijetDijet} & \tab~\ref{tab:CorrDijetTrijet} & -- & -- \\
   $\sT(\Qsq,\meanpttri)$&  \tab~\ref{tab:CorrInclTrijet} & \tab~\ref{tab:CorrDijetTrijet} & \tab~\ref{tab:CorrTrijetTrijet} & -- & -- \\
   $\sD(\Qsq,\xidi)$     &  \tab~\ref{tab:CorrInclDijetXi}&  -- & -- & \tab~\ref{tab:CorrDijetXiDijetXi} & \tab~\ref{tab:CorrDijetXiTrijetXi} \\
   $\sT(\Qsq,\xitri)$     &  \tab~\ref{tab:CorrInclTrijetXi} & -- & -- & \tab~\ref{tab:CorrDijetXiTrijetXi} & \tab~\ref{tab:CorrTrijetXiTrijetXi} \\
  \hline
\end{tabular}
\caption{Overview of the tables of correlation coefficients. The correlation coefficients between the \meanpt\ and $\xi$ measurements are not available.}
\label{tab:TableOfCorrelations}
\end{table}

\begin{figure}[p]
  \centering
  \includegraphics[width=0.99\textwidth]{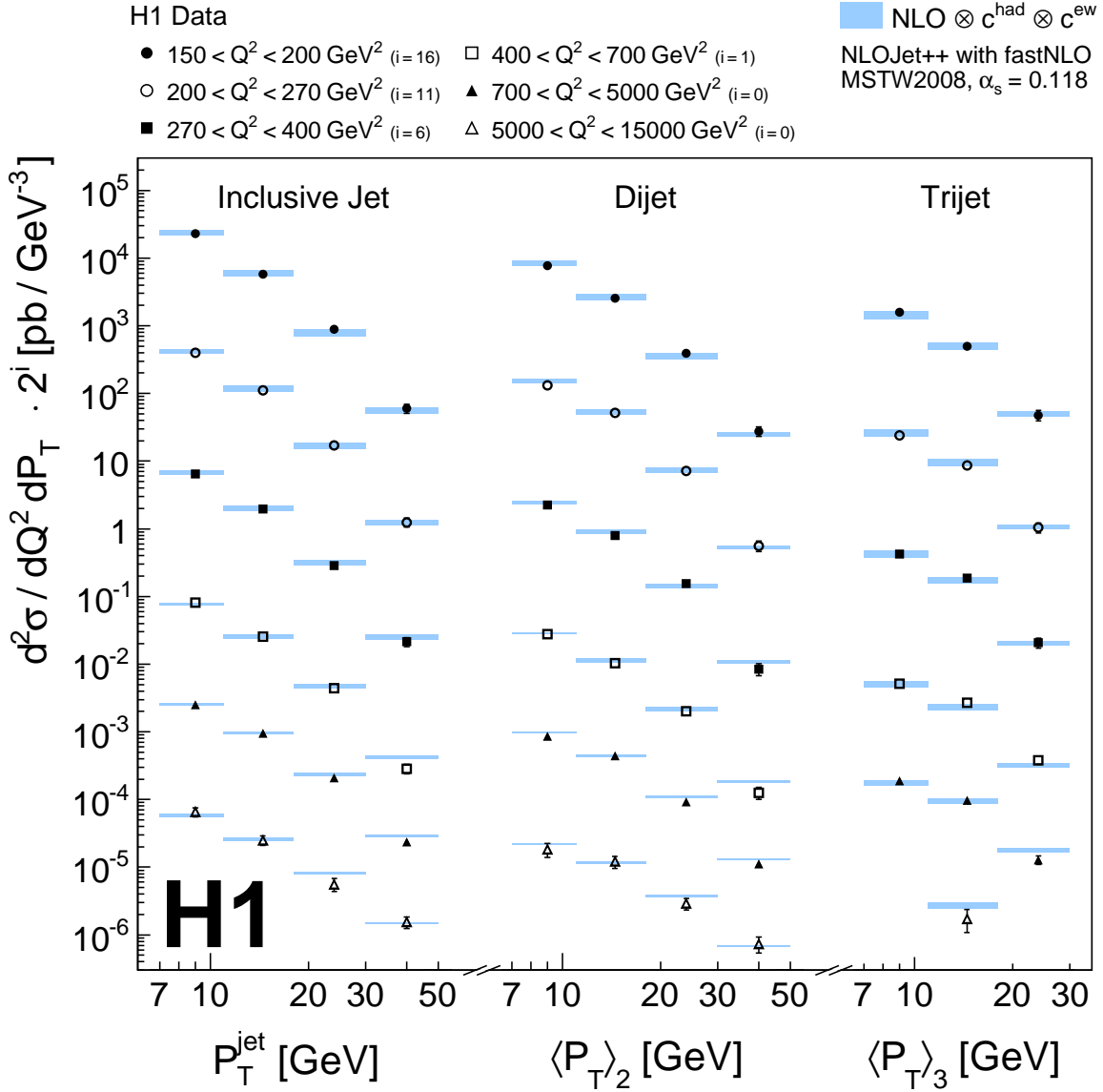}
  \caption{Double-differential cross sections for jet production in DIS as a function of \Qsq\ and \pt.
     The  inner and outer error bars indicate the statistical uncertainties and the statistical and systematic uncertainties added in quadrature.
     The NLO QCD predictions, corrected for hadronisation and electroweak effects, together with their uncertainties are shown by the shaded band.
The cross sections for individual \Qsq~bins are multiplied by a factor of $10^i$ for better readability.
  }
  \label{fig:CrossSections}
\end{figure}

\begin{figure}
  \centering
  \includegraphics[width=0.95\textwidth]{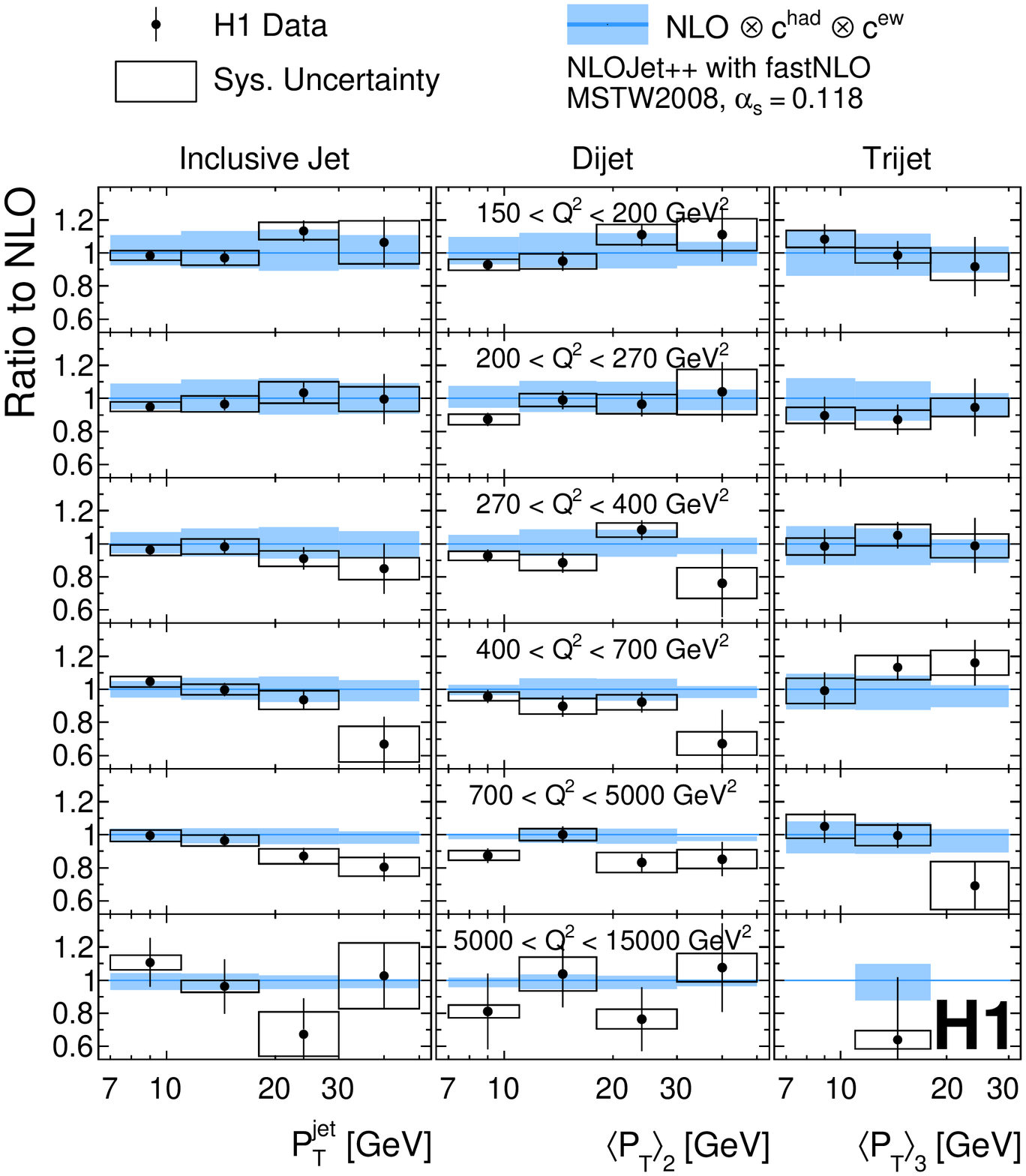}
  \caption{
     Ratio of jet cross sections to NLO predictions as function of \Qsq\ and \pt.
     The error bars on the data indicate the statistical uncertainties of the measurements, while the total systematic uncertainties are given by the open boxes.
The shaded bands show the theory uncertainty.
  }
  \label{fig:CrossSectionRatioMSTW}
\end{figure}

The measured cross sections for the \kt jet algorithm as a function of \pt\ 
(\tab{}s~\ref{tab:IncJet}--\ref{tab:DijetXi}) are displayed in different \Qsq\ bins in \fig~\ref{fig:CrossSections}, together with the NLO predictions.
A detailed comparison of the predictions to the measured cross sections is provided by the 
ratio of data to NLO in \fig~\ref{fig:CrossSectionRatioMSTW}. 
The theory uncertainties from scale variations dominate over the sum of the experimental uncertainties in most bins.

\begin{figure}
  \centering
  \includegraphics[width=0.95\textwidth]{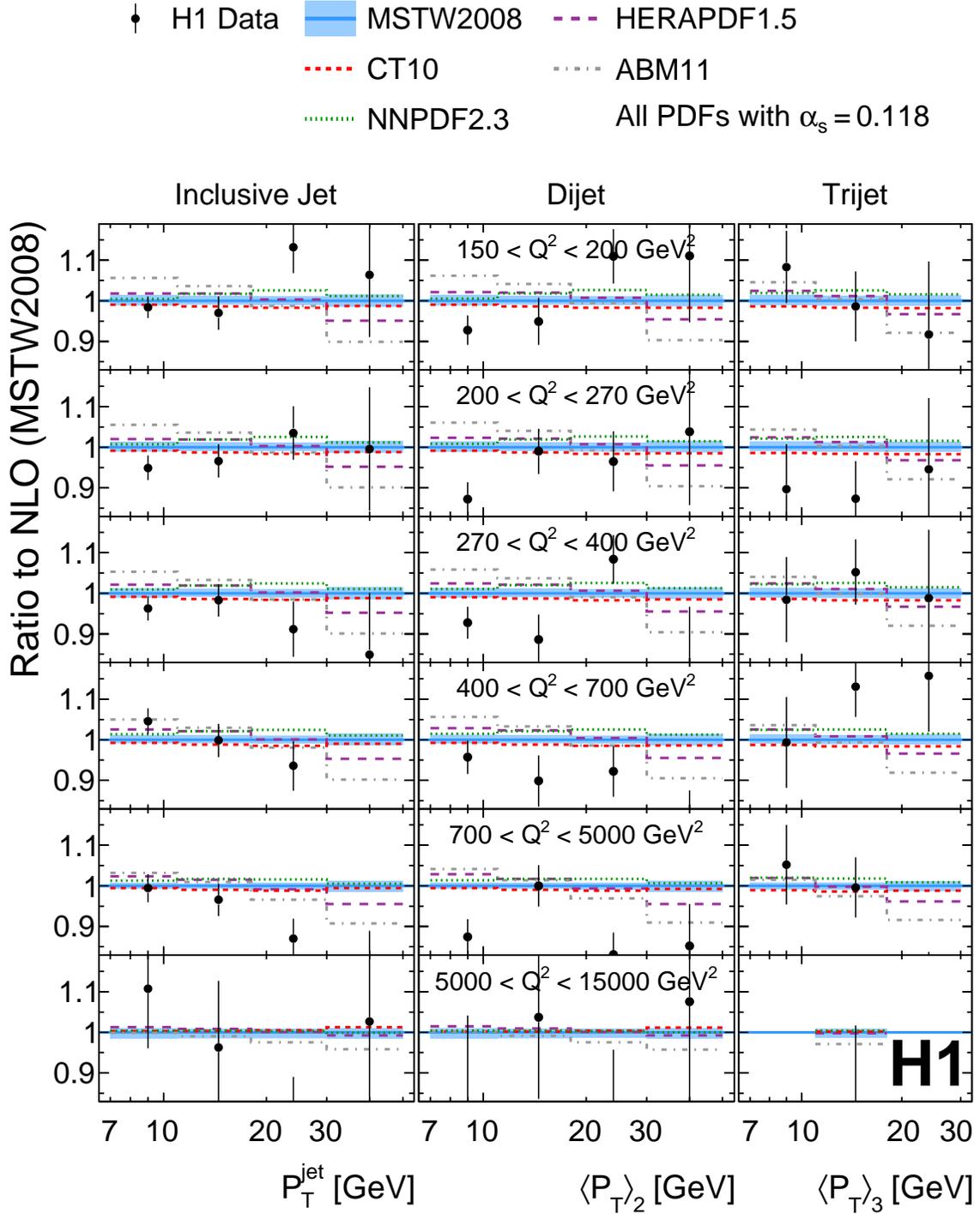}
  \caption{
     Ratio of NLO predictions with various PDF sets to predictions using the MSTW2008 PDF set as a function of \Qsq and \pt.
     For comparison, the data points are displayed together with their statistical uncertainty, which are often outside of the displayed range in this enlarged presentation.
     All PDF sets used are determined at NLO and with a value of $\asmz = 0.118$.
     The shaded bands show the PDF uncertainties of the NLO calculations obtained from the MSTW2008 eigenvector set at a confidence level of $\unit[68]{\%}$.
  }
  \label{fig:CrossSectionRatioPDFComp}
\end{figure}

The data are in general well described by the theoretical predictions.
The predictions are slightly above the measured cross sections for inclusive jet and dijet production, at medium \Qsq\ and at high \Pt. A detailed comparison of NLO predictions using different PDF sets with the measured jet cross sections is shown in 
\fig~\ref{fig:CrossSectionRatioPDFComp}. 
Only small differences are observed between predictions for different choices of PDF sets compared to the theory uncertainty from scale variations shown in 
\fig~\ref{fig:CrossSectionRatioMSTW}.
Predictions using the CT10 PDF set~\cite{Lai:2010vv} are approximately $1$ to $\unit[2]{\%}$ below those using the MSTW2008 PDF set, and predictions using the NNPDF2.3 set~\cite{Ball:2012cx} are about $\unit[2]{\%}$ above the latter.
The calculation using the HERAPDF1.5 set~\cite{Aaron:2009aa, HERAPDF15, Radescu:2010zz} is $\unit[2]{\%}$ above the calculation using MSTW2008 at low \Pt, while at the highest \Pt values it is around $\unit[5]{\%}$ below. The reason for this behaviour is the softer valence quark density at high $x$ of the HERAPDF1.5 set compared to the other PDF sets.
Predictions using the ABM11 PDF set~\cite{Alekhin:2012ig} show larger differences compared to the other PDF sets.

\begin{figure}
  \centering
  \includegraphics[width=0.99\textwidth]{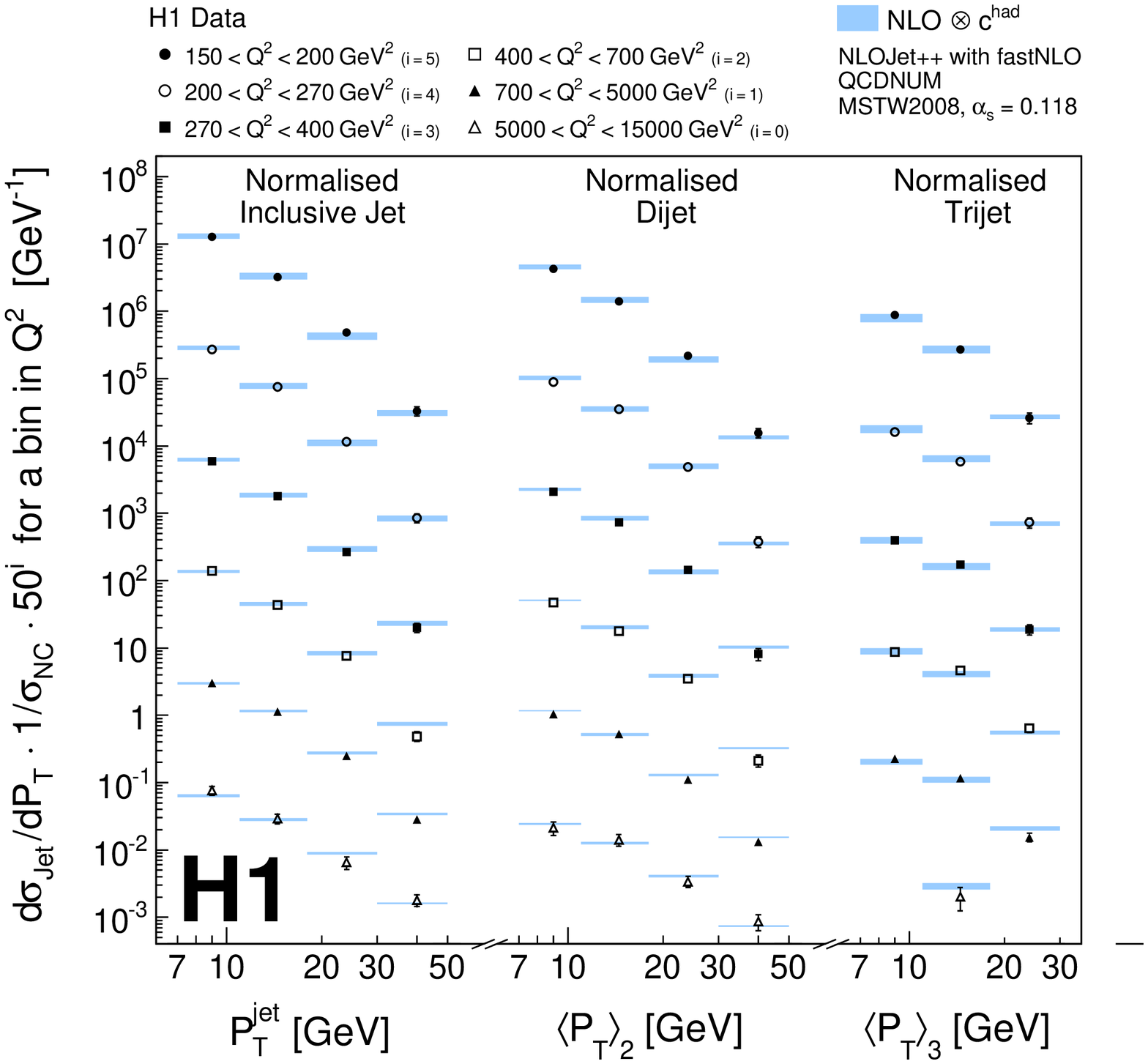}
  \caption{
     Double-differential normalised cross sections for jet production in DIS as a function of \Qsq\ and \pt.
     The NLO predictions, corrected for hadronisation effects, together with their uncertainties are shown by the shaded bands.
     Further details can be found in the caption of \fig~\ref{fig:CrossSections}.
  }
  \label{fig:CSNorm}
\end{figure}
\begin{figure}
  \centering
  \includegraphics[width=0.95\textwidth]{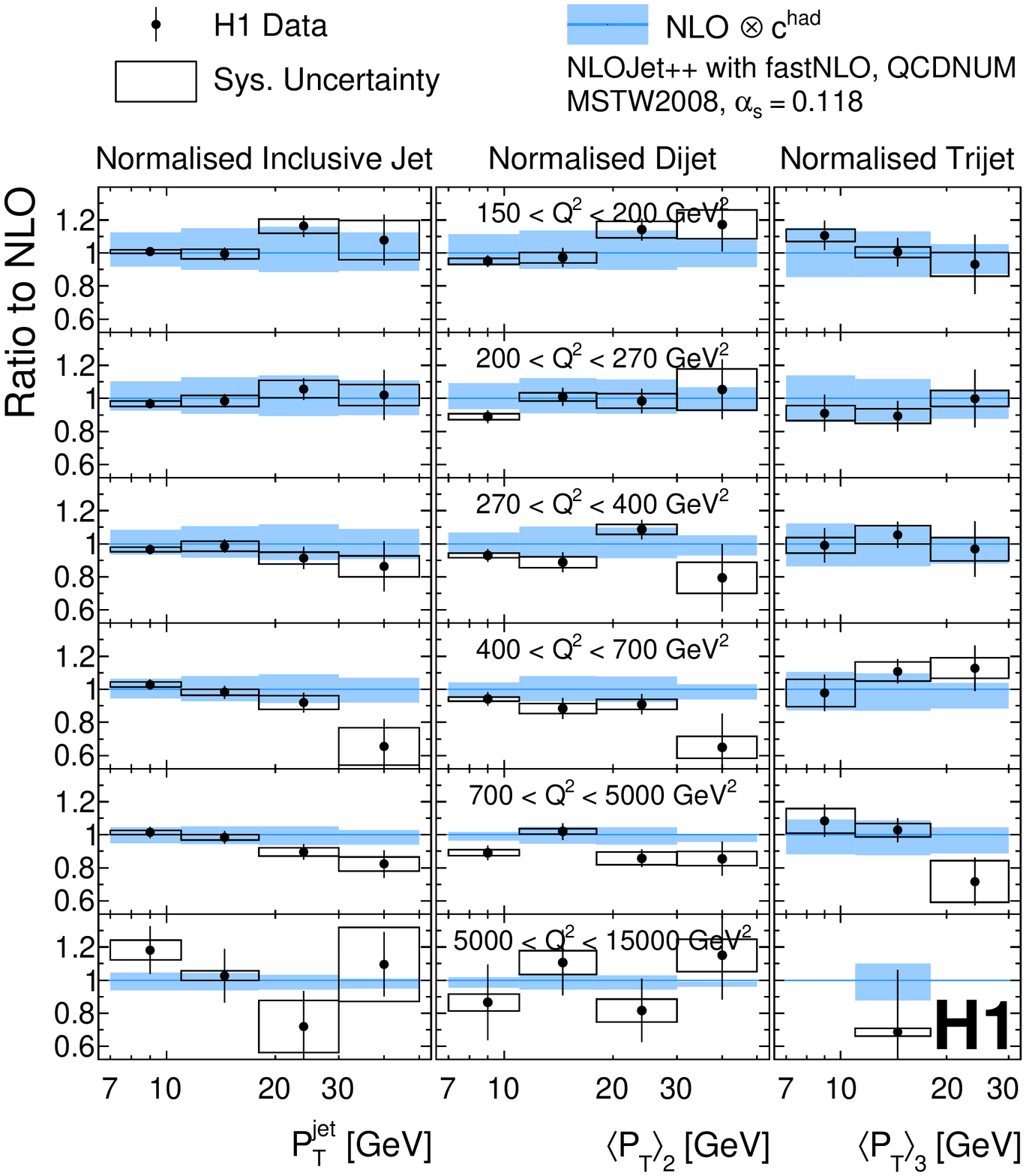}
  \caption{
     Ratio of normalised jet cross sections to NLO predictions as a function of \Qsq\ and \pt.
     The error bars on the data indicate the statistical uncertainties of the measurements, while the total systematic uncertainties are given by the open boxes.
     The shaded bands show the theory uncertainty.
   }
  \label{fig:NormCrossSectionRatio}
\end{figure}

The normalised cross sections using the \kt jet algorithm are displayed in 
\fig~\ref{fig:CSNorm} as a function of \pt\ in different \Qsq\ bins together with the NLO calculations.
The ratio of data to the predictions is shown in \fig~\ref{fig:NormCrossSectionRatio}. 
The comparison is qualitatively similar to the results from the absolute jet cross sections.
Similar to the case of absolute cross sections, the theory uncertainty from scale variations is significantly larger than the total experimental uncertainty in almost all bins.
For the normalised jet cross sections PDF dependencies do not cancel. This is due to the different $x$-dependencies and parton contributions to NC DIS compared to jet production. 
The systematic uncertainties are reduced for normalised cross sections compared to absolute jet cross section, since all normalisation uncertainties cancel fully, and uncertainties on the electron reconstruction and the HFS cancel partly.
The experimental uncertainty is dominated by the statistical, the model and the jet energy scale uncertainties.
Given the high experimental precision, in comparison to the absolute jet cross sections, one observes that the normalised dijet cross sections are below the theory predictions for many data points.

\begin{figure}
  \centering
  \includegraphics[width=0.99\textwidth]{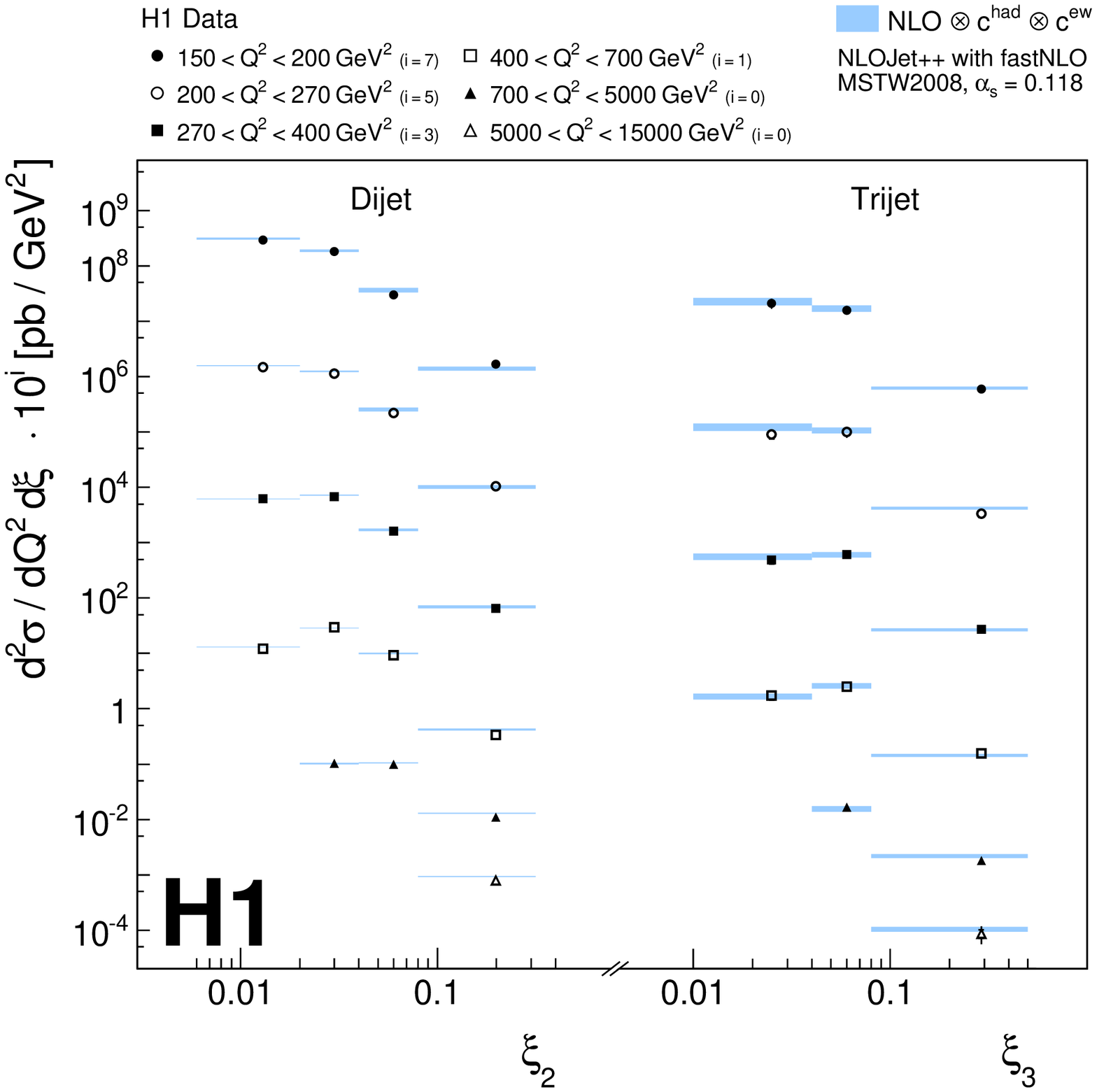}
  \caption{
     Double-differential cross sections for dijet and trijet production in DIS as a function of \Qsq\ and $\xi$.
     The NLO predictions, corrected for hadronisation and electroweak effects, together with their uncertainties are shown by the shaded bands.
     Further details can be found in the caption of \fig~\ref{fig:CrossSections}.
  }
  \label{fig:CrossSectionsXi}
\end{figure}

\begin{figure}
  \centering
  \includegraphics[width=0.99\textwidth]{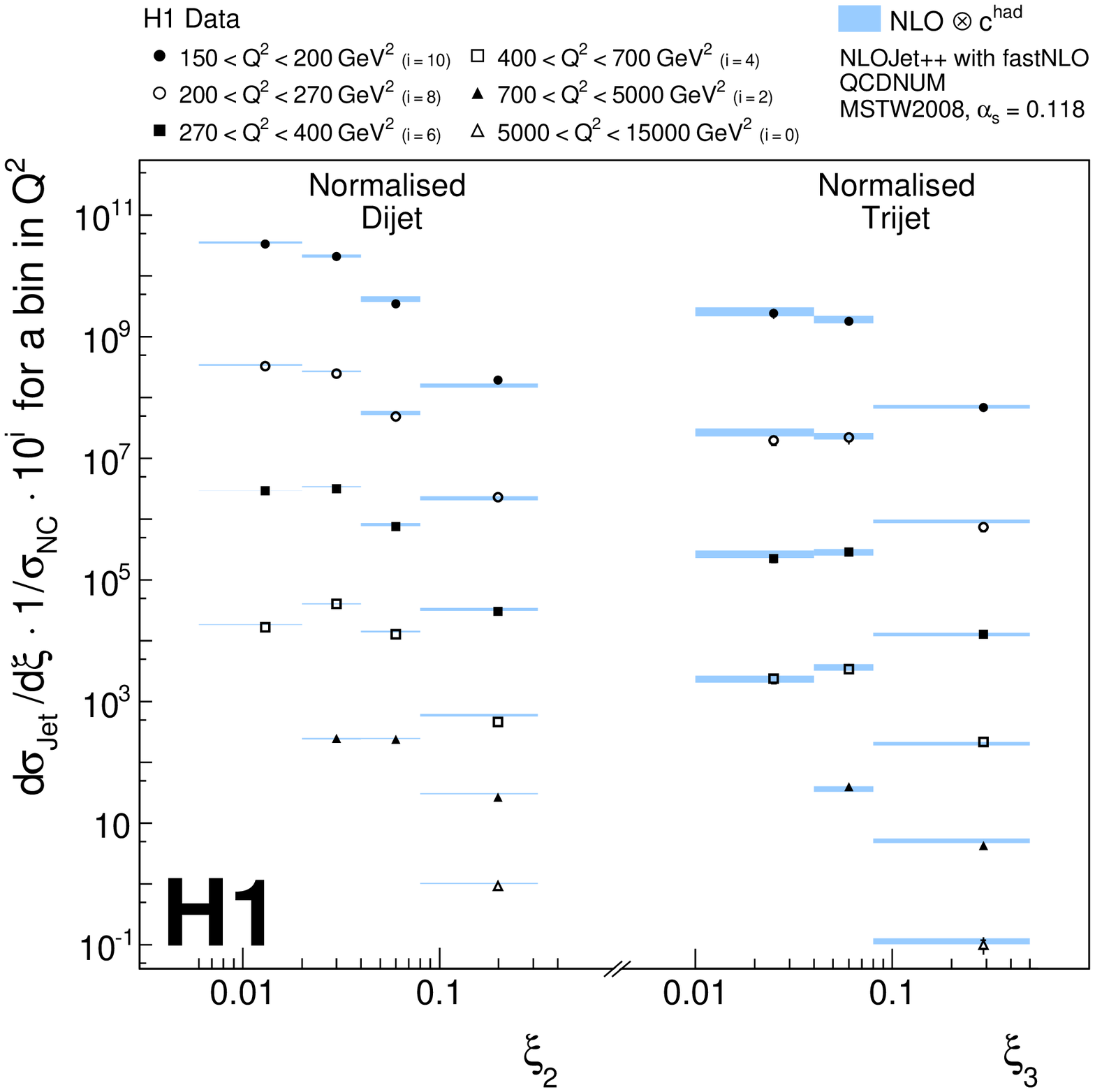}
  \caption{
     Double-differential normalised cross sections for dijet and trijet production in DIS as a function of \Qsq\ and $\xi$.
     The NLO predictions, corrected for hadronisation effects, together with their uncertainties are shown by the shaded bands.
     Further details can be found in the caption of \fig~\ref{fig:CrossSections}.
  }
  \label{fig:CrossSectionsXiNorm}
\end{figure}

\begin{figure}
  \centering
  \includegraphics[width=0.495\textwidth]{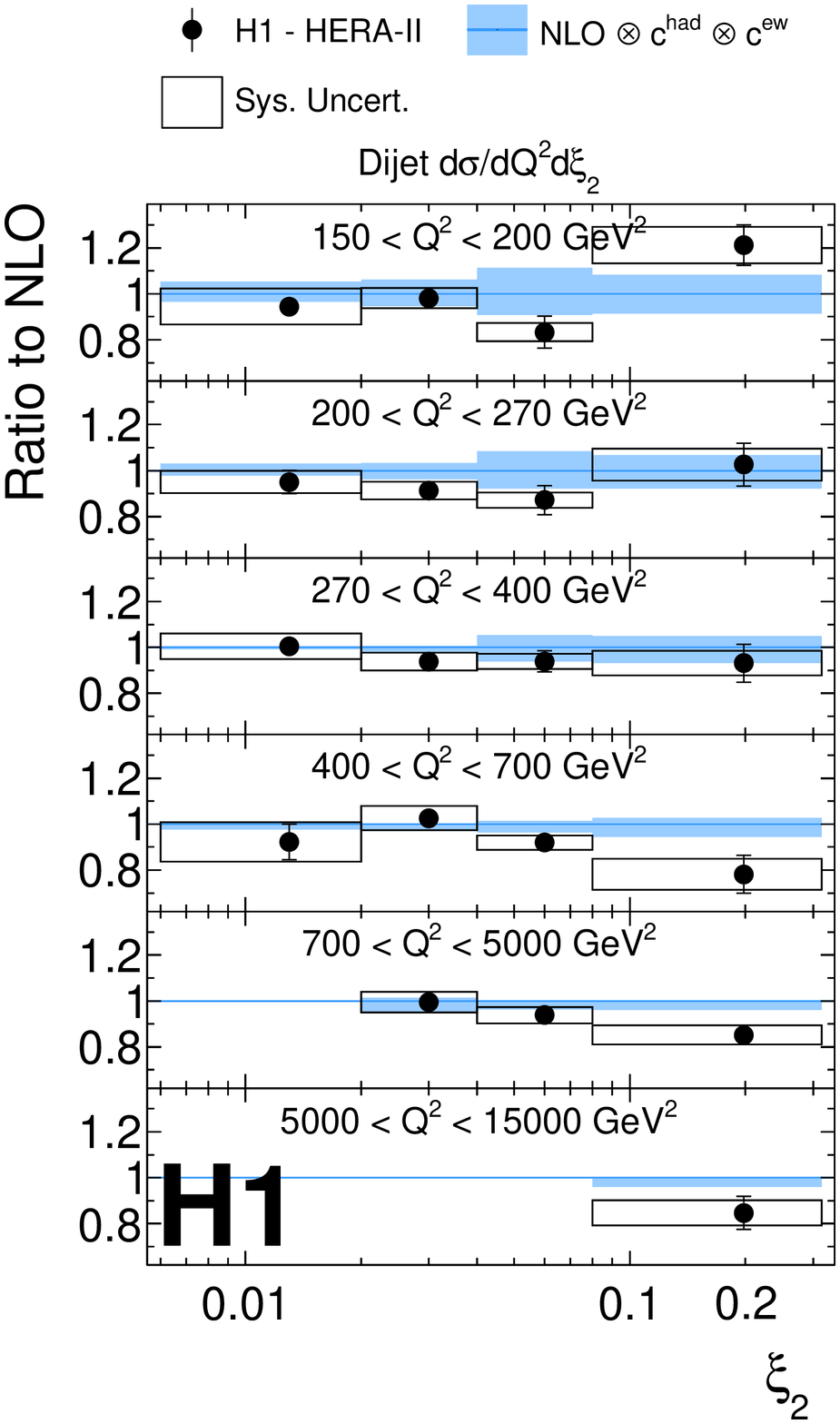}
  \includegraphics[width=0.495\textwidth]{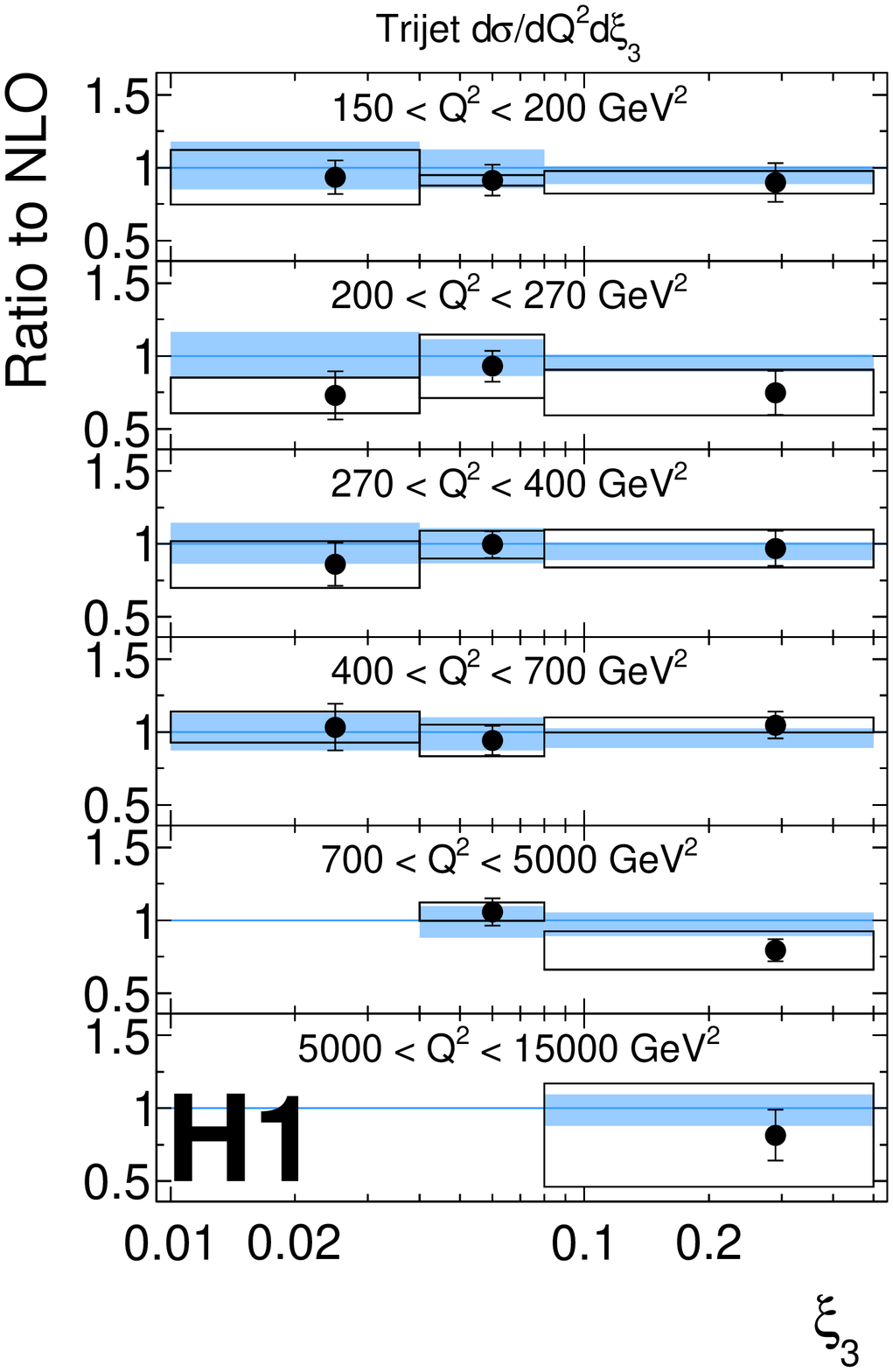}
  \caption{
     Ratio of the dijet and trijet cross sections to NLO QCD predictions as a function of \Qsq\ and $\xi$.
     The error bars on the data indicate the statistical uncertainties of the measurements while the total experimental systematic uncertainties are given by the open boxes.
     The shaded bands show the theory uncertainties.
  }
  \label{fig:RatiosXi}
\end{figure}

The measurements of absolute dijet and trijet cross sections are displayed in \fig~\ref{fig:CrossSectionsXi} as a function of \xidi\ and \xitri\ in different \Qsq\ bins, together with NLO predictions. 
The normalised jet cross sections are shown in \fig~\ref{fig:CrossSectionsXiNorm}.
The ratio of absolute jet cross sections to NLO predictions as a function of $\xi$ in bins of 
\Qsq\ is shown in \fig~\ref{fig:RatiosXi}. 
Good overall agreement between predictions and the data is observed. 
A similar level of agreement is obtained by using other PDF sets than the employed MSTW2008 set.

\kd{
Also the \antikt\ cross sections agree well with the theory predictions. For inclusive jets and dijets the NLO predictions using the \antikt\ or the \kt\ jet algorithm are identical, for trijets they are not. The hadronisation corrections between \antikt\ and \kt\ jets differ slightly. The \antikt\ trijet cross sections have a tendency of being slightly lower than the \kt\ measurement.}

Of the results presented here, those which can be compared to previous H1 measurements are found to be well compatible.

\section[Determination of the strong coupling constant as(MZ)]{Determination of the strong coupling constant \basmz} 

The jet cross sections presented are used to determine the value of the strong coupling constant\footnote{
In this section, the strong coupling constant \asmz is always quoted at the mass of the $Z$-boson, $M_Z=\unit[91.1876]{\GeV}$~\cite{Beringer:1900zz}. 
For better readability the scale dependence is dropped in the notation and henceforth \as\ is written for \asmz; `\asmz' is only used for explicit highlighting.
}
\as at the scale of the mass of the $Z$-boson, $M_Z$, in the framework of perturbative QCD.
The value of the strong coupling constant \as is determined in an iterative 
\chisq-minimisation procedure using NLO calculations, corrected for hadronisation effects and, if applicable, for electroweak effects. 
The sensitivity of the theory prediction to \as arises from the perturbative expansion of the matrix elements in powers of $\asmur=\as(\mur,\asmz)$.
For the \as-fit, the evolution of $\as(\mur)$ is performed
solving this equation numerically, using the renormalisation group equation in two-loop precision with five massless flavours.

\subsection{Fit strategy}
The value of \as\ is determined using a \chisq-minimisation, where \as\ is a free parameter of the theory calculation. 
The agreement between theory and data is estimated using the \chisq definition~\cite{Beringer:1900zz,Barone:00:243}
\begin{equation}
\chisq = \vec{p}^{\rm T}\V^{-1}\vec{p} + \sum_k^\Nsys \varepsilon^2_k \, , 
\label{eq:Chi2Fit}
\end{equation}
where $\V^{-1}$ is the inverse of the covariance matrix with relative uncertainties. 
The element $i$ of the vector $\vec{p}$ stands for the difference between the logarithm of the 
measurement $m_i$ and the logarithm of the theory prediction $t_i = t_i(\asmz)$:
\begin{equation}
p_i = \log m_i - \log t_i - \sum_k^\Nsys E_{i,k} \, .
\end{equation}
This ansatz is equivalent to assuming that the $m_i$ are log-normal distributed,
with $E_{i,k}$ being defined as
\begin{equation}
E_{i,k} = \sqrt{f_k^{\rm C}}\left( \frac{\drel[{\mathnormal k},+]{m,i}   - \drel[{\mathnormal k},-]{m,i}}{2}\varepsilon_k + \frac{\drel[{\mathnormal k},+]{m,i} + \drel[{\mathnormal k},-]{m,i}}{2}\varepsilon_k^2 \right)~.
\label{eq:RelShiftEik}
\end{equation}
The nuisance parameters $\varepsilon_k$ for each source of systematic uncertainty $k$ are free parameters in the \chisq-minimisation.
Sources indicated as uncorrelated between \Qsq bins in table~\ref{tab:ErrorTreatmentInFit} have several nuisance parameters, one for each \Qsq bin.

The parameters $\drel[{\mathnormal k},+]{m,i}$ and $\drel[{\mathnormal k},-]{m,i}$ denote the relative uncertainty on the measurement $m_i$, due to the `up' and `down' variation of the systematic uncertainty $k$. 
Systematic experimental uncertainties are treated in the fit as either relative correlated or uncorrelated uncertainties or as a mixture of both.
The parameter $\fC$ expresses the fraction of the uncertainty $k$ which is considered as relative correlated uncertainty, and $\fU$ expresses the fraction which is treated as uncorrelated uncertainty with $\fC + \fU = 1$. 
The symmetrised uncorrelated uncertainties squared $\fU_k(\drel[{\mathnormal k},+]{m,i}-\drel[{\mathnormal k},-]{m,i})^2$
are added to the diagonal elements of the covariance matrix $\V$.
The covariance matrix $\V$ thus consists of relative statistical uncertainties, including correlations between the data points of the measurements, correlated background uncertainties and the uncorrelated part of the systematic uncertainties.

\subsection[Experimental uncertainties on as]{Experimental uncertainties on \bas}
The experimental uncertainties are treated in the fit as described in the following. 
\begin{table}
\footnotesize
\center
\begin{tabular}[h]{lccc}
  \hline
  {Source of uncertainties} ${k}$ & {Correlated}   & {Uncorrelated}  & {Uncorrelated}\\[-0.15cm]
  & {fraction} ${\fC}$  & {fraction} ${\fU}$  & {between \Qsq\ bins }\\
  \hline 
  Jet energy scale \DJES{}		& 0.5 & 0.5 & \\
  Rem. cluster energy scale \DHFS{}       	& 0.5 & 0.5 & \\
  LAr Noise \DLAr{}			& 1   & 0 & \\
  Electron energy \DEe{} 			& 1   & 0 & \checkmark \\
  Electron polar angle \DThe{}            & 1   & 0 & \checkmark \\
  Electron ID \DID{}			& 1   & 0 & \checkmark \\
  Normalisation \DNorm{}  		& 1   & 0 & \\
  Model \DMod{} 				& 0.25& 0.75 & \checkmark\\
  \hline
\end{tabular}
\caption{Split-up of systematic uncertainties in the fit of the strong coupling constant \as. }
\label{tab:ErrorTreatmentInFit}
\end{table}
\begin{itemize}
\item
The statistical uncertainties are accounted for by using the covariance matrix obtained from the unfolding process.
It includes all point-to-point correlations due to statistical correlations and detector resolutions.
\item 
The uncertainties due to the reconstruction of the hadronic final state, i.e.\ \DJES{}\ and \DHFS{}, are treated as $\unit[50]{\%}$ correlated and uncorrelated, respectively.
\item
The uncertainty \DLAr{}, due to the LAr noise suppression algorithm, is considered to be fully correlated.
\item
  All uncertainties due to the reconstruction of the scattered electron  (\DEe{}, \DThe{} and \DID{}) are treated as fully correlated for data points belonging to the same \Qsq-bin and uncorrelated between different \Qsq-bins.
\item
The uncertainties on the normalisation (\DLumi{}, \DTrig{}\ and \DTCL{}) are summed in quadrature to form the normalisation uncertainty 
$\DNorm{} = \unit[2.9]{\%}$ which is treated as fully correlated.
\item
The model uncertainties are treated as $\unit[75]{\%}$ uncorrelated, whereby the correlated fraction is treated as uncorrelated between different \Qsq-bins.

\end{itemize}

The uncorrelated parts of the systematic uncertainties are expected to account for local variations, while the correlated parts are introduced to account for procedural uncertainties.
%
A summary is given in \tab~\ref{tab:ErrorTreatmentInFit}, showing the treatment
of each experimental uncertainty in the fit.

\begin{table}
\center
\footnotesize
\begin{tabular}[h!]{l|c|cccc}
\multicolumn{6}{c}{\textbf{Experimental uncertainties on {\boldmath $\as \times 10^4$}} } \tabularnewline 
\hline
 {Measurement} & $\dabs[exp]{\as}$
	& \dNorm{\as}
	& \dHFS{\as}
	& \dJES{\as}
	& \dMod{\as}  \\
\hline
\sI				& 22.2 & 18.5 & 4.8 & 5.5 & 4.5 \\
\sD				& 23.4 & 19.4 & 4.4 & 4.3 & 6.4 \\
\sT				& 16.7 & 11.2 & 5.4 & 4.3 & 4.6 \\
\hline 
 & & \\[-2.8ex]                                                                                                  
\sIN				& ~8.9 &  --  & 1.7 & 4.4 & 2.2 \\[1.2em]
\sDN				& ~9.9 &  --  & 1.6 & 3.3 & 3.6 \\[1.2em]
\sTN				& 11.3 &  --  & 4.0 & 3.5 & 4.2 \\[0.6em]
\hline                                                                                                   
$[\sI, \sD, \sT]$               & 16.0 & ~9.6 & 5.9 & 3.2 & 5.0 \\[0.4em]
$\left[\sIN,\sDN,\sTN\right]$   & ~7.6 &  --  & 2.4 & 2.8 & 1.8 \\[0.7em]
\hline
\end{tabular}
\caption{The total experimental uncertainty on \as\ from fits to different jet cross sections, and the contributions from the most relevant sources of uncertainties. These are the normalisation uncertainty, the uncertainties on the reconstruction of the HFS (\dHFS{\as} and \dJES{\as}) and the model uncertainty. }
\label{tab:SysUncAs}
\end{table}
Table~\ref{tab:SysUncAs} lists the size of the most relevant contributions to 
the experimental uncertainty on the \as-value obtained. 
\kd{They are determined using linear error propagation applying an analogous formula as for the theoretical uncertainties (see equation~\ref{eq:SplitedErrorProp}).}
For \as-values determined from the absolute jet cross sections, the dominant uncertainty is the normalisation uncertainty, since it is highly correlated with the value of \asmz\ in the fit.
The errors on the fit parameters, \as\ and $\varepsilon_k$, are determined as the square 
root of the diagonal elements of the inverse of the Hessian matrix.

\subsection[Theoretical uncertainties on as]{Theoretical uncertainties on \bas}
Uncertainties on \as\ from uncertainties on the theory predictions are often determined using the 
offset method \kd{\footnote{ In this procedure, parameters are changed one at a time, the fit is repeated and the difference with respect to the central fit result is calculated.}}.
In this analysis a different approach is taken.
The theory uncertainties are determined for each source separately using linear error propagation~\cite{Britzger:13}.
Uncertainties on \as originating from a specific source of theory uncertainty are calculated as:
\begin{equation}
\left(\dabs[t]{\as}\right)^2 = 
\fC\Bigg(\left.\sum^{N_{\rm bins}}_i \frac{\partial\alpha_s}{\partial t_i}\right|_{\alpha_0}\dabs[]{t_{i}}\Bigg)^2+ \fU \sum^{N_{\rm bins}}_i \Bigg(\left. \frac{\partial\alpha_s}{\partial t_i}\right|_{\alpha_0}\dabs[]{t_{i}}\Bigg)^2 \; ,
\label{eq:SplitedErrorProp}
\end{equation}
where $t_i$ is the prediction in bin $i$, $\dabs[]{t_{i}}$ is the uncertainty of the theory in bin $i$ and $\fC$ ($\fU$) are the correlated (uncorrelated) fractions of the uncertainty source under investigation.
The partial derivatives are calculated numerically at the \as-value, $\alpha_0$, obtained from the fit. 
The uncertainties on \as obtained this way are found to be of comparable size as the uncertainties obtained with other methods, like the offset method~\cite{Aaron:10:363, Aktas:07:134}.
Because equation~\ref{eq:SplitedErrorProp} is linear, the theory uncertainties are symmetric. 

Theoretical uncertainties in the determination of \as arise from unknown higher order corrections beyond NLO, from uncertainties on the hadronisation corrections and from uncertainties on the PDFs. 
Three distinct sources of uncertainties from the PDFs are considered.
These are uncertainties due to the limited precision of the input data in the determination of the PDFs, the uncertainty of the value of \asmz, which was used for obtaining the PDFs, and procedural uncertainties in the PDF fit. 
Details for all theoretical uncertainties considered are given below.

\begin{figure}
  \centering
  \includegraphics[height=0.68\textwidth]{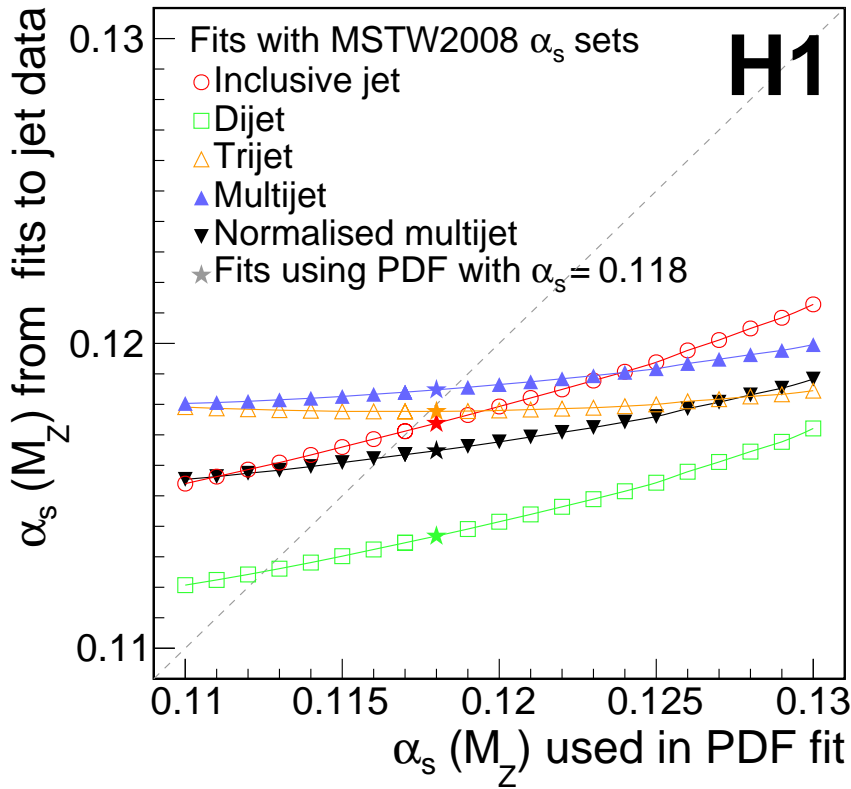}
  \caption{
     Values of \asmz\ extracted from fits of the NLO QCD predictions to the jet cross section measurements.
     Shown are the values of \asmz\ obtained with the inclusive jet, dijet and trijet data separately, and for fits either to the multijet or to the normalized multijet measurements.
     Each point stands for a value of \asmz\ obtained
     using a PDF set which has been determined assuming a fixed values of \asmz as indicated.
  }
  \label{fig:AsFromPDFas}
\end{figure}

\begin{figure}
  \begin{minipage}{0.68\linewidth}
  \centering
    \includegraphics[width=1.0\textwidth]{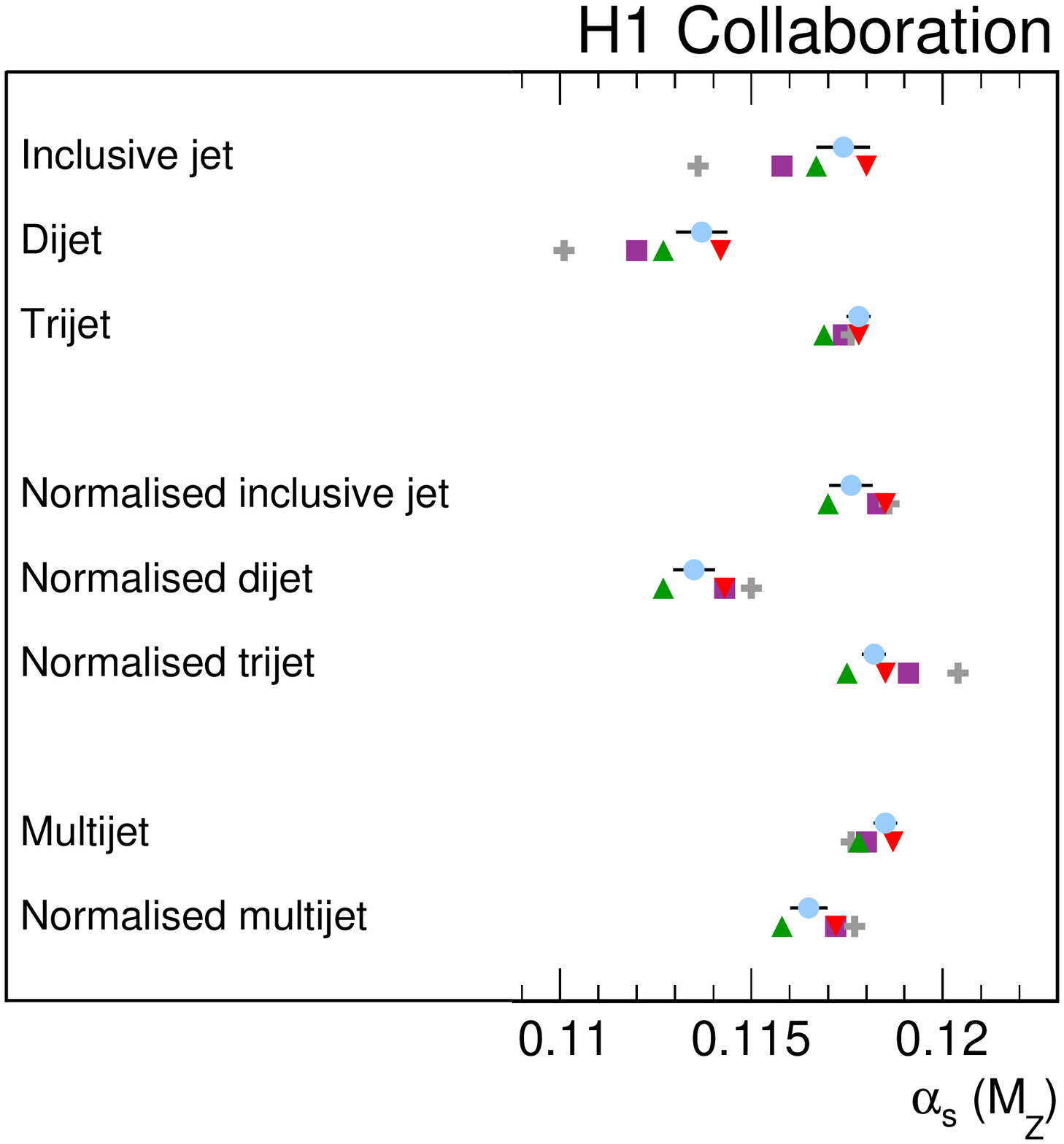}
  \end{minipage}
  \begin{minipage}{0.31\linewidth}
  \includegraphics[width=1.0\textwidth]{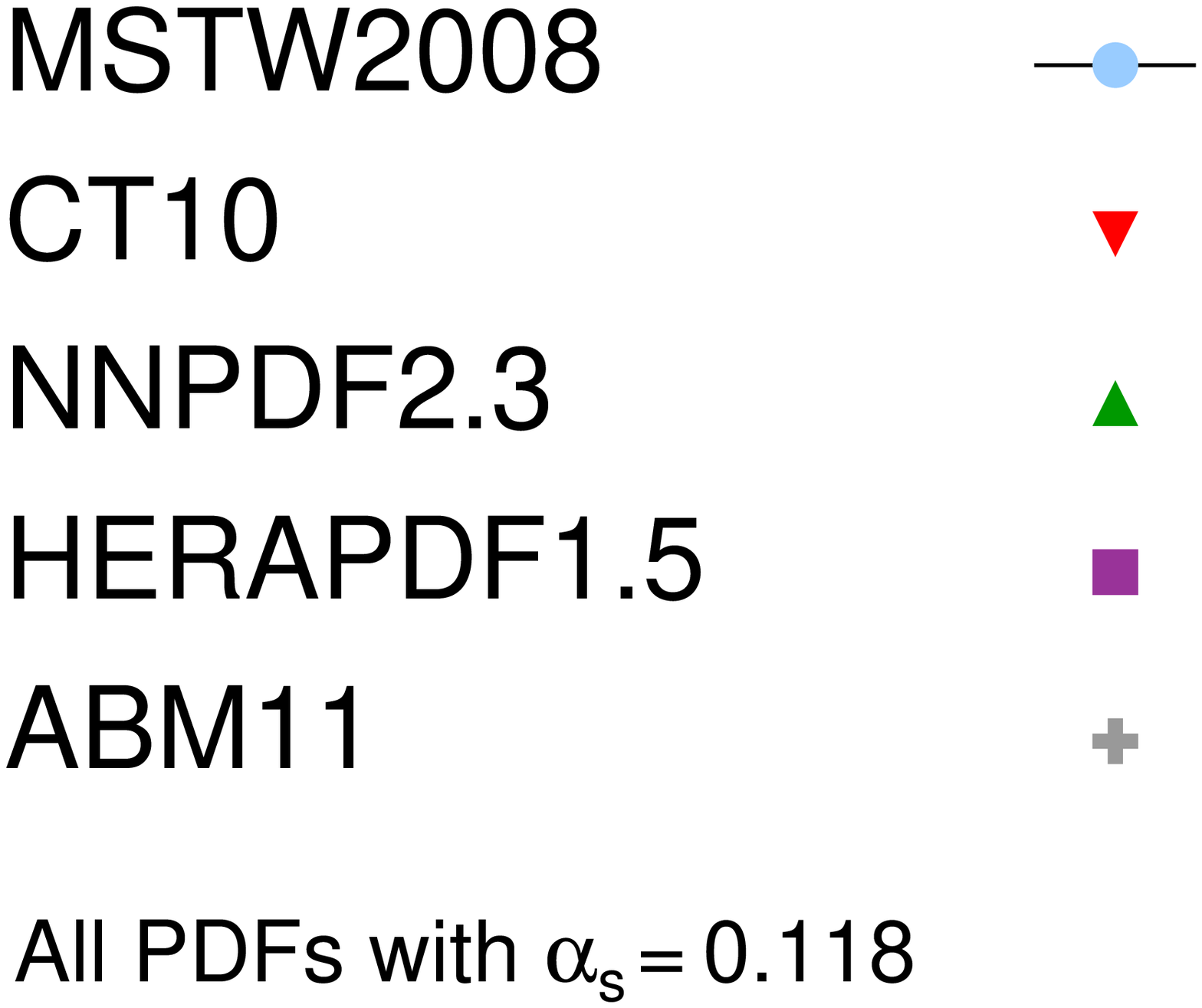}
  \vspace{5.4cm}
  \end{minipage}
  \caption{
     Values of \asmz\ extracted from fits of NLO QCD predictions to the absolute and normalised jet cross sections  using different PDF sets: MSTW2008, CT10, NNPDF2.3, HERAPDF1.5 and ABM11.
     For the MSTW2008 PDF set the PDF uncertainty on \asmz\ as determined from the MSTW2008 eigenvectors is shown as horizontal error bar.
  }
  \label{fig:AsFromPDFs}
\end{figure}

\begin{itemize}
\item \textbf{Uncertainties resulting from truncation of the perturbative series:}
The uncertainty due to missing higher orders is conventionally determined by a variation of \mur and \muf. 
In order to obtain conservative estimates from equation~\ref{eq:SplitedErrorProp},
the uncertainty from scale variations 
on the theory predictions is defined by~\cite{Soper:97}
\begin{equation}
\dabs[\mu]{t_{i}} := \max\left( \left| t_{i}(\mu=c_\mu\mu_0) - t_{i}(\mu=\mu_0) \right| \right)_{0.5\leq c_{\mu}\leq 2} \, ,
\end{equation}
using a continuous variation of the scale in the interval $0.5\leq c_\mu\leq 2$. 
The uncertainty from scale variations on \as, $\dabs[\mu]{\as}$, is then given by 
equation~\ref{eq:SplitedErrorProp} using $\dabs[\mu]{t_{i}}$.
The correlated and uncorrelated fractions of $\dabs[\mu]{t_i}$ are set to $0.5$ each. 
In case of normalised jet cross sections, the uncertainty $\dabs[\mu]{t_{i}}$ is determined 
by a simultaneous variation of the scales in the numerator and denominator.
The scale dependence of the inclusive NC DIS calculation is small compared to the scale dependence of the jet cross sections, since it is in LO of $\Ord(\as^0(\mur))$.
\kd{Changing the renormalisation scale for the jet cross sections to $\mu_r = Q$ or $\mu_r = P_T$ results in changes in \asmz which are typically of similar size as the experimental uncertainty and always smaller than the renormalisation scale uncertainty.}
The uncertainty from the variation of the renormalisation scale is by far the largest uncertainty of all theoretical and experimental uncertainties considered.
Calculations beyond NLO are therefore mandatory for a more precise determination of \as from jet cross sections in DIS.
\item \textbf{Hadronisation uncertainties:}
The uncertainties of the hadronisation correction $\dabs[had]{t}$ on the theory predictions 
are obtained using half the difference of the hadronisation corrections calculated with the Lund string model and the cluster fragmentation model (see section~\ref{sect:xsec_theo_unc}).
The resulting uncertainties on \as are determined using the linear error propagation described above. 
The uncertainty is taken to be half correlated and half uncorrelated.
%
%
\item \textbf{PDF uncertainty:}
  %
PDF uncertainties on \as, $\dabs[PDF]{\as}$ are estimated by propagating 
the uncertainty eigenvectors of the MSTW2008 PDF set. 
Details are described in~\cite{Britzger:13}.
\item \textbf{Uncertainty due to the limited precision of \basmz\ in the PDF fit:}
The PDFs depend on the \asmz\ value used for their determination.
This leads to an additional uncertainty on the PDFs and thus to an additional uncertainty on the \as-value extracted from the jet cross sections.
This uncertainty, $\dabs[PDF(\as)]{\as}$, is conventionally defined as a variation of 
$\pm 0.002$ around the nominal value of $\asmz=0.118$ (see e.g.~\cite{Abramowicz:1900rp}).
For the full range of available MSTW2008 PDF sets with different fixed values of \asmz, the
resulting values of \as\ from fits to jet data are displayed in \fig~\ref{fig:AsFromPDFas}. 
While some dependence on the value of \asmz\ used in the PDF fit is observed for the \as\ 
values obtained from inclusive jet and dijet cross sections, the \as-value obtained from the
trijet cross sections shows only a very weak dependence on \asmz.
This is due to the high sensitivity of the trijet cross sections to \as, where the calculation is of $\Ord(\as^2)$ already at LO. 
Consequently, due to the inclusion of the trijet cross sections, the dependence on \asmz\ as 
used in the PDF fit is reduced for the fit to the multijet dataset. 
  %
  %
  %
  %
\item \textbf{Procedural and theory uncertainties on the PDFs:}
In order to estimate the uncertainty due to the procedure used to extract PDFs, all \as\ fits are repeated using PDF sets from different groups.
The \as-values obtained are displayed in \fig~\ref{fig:AsFromPDFs} and are listed in 
\tab~\ref{tab:AsPDFGroups}.
  %
Half the difference between the \as-values obtained using the NNPDF2.3 and CT10 PDF sets is assigned as PDF set uncertainty, $\dabs[PDFset]{\as}$.
The values for $\dabs[PDFset]{\as}$ are in the range from $0.0007$ to $0.0012$. 
%
\end{itemize}

\subsection[Results of the alphas-fit]{Results of the \bas-fit}
The strong coupling constant is determined from each of the jet measurements, i.e.\ from the absolute and normalised inclusive jet, dijet and trijet cross sections as a function of \Qsq\ and \Pt, as well as from the three absolute and three normalised jet cross sections simultaneously. 
The statistical correlations (\tab{}s~\ref{tab:CorrInclIncl}--\ref{tab:CorrDijetTrijet}) are taken into account.
The \as-values obtained from measurements using the \kt\ jet algorithm are 
compared to those using the \antikt\ jet algorithm with the corresponding NLO calculations. 

The NLO correction to the LO cross section is below $\unit[50]{\%}$ for all of the data points and below $\unit[30]{\%}$ for $\unit[64]{\%}$ of the data points. It is assumed that the perturbative series is converging sufficiently fast, such that NLO calculations are applicable, and that the uncertainty from the variation of the renormalisation and factorisation scales accounts for the not yet calculated contributions beyond NLO.

The \as results, determined from fits to the individual absolute and normalised jet cross sections as well as to the absolute and normalised multijet cross sections using either the \kt\ or the \antikt\ jet algorithms, are summarised in \tab~\ref{tab:AsValuesAll}, together with the split-up of the contributions to the theoretical uncertainty. The largest contribution is due to the variation of the renormalisation scale.
The fits yield, for the \kt-jets taken as an example, the following values of $\chisq/\ndf$ for the absolute (normalised) inclusive jet, dijet and trijet measurements, $24.8 / 23 \, (26.8 / 23)$, $25.1 / 23 \, (31.0 / 23)$ and $13.6 / 15\, (11.8 / 15)$, respectively. 
For the absolute (normalised) multijet measurements the value of $75.2 / 63 \, (89.8 / 63)$ is obtained.
%
%
Note that the theoretical uncertainties on \as are not considered in the calculation of $\chisq/\ndf$.
The fact that $\chisq/\ndf$ degrades as more data are included (multijets as compared to individual data sets) or as the experimental precision is improved (normalised as compared to absolute cross sections) indicates a problem with the theory, possibly related to higher order corrections.
Similarly, the fact that \as extracted from the dijet data is below the values obtained from inclusive jet or trijet data may be due to unknown higher order effects.

All \as-values extracted are compatible within the theoretical uncertainty obtained by the scale variations. 
The values of \as\ extracted using \kt\ or \antikt\-jet cross sections are quite consistent.
Among the absolute cross sections, not considering the multijet fit, the trijet data yield 
values of \as\ with the highest experimental precision, because the LO trijet cross section is proportional to $\as^2$, whereas the inclusive or dijet cross section at LO are proportional to \as\ only.

The best experimental precision on \as\ is achieved for normalised jet cross sections, due to the full cancellation of all normalisation uncertainties, which are highly correlated with the value of \asmz\ in the fit.
A breakdown of the individual uncertainties contributing to the total experimental uncertainty is given in \tab~\ref{tab:SysUncAs}.
For the \as\ extraction using absolute cross sections, the normalisation uncertainty is the dominant uncertainty.
The jet energy scale, the remaining cluster energy scale and the model uncertainty contribute with similar size to the experimental uncertainty.
All other experimental uncertainties are negligible with respect to these uncertainties.
The uncertainties from scale variations are somewhat reduced for normalised jet cross sections, due to the simultaneous variation of the scales in the numerator and the denominator.
%
%
%
The uncertainties from PDFs are of similar size when comparing absolute and normalised jet cross sections. 
The residual differences are well understood~\cite{Britzger:13}.

The absolute and normalised dijet cross sections yield a significantly smaller value of \as\ than the corresponding values from inclusive jet cross sections, considering the experimental uncertainty only.
This is attributed to missing higher order contributions in the calculations, which may be different in the inclusive jet phase space region which is not part of the dijet phase space.
These are, for instance, the dijet topologies with $\Mjj<\unit[16]{\GeV}$, or events where one jet is outside the acceptance in $\etalab$. 
In order to test the influence of the phase space, an inclusive jet measurement is performed in the phase space of the dijet measurement, i.e.\ with the requirement of two jets, 
$\Mjj>\unit[16]{\GeV}$ and $7<\meanptdi<\unit[50]{\GeV}$. 
When using the identical scale $\mu_r^2 = \Qsq$ for the \as-fit to this inclusive jet and the dijet measurement, the difference in \as is only $0.0003$. With the nominal scales, 
$\mur^2 = (\Qsq+(\ptjet)^2)/2$ for this inclusive jet measurement and 
$\mur^2 = (\Qsq+(\meanptdi)^2)/2$ for the dijet measurement, the difference in \as increases to $0.0007$.  Since the \as values obtained are rather similar, this lends some support to the argument given above.
%

The best experimental precision on \as\ is obtained from a fit to normalised multijet cross sections, yielding: 
\begin{flalign}
\asmz|_{\kt} & = 0.1165 \;\,(8)_{\rm exp}\;\,(5)_{\rm PDF}\;\,(7)_{\rm PDFset}\;\,(3)_{\rm PDF(\as)}\;\,(8)_{\rm had}\;\,(36)_{\mur}\;\,(5)_{\muf} \\
& = 0.1165 \;\,(8)_{\rm exp} \;\,(38)_{\rm pdf,theo} ~.
\nonumber
\end{flalign}
Here, we quote the value obtained for jets reconstructed with the \kt\ algorithm. As can be seen in \tab~\ref{tab:AsValuesAll}, it is fully consistent with the \as-value found for jets using the \antikt\ algorithm.

The uncertainties on \asmz\ are dominated by theory uncertainties from missing higher orders and allow a determination of \asmz\ with a precision of $\unit[3.4]{\%}$ only, while an experimental precision of $\unit[0.7]{\%}$ is reached.
Complete next-to-next-to-leading order calculations of jet production in DIS are required to reduce this mismatch in precision between experiment and theory.

\begin{figure}
  \centering
  \includegraphics[width=0.68\textwidth]{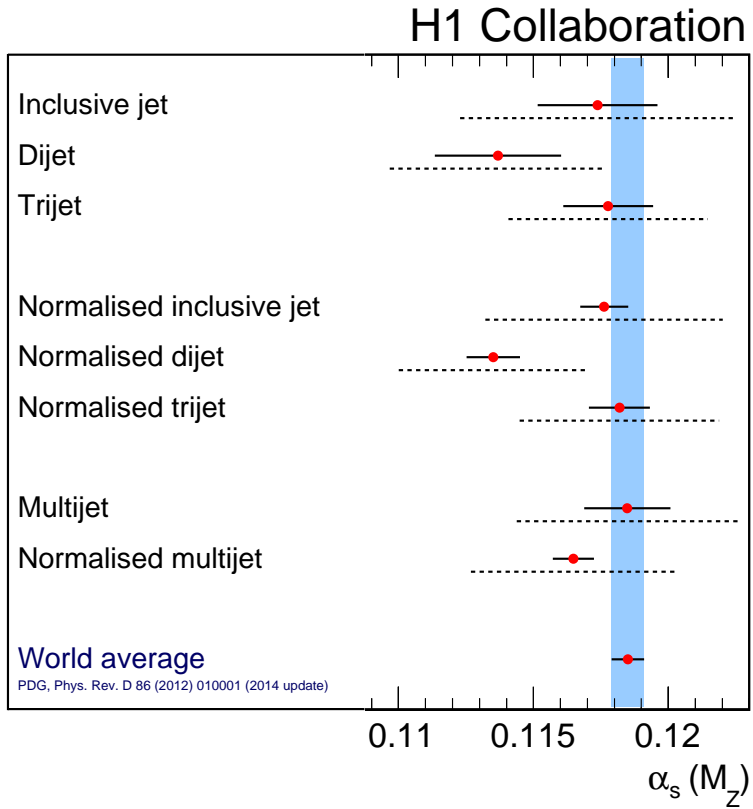}
  \caption{
     Comparison of \as-values extracted from different jet cross section measurements, separately and simultaneously, to the world average value of \asmz. The full line indicates the experimental uncertainty and the dashed line the theoretical uncertainty. The band indicates the uncertainty of the world average value of \asmz.
  }
  \label{fig:AsComparisonPaper}
\end{figure}

The \as-values determined are compatible with the world average~\cite{Beringer:1900zz,Bethke:2012jm} 
value of $\asmz=0.1185\;(6)$ within the experimental and particularly the theoretical uncertainties. 
The \as-values extracted from the \kt-jet cross sections are compared to the world average value in \fig~\ref{fig:AsComparisonPaper}.

The value of \asmz\ with the highest overall precision is obtained from fits to a reduced phase space region, in which the dominant theoretical uncertainty, 
estimated from variations of the renormalisation and factorisation scales, 
are reduced at the expense of an increased experimental uncertainty.
For photon virtualities of $\Qsq>\unit[400]{\GeVsq}$ a total uncertainty of $\unit[2.9]{\%}$  
on the \as-value is obtained, with a value of 
$$\asmz|_\kt= 0.1160\;\,(11)_{\rm exp}\;\,(32)_{\rm pdf,theo}\; .$$ 
The value of \asmz\ is the most precise value ever derived at NLO from jet data recorded in a single experiment.

\begin{figure}
  \centering
  \includegraphics[width=0.99\textwidth]{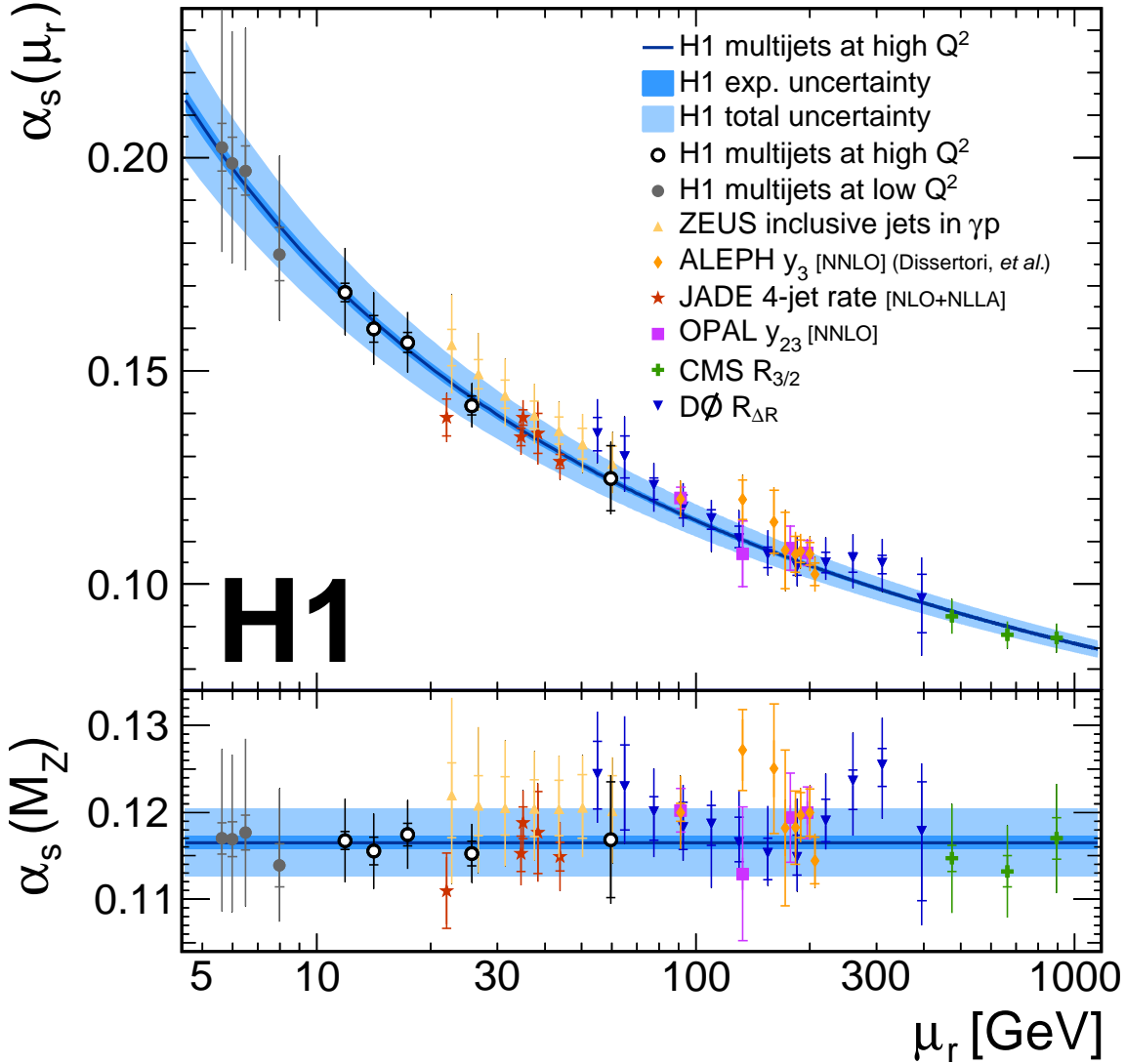}
  \caption{
     The upper panel shows the values of the strong coupling \asmur\ as determined from the normalized multijet measurement (open dots) at different scales \mur.
     \kd{The inner error bars indicate the experimental uncertainty, while the full error bars indicate the total uncertainty, including the experimental and theoretical contributions.}
     The solid line shows the NLO QCD prediction calculated using the renormalisation group equation with $\asmz=0.1165$ as determined from the simultaneous fit to all normalized multijet measurements.
     The dark shaded band around this line indicates the experimental uncertainty on \asmur, while the light shaded band shows the total uncertainty.
     \kd{
     Also shown are the values of \as\ 
      from multijet measurement at low values of \Qsq\ by H1 (circles),
      from inclusive jet measurements in photoproduction by the ZEUS experiment (upper triangles),
      from the 3-jet rate $y_3$ in a fit of NNLO calculations to ALEPH data taken at LEP (diamonds), 
      from the 4-jet rate measured by the JADE experiment at PETRA (stars),
      from the jet transition value $y_{23}$ measured by OPAL at LEP (squares), 
      from the ratio of trijet to dijet cross sections $R_{3/2}$ as measured by the CMS experiment at the LHC (crosses),
      and 
      from jet angular correlations $R_{\Delta R}$ by the D0 experiment at the Tevatron (lower triangles).
     }
      In the lower panel the equivalent values of \asmz\ for all measurements are shown.
  }
  \label{fig:AsRunning}
\end{figure}

The running of \asmur is determined from five fits using the normalised multijet cross sections, each based on a set of measurements with comparable values of the renormalisation scale \mur. 
The values of \asmz\ 
\kd{and \asmur} 
extracted are listed in \tab~\ref{tab:AsMu} together with the cross section weighted average values of \mur.
\kd{The values of \asmur\ are obtained from the values of \asmz\ by applying the 2-loop solution for the evolution equation of \asmur.}
The values of \asmz and \asmur obtained from the \kt-jets are displayed in 
\fig~\ref{fig:AsRunning} together with results from other \kd{recent and precise} jet data\footnote{The values \asmur\ given 
in~\kd{\cite{Dissertori:2007xa, Schieck:2006tc, OPAL:2011aa, Abazov:2012lua} are evolved to \asmz for this comparison, whereas the values of both \asmur\ and \asmz\ are given in~\cite{Chatrchyan:2013txa}.}}~\kd{\cite{Aaron:10:1,Dissertori:2007xa, Schieck:2006tc, OPAL:2011aa, Chatrchyan:2013txa, Abazov:2012lua}}.
Within the small experimental uncertainties the values of \asmz of the present analysis are consistent and independent of \mur. Good agreement is found with H1 data~\cite{Aaron:10:1} at low scales and other jet data~\kd{\cite{Dissertori:2007xa,Schieck:2006tc,OPAL:2011aa, Chatrchyan:2013txa, Abazov:2012lua} at medium} and high scales.
The prediction for the running of \asmur using $\asmz = 0.1165 \;\,(8)_{\rm exp}\;\,(38)_{\rm pdf,theo}$, as extracted from the normalised multijet cross sections, is also shown in \fig~\ref{fig:AsRunning}, together with its experimental and total uncertainty. The prediction is in good agreement with the measured values of \asmur.


\section{Summary}

Measurements of inclusive jet, dijet and trijet cross sections in the Breit frame in deep-inelastic electron-proton scattering in the kinematical range $150 < \Qsq < \unit[\fifteen]{\GeVsq}$ and $0.2 < y < 0.7$ are presented, using H1 data corresponding to an integrated luminosity of $\unit[351]{\invpb}$. 
The measurements consist of absolute jet cross sections as well as jet cross sections normalised to the neutral current DIS cross sections. 
Jets are determined using the \kt and the \antikt jet algorithm.
Compared to previous jet measurements by H1, this analysis makes use of an improved electron calibration and further development of the energy flow algorithm, which combines information from tracking and calorimetric measurements, by including a better separation of electromagnetic and hadronic components of showers. The sum of these improvements, together with a new method to calibrate the hadronic final state, reduces the hadronic energy scale uncertainty by a factor of two to $\unit[1]{\%}$ for \ptlab down to $\unit[5]{GeV}$.


The jet cross section measurements are performed using a regularised unfolding procedure to correct the neutral current DIS, the inclusive jet, the dijet and the trijet measurements simultaneously for detector effects. 
It considers up to seven different observables per measurement for the description
of kinematical migrations due to the limited detector resolution.
This approach provides a reliable treatment of migration effects and enables the determination of the statistical correlations between the three jet measurements and the neutral current DIS measurement.

Theoretical QCD calculations at NLO, corrected for hadronisation and electroweak effects, provide a good description of the measured double-differential jet cross sections as a function of the exchanged boson virtuality \Qsq, the jet transverse momentum \ptjet, the mean transverse momentum \meanptdi and \meanpttri in case of dijets and trijets, as well as of the longitudinal proton momentum fractions \xidi and \xitri. 
In general, the precision of the data is considerably better than that of the NLO calculations.

The measurements of the inclusive, the dijet and the trijet cross section 
are used 
separately and also simultaneously 
to extract values for the strong coupling constant \asmz. 
The best experimental precision of $\unit[0.7]{\%}$ is obtained when using the normalised 
multijet cross sections. 
The simultaneous extraction of the strong coupling constant \asmz\ from the normalised
inclusive jet, the dijet and the trijet samples using the \kt jet algorithm yields:
\begin{flalign}
\asmz|_{\kt} &= 0.1165 \;\,(8)_{\rm exp}\;\,(5)_{\rm PDF}\;\,(7)_{\rm PDFset}\;\,(3)_{\rm PDF(\as)}\;\,(8)_{\rm had}\;\,(36)_{\mur}\;\,(5)_{\muf} \\
&= 0.1165 \;\,(8)_{\rm exp} \;\,(38)_{\rm pdf,theo} ~.
\nonumber
\end{flalign}
A very similar result is obtained when using the \antikt\ jet algorithm.
The values and uncertainties of \asmz obtained using absolute jet cross sections are consistent with the results from the corresponding normalised jet cross sections, albeit with larger experimental uncertainties.
A tension is observed between the value of \asmz\ extracted from the dijet sample
and the similar values obtained from the inclusive jet and the trijet samples. 
This may be caused by missing higher orders in the calculations, which can be different in the inclusive jet phase space region which is not part of the dijet phase space.

When restricting the measurement to regions of higher \Qsq, where the scale uncertainties are reduced, the smallest total uncertainty on the extracted  \asmz\ is found for $\Qsq>\unit[400]{\GeVsq}$.
For this region the loss in experimental precision is compensated by the reduced theory uncertainty, yielding
\begin{equation}
\asmz|_\kt = 0.1160 \;\, (11)_{\rm exp} \;\, (32)_{\rm pdf,theo}~.
\nonumber
\end{equation}
%
%
The extracted \asmz-values are compatible within uncertainties with the world average value of $\asmz=0.1185\;(6)$ and with \as-values from other jet data.
Calculations in NNLO are needed to benefit from the superior experimental precision of the DIS jet data.

The running of \asmur, determined from the normalised multijet cross sections, is shown to be consistent with the expectation from the renormalisation group equation and with values of \asmur from other jet measurements.

\section*{Acknowledgements}

We are grateful to the HERA machine group whose outstanding efforts have made this experiment possible. We thank the engineers and technicians for their work in constructing and maintaining the H1 detector, our funding agencies for financial support, the DESY technical staff for continual assistance and the DESY directorate for support and for the hospitality which they extend to the non DESY members of the collaboration. We would like to give credit to all partners contributing to the EGI computing infrastructure for their support for the H1 Collaboration.
Furthermore, we thank Zolt{\'a}n Nagy and Stefan H{\"o}che for fruitful discussions and for help with their computer programs.

\providecommand{\href}[2]{#2}\begingroup\raggedright\endgroup

\clearpage


\clearpage


\begin{table}[htop]
\center
\footnotesize
\begin{tabular}{cr@{$\;\, \Qsq \; $} l}
\multicolumn{3}{c}{\textbf{Bin labels \bQsq}} \\
\hline
Bin number $q$ &
\multicolumn{2}{c}{\Qsq\ range in \GeVsq} \\
\hline
1             &  $150 \le $ & $ < 200$ \\
2             &  $200 \le $ & $ < 270$ \\
3             &  $270 \le $ & $ < 400$ \\
4             &  $400 \le $ & $ < 700$ \\
5             &  $700 \le $ & $ < 5000$ \\
6             &  $5000 \le $ & $ < \fifteen$\\
\hline
   \multicolumn{3}{c}{} \\
   \multicolumn{3}{c}{} \\
\end{tabular}
\hspace{0.05\linewidth}
\begin{tabular}{cr@{$\;\le\;$}c@{$<\;$}l}
\multicolumn{4}{c}{\textbf{Bin labels \bpt}} \\
\hline
 Label  &  \multicolumn{3}{c}{$\pt$ range in GeV} \\
\hline
 $\alpha$         &    $7  $ &\pt & $ 11$ \\
 $\beta$          &    $11 $ &\pt & $ 18$ \\
 $\gamma$         &    $18 $ &\pt & $  30$ \\
 $\delta$         &    $30 $ &\pt & $ 50$ \\
\hline
\multicolumn{4}{c}{} \\
\multicolumn{4}{c}{} \\

\multicolumn{4}{c}{\textbf{Bin labels \bxidi\ dijet}} \\
\hline
  Label  &  \multicolumn{3}{c}{$\xidi$ range} \\
\hline
 a         &    $0.006  $ & \xidi & $ 0.02~$ \\
 b         &    $0.02~  $ & \xidi & $ 0.04~$ \\
 c         &    $0.04~  $ & \xidi & $ 0.08~$ \\
 d         &    $0.08~  $ & \xidi & $ 0.316$ \\
\hline
\multicolumn{4}{c}{} \\
\multicolumn{4}{c}{} \\

\multicolumn{4}{c}{\textbf{Bin labels \bxitri\ trijet}} \\
\hline
  Label  &  \multicolumn{3}{c}{$\xitri$ range} \\
\hline
 A         &    $0.01 $ & \xitri & $ 0.04$ \\
 B         &    $0.04 $ & \xitri & $ 0.08$ \\
 C         &    $0.08 $ & \xitri & $ 0.5~$ \\
\hline
\end{tabular}
\caption{Bin numbering scheme for \Qsq, \pt, and $\xi$-bins. Bins of the double-differential
measurements are for instance referred to as $3\gamma$ for the bin in the range $270 < \Qsq < \unit[400]{\GeVsq}$
and $18<\ptjet<\unit[30]{\GeV}$.
\label{tab:BinNumbering}
}
\end{table}


\begin{table}
\vspace*{-0.5cm}
\footnotesize
\center
\begin{tabular}{c c r r r c c c c c | c c c}
\multicolumn{13}{c}{ \textbf{Inclusive jet cross sections in bins of $\bQsq$ and $\bptjet$ using the \bkt\ jet algorithm}} \\
\hline
Bin & \multicolumn{1}{c}{\CS} & \multicolumn{1}{c}{\DStat{\csdsub}} & \multicolumn{1}{c}{\DSys{\csdsub}} & \multicolumn{1}{c}{\DMod{\csdsub}} & \DJES{\csdsub} & \DHFS{\csdsub} & \DEe{\csdsub} & \DThe{\csdsub} & \DID{\csdsub} & \cHad & \DHad & \cEW \\
label & \multicolumn{1}{c}{[pb]} & \multicolumn{1}{c}{[\%]} & \multicolumn{1}{c}{[\%]} & \multicolumn{1}{c}{[\%]} & [\%] & [\%] & [\%] & [\%] & [\%] &  & [\%] &  \\
\hline
1$\alpha$ & $7.06\trenn 10^{1}$ & 2.7~ & 2.9~ & $+1.0$~ & $^{+0.9~}_{~-1.1}$ & $^{+0.9~}_{~-1.0}$ & $^{-0.4~}_{~+0.3}$ & $^{-0.4~}_{~+0.3}$ & $^{+0.5~}_{~-0.5}$ & 0.93 & 2.2 & 1.00 \\
1$\beta$ & $3.10\trenn 10^{1}$ & 4.1~ & 4.4~ & $+2.8$~ & $^{+2.4~}_{~-2.5}$ & $^{+0.6~}_{~-0.5}$ & $^{-0.7~}_{~+0.5}$ & $^{-0.3~}_{~+0.2}$ & $^{+0.5~}_{~-0.5}$ & 0.97 & 1.7 & 1.00 \\
1$\gamma$ & $8.07\trenn 10^{0}$ & 6.4~ & 5.3~ & $+3.5$~ & $^{+3.4~}_{~-3.4}$ & $^{+0.3~}_{~-0.1}$ & $^{-0.4~}_{~+0.5}$ & $^{-0.1~}_{~+0.1}$ & $^{+0.5~}_{~-0.5}$ & 0.96 & 1.1 & 1.00 \\
1$\delta$ & $9.18\trenn 10^{-1}$ & 15.3~ & 12.9~ & $+11.7$~ & $^{+4.9~}_{~-5.3}$ & $^{+0.2~}_{~-0.1}$ & $^{-0.1~}_{~-0.5}$ & $^{-0.2~}_{~-0.1}$ & $^{+0.5~}_{~-0.5}$ & 0.95 & 0.7 & 1.00 \\
2$\alpha$ & $5.48\trenn 10^{1}$ & 3.0~ & 2.9~ & $-0.6$~ & $^{+0.9~}_{~-1.0}$ & $^{+1.2~}_{~-1.0}$ & $^{-0.6~}_{~+0.9}$ & $^{-0.3~}_{~+0.4}$ & $^{+0.5~}_{~-0.5}$ & 0.93 & 2.1 & 1.00 \\
2$\beta$ & $2.68\trenn 10^{1}$ & 4.1~ & 4.8~ & $+3.4$~ & $^{+2.4~}_{~-2.4}$ & $^{+0.4~}_{~-0.4}$ & $^{-0.6~}_{~+0.6}$ & $^{-0.3~}_{~+0.3}$ & $^{+0.5~}_{~-0.5}$ & 0.97 & 1.7 & 1.00 \\
2$\gamma$ & $7.01\trenn 10^{0}$ & 6.6~ & 6.4~ & $+4.8$~ & $^{+3.7~}_{~-3.4}$ & $^{+0.2~}_{~-0.2}$ & $^{-0.6~}_{~+0.5}$ & $^{-0.4~}_{~+0.3}$ & $^{+0.5~}_{~-0.5}$ & 0.97 & 1.3 & 1.00 \\
2$\delta$ & $8.52\trenn 10^{-1}$ & 15.2~ & 7.4~ & $+4.6$~ & $^{+5.7~}_{~-4.8}$ & $^{-0.2~}_{~-0.1}$ & $^{+0.0~}_{~+0.1}$ & $^{-0.3~}_{~+0.3}$ & $^{+0.5~}_{~-0.5}$ & 0.96 & 1.2 & 1.00 \\
3$\alpha$ & $5.22\trenn 10^{1}$ & 3.0~ & 3.2~ & $+1.5$~ & $^{+0.9~}_{~-1.0}$ & $^{+1.0~}_{~-1.0}$ & $^{-1.0~}_{~+0.7}$ & $^{-0.3~}_{~+0.3}$ & $^{+0.5~}_{~-0.5}$ & 0.93 & 1.5 & 1.00 \\
3$\beta$ & $2.78\trenn 10^{1}$ & 4.0~ & 4.5~ & $+3.1$~ & $^{+2.3~}_{~-2.2}$ & $^{+0.4~}_{~-0.4}$ & $^{-0.7~}_{~+0.9}$ & $^{-0.2~}_{~+0.3}$ & $^{+0.4~}_{~-0.4}$ & 0.97 & 1.1 & 1.00 \\
3$\gamma$ & $6.99\trenn 10^{0}$ & 6.8~ & 4.7~ & $+1.9$~ & $^{+3.5~}_{~-3.7}$ & $^{+0.2~}_{~-0.1}$ & $^{-1.0~}_{~+0.6}$ & $^{-0.0~}_{~-0.3}$ & $^{+0.4~}_{~-0.4}$ & 0.97 & 0.9 & 1.00 \\
3$\delta$ & $8.69\trenn 10^{-1}$ & 15.1~ & 6.7~ & $-3.0$~ & $^{+5.4~}_{~-5.7}$ & $^{-0.0~}_{~-0.2}$ & $^{+0.8~}_{~-0.3}$ & $^{-0.1~}_{~+0.4}$ & $^{+0.4~}_{~-0.4}$ & 0.95 & 0.5 & 1.00 \\
4$\alpha$ & $4.88\trenn 10^{1}$ & 3.2~ & 3.3~ & $+1.5$~ & $^{+1.2~}_{~-1.4}$ & $^{+0.7~}_{~-0.7}$ & $^{-1.1~}_{~+1.2}$ & $^{-0.2~}_{~+0.2}$ & $^{+0.4~}_{~-0.4}$ & 0.93 & 1.2 & 1.00 \\
4$\beta$ & $2.69\trenn 10^{1}$ & 4.1~ & 3.3~ & $+1.2$~ & $^{+2.0~}_{~-2.0}$ & $^{+0.4~}_{~-0.4}$ & $^{-0.7~}_{~+0.7}$ & $^{-0.1~}_{~+0.1}$ & $^{+0.4~}_{~-0.4}$ & 0.97 & 1.0 & 1.00 \\
4$\gamma$ & $7.95\trenn 10^{0}$ & 6.1~ & 5.6~ & $+3.5$~ & $^{+3.8~}_{~-3.6}$ & $^{+0.2~}_{~-0.4}$ & $^{-0.8~}_{~+0.8}$ & $^{-0.1~}_{~+0.1}$ & $^{+0.3~}_{~-0.3}$ & 0.97 & 0.5 & 1.00 \\
4$\delta$ & $8.57\trenn 10^{-1}$ & 16.5~ & 10.8~ & $-8.9$~ & $^{+5.7~}_{~-5.5}$ & $^{-0.1~}_{~-0.1}$ & $^{-0.1~}_{~-0.1}$ & $^{+0.1~}_{~-0.1}$ & $^{+0.2~}_{~-0.2}$ & 0.96 & 0.4 & 1.00 \\
5$\alpha$ & $4.33\trenn 10^{1}$ & 3.5~ & 3.5~ & $+2.2$~ & $^{+1.0~}_{~-1.2}$ & $^{+0.5~}_{~-0.4}$ & $^{-0.4~}_{~+0.5}$ & $^{-0.5~}_{~+0.5}$ & $^{+1.1~}_{~-1.1}$ & 0.92 & 0.9 & 1.02 \\
5$\beta$ & $2.85\trenn 10^{1}$ & 4.0~ & 3.3~ & $+1.4$~ & $^{+1.6~}_{~-1.5}$ & $^{+0.1~}_{~-0.1}$ & $^{-0.5~}_{~+0.6}$ & $^{-0.6~}_{~+0.6}$ & $^{+1.1~}_{~-1.1}$ & 0.97 & 0.5 & 1.02 \\
5$\gamma$ & $1.07\trenn 10^{1}$ & 4.9~ & 4.6~ & $+2.7$~ & $^{+2.7~}_{~-2.8}$ & $^{+0.1~}_{~-0.1}$ & $^{-0.5~}_{~+0.6}$ & $^{-0.4~}_{~+0.4}$ & $^{+1.1~}_{~-1.1}$ & 0.97 & 0.4 & 1.03 \\
5$\delta$ & $2.04\trenn 10^{0}$ & 8.5~ & 5.7~ & $+2.1$~ & $^{+4.8~}_{~-4.5}$ & $^{+0.1~}_{~-0.0}$ & $^{-0.3~}_{~+0.3}$ & $^{-0.2~}_{~+0.2}$ & $^{+1.0~}_{~-1.0}$ & 0.96 & 0.3 & 1.02 \\
6$\alpha$ & $2.60\trenn 10^{0}$ & 14.7~ & 4.4~ & $-3.0$~ & $^{+0.8~}_{~-0.9}$ & $^{+0.3~}_{~-0.5}$ & $^{-0.6~}_{~-1.6}$ & $^{-0.3~}_{~+0.6}$ & $^{+1.9~}_{~-1.9}$ & 0.91 & 0.6 & 1.11 \\
6$\beta$ & $1.74\trenn 10^{0}$ & 16.4~ & 3.5~ & $+1.1$~ & $^{+1.6~}_{~-1.2}$ & $^{+0.1~}_{~+0.0}$ & $^{+0.2~}_{~+1.2}$ & $^{-0.4~}_{~+0.9}$ & $^{+1.8~}_{~-1.8}$ & 0.96 & 0.6 & 1.11 \\
6$\gamma$ & $6.71\trenn 10^{-1}$ & 21.6~ & 13.4~ & $-12.9$~ & $^{+2.2~}_{~-2.0}$ & $^{+0.2~}_{~-0.3}$ & $^{-0.2~}_{~-0.0}$ & $^{-0.5~}_{~+0.6}$ & $^{+1.8~}_{~-1.8}$ & 0.99 & 1.1 & 1.11 \\
6$\delta$ & $3.09\trenn 10^{-1}$ & 19.7~ & 20.0~ & $-19.5$~ & $^{+2.9~}_{~-2.8}$ & $^{+0.1~}_{~+0.0}$ & $^{+0.3~}_{~-0.9}$ & $^{+0.0~}_{~+0.1}$ & $^{+1.8~}_{~-1.8}$ & 0.98 & 0.8 & 1.11 \\
\hline
\end{tabular}
\caption{Double-differential inclusive jet cross sections measured as a function of \Qsq\ and \ptjet using the \kt\ jet algorithm. The bin labels are defined in \tab~\ref{tab:BinNumbering}. The data points are statistically correlated, and the bin-to-bin correlations are given in the correlation matrix in \tab~\ref{tab:CorrInclIncl}. The correlation with the dijet measurements as a function of \meanptdi\ and $\xidi$ are given in \tab{}s~\ref{tab:CorrInclDijet} and~\ref{tab:CorrInclDijetXi}, respectively. The correlations with the trijet measurements as a function of \meanpttri\ and $\xitri$ are shown in \tab{}s~\ref{tab:CorrInclTrijet} and~\ref{tab:CorrInclTrijetXi}, respectively. The experimental uncertaintes quoted are defined in section~\ref{sec:exp_unc}.
The total systematic uncertainty, \DSys{\csdsub}, sums all systematic uncertainties in quadrature, including the uncertainty due to the LAr noise of $\DLAr{\csdsub} = \unit[0.5]{\%}$ and the total normalisation uncertainty of $\DNorm{\csdsub} =\unit[2.9]{\%}$.
The contributions to the correlated systematic uncertainty from a positive variation of one standard deviation of the model variation (\DMod{\csdsub}), of the jet energy scale (\DJES{\csdsub}), of the remaining cluster energy scale (\DHFS{\csdsub}), of the scattered electron energy (\DEe{\csdsub}), of the polar electron angle (\DThe{\csdsub}) and of the Electron ID (\DID{}) are also given. 
In case of asymmetric uncertainties, the effect due to the positive variation of the underlying error source is given by the upper value for the corresponding table entry.
The correction factors on the theoretical cross sections $\cHad$ and $\cEW$ are listed in the rightmost columns together with the uncertainties $\DHad{}$.
}
\label{tab:IncJet}
\end{table}


\clearpage

\begin{table}
\footnotesize
\center
\begin{tabular}{c c r r r c c c c c | c c c}
\multicolumn{13}{c}{ \textbf{Dijet cross sections in bins of $\bQsq$ and $\bmeanptdi$ using the \bkt\ jet algorithm}} \\
\hline
Bin & \multicolumn{1}{c}{\CS} & \multicolumn{1}{c}{\DStat{\csdsub}} & \multicolumn{1}{c}{\DSys{\csdsub}} & \multicolumn{1}{c}{\DMod{\csdsub}} & \DJES{\csdsub} & \DHFS{\csdsub} & \DEe{\csdsub} & \DThe{\csdsub} & \DID{\csdsub} & \cHad & \DHad & \cEW \\
label & \multicolumn{1}{c}{[pb]} & \multicolumn{1}{c}{[\%]} & \multicolumn{1}{c}{[\%]} & \multicolumn{1}{c}{[\%]} & [\%] & [\%] & [\%] & [\%] & [\%] &  & [\%] &  \\
\hline
1$\alpha$ & $2.34\trenn 10^{1}$ & 3.6~ & 3.4~ & $+2.1$~ & $^{+0.1~}_{~-0.3}$ & $^{+1.3~}_{~-1.3}$ & $^{-0.5~}_{~+0.2}$ & $^{-0.4~}_{~+0.3}$ & $^{+0.5~}_{~-0.5}$ & 0.94 & 2.0 & 1.00 \\
1$\beta$ & $1.36\trenn 10^{1}$ & 5.8~ & 4.5~ & $+3.5$~ & $^{+1.8~}_{~-1.9}$ & $^{+0.2~}_{~-0.3}$ & $^{-0.2~}_{~+0.2}$ & $^{-0.2~}_{~+0.4}$ & $^{+0.5~}_{~-0.5}$ & 0.97 & 1.4 & 1.00 \\
1$\gamma$ & $3.57\trenn 10^{0}$ & 6.7~ & 6.1~ & $+4.0$~ & $^{+4.0~}_{~-3.9}$ & $^{+0.2~}_{~-0.0}$ & $^{-0.4~}_{~+0.2}$ & $^{-0.2~}_{~+0.1}$ & $^{+0.5~}_{~-0.5}$ & 0.96 & 1.0 & 1.00 \\
1$\delta$ & $4.20\trenn 10^{-1}$ & 16.4~ & 9.6~ & $+7.8$~ & $^{+5.4~}_{~-4.9}$ & $^{+0.1~}_{~+0.1}$ & $^{-0.6~}_{~-0.4}$ & $^{-0.2~}_{~-0.1}$ & $^{+0.5~}_{~-0.5}$ & 0.96 & 1.2 & 1.00 \\
2$\alpha$ & $1.81\trenn 10^{1}$ & 4.1~ & 3.3~ & $+2.0$~ & $^{+0.1~}_{~-0.0}$ & $^{+1.4~}_{~-1.2}$ & $^{-0.4~}_{~+0.6}$ & $^{-0.4~}_{~+0.5}$ & $^{+0.5~}_{~-0.5}$ & 0.94 & 1.7 & 1.00 \\
2$\beta$ & $1.24\trenn 10^{1}$ & 5.6~ & 3.9~ & $+2.2$~ & $^{+2.0~}_{~-2.4}$ & $^{+0.4~}_{~-0.4}$ & $^{-0.6~}_{~+0.7}$ & $^{-0.3~}_{~+0.3}$ & $^{+0.5~}_{~-0.5}$ & 0.98 & 1.6 & 1.00 \\
2$\gamma$ & $2.95\trenn 10^{0}$ & 7.4~ & 5.8~ & $+4.0$~ & $^{+3.7~}_{~-3.4}$ & $^{+0.1~}_{~-0.2}$ & $^{-0.1~}_{~+0.1}$ & $^{-0.3~}_{~+0.2}$ & $^{+0.5~}_{~-0.5}$ & 0.97 & 1.0 & 1.00 \\
2$\delta$ & $3.82\trenn 10^{-1}$ & 18.1~ & 13.7~ & $+12.4$~ & $^{+6.2~}_{~-4.3}$ & $^{-0.2~}_{~-0.1}$ & $^{-0.0~}_{~+0.1}$ & $^{-0.4~}_{~+0.2}$ & $^{+0.5~}_{~-0.5}$ & 0.95 & 1.9 & 1.00 \\
3$\alpha$ & $1.83\trenn 10^{1}$ & 3.9~ & 2.8~ & $+1.0$~ & $^{-0.0~}_{~-0.0}$ & $^{+1.1~}_{~-1.1}$ & $^{-0.5~}_{~+0.5}$ & $^{-0.3~}_{~+0.2}$ & $^{+0.4~}_{~-0.4}$ & 0.93 & 1.2 & 1.00 \\
3$\beta$ & $1.13\trenn 10^{1}$ & 6.1~ & 4.9~ & $+3.7$~ & $^{+2.2~}_{~-2.2}$ & $^{+0.3~}_{~-0.3}$ & $^{-0.6~}_{~+0.5}$ & $^{-0.3~}_{~+0.3}$ & $^{+0.4~}_{~-0.4}$ & 0.98 & 0.9 & 1.00 \\
3$\gamma$ & $3.80\trenn 10^{0}$ & 6.0~ & 4.3~ & $+1.2$~ & $^{+3.3~}_{~-3.6}$ & $^{+0.1~}_{~-0.1}$ & $^{-0.4~}_{~+0.1}$ & $^{-0.1~}_{~-0.1}$ & $^{+0.4~}_{~-0.4}$ & 0.97 & 0.8 & 1.00 \\
3$\delta$ & $3.44\trenn 10^{-1}$ & 20.5~ & 9.3~ & $-7.0$~ & $^{+4.9~}_{~-6.4}$ & $^{+0.0~}_{~-0.3}$ & $^{-0.2~}_{~-0.6}$ & $^{-0.2~}_{~-0.2}$ & $^{+0.4~}_{~-0.4}$ & 0.96 & 0.4 & 1.00 \\
4$\alpha$ & $1.67\trenn 10^{1}$ & 4.1~ & 2.5~ & $+0.7$~ & $^{+0.1~}_{~+0.1}$ & $^{+0.9~}_{~-0.8}$ & $^{-0.3~}_{~+0.4}$ & $^{-0.2~}_{~+0.2}$ & $^{+0.4~}_{~-0.4}$ & 0.92 & 1.1 & 1.00 \\
4$\beta$ & $1.08\trenn 10^{1}$ & 6.3~ & 4.7~ & $+3.5$~ & $^{+1.9~}_{~-2.2}$ & $^{+0.3~}_{~-0.4}$ & $^{-0.5~}_{~+0.6}$ & $^{-0.1~}_{~+0.1}$ & $^{+0.4~}_{~-0.4}$ & 0.97 & 0.9 & 1.00 \\
4$\gamma$ & $3.65\trenn 10^{0}$ & 6.2~ & 4.5~ & $+2.2$~ & $^{+3.2~}_{~-3.3}$ & $^{+0.1~}_{~-0.1}$ & $^{-0.3~}_{~+0.4}$ & $^{-0.1~}_{~+0.2}$ & $^{+0.3~}_{~-0.3}$ & 0.98 & 0.5 & 1.00 \\
4$\delta$ & $3.79\trenn 10^{-1}$ & 20.4~ & 7.1~ & $-3.7$~ & $^{+5.5~}_{~-5.8}$ & $^{+0.0~}_{~-0.1}$ & $^{-0.3~}_{~+0.3}$ & $^{+0.0~}_{~-0.1}$ & $^{+0.2~}_{~-0.2}$ & 0.96 & 0.3 & 1.00 \\
5$\alpha$ & $1.49\trenn 10^{1}$ & 4.4~ & 2.9~ & $+1.0$~ & $^{-0.4~}_{~+0.5}$ & $^{+0.6~}_{~-0.5}$ & $^{+0.8~}_{~-0.6}$ & $^{-0.4~}_{~+0.4}$ & $^{+1.2~}_{~-1.2}$ & 0.92 & 0.6 & 1.02 \\
5$\beta$ & $1.32\trenn 10^{1}$ & 5.1~ & 3.6~ & $+2.1$~ & $^{+1.5~}_{~-1.5}$ & $^{+0.2~}_{~-0.1}$ & $^{-0.3~}_{~+0.3}$ & $^{-0.5~}_{~+0.5}$ & $^{+1.1~}_{~-1.1}$ & 0.96 & 0.3 & 1.02 \\
5$\gamma$ & $4.77\trenn 10^{0}$ & 5.4~ & 6.1~ & $+5.0$~ & $^{+2.5~}_{~-2.6}$ & $^{+0.2~}_{~-0.1}$ & $^{-0.2~}_{~+0.3}$ & $^{-0.4~}_{~+0.3}$ & $^{+1.1~}_{~-1.1}$ & 0.98 & 0.4 & 1.03 \\
5$\delta$ & $9.57\trenn 10^{-1}$ & 10.3~ & 5.6~ & $+2.0$~ & $^{+4.7~}_{~-4.5}$ & $^{+0.0~}_{~+0.2}$ & $^{-0.4~}_{~+0.1}$ & $^{-0.1~}_{~+0.1}$ & $^{+1.0~}_{~-1.0}$ & 0.96 & 0.7 & 1.01 \\
6$\alpha$ & $7.29\trenn 10^{-1}$ & 23.0~ & 4.0~ & $-2.2$~ & $^{-0.3~}_{~+0.8}$ & $^{+0.1~}_{~-0.5}$ & $^{+1.1~}_{~-1.4}$ & $^{-0.1~}_{~+0.7}$ & $^{+2.1~}_{~-2.1}$ & 0.89 & 0.2 & 1.11 \\
6$\beta$ & $8.45\trenn 10^{-1}$ & 20.1~ & 10.2~ & $+9.5$~ & $^{+2.8~}_{~-0.6}$ & $^{+0.2~}_{~-0.1}$ & $^{-0.1~}_{~+2.4}$ & $^{-0.4~}_{~+1.8}$ & $^{+1.8~}_{~-1.8}$ & 0.95 & 0.5 & 1.11 \\
6$\gamma$ & $3.49\trenn 10^{-1}$ & 19.3~ & 6.0~ & $-4.8$~ & $^{+1.4~}_{~-2.6}$ & $^{+0.2~}_{~-0.4}$ & $^{+0.1~}_{~-1.1}$ & $^{-1.2~}_{~+0.3}$ & $^{+1.9~}_{~-1.9}$ & 0.97 & 0.8 & 1.11 \\
6$\delta$ & $1.47\trenn 10^{-1}$ & 26.9~ & 8.5~ & $-7.5$~ & $^{+3.1~}_{~-1.7}$ & $^{-0.0~}_{~+0.2}$ & $^{+1.7~}_{~-0.4}$ & $^{+1.0~}_{~-0.3}$ & $^{+1.8~}_{~-1.8}$ & 0.98 & 1.0 & 1.11 \\
\hline
\end{tabular}
\caption{Double-differential dijet cross sections measured as a function of \Qsq\ and \meanptdi\ using the \kt\ jet algorithm.
The total systematic uncertainty, \DSys{\csdsub}, sums all systematic uncertainties in quadrature, including the uncertainty due to the LAr noise of $\DLAr{\csdsub} = \unit[0.6]{\%}$ and the total normalisation uncertainty of $\DNorm{\csdsub} =\unit[2.9]{\%}$.
The correlations between the data points are listed in \tab~\ref{tab:CorrDijetDijet}. The statistical correlations with the trijet measurement as a function of \meanpt\ are listed in \tab~\ref{tab:CorrDijetTrijet}. Further details are given in the caption of \tab~\ref{tab:IncJet}.}
\label{tab:Dijet}
\end{table}

\begin{table}
\footnotesize
\center
\begin{tabular}{c c r r r c c c c c | c c c}
\multicolumn{13}{c}{ \textbf{Dijet cross sections in bins of $\bQsq$ and $\bxidi$ using the \bkt\ jet algorithm}} \\
\hline
Bin & \multicolumn{1}{c}{\CS} & \multicolumn{1}{c}{\DStat{\csdsub}} & \multicolumn{1}{c}{\DSys{\csdsub}} & \multicolumn{1}{c}{\DMod{\csdsub}} & \DJES{\csdsub} & \DHFS{\csdsub} & \DEe{\csdsub} & \DThe{\csdsub} & \DID{\csdsub} & \cHad & \DHad & \cEW \\
label & \multicolumn{1}{c}{[pb]} & \multicolumn{1}{c}{[\%]} & \multicolumn{1}{c}{[\%]} & \multicolumn{1}{c}{[\%]} & [\%] & [\%] & [\%] & [\%] & [\%] &  & [\%] &  \\
\hline
1a & $2.04\trenn 10^{1}$ & 4.2~ & 7.7~ & $+7.2$~ & $^{+1.0~}_{~-1.1}$ & $^{+1.4~}_{~-1.4}$ & $^{+0.3~}_{~-0.5}$ & $^{-0.4~}_{~+0.3}$ & $^{+0.5~}_{~-0.5}$ & 0.94 & 2.1 & 1.00 \\
1b & $1.82\trenn 10^{1}$ & 3.4~ & 4.4~ & $+3.4$~ & $^{+1.2~}_{~-1.5}$ & $^{+1.0~}_{~-1.0}$ & $^{+0.2~}_{~-0.2}$ & $^{-0.2~}_{~+0.2}$ & $^{+0.5~}_{~-0.5}$ & 0.94 & 1.7 & 1.00 \\
1c & $6.01\trenn 10^{0}$ & 7.0~ & 4.0~ & $+2.3$~ & $^{+2.5~}_{~-2.2}$ & $^{+0.1~}_{~-0.1}$ & $^{+0.2~}_{~-0.2}$ & $^{-0.4~}_{~+0.3}$ & $^{+0.5~}_{~-0.5}$ & 0.94 & 1.3 & 1.00 \\
1d & $1.98\trenn 10^{0}$ & 8.8~ & 7.9~ & $+6.7$~ & $^{+3.4~}_{~-3.1}$ & $^{-0.2~}_{~+0.1}$ & $^{-1.7~}_{~+1.3}$ & $^{-0.2~}_{~+0.2}$ & $^{+0.5~}_{~-0.5}$ & 0.92 & 0.7 & 1.00 \\
2a & $1.45\trenn 10^{1}$ & 5.0~ & 4.9~ & $+4.1$~ & $^{+0.8~}_{~-0.9}$ & $^{+1.2~}_{~-1.2}$ & $^{+0.0~}_{~+0.2}$ & $^{-0.3~}_{~+0.4}$ & $^{+0.5~}_{~-0.5}$ & 0.94 & 1.8 & 1.00 \\
2b & $1.58\trenn 10^{1}$ & 3.6~ & 3.9~ & $+2.7$~ & $^{+1.1~}_{~-1.3}$ & $^{+1.0~}_{~-1.0}$ & $^{+0.6~}_{~-0.7}$ & $^{-0.3~}_{~+0.4}$ & $^{+0.5~}_{~-0.5}$ & 0.94 & 1.7 & 1.00 \\
2c & $6.19\trenn 10^{0}$ & 6.3~ & 3.4~ & $+0.7$~ & $^{+2.4~}_{~-2.5}$ & $^{+0.1~}_{~-0.1}$ & $^{+0.3~}_{~-0.1}$ & $^{-0.2~}_{~+0.3}$ & $^{+0.5~}_{~-0.5}$ & 0.94 & 1.1 & 1.00 \\
2d & $1.71\trenn 10^{0}$ & 9.4~ & 7.0~ & $+5.8$~ & $^{+3.2~}_{~-2.8}$ & $^{-0.3~}_{~+0.3}$ & $^{-1.1~}_{~+1.2}$ & $^{-0.3~}_{~+0.5}$ & $^{+0.5~}_{~-0.5}$ & 0.93 & 0.6 & 1.00 \\
3a & $1.13\trenn 10^{1}$ & 4.2~ & 5.5~ & $+4.9$~ & $^{+0.7~}_{~-0.8}$ & $^{+1.0~}_{~-1.1}$ & $^{-0.2~}_{~+0.1}$ & $^{-0.2~}_{~+0.2}$ & $^{+0.4~}_{~-0.4}$ & 0.93 & 1.4 & 1.00 \\
3b & $1.76\trenn 10^{1}$ & 3.0~ & 3.9~ & $+2.8$~ & $^{+1.0~}_{~-1.1}$ & $^{+0.9~}_{~-0.9}$ & $^{+0.5~}_{~-0.6}$ & $^{-0.3~}_{~+0.2}$ & $^{+0.5~}_{~-0.5}$ & 0.94 & 1.2 & 1.00 \\
3c & $8.32\trenn 10^{0}$ & 4.6~ & 3.4~ & $+1.4$~ & $^{+2.0~}_{~-2.1}$ & $^{+0.3~}_{~-0.2}$ & $^{+0.3~}_{~-0.3}$ & $^{-0.2~}_{~+0.3}$ & $^{+0.4~}_{~-0.4}$ & 0.94 & 0.9 & 1.00 \\
3d & $1.99\trenn 10^{0}$ & 8.3~ & 5.3~ & $+3.4$~ & $^{+3.4~}_{~-3.3}$ & $^{-0.2~}_{~+0.1}$ & $^{-0.6~}_{~+0.3}$ & $^{-0.4~}_{~+0.1}$ & $^{+0.5~}_{~-0.5}$ & 0.94 & 0.4 & 1.00 \\
4a & $5.12\trenn 10^{0}$ & 7.7~ & 8.6~ & $+8.2$~ & $^{+0.3~}_{~-0.8}$ & $^{+0.8~}_{~-0.9}$ & $^{+0.4~}_{~-0.5}$ & $^{-0.2~}_{~+0.2}$ & $^{+0.2~}_{~-0.2}$ & 0.92 & 1.4 & 1.00 \\
4b & $1.78\trenn 10^{1}$ & 3.2~ & 5.2~ & $+4.6$~ & $^{+0.8~}_{~-1.0}$ & $^{+0.7~}_{~-0.8}$ & $^{+0.1~}_{~-0.3}$ & $^{-0.1~}_{~+0.1}$ & $^{+0.4~}_{~-0.4}$ & 0.93 & 1.2 & 1.00 \\
4c & $1.12\trenn 10^{1}$ & 3.8~ & 3.1~ & $+1.3$~ & $^{+1.5~}_{~-1.4}$ & $^{+0.5~}_{~-0.5}$ & $^{+0.7~}_{~-0.8}$ & $^{-0.1~}_{~+0.1}$ & $^{+0.4~}_{~-0.4}$ & 0.94 & 0.8 & 1.00 \\
4d & $2.37\trenn 10^{0}$ & 8.2~ & 6.8~ & $+5.6$~ & $^{+3.3~}_{~-3.2}$ & $^{+0.0~}_{~-0.0}$ & $^{-0.0~}_{~+0.3}$ & $^{+0.1~}_{~+0.1}$ & $^{+0.3~}_{~-0.3}$ & 0.95 & 0.5 & 1.00 \\
5b & $8.89\trenn 10^{0}$ & 3.7~ & 4.5~ & $+3.6$~ & $^{+0.7~}_{~-0.9}$ & $^{+0.5~}_{~-0.6}$ & $^{-0.0~}_{~-0.1}$ & $^{-0.5~}_{~+0.4}$ & $^{+1.2~}_{~-1.2}$ & 0.92 & 0.5 & 1.01 \\
5c & $1.71\trenn 10^{1}$ & 2.9~ & 3.5~ & $+2.1$~ & $^{+0.9~}_{~-0.9}$ & $^{+0.6~}_{~-0.6}$ & $^{+0.5~}_{~-0.5}$ & $^{-0.4~}_{~+0.3}$ & $^{+1.0~}_{~-1.0}$ & 0.93 & 0.5 & 1.02 \\
5d & $1.12\trenn 10^{1}$ & 3.0~ & 4.2~ & $+3.1$~ & $^{+1.3~}_{~-1.3}$ & $^{+0.3~}_{~-0.3}$ & $^{-0.0~}_{~-0.2}$ & $^{-0.4~}_{~+0.3}$ & $^{+1.1~}_{~-1.1}$ & 0.94 & 0.4 & 1.03 \\
6d & $1.86\trenn 10^{0}$ & 7.2~ & 5.5~ & $+4.6$~ & $^{+0.6~}_{~-0.6}$ & $^{+0.2~}_{~-0.3}$ & $^{+0.5~}_{~-0.8}$ & $^{-0.3~}_{~+0.4}$ & $^{+1.9~}_{~-1.9}$ & 0.93 & 0.8 & 1.11 \\
\hline
\end{tabular}
\caption{Double-differential dijet cross sections measured as a function of \Qsq\ and $\xidi$ using the \kt\ jet algorithm.
The total systematic uncertainty, \DSys{\csdsub}, sums all systematic uncertainties in quadrature, including the uncertainty due to the LAr noise of $\DLAr{\csdsub} = \unit[0.6]{\%}$ and the total normalisation uncertainty of $\DNorm{\csdsub} =\unit[2.9]{\%}$.
 The correlations between the data points are listed in \tab~\ref{tab:CorrDijetXiDijetXi}. The statistical correlations with the trijet measurement as a function of $\xitri$ are listed in \tab~\ref{tab:CorrDijetXiTrijetXi}.
 Further details are given in the caption of \tab~\ref{tab:IncJet}.
}
\label{tab:DijetXi}
\end{table}

\clearpage
\begin{table}
\footnotesize
\center
\begin{tabular}{c c r r r c c c c c | c c c}
\multicolumn{13}{c}{ \textbf{Trijet cross sections in bins of $\bQsq$ and $\bmeanpttri$ using the \bkt\ jet algorithm}} \\
\hline
Bin & \multicolumn{1}{c}{\CS} & \multicolumn{1}{c}{\DStat{\csdsub}} & \multicolumn{1}{c}{\DSys{\csdsub}} & \multicolumn{1}{c}{\DMod{\csdsub}} & \DJES{\csdsub} & \DHFS{\csdsub} & \DEe{\csdsub} & \DThe{\csdsub} & \DID{\csdsub} & \cHad & \DHad & \cEW \\
label & \multicolumn{1}{c}{[pb]} & \multicolumn{1}{c}{[\%]} & \multicolumn{1}{c}{[\%]} & \multicolumn{1}{c}{[\%]} & [\%] & [\%] & [\%] & [\%] & [\%] &  & [\%] &  \\
\hline
1$\alpha$ & $4.86\trenn 10^{0}$ & 8.9~ & 5.1~ & $+2.9$~ & $^{-0.9~}_{~+1.2}$ & $^{+3.5~}_{~-3.3}$ & $^{-0.2~}_{~+0.3}$ & $^{-0.2~}_{~+0.3}$ & $^{+0.5~}_{~-0.5}$ & 0.79 & 5.3 & 1.00 \\
1$\beta$ & $2.65\trenn 10^{0}$ & 8.6~ & 4.5~ & $+1.8$~ & $^{+3.0~}_{~-3.3}$ & $^{+1.0~}_{~-1.2}$ & $^{+0.2~}_{~+0.0}$ & $^{-0.4~}_{~+0.3}$ & $^{+0.5~}_{~-0.5}$ & 0.85 & 4.3 & 1.00 \\
1$\gamma$ & $4.37\trenn 10^{-1}$ & 18.0~ & 8.4~ & $+6.7$~ & $^{+4.4~}_{~-4.8}$ & $^{+0.4~}_{~-0.1}$ & $^{-1.0~}_{~-0.0}$ & $^{+0.3~}_{~-0.1}$ & $^{+0.5~}_{~-0.5}$ & 0.89 & 3.6 & 1.00 \\
2$\alpha$ & $3.28\trenn 10^{0}$ & 11.1~ & 4.9~ & $-2.0$~ & $^{-1.5~}_{~+1.0}$ & $^{+3.3~}_{~-3.8}$ & $^{-0.2~}_{~+0.3}$ & $^{-0.4~}_{~+0.4}$ & $^{+0.5~}_{~-0.5}$ & 0.78 & 5.0 & 1.00 \\
2$\beta$ & $2.06\trenn 10^{0}$ & 9.2~ & 5.7~ & $+4.0$~ & $^{+2.9~}_{~-3.0}$ & $^{+1.4~}_{~-1.4}$ & $^{-0.3~}_{~+0.1}$ & $^{-0.2~}_{~+0.2}$ & $^{+0.5~}_{~-0.5}$ & 0.84 & 4.4 & 1.00 \\
2$\gamma$ & $4.28\trenn 10^{-1}$ & 17.5~ & 5.5~ & $-1.2$~ & $^{+4.8~}_{~-4.5}$ & $^{+0.7~}_{~-0.6}$ & $^{+1.1~}_{~-0.3}$ & $^{-0.7~}_{~+0.5}$ & $^{+0.5~}_{~-0.5}$ & 0.89 & 2.7 & 1.00 \\
3$\alpha$ & $3.46\trenn 10^{0}$ & 10.5~ & 5.1~ & $-2.5$~ & $^{-1.2~}_{~+1.2}$ & $^{+3.5~}_{~-3.6}$ & $^{-0.2~}_{~+0.5}$ & $^{-0.2~}_{~+0.2}$ & $^{+0.4~}_{~-0.4}$ & 0.78 & 4.6 & 1.00 \\
3$\beta$ & $2.65\trenn 10^{0}$ & 8.0~ & 6.5~ & $+5.3$~ & $^{+2.5~}_{~-2.8}$ & $^{+1.3~}_{~-1.4}$ & $^{-0.7~}_{~+0.5}$ & $^{-0.0~}_{~+0.1}$ & $^{+0.4~}_{~-0.4}$ & 0.85 & 3.7 & 1.00 \\
3$\gamma$ & $5.07\trenn 10^{-1}$ & 16.8~ & 7.2~ & $-3.8$~ & $^{+5.9~}_{~-5.4}$ & $^{+0.7~}_{~-0.6}$ & $^{-1.0~}_{~+0.1}$ & $^{+0.1~}_{~-0.6}$ & $^{+0.4~}_{~-0.4}$ & 0.87 & 2.3 & 1.00 \\
4$\alpha$ & $3.06\trenn 10^{0}$ & 11.2~ & 7.6~ & $-6.5$~ & $^{-0.9~}_{~+0.8}$ & $^{+3.3~}_{~-3.0}$ & $^{-0.4~}_{~+0.3}$ & $^{-0.1~}_{~+0.0}$ & $^{+0.3~}_{~-0.3}$ & 0.77 & 4.1 & 1.00 \\
4$\beta$ & $2.83\trenn 10^{0}$ & 7.4~ & 7.3~ & $+6.4$~ & $^{+2.4~}_{~-2.4}$ & $^{+1.2~}_{~-1.3}$ & $^{-0.8~}_{~+0.9}$ & $^{-0.1~}_{~+0.1}$ & $^{+0.3~}_{~-0.3}$ & 0.85 & 3.6 & 1.00 \\
4$\gamma$ & $6.86\trenn 10^{-1}$ & 13.8~ & 7.5~ & $+3.8$~ & $^{+6.0~}_{~-6.0}$ & $^{+0.9~}_{~-0.4}$ & $^{-0.3~}_{~+0.5}$ & $^{+0.1~}_{~+0.1}$ & $^{+0.1~}_{~-0.1}$ & 0.87 & 2.3 & 1.00 \\
5$\alpha$ & $3.23\trenn 10^{0}$ & 9.8~ & 7.1~ & $-5.9$~ & $^{-1.6~}_{~+1.6}$ & $^{+2.0~}_{~-2.0}$ & $^{+1.3~}_{~-1.1}$ & $^{-0.3~}_{~+0.4}$ & $^{+1.4~}_{~-1.4}$ & 0.77 & 3.5 & 1.03 \\
5$\beta$ & $2.91\trenn 10^{0}$ & 7.4~ & 6.2~ & $+5.3$~ & $^{+1.5~}_{~-1.6}$ & $^{+1.0~}_{~-0.9}$ & $^{-0.2~}_{~+0.5}$ & $^{-0.4~}_{~+0.3}$ & $^{+1.3~}_{~-1.3}$ & 0.83 & 2.9 & 1.03 \\
5$\gamma$ & $6.61\trenn 10^{-1}$ & 14.5~ & 14.5~ & $+13.5$~ & $^{+4.8~}_{~-4.6}$ & $^{+0.5~}_{~-0.6}$ & $^{-1.0~}_{~+0.6}$ & $^{-0.0~}_{~+0.0}$ & $^{+1.1~}_{~-1.1}$ & 0.86 & 2.2 & 1.03 \\
6$\beta$ & $1.21\trenn 10^{-1}$ & 37.9~ & 5.5~ & $+4.2$~ & $^{+0.0~}_{~+0.0}$ & $^{+1.1~}_{~-0.9}$ & $^{+1.4~}_{~-0.5}$ & $^{-0.2~}_{~+0.8}$ & $^{+2.2~}_{~-2.2}$ & 0.82 & 0.8 & 1.12 \\
\hline
\end{tabular}
\caption{Double-differential trijet cross sections measured as a function of \Qsq\ and \meanpttri\ using the \kt\ jet algorithm.
The total systematic uncertainty, \DSys{\csdsub}, sums all systematic uncertainties in quadrature, including the uncertainty due to the LAr noise of $\DLAr{\csdsub} = \unit[0.9]{\%}$ and the total normalisation uncertainty of $\DNorm{\csdsub} =\unit[2.9]{\%}$.
The correlations between the data points are listed in \tab~\ref{tab:CorrTrijetTrijet}. Further details are given in the caption of \tab~\ref{tab:IncJet}.}
\label{tab:Trijet}
\end{table}

\begin{table}
\footnotesize
\center
\begin{tabular}{c c r r r c c c c c | c c c}
\multicolumn{13}{c}{ \textbf{Trijet cross sections in bins of $\bQsq$ and $\bxitri$ using the \bkt\ jet algorithm}} \\
\hline
Bin & \multicolumn{1}{c}{\CS} & \multicolumn{1}{c}{\DStat{\csdsub}} & \multicolumn{1}{c}{\DSys{\csdsub}} & \multicolumn{1}{c}{\DMod{\csdsub}} & \DJES{\csdsub} & \DHFS{\csdsub} & \DEe{\csdsub} & \DThe{\csdsub} & \DID{\csdsub} & \cHad & \DHad & \cEW \\
label & \multicolumn{1}{c}{[pb]} & \multicolumn{1}{c}{[\%]} & \multicolumn{1}{c}{[\%]} & \multicolumn{1}{c}{[\%]} & [\%] & [\%] & [\%] & [\%] & [\%] &  & [\%] &  \\
\hline
1A & $3.15\trenn 10^{0}$ & 11.4~ & 18.7~ & $+18.1$~ & $^{-0.2~}_{~+0.2}$ & $^{+4.1~}_{~-4.2}$ & $^{+0.5~}_{~-1.0}$ & $^{-0.4~}_{~+0.3}$ & $^{+0.5~}_{~-0.5}$ & 0.81 & 6.5 & 1.00 \\
1B & $3.12\trenn 10^{0}$ & 10.6~ & 3.8~ & $+2.2$~ & $^{+1.2~}_{~-1.6}$ & $^{+1.5~}_{~-1.3}$ & $^{-0.2~}_{~+0.3}$ & $^{-0.2~}_{~+0.1}$ & $^{+0.4~}_{~-0.4}$ & 0.81 & 5.3 & 1.00 \\
1C & $1.24\trenn 10^{0}$ & 13.2~ & 7.7~ & $-5.8$~ & $^{+4.6~}_{~-4.3}$ & $^{+0.4~}_{~-0.6}$ & $^{+0.5~}_{~-0.6}$ & $^{-0.4~}_{~+0.3}$ & $^{+0.4~}_{~-0.4}$ & 0.81 & 3.7 & 1.00 \\
2A & $1.87\trenn 10^{0}$ & 16.5~ & 12.2~ & $+11.3$~ & $^{+0.5~}_{~-0.2}$ & $^{+3.7~}_{~-3.7}$ & $^{+0.8~}_{~-0.8}$ & $^{-0.4~}_{~+0.3}$ & $^{+0.5~}_{~-0.5}$ & 0.80 & 5.7 & 1.00 \\
2B & $2.80\trenn 10^{0}$ & 10.7~ & 21.7~ & $-21.4$~ & $^{+1.2~}_{~-1.9}$ & $^{+1.6~}_{~-1.8}$ & $^{+0.5~}_{~-0.6}$ & $^{-0.4~}_{~+0.3}$ & $^{+0.5~}_{~-0.5}$ & 0.81 & 4.9 & 1.00 \\
2C & $9.74\trenn 10^{-1}$ & 15.0~ & 15.6~ & $+15.0$~ & $^{+3.9~}_{~-3.0}$ & $^{+0.6~}_{~-0.4}$ & $^{+0.7~}_{~-0.2}$ & $^{-0.3~}_{~+0.4}$ & $^{+0.4~}_{~-0.4}$ & 0.80 & 3.5 & 1.00 \\
3A & $1.88\trenn 10^{0}$ & 14.7~ & 16.0~ & $+15.4$~ & $^{-0.1~}_{~-0.1}$ & $^{+3.4~}_{~-3.5}$ & $^{+1.0~}_{~-0.6}$ & $^{+0.1~}_{~+0.1}$ & $^{+0.4~}_{~-0.4}$ & 0.80 & 5.1 & 1.00 \\
3B & $3.19\trenn 10^{0}$ & 9.3~ & 9.4~ & $+8.8$~ & $^{+0.6~}_{~-0.8}$ & $^{+2.1~}_{~-2.0}$ & $^{+0.0~}_{~-0.2}$ & $^{-0.3~}_{~+0.3}$ & $^{+0.4~}_{~-0.4}$ & 0.81 & 4.5 & 1.00 \\
3C & $1.48\trenn 10^{0}$ & 12.0~ & 13.0~ & $-12.2$~ & $^{+3.9~}_{~-2.9}$ & $^{+0.9~}_{~-0.8}$ & $^{+0.8~}_{~-0.3}$ & $^{+0.1~}_{~+0.2}$ & $^{+0.4~}_{~-0.4}$ & 0.80 & 3.0 & 1.00 \\
4A & $1.55\trenn 10^{0}$ & 16.0~ & 10.7~ & $+10.0$~ & $^{-1.0~}_{~+1.1}$ & $^{+2.3~}_{~-2.7}$ & $^{+0.8~}_{~-1.5}$ & $^{-0.4~}_{~+0.2}$ & $^{+0.1~}_{~-0.1}$ & 0.80 & 5.1 & 1.00 \\
4B & $2.99\trenn 10^{0}$ & 10.1~ & 10.9~ & $+10.4$~ & $^{+0.4~}_{~-0.4}$ & $^{+2.1~}_{~-2.0}$ & $^{+0.4~}_{~-0.1}$ & $^{+0.1~}_{~-0.0}$ & $^{+0.4~}_{~-0.4}$ & 0.81 & 4.5 & 1.00 \\
4C & $1.98\trenn 10^{0}$ & 9.2~ & 5.3~ & $-3.6$~ & $^{+3.1~}_{~-3.1}$ & $^{+0.6~}_{~-0.5}$ & $^{+0.1~}_{~-0.3}$ & $^{-0.2~}_{~+0.0}$ & $^{+0.3~}_{~-0.3}$ & 0.81 & 3.1 & 1.00 \\
5B & $2.86\trenn 10^{0}$ & 9.4~ & 6.3~ & $+5.5$~ & $^{-0.0~}_{~+0.0}$ & $^{+1.4~}_{~-1.3}$ & $^{+0.1~}_{~+0.1}$ & $^{-0.6~}_{~+0.6}$ & $^{+1.4~}_{~-1.4}$ & 0.80 & 2.9 & 1.03 \\
5C & $3.26\trenn 10^{0}$ & 7.6~ & 13.1~ & $+12.8$~ & $^{+1.3~}_{~-1.5}$ & $^{+1.2~}_{~-1.2}$ & $^{+0.2~}_{~-0.2}$ & $^{-0.1~}_{~+0.1}$ & $^{+1.2~}_{~-1.2}$ & 0.80 & 2.8 & 1.04 \\
6C & $3.63\trenn 10^{-1}$ & 17.4~ & 35.5~ & $+35.3$~ & $^{+1.0~}_{~-0.6}$ & $^{+1.2~}_{~-1.1}$ & $^{+1.6~}_{~-1.0}$ & $^{+0.3~}_{~+0.3}$ & $^{+2.2~}_{~-2.2}$ & 0.79 & 1.1 & 1.11 \\
\hline
\end{tabular}
\caption{Double-differential trijet cross sections measured as a function of \Qsq\ and \xitri using the \kt\ jet algorithm.
The total systematic uncertainty, \DSys{\csdsub}, sums all systematic uncertainties in quadrature, including the uncertainty due to the LAr noise of $\DLAr{\csdsub} = \unit[0.9]{\%}$ and the total normalisation uncertainty of $\DNorm{\csdsub} =\unit[2.9]{\%}$.
The correlations between the data points are listed in \tab~\ref{tab:CorrTrijetXiTrijetXi}. Further details are given in the captions of the \tab 
\ref{tab:IncJet}.
}
\label{tab:TrijetXi}
\end{table}

\clearpage
\begin{table}
\footnotesize
\center
\begin{tabular}{c c r r r c c c c c | c c }
\multicolumn{12}{c}{ \textbf{Inclusive jet cross sections in bins of $\bQsq$ and $\bptjet$ using the \bantikt\ jet algorithm}} \\
\hline
Bin & \multicolumn{1}{c}{\CS} & \multicolumn{1}{c}{\DStat{\csdsub}} & \multicolumn{1}{c}{\DSys{\csdsub}} & \multicolumn{1}{c}{\DMod{\csdsub}} & \DJES{\csdsub} & \DHFS{\csdsub} & \DEe{\csdsub} & \DThe{\csdsub} & \DID{\csdsub} & \cHad & \cEW \\
label & \multicolumn{1}{c}{[pb]} & \multicolumn{1}{c}{[\%]} & \multicolumn{1}{c}{[\%]} & \multicolumn{1}{c}{[\%]} & [\%] & [\%] & [\%] & [\%] & [\%] &  &  \\
\hline
1$\alpha$ & $6.99\trenn 10^{1}$ & 2.3~ & 2.9~ & $+0.8$~ & $^{+1.0~}_{~-1.1}$ & $^{+1.1~}_{~-1.1}$ & $^{-0.5~}_{~+0.3}$ & $^{-0.4~}_{~+0.3}$ & $^{+0.5~}_{~-0.5}$ & 0.93 & 1.00 \\
1$\beta$ & $3.11\trenn 10^{1}$ & 3.4~ & 4.6~ & $+3.0$~ & $^{+2.5~}_{~-2.6}$ & $^{+0.6~}_{~-0.5}$ & $^{-0.5~}_{~+0.6}$ & $^{-0.3~}_{~+0.2}$ & $^{+0.5~}_{~-0.5}$ & 0.94 & 1.00 \\
1$\gamma$ & $7.28\trenn 10^{0}$ & 6.3~ & 6.2~ & $+4.5$~ & $^{+3.8~}_{~-3.4}$ & $^{+0.3~}_{~-0.3}$ & $^{-0.3~}_{~+0.7}$ & $^{-0.1~}_{~+0.1}$ & $^{+0.5~}_{~-0.5}$ & 0.93 & 1.00 \\
1$\delta$ & $8.68\trenn 10^{-1}$ & 16.2~ & 6.9~ & $+4.4$~ & $^{+4.9~}_{~-4.9}$ & $^{-0.0~}_{~-0.2}$ & $^{-0.2~}_{~-0.4}$ & $^{-0.2~}_{~+0.0}$ & $^{+0.5~}_{~-0.5}$ & 0.93 & 1.00 \\
2$\alpha$ & $5.57\trenn 10^{1}$ & 2.5~ & 2.8~ & $+0.5$~ & $^{+0.9~}_{~-0.9}$ & $^{+1.1~}_{~-1.1}$ & $^{-0.5~}_{~+0.7}$ & $^{-0.3~}_{~+0.4}$ & $^{+0.5~}_{~-0.5}$ & 0.93 & 1.00 \\
\hline
\end{tabular}
\caption{
Double-differential inclusive jet cross sections measured as a function of \Qsq\ and \ptjet\ using the \antikt\ jet algorithm.
The uncertainties $\DHad{}$ are identical to those in \tab~\ref{tab:IncJet} and are not repeated here.
Further details are given in the caption of \tab~\ref{tab:IncJet}.}
\label{tab:IncJetAntiKt}
\end{table}

\clearpage
\begin{table}
\footnotesize
\center
\begin{tabular}{c c r r r c c c c c | c c }
\multicolumn{12}{c}{ \textbf{Dijet cross sections in bins of $\bQsq$ and $\bmeanptdi$ using the \bantikt\ jet algorithm}} \\
\hline
Bin & \multicolumn{1}{c}{\CS} & \multicolumn{1}{c}{\DStat{\csdsub}} & \multicolumn{1}{c}{\DSys{\csdsub}} & \multicolumn{1}{c}{\DMod{\csdsub}} & \DJES{\csdsub} & \DHFS{\csdsub} & \DEe{\csdsub} & \DThe{\csdsub} & \DID{\csdsub} & \cHad & \cEW \\
label & \multicolumn{1}{c}{[pb]} & \multicolumn{1}{c}{[\%]} & \multicolumn{1}{c}{[\%]} & \multicolumn{1}{c}{[\%]} & [\%] & [\%] & [\%] & [\%] & [\%] &  &  \\
\hline
1$\alpha$ & $2.36\trenn 10^{1}$ & 3.2~ & 3.5~ & $+2.4$~ & $^{+0.2~}_{~-0.4}$ & $^{+1.2~}_{~-1.1}$ & $^{-0.5~}_{~+0.3}$ & $^{-0.4~}_{~+0.3}$ & $^{+0.5~}_{~-0.5}$ & 0.95 & 1.00 \\
1$\beta$ & $1.43\trenn 10^{1}$ & 4.5~ & 5.2~ & $+4.1$~ & $^{+2.0~}_{~-2.2}$ & $^{+0.3~}_{~-0.4}$ & $^{-0.3~}_{~+0.2}$ & $^{-0.3~}_{~+0.3}$ & $^{+0.5~}_{~-0.5}$ & 0.95 & 1.00 \\
1$\gamma$ & $3.19\trenn 10^{0}$ & 6.7~ & 6.3~ & $+4.5$~ & $^{+3.8~}_{~-3.7}$ & $^{+0.2~}_{~-0.2}$ & $^{-0.4~}_{~+0.1}$ & $^{-0.2~}_{~+0.1}$ & $^{+0.5~}_{~-0.5}$ & 0.94 & 1.00 \\
1$\delta$ & $3.96\trenn 10^{-1}$ & 17.1~ & 7.4~ & $-5.1$~ & $^{+5.5~}_{~-4.2}$ & $^{-0.1~}_{~-0.0}$ & $^{-0.4~}_{~-0.3}$ & $^{-0.3~}_{~+0.0}$ & $^{+0.5~}_{~-0.5}$ & 0.94 & 1.00 \\
2$\alpha$ & $1.98\trenn 10^{1}$ & 3.3~ & 3.3~ & $+2.1$~ & $^{+0.2~}_{~-0.2}$ & $^{+1.2~}_{~-1.0}$ & $^{-0.4~}_{~+0.6}$ & $^{-0.4~}_{~+0.5}$ & $^{+0.5~}_{~-0.5}$ & 0.95 & 1.00 \\
\hline
\end{tabular}
\caption{Double-differential dijet cross sections measured as a function of \Qsq\ and \meanptdi\ using the \antikt\ jet algorithm.
The uncertainties $\DHad{}$ are identical to those in \tab~\ref{tab:Dijet} and are not repeated here.
Further details are given in the caption of \tab~\ref{tab:Dijet}.
}
\label{tab:DijetAntiKt}
\end{table}

\begin{table}
\footnotesize
\center
\begin{tabular}{c c r r r c c c c c | c c }
\multicolumn{12}{c}{ \textbf{Dijet cross sections in bins of $\bQsq$ and $\bxidi$ using the \bantikt\ jet algorithm}} \\
\hline
Bin & \multicolumn{1}{c}{\CS} & \multicolumn{1}{c}{\DStat{\csdsub}} & \multicolumn{1}{c}{\DSys{\csdsub}} & \multicolumn{1}{c}{\DMod{\csdsub}} & \DJES{\csdsub} & \DHFS{\csdsub} & \DEe{\csdsub} & \DThe{\csdsub} & \DID{\csdsub} & \cHad & \cEW \\
label & \multicolumn{1}{c}{[pb]} & \multicolumn{1}{c}{[\%]} & \multicolumn{1}{c}{[\%]} & \multicolumn{1}{c}{[\%]} & [\%] & [\%] & [\%] & [\%] & [\%] &  &  \\
\hline
1a & $2.16\trenn 10^{1}$ & 3.0~ & 9.5~ & $+9.0$~ & $^{+1.0~}_{~-1.1}$ & $^{+1.4~}_{~-1.4}$ & $^{+0.1~}_{~-0.2}$ & $^{-0.3~}_{~+0.3}$ & $^{+0.5~}_{~-0.5}$ & 0.96 & 1.00 \\
1b & $1.86\trenn 10^{1}$ & 3.1~ & 4.3~ & $+3.2$~ & $^{+1.3~}_{~-1.5}$ & $^{+1.0~}_{~-1.1}$ & $^{+0.3~}_{~-0.5}$ & $^{-0.3~}_{~+0.2}$ & $^{+0.5~}_{~-0.5}$ & 0.95 & 1.00 \\
1c & $6.08\trenn 10^{0}$ & 6.5~ & 9.1~ & $-8.5$~ & $^{+2.3~}_{~-2.1}$ & $^{+0.2~}_{~-0.2}$ & $^{+0.2~}_{~+0.2}$ & $^{-0.2~}_{~+0.2}$ & $^{+0.5~}_{~-0.5}$ & 0.92 & 1.00 \\
1d & $1.75\trenn 10^{0}$ & 8.9~ & 7.7~ & $+6.5$~ & $^{+3.1~}_{~-3.5}$ & $^{-0.0~}_{~-0.0}$ & $^{-1.4~}_{~+0.9}$ & $^{-0.4~}_{~+0.2}$ & $^{+0.5~}_{~-0.5}$ & 0.90 & 1.00 \\
2a & $1.46\trenn 10^{1}$ & 5.3~ & 10.3~ & $+10.0$~ & $^{+0.8~}_{~-0.7}$ & $^{+1.2~}_{~-1.2}$ & $^{-0.1~}_{~+0.2}$ & $^{-0.4~}_{~+0.5}$ & $^{+0.5~}_{~-0.5}$ & 0.96 & 1.00 \\
\hline
\end{tabular}
\caption{Double-differential dijet cross sections measured as a function of \Qsq\ and $\xidi$ using the \antikt\ jet algorithm.
The uncertainties $\DHad{}$ are identical to those in \tab~\ref{tab:DijetXi} and are not repeated here.
Further details are given in the caption of \tab~\ref{tab:DijetXi}.
}
\label{tab:DijetXiAntiKt}
\end{table}

\clearpage
\begin{table}
\footnotesize
\center
\begin{tabular}{c c r r r c c c c c | c c }
\multicolumn{12}{c}{ \textbf{Trijet cross sections in bins of $\bQsq$ and $\bmeanpttri$ using the \bantikt\ jet algorithm}} \\
\hline
Bin & \multicolumn{1}{c}{\CS} & \multicolumn{1}{c}{\DStat{\csdsub}} & \multicolumn{1}{c}{\DSys{\csdsub}} & \multicolumn{1}{c}{\DMod{\csdsub}} & \DJES{\csdsub} & \DHFS{\csdsub} & \DEe{\csdsub} & \DThe{\csdsub} & \DID{\csdsub} & \cHad & \cEW \\
label & \multicolumn{1}{c}{[pb]} & \multicolumn{1}{c}{[\%]} & \multicolumn{1}{c}{[\%]} & \multicolumn{1}{c}{[\%]} & [\%] & [\%] & [\%] & [\%] & [\%] &  &  \\
\hline
1$\alpha$ & $4.21\trenn 10^{0}$ & 8.9~ & 7.1~ & $+5.6$~ & $^{-0.8~}_{~+0.6}$ & $^{+3.4~}_{~-3.8}$ & $^{-0.2~}_{~+0.2}$ & $^{-0.4~}_{~+0.3}$ & $^{+0.4~}_{~-0.4}$ & 0.75 & 1.00 \\
1$\beta$ & $2.57\trenn 10^{0}$ & 8.2~ & 6.0~ & $+4.4$~ & $^{+3.2~}_{~-3.2}$ & $^{+1.2~}_{~-1.0}$ & $^{+0.4~}_{~-0.1}$ & $^{-0.4~}_{~+0.3}$ & $^{+0.5~}_{~-0.5}$ & 0.78 & 1.00 \\
1$\gamma$ & $3.10\trenn 10^{-1}$ & 24.0~ & 19.2~ & $+18.4$~ & $^{+4.8~}_{~-4.7}$ & $^{+0.2~}_{~-0.9}$ & $^{-1.2~}_{~+0.3}$ & $^{+0.1~}_{~-0.3}$ & $^{+0.5~}_{~-0.5}$ & 0.81 & 1.00 \\
2$\alpha$ & $3.12\trenn 10^{0}$ & 10.0~ & 6.1~ & $+4.3$~ & $^{-0.5~}_{~+1.3}$ & $^{+3.6~}_{~-3.5}$ & $^{-0.2~}_{~+0.6}$ & $^{-0.3~}_{~+0.4}$ & $^{+0.5~}_{~-0.5}$ & 0.74 & 1.00 \\
2$\beta$ & $1.77\trenn 10^{0}$ & 9.7~ & 6.2~ & $+4.8$~ & $^{+2.5~}_{~-3.3}$ & $^{+1.2~}_{~-1.5}$ & $^{-0.4~}_{~-0.4}$ & $^{-0.3~}_{~+0.1}$ & $^{+0.5~}_{~-0.5}$ & 0.78 & 1.00 \\
\hline
\end{tabular}
\caption{Double-differential trijet cross sections measured as a function of \Qsq\ and \meanpttri\ using the \antikt\ jet algorithm.
The uncertainties $\DHad{}$ are identical to those in \tab~\ref{tab:Trijet} and are not repeated here.
Further details are given in the caption of \tab~\ref{tab:Trijet}.}
\label{tab:TrijetAntiKt}
\end{table}

\begin{table}
\footnotesize
\center
\begin{tabular}{c c r r r c c c c c | c c }
\multicolumn{12}{c}{ \textbf{Trijet cross sections in bins of $\bQsq$ and $\bxitri$ using the \bantikt\ jet algorithm}} \\
\hline
Bin & \multicolumn{1}{c}{\CS} & \multicolumn{1}{c}{\DStat{\csdsub}} & \multicolumn{1}{c}{\DSys{\csdsub}} & \multicolumn{1}{c}{\DMod{\csdsub}} & \DJES{\csdsub} & \DHFS{\csdsub} & \DEe{\csdsub} & \DThe{\csdsub} & \DID{\csdsub} & \cHad & \cEW \\
label & \multicolumn{1}{c}{[pb]} & \multicolumn{1}{c}{[\%]} & \multicolumn{1}{c}{[\%]} & \multicolumn{1}{c}{[\%]} & [\%] & [\%] & [\%] & [\%] & [\%] &  &  \\
\hline
1A & $2.71\trenn 10^{0}$ & 11.5~ & 18.3~ & $+17.7$~ & $^{-0.2~}_{~+0.2}$ & $^{+4.0~}_{~-3.9}$ & $^{+0.5~}_{~-0.9}$ & $^{-0.4~}_{~+0.3}$ & $^{+0.5~}_{~-0.5}$ & 0.76 & 1.00 \\
1B & $3.04\trenn 10^{0}$ & 9.0~ & 14.0~ & $+13.7$~ & $^{+1.2~}_{~-1.4}$ & $^{+1.6~}_{~-1.6}$ & $^{-0.2~}_{~+0.3}$ & $^{-0.3~}_{~+0.2}$ & $^{+0.4~}_{~-0.4}$ & 0.76 & 1.00 \\
1C & $1.00\trenn 10^{0}$ & 13.5~ & 16.2~ & $-15.4$~ & $^{+4.2~}_{~-4.5}$ & $^{+0.5~}_{~-0.8}$ & $^{+0.5~}_{~-0.6}$ & $^{-0.1~}_{~+0.4}$ & $^{+0.4~}_{~-0.4}$ & 0.74 & 1.00 \\
2A & $1.68\trenn 10^{0}$ & 16.6~ & 17.8~ & $+17.2$~ & $^{+0.6~}_{~-0.2}$ & $^{+4.0~}_{~-3.5}$ & $^{+1.1~}_{~-1.0}$ & $^{-0.4~}_{~+0.4}$ & $^{+0.5~}_{~-0.5}$ & 0.75 & 1.00 \\
2B & $2.56\trenn 10^{0}$ & 9.7~ & 12.5~ & $+12.1$~ & $^{+0.9~}_{~-1.0}$ & $^{+1.6~}_{~-1.6}$ & $^{+0.1~}_{~-0.1}$ & $^{-0.3~}_{~+0.3}$ & $^{+0.5~}_{~-0.5}$ & 0.76 & 1.00 \\
\hline
\end{tabular}
\caption{Double-differential normalised trijet cross sections measured as a function of \Qsq\ and $\xitri$ using the \antikt\ jet algorithm.
The uncertainties $\DHad{}$ are identical to those in \tab~\ref{tab:TrijetXi} and are not repeated here.
Further details are given in the caption of \tab~\ref{tab:TrijetXi}.}
\label{tab:TrijetXiAntiKt}
\end{table}

\clearpage
\begin{table}
\footnotesize
\center
\begin{tabular}{c c r r r c c c c | c c }
\multicolumn{11}{c}{ \textbf{Normalised inclusive jet cross sections in bins of $\bQsq$ and $\bptjet$ using the \bkt\ jet algorithm}} \\
\hline
Bin & \multicolumn{1}{c}{\CSN} & \multicolumn{1}{c}{\DStat{\csdsubn}} & \multicolumn{1}{c}{\DSys{\csdsubn}} & \multicolumn{1}{c}{\DMod{\csdsubn}} & \DJES{\csdsubn} & \DHFS{\csdsubn} & \DEe{\csdsubn} & \DThe{\csdsubn} & \cHad & \DHad  \\
label & \multicolumn{1}{c}{} & \multicolumn{1}{c}{[\%]} & \multicolumn{1}{c}{[\%]} & \multicolumn{1}{c}{[\%]} & [\%] & [\%] & [\%] & [\%] &  & [\%]  \\
\hline
1$\alpha$ & $1.63\trenn 10^{-1}$ & 2.7~ & 1.1~ & $-0.6$~ & $^{+0.6~}_{~-0.8}$ & $^{+0.3~}_{~-0.3}$ & $^{-0.4~}_{~+0.3}$ & $^{+0.1~}_{~-0.1}$ & 0.93 & 2.2 \\
1$\beta$ & $7.16\trenn 10^{-2}$ & 4.1~ & 3.0~ & $+2.0$~ & $^{+2.0~}_{~-2.2}$ & $^{-0.1~}_{~+0.2}$ & $^{-0.7~}_{~+0.5}$ & $^{+0.2~}_{~-0.3}$ & 0.97 & 1.7 \\
1$\gamma$ & $1.87\trenn 10^{-2}$ & 6.4~ & 4.2~ & $+2.9$~ & $^{+3.0~}_{~-3.0}$ & $^{-0.3~}_{~+0.6}$ & $^{-0.4~}_{~+0.5}$ & $^{+0.3~}_{~-0.4}$ & 0.96 & 1.1 \\
1$\delta$ & $2.10\trenn 10^{-3}$ & 15.3~ & 11.9~ & $+10.9$~ & $^{+4.5~}_{~-4.9}$ & $^{-0.5~}_{~+0.6}$ & $^{-0.1~}_{~-0.5}$ & $^{+0.2~}_{~-0.5}$ & 0.95 & 0.7 \\
2$\alpha$ & $1.73\trenn 10^{-1}$ & 2.9~ & 1.7~ & $-1.2$~ & $^{+0.5~}_{~-0.6}$ & $^{+0.6~}_{~-0.4}$ & $^{-0.6~}_{~+0.9}$ & $^{+0.1~}_{~-0.1}$ & 0.93 & 2.1 \\
\hline
\end{tabular}
\caption{Double-differential normalised inclusive jet cross sections measured as a function of \Qsq\ and \ptjet using the \kt\ jet algorithm. 
The total systematic uncertainty, \DSys{\csdsub}, sums all systematic uncertainties in quadrature, including the uncertainty due to the LAr noise of $\DLAr{\csdsub} = \unit[0.5]{\%}$.
Further details are given in the caption of \tab~\ref{tab:IncJet}.
}
\label{tab:NormIncJet}
\end{table}

\begin{table}
\footnotesize
\center
\begin{tabular}{c c r r r c c c c | c c }
\multicolumn{11}{c}{ \textbf{Normalised dijet cross sections in bins of $\bQsq$ and $\bmeanptdi$ using the \bkt\ jet algorithm}} \\
\hline
Bin & \multicolumn{1}{c}{\CSN} & \multicolumn{1}{c}{\DStat{\csdsubn}} & \multicolumn{1}{c}{\DSys{\csdsubn}} & \multicolumn{1}{c}{\DMod{\csdsubn}} & \DJES{\csdsubn} & \DHFS{\csdsubn} & \DEe{\csdsubn} & \DThe{\csdsubn} & \cHad & \DHad  \\
label & \multicolumn{1}{c}{} & \multicolumn{1}{c}{[\%]} & \multicolumn{1}{c}{[\%]} & \multicolumn{1}{c}{[\%]} & [\%] & [\%] & [\%] & [\%] &  & [\%]  \\
\hline
1$\alpha$ & $5.42\trenn 10^{-2}$ & 3.6~ & 1.8~ & $+1.5$~ & $^{-0.2~}_{~+0.1}$ & $^{+0.6~}_{~-0.5}$ & $^{-0.5~}_{~+0.3}$ & $^{+0.1~}_{~-0.2}$ & 0.94 & 2.0 \\
1$\beta$ & $3.13\trenn 10^{-2}$ & 5.8~ & 3.1~ & $+2.6$~ & $^{+1.4~}_{~-1.5}$ & $^{-0.4~}_{~+0.4}$ & $^{-0.1~}_{~+0.3}$ & $^{+0.2~}_{~-0.1}$ & 0.97 & 1.4 \\
1$\gamma$ & $8.30\trenn 10^{-3}$ & 6.6~ & 5.0~ & $+3.3$~ & $^{+3.7~}_{~-3.5}$ & $^{-0.4~}_{~+0.7}$ & $^{-0.4~}_{~+0.2}$ & $^{+0.3~}_{~-0.4}$ & 0.96 & 1.0 \\
1$\delta$ & $1.00\trenn 10^{-3}$ & 16.4~ & 8.7~ & $+7.2$~ & $^{+5.0~}_{~-4.5}$ & $^{-0.6~}_{~+0.8}$ & $^{-0.6~}_{~-0.4}$ & $^{+0.3~}_{~-0.6}$ & 0.96 & 1.2 \\
2$\alpha$ & $5.71\trenn 10^{-2}$ & 4.0~ & 1.8~ & $+1.4$~ & $^{-0.3~}_{~+0.4}$ & $^{+0.8~}_{~-0.6}$ & $^{-0.4~}_{~+0.6}$ & $^{+0.1~}_{~-0.0}$ & 0.94 & 1.7 \\
2$\beta$ & $3.92\trenn 10^{-2}$ & 5.5~ & 2.6~ & $+1.6$~ & $^{+1.6~}_{~-2.0}$ & $^{-0.2~}_{~+0.2}$ & $^{-0.7~}_{~+0.7}$ & $^{+0.2~}_{~-0.1}$ & 0.98 & 1.6 \\
2$\gamma$ & $9.30\trenn 10^{-3}$ & 7.4~ & 4.4~ & $+3.0$~ & $^{+3.3~}_{~-3.0}$ & $^{-0.5~}_{~+0.4}$ & $^{-0.2~}_{~+0.1}$ & $^{+0.2~}_{~-0.3}$ & 0.97 & 1.0 \\
2$\delta$ & $1.20\trenn 10^{-3}$ & 18.1~ & 12.5~ & $+11.5$~ & $^{+5.8~}_{~-3.9}$ & $^{-0.8~}_{~+0.5}$ & $^{-0.1~}_{~+0.1}$ & $^{+0.1~}_{~-0.3}$ & 0.95 & 1.9 \\
3$\alpha$ & $6.66\trenn 10^{-2}$ & 3.9~ & 1.4~ & $+0.6$~ & $^{-0.5~}_{~+0.5}$ & $^{+0.8~}_{~-0.7}$ & $^{-0.5~}_{~+0.6}$ & $^{+0.1~}_{~-0.1}$ & 0.93 & 1.2 \\
3$\beta$ & $4.11\trenn 10^{-2}$ & 6.1~ & 3.4~ & $+2.8$~ & $^{+1.7~}_{~-1.7}$ & $^{-0.0~}_{~+0.1}$ & $^{-0.6~}_{~+0.5}$ & $^{+0.1~}_{~-0.1}$ & 0.98 & 0.9 \\
3$\gamma$ & $1.38\trenn 10^{-2}$ & 5.9~ & 3.1~ & $+0.5$~ & $^{+2.9~}_{~-3.1}$ & $^{-0.3~}_{~+0.3}$ & $^{-0.4~}_{~+0.1}$ & $^{+0.2~}_{~-0.4}$ & 0.97 & 0.8 \\
3$\delta$ & $1.30\trenn 10^{-3}$ & 20.5~ & 9.5~ & $-8.0$~ & $^{+4.4~}_{~-5.9}$ & $^{-0.4~}_{~+0.1}$ & $^{-0.2~}_{~-0.6}$ & $^{+0.1~}_{~-0.5}$ & 0.96 & 0.4 \\
4$\alpha$ & $7.61\trenn 10^{-2}$ & 4.1~ & 1.3~ & $-0.7$~ & $^{-0.4~}_{~+0.6}$ & $^{+0.7~}_{~-0.6}$ & $^{-0.2~}_{~+0.3}$ & $^{+0.1~}_{~-0.0}$ & 0.92 & 1.1 \\
4$\beta$ & $4.95\trenn 10^{-2}$ & 6.3~ & 3.0~ & $+2.5$~ & $^{+1.3~}_{~-1.6}$ & $^{+0.1~}_{~-0.2}$ & $^{-0.5~}_{~+0.5}$ & $^{+0.1~}_{~-0.1}$ & 0.97 & 0.9 \\
4$\gamma$ & $1.67\trenn 10^{-2}$ & 6.2~ & 3.1~ & $+1.3$~ & $^{+2.7~}_{~-2.7}$ & $^{-0.1~}_{~+0.2}$ & $^{-0.3~}_{~+0.3}$ & $^{+0.1~}_{~-0.1}$ & 0.98 & 0.5 \\
4$\delta$ & $1.70\trenn 10^{-3}$ & 20.4~ & 6.6~ & $-4.2$~ & $^{+4.9~}_{~-5.2}$ & $^{-0.2~}_{~+0.2}$ & $^{-0.3~}_{~+0.3}$ & $^{+0.2~}_{~-0.3}$ & 0.96 & 0.3 \\
5$\alpha$ & $8.27\trenn 10^{-2}$ & 4.4~ & 1.8~ & $-1.2$~ & $^{-0.9~}_{~+1.0}$ & $^{+0.5~}_{~-0.5}$ & $^{+0.6~}_{~-0.5}$ & $^{+0.1~}_{~-0.1}$ & 0.92 & 0.6 \\
5$\beta$ & $7.37\trenn 10^{-2}$ & 5.1~ & 1.6~ & $+1.0$~ & $^{+1.0~}_{~-1.0}$ & $^{+0.1~}_{~-0.1}$ & $^{-0.5~}_{~+0.4}$ & $^{+0.0~}_{~+0.0}$ & 0.96 & 0.3 \\
5$\gamma$ & $2.66\trenn 10^{-2}$ & 5.4~ & 3.9~ & $+3.3$~ & $^{+2.0~}_{~-2.0}$ & $^{+0.1~}_{~-0.0}$ & $^{-0.4~}_{~+0.5}$ & $^{+0.1~}_{~-0.1}$ & 0.98 & 0.4 \\
5$\delta$ & $5.30\trenn 10^{-3}$ & 10.3~ & 4.3~ & $+0.9$~ & $^{+4.2~}_{~-4.0}$ & $^{-0.0~}_{~+0.2}$ & $^{-0.5~}_{~+0.2}$ & $^{+0.3~}_{~-0.4}$ & 0.96 & 0.7 \\
6$\alpha$ & $8.53\trenn 10^{-2}$ & 22.9~ & 5.1~ & $-4.9$~ & $^{-0.9~}_{~+1.3}$ & $^{+0.1~}_{~-0.5}$ & $^{+0.5~}_{~-0.9}$ & $^{+0.2~}_{~+0.4}$ & 0.89 & 0.2 \\
6$\beta$ & $9.88\trenn 10^{-2}$ & 20.0~ & 7.2~ & $+6.8$~ & $^{+2.2~}_{~-0.1}$ & $^{+0.2~}_{~-0.1}$ & $^{-0.7~}_{~+3.0}$ & $^{-0.1~}_{~+1.5}$ & 0.95 & 0.5 \\
6$\gamma$ & $4.08\trenn 10^{-2}$ & 19.2~ & 7.0~ & $-6.8$~ & $^{+0.8~}_{~-2.1}$ & $^{+0.2~}_{~-0.4}$ & $^{-0.5~}_{~-0.6}$ & $^{-0.9~}_{~-0.0}$ & 0.97 & 0.8 \\
6$\delta$ & $1.72\trenn 10^{-2}$ & 26.7~ & 9.8~ & $-9.6$~ & $^{+2.5~}_{~-1.2}$ & $^{-0.0~}_{~+0.2}$ & $^{+1.2~}_{~+0.2}$ & $^{+1.3~}_{~-0.6}$ & 0.98 & 1.0 \\
\hline
\end{tabular}
\caption{Double-differential normalised dijet cross sections measured as a function of \Qsq\ and \meanptdi using the \kt\ jet algorithm.
The total systematic uncertainty, \DSys{\csdsub}, sums all systematic uncertainties in quadrature, including the uncertainty due to the LAr noise of $\DLAr{\csdsub} = \unit[0.6]{\%}$.
Further details are given in the caption of \tab~\ref{tab:Dijet}.
}
\label{tab:NormDijet}
\end{table}

\begin{table}
\footnotesize
\center
\begin{tabular}{c c r r r c c c c | c c }
\multicolumn{11}{c}{ \textbf{Normalised dijet cross sections in bins of $\bQsq$ and $\bxidi$ using the \bkt\ jet algorithm}} \\
\hline
Bin & \multicolumn{1}{c}{\CSN} & \multicolumn{1}{c}{\DStat{\csdsubn}} & \multicolumn{1}{c}{\DSys{\csdsubn}} & \multicolumn{1}{c}{\DMod{\csdsubn}} & \DJES{\csdsubn} & \DHFS{\csdsubn} & \DEe{\csdsubn} & \DThe{\csdsubn} & \cHad & \DHad  \\
label & \multicolumn{1}{c}{} & \multicolumn{1}{c}{[\%]} & \multicolumn{1}{c}{[\%]} & \multicolumn{1}{c}{[\%]} & [\%] & [\%] & [\%] & [\%] &  & [\%]  \\
\hline
1a & $4.72\trenn 10^{-2}$ & 4.2~ & 6.7~ & $+6.5$~ & $^{+0.6~}_{~-0.7}$ & $^{+1.0~}_{~-1.0}$ & $^{+0.2~}_{~-0.4}$ & $^{+0.1~}_{~-0.2}$ & 0.94 & 2.1 \\
1b & $4.23\trenn 10^{-2}$ & 3.4~ & 3.1~ & $+2.7$~ & $^{+0.9~}_{~-1.0}$ & $^{+0.6~}_{~-0.6}$ & $^{+0.2~}_{~-0.2}$ & $^{+0.3~}_{~-0.3}$ & 0.94 & 1.7 \\
1c & $1.39\trenn 10^{-2}$ & 7.0~ & 2.7~ & $+1.7$~ & $^{+2.1~}_{~-1.8}$ & $^{-0.2~}_{~+0.3}$ & $^{+0.2~}_{~-0.1}$ & $^{+0.1~}_{~-0.2}$ & 0.94 & 1.3 \\
1d & $4.60\trenn 10^{-3}$ & 8.8~ & 6.9~ & $+6.0$~ & $^{+3.1~}_{~-2.7}$ & $^{-0.6~}_{~+0.5}$ & $^{-1.7~}_{~+1.4}$ & $^{+0.3~}_{~-0.3}$ & 0.92 & 0.7 \\
2a & $4.59\trenn 10^{-2}$ & 4.9~ & 3.7~ & $+3.5$~ & $^{+0.3~}_{~-0.4}$ & $^{+0.9~}_{~-0.9}$ & $^{+0.0~}_{~+0.1}$ & $^{+0.2~}_{~-0.1}$ & 0.94 & 1.8 \\
2b & $4.99\trenn 10^{-2}$ & 3.5~ & 2.5~ & $+2.0$~ & $^{+0.6~}_{~-0.8}$ & $^{+0.7~}_{~-0.7}$ & $^{+0.6~}_{~-0.7}$ & $^{+0.1~}_{~-0.1}$ & 0.94 & 1.7 \\
2c & $1.95\trenn 10^{-2}$ & 6.3~ & 2.4~ & $-1.0$~ & $^{+2.0~}_{~-2.1}$ & $^{-0.2~}_{~+0.2}$ & $^{+0.3~}_{~-0.2}$ & $^{+0.2~}_{~-0.2}$ & 0.94 & 1.1 \\
2d & $5.40\trenn 10^{-3}$ & 9.4~ & 5.9~ & $+5.2$~ & $^{+2.7~}_{~-2.3}$ & $^{-0.6~}_{~+0.6}$ & $^{-1.1~}_{~+1.1}$ & $^{+0.1~}_{~+0.0}$ & 0.93 & 0.6 \\
3a & $4.10\trenn 10^{-2}$ & 4.1~ & 4.3~ & $+4.2$~ & $^{+0.3~}_{~-0.3}$ & $^{+0.9~}_{~-0.9}$ & $^{-0.2~}_{~+0.1}$ & $^{+0.1~}_{~-0.1}$ & 0.93 & 1.4 \\
3b & $6.40\trenn 10^{-2}$ & 3.0~ & 2.5~ & $+2.1$~ & $^{+0.5~}_{~-0.6}$ & $^{+0.7~}_{~-0.7}$ & $^{+0.5~}_{~-0.6}$ & $^{+0.1~}_{~-0.1}$ & 0.94 & 1.2 \\
3c & $3.02\trenn 10^{-2}$ & 4.6~ & 2.1~ & $-1.2$~ & $^{+1.5~}_{~-1.6}$ & $^{+0.1~}_{~-0.1}$ & $^{+0.3~}_{~-0.3}$ & $^{+0.2~}_{~-0.1}$ & 0.94 & 0.9 \\
3d & $7.20\trenn 10^{-3}$ & 8.3~ & 4.0~ & $+2.7$~ & $^{+3.0~}_{~-2.9}$ & $^{-0.4~}_{~+0.3}$ & $^{-0.6~}_{~+0.3}$ & $^{-0.1~}_{~-0.2}$ & 0.94 & 0.4 \\
4a & $2.34\trenn 10^{-2}$ & 7.7~ & 7.3~ & $+7.2$~ & $^{-0.1~}_{~-0.3}$ & $^{+0.7~}_{~-0.8}$ & $^{+0.4~}_{~-0.4}$ & $^{-0.0~}_{~-0.0}$ & 0.92 & 1.4 \\
4b & $8.14\trenn 10^{-2}$ & 3.2~ & 3.8~ & $+3.6$~ & $^{+0.3~}_{~-0.5}$ & $^{+0.6~}_{~-0.7}$ & $^{+0.1~}_{~-0.2}$ & $^{+0.1~}_{~-0.1}$ & 0.93 & 1.2 \\
4c & $5.11\trenn 10^{-2}$ & 3.7~ & 2.0~ & $-1.4$~ & $^{+1.0~}_{~-1.0}$ & $^{+0.4~}_{~-0.4}$ & $^{+0.7~}_{~-0.8}$ & $^{+0.1~}_{~-0.1}$ & 0.94 & 0.8 \\
4d & $1.08\trenn 10^{-2}$ & 8.2~ & 5.4~ & $+4.7$~ & $^{+2.8~}_{~-2.7}$ & $^{-0.1~}_{~+0.1}$ & $^{-0.0~}_{~+0.4}$ & $^{+0.2~}_{~-0.1}$ & 0.95 & 0.5 \\
5b & $4.94\trenn 10^{-2}$ & 3.6~ & 2.1~ & $+1.9$~ & $^{+0.3~}_{~-0.5}$ & $^{+0.5~}_{~-0.6}$ & $^{-0.1~}_{~+0.1}$ & $^{+0.0~}_{~-0.1}$ & 0.92 & 0.5 \\
5c & $9.52\trenn 10^{-2}$ & 2.9~ & 1.1~ & $+0.4$~ & $^{+0.5~}_{~-0.5}$ & $^{+0.6~}_{~-0.6}$ & $^{+0.4~}_{~-0.3}$ & $^{+0.1~}_{~-0.1}$ & 0.93 & 0.5 \\
5d & $6.25\trenn 10^{-2}$ & 2.9~ & 1.9~ & $+1.4$~ & $^{+1.0~}_{~-0.9}$ & $^{+0.3~}_{~-0.3}$ & $^{-0.2~}_{~-0.0}$ & $^{+0.1~}_{~-0.1}$ & 0.94 & 0.4 \\
6d & $2.17\trenn 10^{-1}$ & 6.7~ & 2.3~ & $+2.2$~ & $^{+0.2~}_{~-0.3}$ & $^{+0.3~}_{~-0.4}$ & $^{-0.1~}_{~-0.2}$ & $^{-0.0~}_{~+0.1}$ & 0.93 & 0.8 \\
\hline
\end{tabular}
\caption{Double-differential normalised inclusive dijet cross sections measured as a function of \Qsq\ and $\xidi$ using the \kt\ jet algorithm.
The total systematic uncertainty, \DSys{\csdsub}, sums all systematic uncertainties in quadrature, including the uncertainty due to the LAr noise of $\DLAr{\csdsub} = \unit[0.6]{\%}$.
Further details are given in the caption of \tab~\ref{tab:DijetXi}.
}
\label{tab:NormDijetXi}
\end{table}

\clearpage
\begin{table}
\footnotesize
\center
\begin{tabular}{c c r r r c c c c | c c }
\multicolumn{11}{c}{ \textbf{Normalised trijet cross sections in bins of $\bQsq$ and $\bmeanpttri$ using the \bkt\ jet algorithm}} \\
\hline
Bin & \multicolumn{1}{c}{\CSN} & \multicolumn{1}{c}{\DStat{\csdsubn}} & \multicolumn{1}{c}{\DSys{\csdsubn}} & \multicolumn{1}{c}{\DMod{\csdsubn}} & \DJES{\csdsubn} & \DHFS{\csdsubn} & \DEe{\csdsubn} & \DThe{\csdsubn} & \cHad & \DHad  \\
label & \multicolumn{1}{c}{} & \multicolumn{1}{c}{[\%]} & \multicolumn{1}{c}{[\%]} & \multicolumn{1}{c}{[\%]} & [\%] & [\%] & [\%] & [\%] &  & [\%]  \\
\hline
1$\alpha$ & $1.12\trenn 10^{-2}$ & 8.9~ & 3.8~ & $+2.1$~ & $^{-1.3~}_{~+1.6}$ & $^{+2.8~}_{~-2.6}$ & $^{-0.2~}_{~+0.3}$ & $^{+0.3~}_{~-0.1}$ & 0.79 & 5.3 \\
1$\beta$ & $6.10\trenn 10^{-3}$ & 8.6~ & 3.1~ & $+1.1$~ & $^{+2.6~}_{~-2.9}$ & $^{+0.3~}_{~-0.4}$ & $^{+0.2~}_{~+0.1}$ & $^{+0.0~}_{~-0.2}$ & 0.85 & 4.3 \\
1$\gamma$ & $1.00\trenn 10^{-3}$ & 18.0~ & 7.3~ & $+5.8$~ & $^{+4.1~}_{~-4.4}$ & $^{-0.3~}_{~+0.6}$ & $^{-0.9~}_{~+0.0}$ & $^{+0.8~}_{~-0.6}$ & 0.89 & 3.6 \\
2$\alpha$ & $1.03\trenn 10^{-2}$ & 11.1~ & 4.6~ & $-2.9$~ & $^{-1.9~}_{~+1.5}$ & $^{+2.7~}_{~-3.3}$ & $^{-0.2~}_{~+0.3}$ & $^{+0.0~}_{~-0.1}$ & 0.78 & 5.0 \\
2$\beta$ & $6.50\trenn 10^{-3}$ & 9.2~ & 4.5~ & $+3.5$~ & $^{+2.5~}_{~-2.6}$ & $^{+0.8~}_{~-0.8}$ & $^{-0.3~}_{~+0.1}$ & $^{+0.3~}_{~-0.3}$ & 0.84 & 4.4 \\
2$\gamma$ & $1.40\trenn 10^{-3}$ & 17.5~ & 4.9~ & $-2.2$~ & $^{+4.4~}_{~-4.1}$ & $^{+0.2~}_{~-0.0}$ & $^{+1.1~}_{~-0.3}$ & $^{-0.2~}_{~+0.0}$ & 0.89 & 2.7 \\
3$\alpha$ & $1.26\trenn 10^{-2}$ & 10.5~ & 4.7~ & $-2.9$~ & $^{-1.6~}_{~+1.6}$ & $^{+3.2~}_{~-3.2}$ & $^{-0.2~}_{~+0.5}$ & $^{+0.1~}_{~-0.2}$ & 0.78 & 4.6 \\
3$\beta$ & $9.60\trenn 10^{-3}$ & 8.0~ & 5.5~ & $+4.8$~ & $^{+2.0~}_{~-2.3}$ & $^{+0.9~}_{~-1.1}$ & $^{-0.7~}_{~+0.5}$ & $^{+0.4~}_{~-0.2}$ & 0.85 & 3.7 \\
3$\gamma$ & $1.80\trenn 10^{-3}$ & 16.8~ & 7.0~ & $-4.5$~ & $^{+5.4~}_{~-5.0}$ & $^{+0.3~}_{~-0.2}$ & $^{-1.0~}_{~+0.1}$ & $^{+0.4~}_{~-1.0}$ & 0.87 & 2.3 \\
4$\alpha$ & $1.40\trenn 10^{-2}$ & 11.1~ & 8.2~ & $-7.5$~ & $^{-1.5~}_{~+1.4}$ & $^{+3.1~}_{~-2.8}$ & $^{-0.4~}_{~+0.2}$ & $^{+0.1~}_{~-0.2}$ & 0.77 & 4.1 \\
4$\beta$ & $1.29\trenn 10^{-2}$ & 7.4~ & 5.8~ & $+5.3$~ & $^{+1.8~}_{~-1.8}$ & $^{+1.0~}_{~-1.1}$ & $^{-0.8~}_{~+0.8}$ & $^{+0.2~}_{~-0.1}$ & 0.85 & 3.6 \\
4$\gamma$ & $3.10\trenn 10^{-3}$ & 13.8~ & 6.2~ & $+2.8$~ & $^{+5.4~}_{~-5.4}$ & $^{+0.6~}_{~-0.2}$ & $^{-0.2~}_{~+0.4}$ & $^{+0.3~}_{~-0.1}$ & 0.87 & 2.3 \\
5$\alpha$ & $1.80\trenn 10^{-2}$ & 9.8~ & 7.5~ & $-6.8$~ & $^{-2.1~}_{~+2.1}$ & $^{+1.9~}_{~-2.0}$ & $^{+1.2~}_{~-1.0}$ & $^{+0.2~}_{~-0.1}$ & 0.77 & 3.5 \\
5$\beta$ & $1.62\trenn 10^{-2}$ & 7.4~ & 4.0~ & $+3.6$~ & $^{+1.0~}_{~-1.1}$ & $^{+0.9~}_{~-0.8}$ & $^{-0.4~}_{~+0.6}$ & $^{+0.0~}_{~-0.1}$ & 0.83 & 2.9 \\
5$\gamma$ & $3.70\trenn 10^{-3}$ & 14.5~ & 12.6~ & $+11.7$~ & $^{+4.3~}_{~-4.1}$ & $^{+0.4~}_{~-0.6}$ & $^{-1.1~}_{~+0.7}$ & $^{+0.5~}_{~-0.4}$ & 0.86 & 2.2 \\
6$\beta$ & $1.41\trenn 10^{-2}$ & 37.8~ & 2.4~ & $+1.9$~ & $^{-0.5~}_{~+0.5}$ & $^{+1.1~}_{~-0.9}$ & $^{+0.9~}_{~+0.1}$ & $^{+0.1~}_{~+0.5}$ & 0.82 & 0.8 \\
\hline
\end{tabular}
\caption{Double-differential normalised trijet cross sections measured as a function of \Qsq\ and \meanpttri using the \kt\ jet algorithm.
The total systematic uncertainty, \DSys{\csdsub}, sums all systematic uncertainties in quadrature, including the uncertainty due to the LAr noise of $\DLAr{\csdsub} = \unit[0.9]{\%}$.
Further details are given in the caption of \tab~\ref{tab:Trijet}.
}
\label{tab:NormTrijet}
\end{table}

\begin{table}
\footnotesize
\center
\begin{tabular}{c c r r r c c c c | c c }
\multicolumn{11}{c}{ \textbf{Normalised trijet cross sections in bins of $\bQsq$ and $\bxitri$ using the \bkt\ jet algorithm}} \\
\hline
Bin & \multicolumn{1}{c}{\CSN} & \multicolumn{1}{c}{\DStat{\csdsubn}} & \multicolumn{1}{c}{\DSys{\csdsubn}} & \multicolumn{1}{c}{\DMod{\csdsubn}} & \DJES{\csdsubn} & \DHFS{\csdsubn} & \DEe{\csdsubn} & \DThe{\csdsubn} & \cHad & \DHad  \\
label & \multicolumn{1}{c}{} & \multicolumn{1}{c}{[\%]} & \multicolumn{1}{c}{[\%]} & \multicolumn{1}{c}{[\%]} & [\%] & [\%] & [\%] & [\%] &  & [\%]  \\
\hline
1A & $7.30\trenn 10^{-3}$ & 11.4~ & 17.9~ & $+17.4$~ & $^{-0.6~}_{~+0.6}$ & $^{+3.7~}_{~-3.8}$ & $^{+0.5~}_{~-0.9}$ & $^{+0.1~}_{~-0.2}$ & 0.81 & 6.5 \\
1B & $7.20\trenn 10^{-3}$ & 10.6~ & 2.3~ & $+1.5$~ & $^{+0.8~}_{~-1.2}$ & $^{+1.1~}_{~-0.9}$ & $^{-0.2~}_{~+0.4}$ & $^{+0.3~}_{~-0.4}$ & 0.81 & 5.3 \\
1C & $2.90\trenn 10^{-3}$ & 13.2~ & 7.7~ & $-6.4$~ & $^{+4.2~}_{~-3.9}$ & $^{+0.1~}_{~-0.2}$ & $^{+0.5~}_{~-0.6}$ & $^{+0.1~}_{~-0.2}$ & 0.81 & 3.7 \\
2A & $5.90\trenn 10^{-3}$ & 16.5~ & 11.3~ & $+10.7$~ & $^{+0.0~}_{~+0.3}$ & $^{+3.4~}_{~-3.4}$ & $^{+0.7~}_{~-0.8}$ & $^{+0.1~}_{~-0.2}$ & 0.80 & 5.7 \\
2B & $8.80\trenn 10^{-3}$ & 10.6~ & 22.0~ & $-21.9$~ & $^{+0.7~}_{~-1.5}$ & $^{+1.3~}_{~-1.4}$ & $^{+0.5~}_{~-0.6}$ & $^{+0.1~}_{~-0.1}$ & 0.81 & 4.9 \\
2C & $3.10\trenn 10^{-3}$ & 15.0~ & 14.9~ & $+14.5$~ & $^{+3.5~}_{~-2.6}$ & $^{+0.3~}_{~-0.1}$ & $^{+0.7~}_{~-0.3}$ & $^{+0.1~}_{~-0.1}$ & 0.80 & 3.5 \\
3A & $6.80\trenn 10^{-3}$ & 14.6~ & 15.2~ & $+14.7$~ & $^{-0.5~}_{~+0.3}$ & $^{+3.2~}_{~-3.3}$ & $^{+1.0~}_{~-0.6}$ & $^{+0.4~}_{~-0.3}$ & 0.80 & 5.1 \\
3B & $1.16\trenn 10^{-2}$ & 9.3~ & 8.4~ & $+8.1$~ & $^{+0.1~}_{~-0.3}$ & $^{+1.9~}_{~-1.8}$ & $^{+0.0~}_{~-0.2}$ & $^{+0.0~}_{~-0.1}$ & 0.81 & 4.5 \\
3C & $5.40\trenn 10^{-3}$ & 12.0~ & 13.4~ & $-13.0$~ & $^{+3.4~}_{~-2.5}$ & $^{+0.7~}_{~-0.6}$ & $^{+0.8~}_{~-0.3}$ & $^{+0.4~}_{~-0.1}$ & 0.80 & 3.0 \\
4A & $7.10\trenn 10^{-3}$ & 16.0~ & 9.7~ & $+9.2$~ & $^{-1.4~}_{~+1.6}$ & $^{+2.2~}_{~-2.6}$ & $^{+0.8~}_{~-1.5}$ & $^{-0.2~}_{~-0.1}$ & 0.80 & 5.1 \\
4B & $1.36\trenn 10^{-2}$ & 10.1~ & 9.7~ & $+9.4$~ & $^{-0.0~}_{~+0.1}$ & $^{+2.0~}_{~-1.9}$ & $^{+0.3~}_{~-0.0}$ & $^{+0.3~}_{~-0.2}$ & 0.81 & 4.5 \\
4C & $9.10\trenn 10^{-3}$ & 9.2~ & 5.0~ & $-4.1$~ & $^{+2.7~}_{~-2.6}$ & $^{+0.5~}_{~-0.4}$ & $^{+0.1~}_{~-0.3}$ & $^{+0.0~}_{~-0.2}$ & 0.81 & 3.1 \\
5B & $1.59\trenn 10^{-2}$ & 9.4~ & 4.4~ & $+4.1$~ & $^{-0.4~}_{~+0.4}$ & $^{+1.4~}_{~-1.3}$ & $^{-0.0~}_{~+0.2}$ & $^{-0.1~}_{~+0.2}$ & 0.80 & 2.9 \\
5C & $1.81\trenn 10^{-2}$ & 7.6~ & 11.5~ & $+11.3$~ & $^{+0.9~}_{~-1.1}$ & $^{+1.2~}_{~-1.2}$ & $^{+0.1~}_{~-0.1}$ & $^{+0.4~}_{~-0.3}$ & 0.80 & 2.8 \\
6C & $4.23\trenn 10^{-2}$ & 17.1~ & 32.6~ & $+32.5$~ & $^{+0.6~}_{~-0.3}$ & $^{+1.3~}_{~-1.2}$ & $^{+1.0~}_{~-0.4}$ & $^{+0.5~}_{~+0.0}$ & 0.79 & 1.1 \\
\hline
\end{tabular}
\caption{Double-differential normalised trijet cross sections measured as a function of \Qsq\ and $\xitri$ using the \kt\ jet algorithm.
The total systematic uncertainty, \DSys{\csdsub}, sums all systematic uncertainties in quadrature, including the uncertainty due to the LAr noise of $\DLAr{\csdsub} = \unit[0.9]{\%}$.
Further details are given in the caption of \tab~\ref{tab:TrijetXi}.
}
\label{tab:NormTrijetXi}
\end{table}

\clearpage
\begin{table}
\footnotesize
\center
\begin{tabular}{c c r r r c c c c | c }
\multicolumn{10}{c}{ \textbf{Normalised inclusive jet cross sections in bins of $\bQsq$ and $\bptjet$ using the \bantikt\ jet algorithm}} \\
\hline
Bin & \multicolumn{1}{c}{\CSN} & \multicolumn{1}{c}{\DStat{\csdsubn}} & \multicolumn{1}{c}{\DSys{\csdsubn}} & \multicolumn{1}{c}{\DMod{\csdsubn}} & \DJES{\csdsubn} & \DHFS{\csdsubn} & \DEe{\csdsubn} & \DThe{\csdsubn} & \cHad  \\
label & \multicolumn{1}{c}{} & \multicolumn{1}{c}{[\%]} & \multicolumn{1}{c}{[\%]} & \multicolumn{1}{c}{[\%]} & [\%] & [\%] & [\%] & [\%] &   \\
\hline
1$\alpha$ & $1.61\trenn 10^{-1}$ & 2.3~ & 1.2~ & $-0.6$~ & $^{+0.7~}_{~-0.7}$ & $^{+0.4~}_{~-0.4}$ & $^{-0.5~}_{~+0.4}$ & $^{+0.1~}_{~-0.2}$ & 0.93 \\
1$\beta$ & $7.17\trenn 10^{-2}$ & 3.4~ & 3.3~ & $+2.3$~ & $^{+2.1~}_{~-2.3}$ & $^{-0.1~}_{~+0.2}$ & $^{-0.5~}_{~+0.7}$ & $^{+0.2~}_{~-0.3}$ & 0.94 \\
1$\gamma$ & $1.68\trenn 10^{-2}$ & 6.3~ & 4.9~ & $+3.6$~ & $^{+3.4~}_{~-3.0}$ & $^{-0.3~}_{~+0.4}$ & $^{-0.3~}_{~+0.8}$ & $^{+0.4~}_{~-0.4}$ & 0.93 \\
1$\delta$ & $2.00\trenn 10^{-3}$ & 16.2~ & 6.4~ & $-4.4$~ & $^{+4.5~}_{~-4.5}$ & $^{-0.7~}_{~+0.5}$ & $^{-0.3~}_{~-0.3}$ & $^{+0.3~}_{~-0.5}$ & 0.93 \\
2$\alpha$ & $1.75\trenn 10^{-1}$ & 2.5~ & 1.3~ & $-0.8$~ & $^{+0.5~}_{~-0.5}$ & $^{+0.6~}_{~-0.5}$ & $^{-0.6~}_{~+0.7}$ & $^{+0.2~}_{~-0.1}$ & 0.93 \\
2$\beta$ & $8.24\trenn 10^{-2}$ & 3.6~ & 3.1~ & $+2.1$~ & $^{+2.1~}_{~-2.2}$ & $^{-0.1~}_{~+0.2}$ & $^{-0.6~}_{~+0.5}$ & $^{+0.1~}_{~-0.2}$ & 0.95 \\
2$\gamma$ & $2.10\trenn 10^{-2}$ & 6.4~ & 5.6~ & $+4.6$~ & $^{+3.0~}_{~-2.9}$ & $^{-0.5~}_{~+0.2}$ & $^{-0.8~}_{~+0.4}$ & $^{-0.0~}_{~-0.2}$ & 0.94 \\
2$\delta$ & $2.80\trenn 10^{-3}$ & 14.1~ & 7.0~ & $+5.0$~ & $^{+4.8~}_{~-4.7}$ & $^{-0.4~}_{~+0.7}$ & $^{-0.2~}_{~+0.3}$ & $^{+0.2~}_{~-0.3}$ & 0.93 \\
3$\alpha$ & $1.93\trenn 10^{-1}$ & 2.5~ & 1.2~ & $-0.1$~ & $^{+0.4~}_{~-0.3}$ & $^{+0.7~}_{~-0.7}$ & $^{-0.8~}_{~+0.7}$ & $^{+0.0~}_{~-0.1}$ & 0.94 \\
3$\beta$ & $9.90\trenn 10^{-2}$ & 3.4~ & 2.8~ & $+1.8$~ & $^{+1.9~}_{~-1.9}$ & $^{+0.0~}_{~-0.0}$ & $^{-0.8~}_{~+0.8}$ & $^{+0.2~}_{~-0.2}$ & 0.95 \\
3$\gamma$ & $2.73\trenn 10^{-2}$ & 5.7~ & 3.8~ & $+1.9$~ & $^{+3.1~}_{~-3.2}$ & $^{-0.2~}_{~+0.2}$ & $^{-0.7~}_{~+0.7}$ & $^{+0.3~}_{~-0.3}$ & 0.95 \\
3$\delta$ & $3.30\trenn 10^{-3}$ & 14.7~ & 7.5~ & $+5.3$~ & $^{+4.9~}_{~-5.6}$ & $^{-0.3~}_{~+0.4}$ & $^{+0.3~}_{~-0.3}$ & $^{+0.4~}_{~-0.2}$ & 0.93 \\
4$\alpha$ & $2.11\trenn 10^{-1}$ & 2.8~ & 1.6~ & $+0.7$~ & $^{+0.6~}_{~-0.7}$ & $^{+0.6~}_{~-0.6}$ & $^{-1.0~}_{~+1.1}$ & $^{+0.0~}_{~-0.0}$ & 0.94 \\
4$\beta$ & $1.23\trenn 10^{-1}$ & 3.5~ & 1.6~ & $+0.3$~ & $^{+1.3~}_{~-1.3}$ & $^{+0.2~}_{~-0.0}$ & $^{-0.7~}_{~+0.6}$ & $^{+0.1~}_{~-0.1}$ & 0.95 \\
4$\gamma$ & $3.58\trenn 10^{-2}$ & 5.5~ & 4.2~ & $+2.9$~ & $^{+3.0~}_{~-2.8}$ & $^{+0.1~}_{~-0.0}$ & $^{-0.6~}_{~+0.7}$ & $^{+0.2~}_{~-0.1}$ & 0.96 \\
4$\delta$ & $3.80\trenn 10^{-3}$ & 16.3~ & 9.6~ & $-8.2$~ & $^{+5.3~}_{~-4.9}$ & $^{-0.3~}_{~+0.4}$ & $^{-0.0~}_{~-0.0}$ & $^{+0.3~}_{~-0.2}$ & 0.93 \\
5$\alpha$ & $2.36\trenn 10^{-1}$ & 3.0~ & 1.0~ & $+0.4$~ & $^{+0.5~}_{~-0.5}$ & $^{+0.4~}_{~-0.3}$ & $^{-0.3~}_{~+0.5}$ & $^{-0.0~}_{~+0.0}$ & 0.92 \\
5$\beta$ & $1.58\trenn 10^{-1}$ & 3.4~ & 1.6~ & $-0.6$~ & $^{+1.2~}_{~-1.1}$ & $^{+0.1~}_{~-0.1}$ & $^{-0.7~}_{~+0.6}$ & $^{-0.1~}_{~+0.1}$ & 0.97 \\
5$\gamma$ & $5.96\trenn 10^{-2}$ & 4.3~ & 2.5~ & $+0.8$~ & $^{+2.3~}_{~-2.2}$ & $^{+0.1~}_{~-0.1}$ & $^{-0.8~}_{~+0.8}$ & $^{+0.1~}_{~-0.1}$ & 0.96 \\
5$\delta$ & $1.02\trenn 10^{-2}$ & 9.0~ & 4.6~ & $+1.9$~ & $^{+4.1~}_{~-4.1}$ & $^{-0.0~}_{~-0.1}$ & $^{-0.6~}_{~+0.4}$ & $^{+0.2~}_{~-0.3}$ & 0.95 \\
6$\alpha$ & $2.96\trenn 10^{-1}$ & 12.6~ & 2.2~ & $-2.0$~ & $^{-0.3~}_{~-0.2}$ & $^{+0.5~}_{~-0.3}$ & $^{-0.8~}_{~-0.5}$ & $^{-0.5~}_{~-0.2}$ & 0.90 \\
6$\beta$ & $2.13\trenn 10^{-1}$ & 13.3~ & 4.4~ & $-3.9$~ & $^{+1.1~}_{~-1.1}$ & $^{+0.2~}_{~-0.4}$ & $^{+1.8~}_{~+0.8}$ & $^{+0.9~}_{~-0.0}$ & 0.95 \\
6$\gamma$ & $7.13\trenn 10^{-2}$ & 20.7~ & 8.5~ & $-8.4$~ & $^{+1.6~}_{~-1.3}$ & $^{+0.3~}_{~-0.2}$ & $^{+0.4~}_{~+0.5}$ & $^{+0.2~}_{~+0.2}$ & 0.98 \\
6$\delta$ & $3.18\trenn 10^{-2}$ & 20.3~ & 20.0~ & $-19.9$~ & $^{+2.1~}_{~-2.5}$ & $^{+0.1~}_{~-0.2}$ & $^{-0.2~}_{~+0.3}$ & $^{+0.1~}_{~-0.1}$ & 0.98 \\
\hline
\end{tabular}
\caption{Double-differential normalised inclusive jet cross sections measured as a function of \Qsq\ and \ptjet\ using the \antikt jet algorithm.
The uncertainties $\DHad{}$ are identical to those in \tab~\ref{tab:NormIncJet} and are not repeated here.
Further details are given in the caption of \tab~\ref{tab:NormIncJet}.
}
\label{tab:NormIncJetAntiKt}
\end{table}

\begin{table}
\footnotesize
\center
\begin{tabular}{c c r r r c c c c | c }
\multicolumn{10}{c}{ \textbf{Normalised dijet cross sections in bins of $\bQsq$ and $\bmeanptdi$ using the \bantikt\ jet algorithm}} \\
\hline
Bin & \multicolumn{1}{c}{\CSN} & \multicolumn{1}{c}{\DStat{\csdsubn}} & \multicolumn{1}{c}{\DSys{\csdsubn}} & \multicolumn{1}{c}{\DMod{\csdsubn}} & \DJES{\csdsubn} & \DHFS{\csdsubn} & \DEe{\csdsubn} & \DThe{\csdsubn} & \cHad  \\
label & \multicolumn{1}{c}{} & \multicolumn{1}{c}{[\%]} & \multicolumn{1}{c}{[\%]} & \multicolumn{1}{c}{[\%]} & [\%] & [\%] & [\%] & [\%] &   \\
\hline
1$\alpha$ & $5.45\trenn 10^{-2}$ & 3.1~ & 2.0~ & $+1.8$~ & $^{-0.1~}_{~-0.1}$ & $^{+0.6~}_{~-0.4}$ & $^{-0.5~}_{~+0.4}$ & $^{+0.1~}_{~-0.2}$ & 0.95 \\
1$\beta$ & $3.31\trenn 10^{-2}$ & 4.5~ & 3.7~ & $+3.2$~ & $^{+1.7~}_{~-1.8}$ & $^{-0.3~}_{~+0.3}$ & $^{-0.3~}_{~+0.3}$ & $^{+0.2~}_{~-0.1}$ & 0.95 \\
1$\gamma$ & $7.40\trenn 10^{-3}$ & 6.7~ & 5.2~ & $+3.8$~ & $^{+3.5~}_{~-3.3}$ & $^{-0.4~}_{~+0.5}$ & $^{-0.4~}_{~+0.2}$ & $^{+0.3~}_{~-0.4}$ & 0.94 \\
1$\delta$ & $9.00\trenn 10^{-4}$ & 17.1~ & 7.5~ & $-5.9$~ & $^{+5.2~}_{~-3.9}$ & $^{-0.8~}_{~+0.7}$ & $^{-0.4~}_{~-0.2}$ & $^{+0.2~}_{~-0.5}$ & 0.94 \\
2$\alpha$ & $6.23\trenn 10^{-2}$ & 3.3~ & 1.8~ & $+1.5$~ & $^{-0.2~}_{~+0.2}$ & $^{+0.6~}_{~-0.5}$ & $^{-0.5~}_{~+0.6}$ & $^{+0.1~}_{~+0.0}$ & 0.95 \\
2$\beta$ & $3.60\trenn 10^{-2}$ & 5.1~ & 2.6~ & $+1.9$~ & $^{+1.5~}_{~-1.7}$ & $^{-0.1~}_{~+0.2}$ & $^{-0.8~}_{~+0.6}$ & $^{+0.2~}_{~-0.3}$ & 0.96 \\
2$\gamma$ & $8.90\trenn 10^{-3}$ & 7.1~ & 6.1~ & $+5.2$~ & $^{+3.2~}_{~-3.1}$ & $^{-0.6~}_{~+0.3}$ & $^{-0.2~}_{~+0.3}$ & $^{+0.2~}_{~-0.2}$ & 0.95 \\
2$\delta$ & $1.30\trenn 10^{-3}$ & 16.3~ & 8.2~ & $+5.8$~ & $^{+6.2~}_{~-5.1}$ & $^{-0.4~}_{~+0.8}$ & $^{+0.5~}_{~+0.3}$ & $^{+0.3~}_{~-0.1}$ & 0.94 \\
3$\alpha$ & $6.94\trenn 10^{-2}$ & 3.3~ & 1.4~ & $+0.9$~ & $^{-0.2~}_{~+0.1}$ & $^{+0.7~}_{~-0.7}$ & $^{-0.6~}_{~+0.6}$ & $^{+0.1~}_{~-0.1}$ & 0.94 \\
3$\beta$ & $4.29\trenn 10^{-2}$ & 4.9~ & 3.0~ & $+2.4$~ & $^{+1.6~}_{~-1.8}$ & $^{+0.1~}_{~-0.1}$ & $^{-0.6~}_{~+0.6}$ & $^{+0.1~}_{~-0.1}$ & 0.96 \\
3$\gamma$ & $1.33\trenn 10^{-2}$ & 5.6~ & 3.2~ & $+0.9$~ & $^{+3.0~}_{~-3.0}$ & $^{-0.2~}_{~+0.2}$ & $^{-0.4~}_{~+0.3}$ & $^{+0.2~}_{~-0.3}$ & 0.95 \\
3$\delta$ & $1.00\trenn 10^{-3}$ & 23.3~ & 6.2~ & $-3.6$~ & $^{+5.1~}_{~-5.0}$ & $^{-0.3~}_{~+0.4}$ & $^{-0.2~}_{~-0.1}$ & $^{+0.1~}_{~-0.3}$ & 0.95 \\
4$\alpha$ & $7.69\trenn 10^{-2}$ & 3.5~ & 1.0~ & $+0.4$~ & $^{-0.4~}_{~+0.3}$ & $^{+0.6~}_{~-0.5}$ & $^{-0.3~}_{~+0.3}$ & $^{+0.0~}_{~+0.0}$ & 0.93 \\
4$\beta$ & $5.11\trenn 10^{-2}$ & 5.0~ & 2.8~ & $+2.3$~ & $^{+1.2~}_{~-1.5}$ & $^{+0.2~}_{~-0.2}$ & $^{-0.5~}_{~+0.7}$ & $^{+0.1~}_{~-0.1}$ & 0.96 \\
4$\gamma$ & $1.69\trenn 10^{-2}$ & 5.5~ & 3.8~ & $+2.8$~ & $^{+2.7~}_{~-2.3}$ & $^{-0.1~}_{~+0.1}$ & $^{-0.2~}_{~+0.4}$ & $^{+0.1~}_{~-0.1}$ & 0.96 \\
4$\delta$ & $1.80\trenn 10^{-3}$ & 18.2~ & 8.7~ & $-6.8$~ & $^{+5.4~}_{~-5.4}$ & $^{-0.1~}_{~+0.3}$ & $^{-0.1~}_{~+0.2}$ & $^{+0.3~}_{~-0.3}$ & 0.94 \\
5$\alpha$ & $8.54\trenn 10^{-2}$ & 3.8~ & 1.4~ & $-0.8$~ & $^{-0.8~}_{~+0.8}$ & $^{+0.3~}_{~-0.3}$ & $^{+0.4~}_{~-0.2}$ & $^{+0.1~}_{~-0.0}$ & 0.92 \\
5$\beta$ & $7.38\trenn 10^{-2}$ & 4.3~ & 1.9~ & $+1.4$~ & $^{+1.0~}_{~-1.1}$ & $^{+0.2~}_{~-0.2}$ & $^{-0.4~}_{~+0.3}$ & $^{-0.0~}_{~-0.1}$ & 0.95 \\
5$\gamma$ & $2.62\trenn 10^{-2}$ & 5.0~ & 3.4~ & $+2.7$~ & $^{+2.0~}_{~-1.8}$ & $^{+0.2~}_{~-0.1}$ & $^{-0.5~}_{~+0.5}$ & $^{+0.2~}_{~-0.1}$ & 0.97 \\
5$\delta$ & $4.90\trenn 10^{-3}$ & 10.5~ & 5.1~ & $+2.9$~ & $^{+4.3~}_{~-3.9}$ & $^{+0.0~}_{~+0.0}$ & $^{-0.2~}_{~+0.4}$ & $^{+0.4~}_{~-0.3}$ & 0.96 \\
6$\alpha$ & $9.70\trenn 10^{-2}$ & 17.4~ & 1.2~ & $+0.4$~ & $^{-0.1~}_{~+0.6}$ & $^{+0.6~}_{~-0.4}$ & $^{-0.5~}_{~+1.0}$ & $^{-0.2~}_{~+0.5}$ & 0.91 \\
6$\beta$ & $8.19\trenn 10^{-2}$ & 19.7~ & 1.7~ & $+1.5$~ & $^{+0.5~}_{~-0.4}$ & $^{+0.4~}_{~-0.2}$ & $^{+0.2~}_{~+0.6}$ & $^{+0.1~}_{~+0.3}$ & 0.94 \\
6$\gamma$ & $4.96\trenn 10^{-2}$ & 14.8~ & 5.3~ & $-5.0$~ & $^{+1.5~}_{~-1.9}$ & $^{+0.2~}_{~-0.1}$ & $^{-0.3~}_{~+0.2}$ & $^{-0.3~}_{~+0.4}$ & 0.96 \\
6$\delta$ & $1.45\trenn 10^{-2}$ & 26.9~ & 10.1~ & $-9.8$~ & $^{+2.3~}_{~-2.2}$ & $^{+0.1~}_{~-0.3}$ & $^{-0.2~}_{~+0.9}$ & $^{+0.2~}_{~-0.0}$ & 0.97 \\
\hline
\end{tabular}
\caption{Double-differential normalised dijet cross sections measured as a function of \Qsq\ and \meanptdi using the \antikt\ jet algorithm.
The uncertainties $\DHad{}$ are identical to those in \tab~\ref{tab:NormDijet} and are not repeated here.
Further details are given in the caption of \tab~\ref{tab:NormDijet}.
}
\label{tab:NormDijetAntiKt}
\end{table}

\begin{table}
\footnotesize
\center
\begin{tabular}{c c r r r c c c c | c }
\multicolumn{10}{c}{ \textbf{Normalised dijet cross sections in bins of $\bQsq$ and $\bxidi$ using the \bantikt\ jet algorithm}} \\
\hline
Bin & \multicolumn{1}{c}{\CSN} & \multicolumn{1}{c}{\DStat{\csdsubn}} & \multicolumn{1}{c}{\DSys{\csdsubn}} & \multicolumn{1}{c}{\DMod{\csdsubn}} & \DJES{\csdsubn} & \DHFS{\csdsubn} & \DEe{\csdsubn} & \DThe{\csdsubn} & \cHad  \\
label & \multicolumn{1}{c}{} & \multicolumn{1}{c}{[\%]} & \multicolumn{1}{c}{[\%]} & \multicolumn{1}{c}{[\%]} & [\%] & [\%] & [\%] & [\%] &   \\
\hline
1a & $5.00\trenn 10^{-2}$ & 3.0~ & 8.5~ & $+8.4$~ & $^{+0.7~}_{~-0.7}$ & $^{+1.1~}_{~-1.0}$ & $^{+0.1~}_{~-0.1}$ & $^{+0.2~}_{~-0.2}$ & 0.96 \\
1b & $4.30\trenn 10^{-2}$ & 3.1~ & 3.0~ & $+2.6$~ & $^{+0.9~}_{~-1.1}$ & $^{+0.7~}_{~-0.7}$ & $^{+0.3~}_{~-0.4}$ & $^{+0.2~}_{~-0.2}$ & 0.95 \\
1c & $1.41\trenn 10^{-2}$ & 6.5~ & 9.2~ & $-9.0$~ & $^{+1.9~}_{~-1.7}$ & $^{-0.2~}_{~+0.2}$ & $^{+0.1~}_{~+0.3}$ & $^{+0.3~}_{~-0.3}$ & 0.92 \\
1d & $4.10\trenn 10^{-3}$ & 8.9~ & 6.7~ & $+5.9$~ & $^{+2.7~}_{~-3.1}$ & $^{-0.4~}_{~+0.4}$ & $^{-1.4~}_{~+0.9}$ & $^{+0.1~}_{~-0.3}$ & 0.90 \\
2a & $4.61\trenn 10^{-2}$ & 5.2~ & 9.1~ & $+9.0$~ & $^{+0.3~}_{~-0.2}$ & $^{+0.9~}_{~-0.9}$ & $^{-0.1~}_{~+0.2}$ & $^{+0.1~}_{~-0.0}$ & 0.96 \\
2b & $5.17\trenn 10^{-2}$ & 3.3~ & 2.5~ & $+2.0$~ & $^{+0.7~}_{~-0.9}$ & $^{+0.7~}_{~-0.7}$ & $^{+0.6~}_{~-0.6}$ & $^{+0.1~}_{~-0.1}$ & 0.95 \\
2c & $1.84\trenn 10^{-2}$ & 6.0~ & 2.6~ & $+1.7$~ & $^{+1.8~}_{~-1.8}$ & $^{-0.0~}_{~+0.0}$ & $^{+0.1~}_{~-0.0}$ & $^{+0.2~}_{~-0.1}$ & 0.93 \\
2d & $5.20\trenn 10^{-3}$ & 8.8~ & 5.2~ & $+4.3$~ & $^{+2.9~}_{~-2.3}$ & $^{-0.3~}_{~+0.6}$ & $^{-0.8~}_{~+1.3}$ & $^{+0.3~}_{~-0.1}$ & 0.91 \\
3a & $4.14\trenn 10^{-2}$ & 4.0~ & 8.0~ & $+8.0$~ & $^{+0.3~}_{~-0.3}$ & $^{+0.9~}_{~-0.8}$ & $^{-0.2~}_{~+0.2}$ & $^{+0.2~}_{~-0.2}$ & 0.95 \\
3b & $6.69\trenn 10^{-2}$ & 2.8~ & 2.9~ & $+2.6$~ & $^{+0.7~}_{~-0.7}$ & $^{+0.7~}_{~-0.7}$ & $^{+0.4~}_{~-0.5}$ & $^{+0.0~}_{~-0.1}$ & 0.95 \\
3c & $2.84\trenn 10^{-2}$ & 4.6~ & 2.9~ & $-2.5$~ & $^{+1.4~}_{~-1.5}$ & $^{+0.1~}_{~-0.2}$ & $^{+0.3~}_{~-0.3}$ & $^{+0.2~}_{~-0.2}$ & 0.93 \\
3d & $7.10\trenn 10^{-3}$ & 7.8~ & 5.7~ & $-4.9$~ & $^{+2.8~}_{~-2.8}$ & $^{-0.3~}_{~+0.2}$ & $^{-0.8~}_{~+0.4}$ & $^{-0.1~}_{~-0.3}$ & 0.92 \\
4a & $2.38\trenn 10^{-2}$ & 7.5~ & 3.3~ & $+3.1$~ & $^{+0.2~}_{~-0.2}$ & $^{+0.7~}_{~-0.8}$ & $^{+0.5~}_{~-0.3}$ & $^{+0.1~}_{~-0.0}$ & 0.94 \\
4b & $8.26\trenn 10^{-2}$ & 3.0~ & 2.4~ & $+2.2$~ & $^{+0.6~}_{~-0.6}$ & $^{+0.7~}_{~-0.7}$ & $^{+0.3~}_{~-0.3}$ & $^{+0.1~}_{~-0.1}$ & 0.95 \\
4c & $5.28\trenn 10^{-2}$ & 3.4~ & 6.0~ & $+5.8$~ & $^{+0.9~}_{~-0.9}$ & $^{+0.4~}_{~-0.5}$ & $^{+0.6~}_{~-0.9}$ & $^{+0.1~}_{~-0.1}$ & 0.94 \\
4d & $1.10\trenn 10^{-2}$ & 7.4~ & 13.9~ & $+13.6$~ & $^{+2.8~}_{~-2.8}$ & $^{-0.0~}_{~+0.0}$ & $^{-0.4~}_{~+0.4}$ & $^{+0.2~}_{~-0.1}$ & 0.93 \\
5b & $5.07\trenn 10^{-2}$ & 3.3~ & 1.7~ & $+1.4$~ & $^{+0.4~}_{~-0.4}$ & $^{+0.5~}_{~-0.5}$ & $^{-0.1~}_{~+0.1}$ & $^{-0.0~}_{~-0.1}$ & 0.94 \\
5c & $9.60\trenn 10^{-2}$ & 2.7~ & 6.8~ & $+6.7$~ & $^{+0.5~}_{~-0.6}$ & $^{+0.5~}_{~-0.6}$ & $^{+0.3~}_{~-0.3}$ & $^{+0.2~}_{~-0.2}$ & 0.94 \\
5d & $6.24\trenn 10^{-2}$ & 2.8~ & 8.5~ & $+8.4$~ & $^{+0.9~}_{~-1.0}$ & $^{+0.3~}_{~-0.3}$ & $^{-0.1~}_{~-0.1}$ & $^{+0.1~}_{~-0.1}$ & 0.93 \\
6d & $2.17\trenn 10^{-1}$ & 6.7~ & 4.9~ & $+4.8$~ & $^{+0.2~}_{~-0.5}$ & $^{+0.4~}_{~-0.3}$ & $^{+0.2~}_{~-0.4}$ & $^{-0.1~}_{~-0.1}$ & 0.93 \\
\hline
\end{tabular}
\caption{Double-differential normalised inclusive dijet cross sections measured as a function of \Qsq\ and $\xidi$ using the \antikt\ jet algorithm.
The uncertainties $\DHad{}$ are identical to those in \tab~\ref{tab:NormDijetXi} and are not repeated here.
Further details are given in the caption of \tab~\ref{tab:NormDijetXi}.
}
\label{tab:NormDijetXiAntiKt}
\end{table}

\clearpage
\begin{table}
\footnotesize
\center
\begin{tabular}{c c r r r c c c c | c }
\multicolumn{10}{c}{ \textbf{Normalised trijet cross sections in bins of $\bQsq$ and $\bmeanpttri$ using the \bantikt\ jet algorithm}} \\
\hline
Bin & \multicolumn{1}{c}{\CSN} & \multicolumn{1}{c}{\DStat{\csdsubn}} & \multicolumn{1}{c}{\DSys{\csdsubn}} & \multicolumn{1}{c}{\DMod{\csdsubn}} & \DJES{\csdsubn} & \DHFS{\csdsubn} & \DEe{\csdsubn} & \DThe{\csdsubn} & \cHad  \\
label & \multicolumn{1}{c}{} & \multicolumn{1}{c}{[\%]} & \multicolumn{1}{c}{[\%]} & \multicolumn{1}{c}{[\%]} & [\%] & [\%] & [\%] & [\%] &   \\
\hline
1$\alpha$ & $9.70\trenn 10^{-3}$ & 8.9~ & 5.8~ & $+4.8$~ & $^{-1.1~}_{~+1.0}$ & $^{+2.7~}_{~-3.1}$ & $^{-0.2~}_{~+0.3}$ & $^{+0.1~}_{~-0.2}$ & 0.75 \\
1$\beta$ & $5.90\trenn 10^{-3}$ & 8.1~ & 4.8~ & $+3.8$~ & $^{+2.8~}_{~-2.8}$ & $^{+0.6~}_{~-0.3}$ & $^{+0.4~}_{~-0.1}$ & $^{+0.1~}_{~-0.1}$ & 0.78 \\
1$\gamma$ & $7.00\trenn 10^{-4}$ & 24.0~ & 18.0~ & $+17.5$~ & $^{+4.4~}_{~-4.3}$ & $^{-0.4~}_{~-0.2}$ & $^{-1.2~}_{~+0.3}$ & $^{+0.6~}_{~-0.8}$ & 0.81 \\
2$\alpha$ & $9.80\trenn 10^{-3}$ & 10.0~ & 4.8~ & $+3.3$~ & $^{-1.0~}_{~+1.7}$ & $^{+3.1~}_{~-2.9}$ & $^{-0.2~}_{~+0.6}$ & $^{+0.1~}_{~-0.1}$ & 0.74 \\
2$\beta$ & $5.60\trenn 10^{-3}$ & 9.7~ & 5.1~ & $+4.2$~ & $^{+2.0~}_{~-2.9}$ & $^{+0.7~}_{~-0.9}$ & $^{-0.5~}_{~-0.4}$ & $^{+0.2~}_{~-0.4}$ & 0.78 \\
2$\gamma$ & $1.30\trenn 10^{-3}$ & 17.6~ & 8.0~ & $+6.9$~ & $^{+4.9~}_{~-3.0}$ & $^{+0.2~}_{~+0.2}$ & $^{+1.5~}_{~+0.0}$ & $^{-0.2~}_{~+0.3}$ & 0.81 \\
3$\alpha$ & $1.23\trenn 10^{-2}$ & 9.2~ & 4.0~ & $+2.0$~ & $^{-1.2~}_{~+1.5}$ & $^{+2.9~}_{~-2.9}$ & $^{-0.4~}_{~+1.1}$ & $^{+0.3~}_{~-0.2}$ & 0.73 \\
3$\beta$ & $7.60\trenn 10^{-3}$ & 8.6~ & 7.8~ & $+7.4$~ & $^{+2.0~}_{~-2.4}$ & $^{+1.0~}_{~-1.0}$ & $^{-0.5~}_{~+0.2}$ & $^{+0.3~}_{~-0.1}$ & 0.78 \\
3$\gamma$ & $1.90\trenn 10^{-3}$ & 14.8~ & 5.7~ & $+1.6$~ & $^{+5.4~}_{~-5.1}$ & $^{+0.1~}_{~-0.4}$ & $^{-1.4~}_{~+0.4}$ & $^{+0.4~}_{~-0.9}$ & 0.80 \\
4$\alpha$ & $1.17\trenn 10^{-2}$ & 11.0~ & 3.6~ & $-1.9$~ & $^{-1.5~}_{~+1.3}$ & $^{+2.4~}_{~-2.6}$ & $^{-0.3~}_{~-0.1}$ & $^{-0.1~}_{~-0.0}$ & 0.73 \\
4$\beta$ & $1.13\trenn 10^{-2}$ & 7.4~ & 8.6~ & $+8.2$~ & $^{+2.2~}_{~-1.8}$ & $^{+1.5~}_{~-1.3}$ & $^{-0.6~}_{~+0.9}$ & $^{+0.2~}_{~-0.2}$ & 0.78 \\
4$\gamma$ & $3.00\trenn 10^{-3}$ & 14.0~ & 10.5~ & $+9.0$~ & $^{+5.2~}_{~-5.3}$ & $^{+0.4~}_{~-0.6}$ & $^{-0.2~}_{~+0.6}$ & $^{+0.1~}_{~+0.2}$ & 0.80 \\
5$\alpha$ & $1.46\trenn 10^{-2}$ & 10.2~ & 3.4~ & $-2.0$~ & $^{-1.8~}_{~+1.8}$ & $^{+1.7~}_{~-2.0}$ & $^{+0.9~}_{~-0.6}$ & $^{+0.2~}_{~-0.2}$ & 0.71 \\
5$\beta$ & $1.44\trenn 10^{-2}$ & 7.4~ & 6.7~ & $+6.4$~ & $^{+1.1~}_{~-1.1}$ & $^{+1.1~}_{~-0.9}$ & $^{-0.2~}_{~+0.4}$ & $^{+0.2~}_{~-0.1}$ & 0.77 \\
5$\gamma$ & $3.10\trenn 10^{-3}$ & 18.6~ & 21.3~ & $+20.8$~ & $^{+4.1~}_{~-3.9}$ & $^{+0.6~}_{~-0.8}$ & $^{-1.3~}_{~+1.0}$ & $^{+0.3~}_{~-0.3}$ & 0.79 \\
6$\beta$ & $1.52\trenn 10^{-2}$ & 33.0~ & 9.1~ & $+9.0$~ & $^{+0.6~}_{~-1.1}$ & $^{+1.4~}_{~-0.5}$ & $^{+0.3~}_{~+0.1}$ & $^{-0.2~}_{~+0.2}$ & 0.74 \\
\hline
\end{tabular}
\caption{Double-differential normalised trijet cross sections measured as a function of \Qsq\ and \meanpttri using the \antikt\ jet algorithm.
Further details are given in the caption of \tab~\ref{tab:NormTrijet}.
The uncertainties $\DHad{}$ are identical to those in \tab~\ref{tab:NormTrijet} and are not repeated here.
Further details are given in the caption of \tab~\ref{tab:NormTrijet}. }
\label{tab:NormTrijetAntiKt}
\end{table}

\begin{table}
\footnotesize
\center
\begin{tabular}{c c r r r c c c c | c }
\multicolumn{10}{c}{ \textbf{Normalised trijet cross sections in bins of $\bQsq$ and $\bxitri$ using the \bantikt\ jet algorithm}} \\
\hline
Bin & \multicolumn{1}{c}{\CSN} & \multicolumn{1}{c}{\DStat{\csdsubn}} & \multicolumn{1}{c}{\DSys{\csdsubn}} & \multicolumn{1}{c}{\DMod{\csdsubn}} & \DJES{\csdsubn} & \DHFS{\csdsubn} & \DEe{\csdsubn} & \DThe{\csdsubn} & \cHad  \\
label & \multicolumn{1}{c}{} & \multicolumn{1}{c}{[\%]} & \multicolumn{1}{c}{[\%]} & \multicolumn{1}{c}{[\%]} & [\%] & [\%] & [\%] & [\%] &   \\
\hline
1A & $6.30\trenn 10^{-3}$ & 11.5~ & 17.5~ & $+17.1$~ & $^{-0.5~}_{~+0.6}$ & $^{+3.7~}_{~-3.5}$ & $^{+0.5~}_{~-0.8}$ & $^{+0.1~}_{~-0.2}$ & 0.76 \\
1B & $7.00\trenn 10^{-3}$ & 9.0~ & 13.2~ & $+13.1$~ & $^{+0.8~}_{~-0.9}$ & $^{+1.2~}_{~-1.2}$ & $^{-0.2~}_{~+0.4}$ & $^{+0.2~}_{~-0.2}$ & 0.76 \\
1C & $2.30\trenn 10^{-3}$ & 13.5~ & 16.4~ & $-15.9$~ & $^{+3.8~}_{~-4.1}$ & $^{+0.1~}_{~-0.4}$ & $^{+0.5~}_{~-0.5}$ & $^{+0.4~}_{~-0.1}$ & 0.74 \\
2A & $5.30\trenn 10^{-3}$ & 16.6~ & 16.7~ & $+16.3$~ & $^{+0.1~}_{~+0.2}$ & $^{+3.7~}_{~-3.2}$ & $^{+1.1~}_{~-1.0}$ & $^{+0.1~}_{~-0.0}$ & 0.75 \\
2B & $8.10\trenn 10^{-3}$ & 9.7~ & 11.3~ & $+11.2$~ & $^{+0.4~}_{~-0.6}$ & $^{+1.3~}_{~-1.3}$ & $^{+0.1~}_{~-0.2}$ & $^{+0.1~}_{~-0.2}$ & 0.76 \\
2C & $2.60\trenn 10^{-3}$ & 14.7~ & 47.6~ & $-47.4$~ & $^{+4.0~}_{~-3.7}$ & $^{+0.4~}_{~-0.4}$ & $^{+1.0~}_{~-1.3}$ & $^{-0.0~}_{~-0.2}$ & 0.74 \\
3A & $5.30\trenn 10^{-3}$ & 16.1~ & 18.4~ & $+18.1$~ & $^{-0.7~}_{~+0.6}$ & $^{+3.5~}_{~-3.0}$ & $^{+1.4~}_{~-0.7}$ & $^{+0.4~}_{~-0.3}$ & 0.75 \\
3B & $1.12\trenn 10^{-2}$ & 8.4~ & 17.2~ & $+17.0$~ & $^{+0.7~}_{~-0.2}$ & $^{+1.9~}_{~-2.0}$ & $^{+0.2~}_{~-0.1}$ & $^{+0.2~}_{~-0.0}$ & 0.76 \\
3C & $4.20\trenn 10^{-3}$ & 11.7~ & 11.3~ & $-10.8$~ & $^{+2.9~}_{~-3.6}$ & $^{+0.4~}_{~-0.3}$ & $^{+0.2~}_{~-0.4}$ & $^{+0.3~}_{~-0.4}$ & 0.74 \\
4A & $5.80\trenn 10^{-3}$ & 15.9~ & 16.3~ & $+16.1$~ & $^{-0.9~}_{~+0.2}$ & $^{+2.2~}_{~-2.3}$ & $^{+1.0~}_{~-1.8}$ & $^{-0.1~}_{~-0.1}$ & 0.73 \\
4B & $1.22\trenn 10^{-2}$ & 9.4~ & 20.0~ & $-19.9$~ & $^{+0.0~}_{~+0.4}$ & $^{+2.0~}_{~-1.9}$ & $^{+0.5~}_{~-0.5}$ & $^{+0.3~}_{~-0.1}$ & 0.76 \\
4C & $7.60\trenn 10^{-3}$ & 9.1~ & 17.3~ & $-17.0$~ & $^{+2.7~}_{~-3.0}$ & $^{+0.6~}_{~-0.7}$ & $^{+0.4~}_{~-0.4}$ & $^{+0.1~}_{~-0.2}$ & 0.75 \\
5B & $1.40\trenn 10^{-2}$ & 9.1~ & 7.7~ & $+7.5$~ & $^{-0.3~}_{~+0.6}$ & $^{+1.3~}_{~-1.4}$ & $^{+0.3~}_{~-0.1}$ & $^{+0.0~}_{~+0.1}$ & 0.75 \\
5C & $1.60\trenn 10^{-2}$ & 7.0~ & 52.8~ & $-52.8$~ & $^{+1.0~}_{~-1.0}$ & $^{+1.2~}_{~-1.2}$ & $^{+0.2~}_{~-0.1}$ & $^{+0.3~}_{~-0.3}$ & 0.75 \\
6C & $3.52\trenn 10^{-2}$ & 17.5~ & 94.7~ & $-94.6$~ & $^{+0.3~}_{~-0.1}$ & $^{+1.3~}_{~-1.0}$ & $^{+0.6~}_{~-0.4}$ & $^{+0.3~}_{~-0.2}$ & 0.73 \\
\hline
\end{tabular}
\caption{Double-differential normalised trijet cross sections measured as a function of \Qsq\ and $\xitri$ using the \antikt\ jet algorithm.
The uncertainties $\DHad{}$ are identical to those in \tab~\ref{tab:NormTrijetXi} and are not repeated here.
Further details are given in the caption of \tab~\ref{tab:NormTrijetXi}.}
\label{tab:NormTrijetXiAntiKt}
\end{table}


\clearpage
\begin{table}[htbp]
\footnotesize
\center
\def\arraystretch{1}
\def\tabcolsep{1pt}
\begin{tabular}{cc|cccccccccccccccccccccccc}
 & & \multicolumn{24}{c}{\textbf{Inclusive jet as function of \bQsq and \bptjet}} \\
 &  & 1$\alpha$ & 1$\beta$ & 1$\gamma$ & 1$\delta$ & 2$\alpha$ & 2$\beta$ & 2$\gamma$ & 2$\delta$ & 3$\alpha$ & 3$\beta$ & 3$\gamma$ & 3$\delta$ & 4$\alpha$ & 4$\beta$ & 4$\gamma$ & 4$\delta$ & 5$\alpha$ & 5$\beta$ & 5$\gamma$ & 5$\delta$ & 6$\alpha$ & 6$\beta$ & 6$\gamma$ & 6$\delta$ \\
\hline
\multirow{24}{*}{\begin{sideways}\textbf{Inclusive jet as function of \bQsq and \bptjet}~~~\end{sideways}}  & 1$\alpha$&  100 & -20& -11& -2& -14& 2& 1& & 1& & & & & & & & & & & & & & & \\
 & 1$\beta$& &  100 & 2& -1& 4& -13& & & & 2& & & & & & & & & & & & & & \\
 & 1$\gamma$& & &  100 & 6& 1& & -13& -1& & & 2& & & & & & & & 1& & & & & \\
 & 1$\delta$& & & &  100 & & & & -14& & & & 2& & & & & & & & 1& & & & 1\\
 & 2$\alpha$& & & & &  100 & -21& -10& -2& -11& 2& 1& & -1& & & & -1& & & & & & & \\
 & 2$\beta$& & & & & &  100 & 2& -1& 3& -10& & & & & & & & -1& & & & & & \\
\end{tabular}
\caption{Correlation coefficients between data points of the inclusive jet measurement as a function of \Qsq\ and \ptjet. Since the matrix is symmetric only the upper triangle is given. 
 The bin labels are defined in \tab~\ref{tab:BinNumbering}. 
 All values are multiplied by a factor of $100$.
\label{tab:CorrInclIncl}}
\end{table}

\newpage
\begin{table}[htbp]
\footnotesize
\center
\def\arraystretch{1}
\def\tabcolsep{1pt}
\begin{tabular}{cc|cccccccccccccccccccccccc}
 & & \multicolumn{24}{c}{\textbf{Dijet as function of \bQsq and \bmeanptdi}} \\
 &  & 1$\alpha$ & 1$\beta$ & 1$\gamma$ & 1$\delta$ & 2$\alpha$ & 2$\beta$ & 2$\gamma$ & 2$\delta$ & 3$\alpha$ & 3$\beta$ & 3$\gamma$ & 3$\delta$ & 4$\alpha$ & 4$\beta$ & 4$\gamma$ & 4$\delta$ & 5$\alpha$ & 5$\beta$ & 5$\gamma$ & 5$\delta$ & 6$\alpha$ & 6$\beta$ & 6$\gamma$ & 6$\delta$ \\
\hline
\multirow{24}{*}{\begin{sideways}\textbf{Dijet as function of \bQsq and \bmeanptdi}~~~\end{sideways}}  & 1$\alpha$&  100 & -44& 11& 3& -3& 6& -2& & 11& -1& & & 9& & & & 8& & & & 2& & & \\
 & 1$\beta$& &  100 & -36& -9& 7& -13& 5& 1& -1& 2& -1& & & 1& & & & 1& & & & & & \\
 & 1$\gamma$& & &  100 & 6& -1& 4& -14& & & -1& 2& & & & 1& & & & 1& & & & 1& \\
 & 1$\delta$& & & &  100 & & 1& & -14& & & & 2& & & & & & & & 1& & & & 1\\
 & 2$\alpha$& & & & &  100 & -44& 10& 2& -4& 6& -1& & 4& 1& & & 4& 1& & & 1& & & \\
 & 2$\beta$& & & & & &  100 & -34& -8& 7& -11& 4& 1& 1& -1& & & 2& -1& & & & & & \\
\end{tabular}
\caption{Correlation coefficients between data points of the dijet measurement as a function of \Qsq\ and \meanptdi. Since the matrix is symmetric only the upper triangle is given. 
 The bin labels are defined in \tab~\ref{tab:BinNumbering}. 
All values are multiplied by a factor of $100$.
\label{tab:CorrDijetDijet}}
\end{table}

\begin{table}[htbp]
\footnotesize
\center
\def\arraystretch{1}
\def\tabcolsep{1pt}
\begin{tabular}{cc|cccccccccccccccccc}
 & & \multicolumn{16}{c}{\textbf{Trijet as function of \bQsq and \bmeanpttri}} \\
 &  & 1$\alpha$ & 1$\beta$ & 1$\gamma$ & 2$\alpha$ & 2$\beta$ & 2$\gamma$ & 3$\alpha$ & 3$\beta$ & 3$\gamma$ & 4$\alpha$ & 4$\beta$ & 4$\gamma$ & 5$\alpha$ & 5$\beta$ & 5$\gamma$ & 6$\beta$ \\
\hline
\multirow{16}{*}{\begin{sideways}\textbf{Trijet as function of \bQsq and \bmeanpttri}~~~\end{sideways}}  & 1$\alpha$&  100 & -37& 9& 1& 2& & 14& -4& 1& 12& -3& 1& 12& -3& 1& -1\\
 & 1$\beta$& &  100 & -26& 1& -8& 2& -3& 7& -2& -3& 5& -1& -3& 6& -1& 2\\
 & 1$\gamma$& & &  100 & & 2& -11& 1& -2& 5& 1& -1& 3& & -1& 3& \\
 & 2$\alpha$& & & &  100 & -35& 8& 2& 1& & 11& -3& 1& 10& -2& & -1\\
 & 2$\beta$& & & & &  100 & -24& 1& -5& 2& -3& 6& -1& -2& 6& -1& 2\\
 & 2$\gamma$& & & & & &  100 & & 2& -8& & -1& 3& & -1& 3& \\
 & 3$\alpha$& & & & & & &  100 & -37& 10& 3& 1& & 10& -3& 1& -1\\
 & 3$\beta$& & & & & & & &  100 & -29& & -2& 1& -2& 7& -1& 2\\
 & 3$\gamma$& & & & & & & & &  100 & & 1& -5& & -1& 3& \\
 & 4$\alpha$& & & & & & & & & &  100 & -35& 9& 5& -1& & -1\\
 & 4$\beta$& & & & & & & & & & &  100 & -27& -1& 3& & 2\\
 & 4$\gamma$& & & & & & & & & & & &  100 & & & 1& \\
 & 5$\alpha$& & & & & & & & & & & & &  100 & -35& 9& \\
 & 5$\beta$& & & & & & & & & & & & & &  100 & -28& 2\\
 & 5$\gamma$& & & & & & & & & & & & & & &  100 & \\
 & 6$\beta$& & & & & & & & & & & & & & & &  100 \\
\end{tabular}
\caption{Correlation coefficients between data points of the trijet measurement as a function of \Qsq\ and \meanpttri. Since the matrix is symmetric only the upper triangle is given. 
 The bin labels are defined in \tab~\ref{tab:BinNumbering}. 
All values are multiplied by a factor of $100$.}
\label{tab:CorrTrijetTrijet}
\end{table}

\begin{table}[htbp]
\footnotesize
\center
\def\arraystretch{1}
\def\tabcolsep{1pt}
\begin{tabular}{cc|cccccccccccccccccccccccc}
 & & \multicolumn{24}{c}{\textbf{Dijet as function of \bQsq and \bmeanptdi}} \\
 &  & 1$\alpha$ & 1$\beta$ & 1$\gamma$ & 1$\delta$ & 2$\alpha$ & 2$\beta$ & 2$\gamma$ & 2$\delta$ & 3$\alpha$ & 3$\beta$ & 3$\gamma$ & 3$\delta$ & 4$\alpha$ & 4$\beta$ & 4$\gamma$ & 4$\delta$ & 5$\alpha$ & 5$\beta$ & 5$\gamma$ & 5$\delta$ & 6$\alpha$ & 6$\beta$ & 6$\gamma$ & 6$\delta$ \\
\hline
\multirow{24}{*}{\begin{sideways}\textbf{Inclusive jet as function of \bQsq and \bptjet}~~~\end{sideways}}  & 1$\alpha$& 35& 1& -2& & -5& & & & 1& & & & & & & & & & & & & & & \\
 & 1$\beta$& -6& 25& -1& -1& 1& -3& & & & 1& & & & & & & & & & & & & & \\
 & 1$\gamma$& -1& -3& 48& 1& & & -6& & & & 1& & & & & & & & & & & & & \\
 & 1$\delta$& & 1& -6& 71& & & & -10& & & & 1& & & & & & & & & & & & \\
 & 2$\alpha$& -5& & & & 34& & -1& & -4& & & & -1& & & & -1& & & & & & & \\
 & 2$\beta$& 1& -4& & & -7& 27& -1& & 1& -3& & & & & & & & -1& & & & & & \\
\end{tabular}
\caption{Correlation coefficients between data points of the inclusive jet measurement as a function of \Qsq\ and \ptjet\ and of the dijet measurement as a function of \Qsq\ and \meanptdi. 
 The bin labels are defined in \tab~\ref{tab:BinNumbering}. 
All values are multiplied by a factor of $100$.
\label{tab:CorrInclDijet}}
\end{table}

\begin{table}[htbp]
\footnotesize
\center
\def\arraystretch{1}
\def\tabcolsep{1pt}
\begin{tabular}{cc|cccccccccccccccccc}
 & & \multicolumn{16}{c}{\textbf{Trijet as function of \bQsq and \bmeanpttri}} \\
 &  & 1$\alpha$ & 1$\beta$ & 1$\gamma$ & 2$\alpha$ & 2$\beta$ & 2$\gamma$ & 3$\alpha$ & 3$\beta$ & 3$\gamma$ & 4$\alpha$ & 4$\beta$ & 4$\gamma$ & 5$\alpha$ & 5$\beta$ & 5$\gamma$ & 6$\beta$ \\
\hline
\multirow{24}{*}{\begin{sideways}\textbf{Inclusive jet as function of \bQsq and \bptjet}~~~\end{sideways}}  & 1$\alpha$& 11& 2& 1& -2& & & & & & -1& & & -1& & & \\
 & 1$\beta$& 10& 12& 1& -3& -2& & -1& & & -1& & & -2& & & \\
 & 1$\gamma$& -6& 18& 12& 1& -4& -2& & -1& & & -1& & & -2& & \\
 & 1$\delta$& 2& -4& 18& & 2& -6& & 1& -2& & 1& -2& & & -2& \\
 & 2$\alpha$& -2& & & 10& 1& 1& -2& & & & & & -1& & & \\
 & 2$\beta$& -3& -2& & 10& 13& -1& -1& -1& & & & & -1& & & \\
 & 2$\gamma$& 1& -3& -2& -6& 16& 16& 1& -3& -2& & -1& & & -1& & \\
 & 2$\delta$& & 1& -5& 2& -7& 22& & 1& -2& & & -1& & & -3& \\
 & 3$\alpha$& & & & -2& & & 12& 1& 2& -1& & & & & & \\
 & 3$\beta$& -2& & & -2& -2& & 7& 12& & -1& -1& & -1& & & \\
 & 3$\gamma$& & -1& & 1& -3& -2& -7& 15& 12& 1& -3& -1& & -1& & \\
 & 3$\delta$& & & -1& & 1& -3& 2& -7& 23& & 1& -3& & & -2& \\
 & 4$\alpha$& -1& & & & & & -1& & & 8& 3& 2& -1& & & \\
 & 4$\beta$& -1& & & -1& & & -1& -1& & 8& 11& 1& -1& -1& & \\
 & 4$\gamma$& & -1& & & -1& & 1& -2& -1& -6& 16& 15& & -2& -1& \\
 & 4$\delta$& & & -1& & & & & 1& -3& 2& -7& 23& & & -4& \\
 & 5$\alpha$& -1& & & & & & & & & -1& & & 9& 3& & \\
 & 5$\beta$& -1& & & -1& & & -1& & & -1& -1& & 8& 10& 2& \\
 & 5$\gamma$& & -1& -1& & -1& & & -1& & 1& -2& -1& -5& 14& 13& -1\\
 & 5$\delta$& & & -1& & & -1& & & -1& & 1& -3& 2& -6& 13& \\
 & 6$\alpha$& & & & & & & & & & & & & & & & 3\\
 & 6$\beta$& -1& & & & & & & & & & & & & & & 7\\
 & 6$\gamma$& & & & & & & & & & & & & & & & 14\\
 & 6$\delta$& & & & & & & & & & & & & & & -1& -8\\
\end{tabular}
\caption{Correlation coefficients between data points of the inclusive jet measurement as a function of \Qsq\ and \ptjet\ and of the trijet measurement as a function of \Qsq\ and \meanpttri. 
 The bin labels are defined in \tab~\ref{tab:BinNumbering}. 
All values are multiplied by a factor of $100$.
\label{tab:CorrInclTrijet}}
\end{table}

\begin{table}[htbp]
\footnotesize
\center
\def\arraystretch{1}
\def\tabcolsep{1pt}
\begin{tabular}{cc|cccccccccccccccccc}
 & & \multicolumn{16}{c}{\textbf{Trijet as function of \bQsq and \bmeanpttri}} \\
 &  & 1$\alpha$ & 1$\beta$ & 1$\gamma$ & 2$\alpha$ & 2$\beta$ & 2$\gamma$ & 3$\alpha$ & 3$\beta$ & 3$\gamma$ & 4$\alpha$ & 4$\beta$ & 4$\gamma$ & 5$\alpha$ & 5$\beta$ & 5$\gamma$ & 6$\beta$ \\
\hline
\multirow{24}{*}{\begin{sideways}\textbf{Dijet as function of \bQsq and \bmeanptdi}~~~\end{sideways}}  & 1$\alpha$& 13& -7& 2& -2& 1& & & & & & & & & & & \\
 & 1$\beta$& 5& 16& -5& -2& -3& 1& -1& & & -1& & & -1& & & \\
 & 1$\gamma$& -4& 12& 21& 1& -3& -3& 1& -1& & 1& -1& & 1& -2& & \\
 & 1$\delta$& 2& -5& 15& & 1& -4& & & -2& & & -2& & & -2& \\
 & 2$\alpha$& -2& 1& & 13& -7& 2& -1& 1& & & & & & & & \\
 & 2$\beta$& -2& -2& 1& 5& 17& -5& -1& -2& 1& & & & & & & \\
 & 2$\gamma$& 1& -3& -3& -3& 8& 22& 1& -2& -3& & -1& & & -1& & \\
 & 2$\delta$& & 1& -3& 1& -4& 13& & 1& -2& & & -1& & & -3& \\
 & 3$\alpha$& & & & -1& & & 14& -7& 3& -1& 1& & & & & \\
 & 3$\beta$& -1& & & -1& -2& 1& 3& 17& -7& -1& -1& 1& & & & \\
 & 3$\gamma$& 1& -1& & 1& -2& -3& -4& 12& 23& 1& -2& -2& & -1& & \\
 & 3$\delta$& & & & & 1& -2& 3& -8& 20& & 1& -2& & & -1& \\
 & 4$\alpha$& & & & & & & -1& & & 12& -6& 3& & & & \\
 & 4$\beta$& -1& & & & & & -1& -1& 1& 4& 17& -7& & -1& & \\
 & 4$\gamma$& & -1& -1& & -1& & 1& -2& -2& -4& 9& 29& 1& -1& -1& \\
 & 4$\delta$& & & -1& & & & & 1& -2& 2& -7& 18& & & -4& \\
 & 5$\alpha$& & & & & & & & & & & & & 13& -6& 2& \\
 & 5$\beta$& -1& & & & & & & & & & -1& & 4& 16& -5& \\
 & 5$\gamma$& 1& -1& -1& 1& -1& & 1& -1& & 1& -2& -1& -4& 11& 18& \\
 & 5$\delta$& & & -1& & & & & & & & 1& -2& 2& -5& 11& \\
 & 6$\alpha$& & & & & & & & & & & & & & & & -10\\
 & 6$\beta$& & & & & & & & & & & & & & & & 22\\
 & 6$\gamma$& & -1& & & & & & -1& & & & & & -1& -1& 4\\
 & 6$\delta$& & & & & & & & & & & & & & & -1& -5\\
\end{tabular}
\caption{Correlation coefficients between data points of the dijet measurement as a function of \Qsq\ and \meanptdi\ and of the trijet measurement as a function of \Qsq\ and \meanpttri. 
 The bin labels are defined in \tab~\ref{tab:BinNumbering}. 
All values are multiplied by a factor of $100$.
\label{tab:CorrDijetTrijet}}
\end{table}

\newpage
\begin{table}[htbp]
\footnotesize
\center
\def\arraystretch{1}
\def\tabcolsep{1pt}
\begin{tabular}{cc|cccccccccccccccccccccccc}
 & & \multicolumn{20}{c}{\textbf{Dijet as function of \bQsq and \bxidi}} \\
 &  & 1a & 1b & 1c & 1d & 2a & 2b & 2c & 2d & 3a & 3b & 3c & 3d & 4a & 4b & 4c & 4d & 5b & 5c & 5d & 6d \\
\hline
\multirow{20}{*}{\begin{sideways}\textbf{Dijet as function of \bQsq and \bxidi}~~~\end{sideways}}  & 1a&  100 & -35& 2& -23& -14& 1& & 3& 3& & & & 3& & & & -1& & & \\
 & 1b& &  100 & -39& 14& 5& -10& 4& -1& & 4& -1& & 1& 3& & 1& 3& -1& 2& \\
 & 1c& & &  100 & -25& & 5& -10& 3& & -1& 3& -1& & 1& 2& -1& & 2& -1& \\
 & 1d& & & &  100 & 3& -1& 3& -14& & & -1& 2& & & & 1& & & 1& \\
 & 2a& & & & &  100 & -22& 13& -22& 2& & -1& 1& 1& & -1& & -1& & & -1\\
 & 2b& & & & & &  100 & -38& 8& 5& -7& 3& -1& 1& 3& -1& & 4& -1& 1& \\
 & 2c& & & & & & &  100 & -26& -2& 4& -7& 3& & & 1& -1& & 2& -1& \\
 & 2d& & & & & & & &  100 & -3& -1& 3& -10& & & & 1& & & 1& 1\\
 & 3a& & & & & & & & &  100 & -33& 15& -7& -6& 2& -2& & -2& & & \\
 & 3b& & & & & & & & & &  100 & -34& 9& 2& -3& 1& & 6& -1& 1& 1\\
 & 3c& & & & & & & & & & &  100 & -27& -1& 3& -3& 1& & 4& -1& -1\\
 & 3d& & & & & & & & & & & &  100 & & -1& 2& -5& & & 1& \\
 & 4a& & & & & & & & & & & & &  100 & -13& 12& -17& 3& -2& 1& \\
 & 4b& & & & & & & & & & & & & &  100 & -29& -9& 4& & -1& \\
 & 4c& & & & & & & & & & & & & & &  100 & -20& 3& 6& -2& -1\\
 & 4d& & & & & & & & & & & & & & & &  100 & -2& 1& & \\
 & 5b& & & & & & & & & & & & & & & & &  100 & -16& 21& \\
 & 5c& & & & & & & & & & & & & & & & & &  100 & -24& -3\\
 & 5d& & & & & & & & & & & & & & & & & & &  100 & 14\\
 & 6d& & & & & & & & & & & & & & & & & & & &  100 \\
\end{tabular}
\caption{Correlation coefficients between data points of the dijet measurement as a function of \Qsq\ and $\xidi$. Since the matrix is symmetric only the upper triangle is given.
 The bin labels are defined in \tab~\ref{tab:BinNumbering}. 
 All values are multiplied by a factor of $100$.
\label{tab:CorrDijetXiDijetXi}}
\end{table}

\begin{table}[htbp]
\footnotesize
\center
\def\arraystretch{1}
\def\tabcolsep{1pt}
\begin{tabular}{cc|cccccccccccccccccc}
 & & \multicolumn{15}{c}{\textbf{Trijet as function of \bQsq and \bxitri}} \\
 &  & 1A & 1B & 1C & 2A & 2B & 2C & 3A & 3B & 3C & 4A & 4B & 4C & 5B & 5C & 6C \\
\hline
\multirow{15}{*}{\begin{sideways}\textbf{Trijet as function of \bQsq and \bxitri}~~~\end{sideways}}  & 1A&  100 & -35& 12& -5& 2& -1& 11& -3& & 6& -2& 1& -3& 2& 1\\
 & 1B& &  100 & -43& 3& -8& 1& -2& 6& -1& -1& 6& -1& 8& -5& -2\\
 & 1C& & &  100 & & 4& -8& 1& -1& 2& & -1& 3& -2& 6& 3\\
 & 2A& & & &  100 & -33& 9& -4& 1& 1& 5& -2& 1& -2& 1& \\
 & 2B& & & & &  100 & -43& 3& -7& 3& -1& 6& -1& 6& -3& -2\\
 & 2C& & & & & &  100 & -2& 5& -6& & -1& 3& -2& 6& 3\\
 & 3A& & & & & & &  100 & -37& 8& -4& 1& -1& -2& & \\
 & 3B& & & & & & & &  100 & -36& 2& & & 8& -2& -2\\
 & 3C& & & & & & & & &  100 & -1& 2& -2& & 5& 3\\
 & 4A& & & & & & & & & &  100 & -39& 11& & & 1\\
 & 4B& & & & & & & & & & &  100 & -36& 6& -3& -2\\
 & 4C& & & & & & & & & & & &  100 & -1& 5& 4\\
 & 5B& & & & & & & & & & & & &  100 & -33& -4\\
 & 5C& & & & & & & & & & & & & &  100 & 10\\
 & 6C& & & & & & & & & & & & & & &  100 \\
\end{tabular}
\caption{Correlation coefficients between data points of the trijet measurement as a function of \Qsq\ and $\xidi$. Since the matrix is symmetric only the upper triangle is given. 
 The bin labels are defined in \tab~\ref{tab:BinNumbering}. 
All values are multiplied by a factor of $100$.}
\label{tab:CorrTrijetXiTrijetXi}
\end{table}

\begin{table}[htbp]
\footnotesize
\center
\def\arraystretch{1}
\def\tabcolsep{1pt}
\begin{tabular}{cc|cccccccccccccccccccccccc}
 & & \multicolumn{20}{c}{\textbf{Dijet as function of \bQsq and \bxidi}} \\
 &  & 1a & 1b & 1c & 1d & 2a & 2b & 2c & 2d & 3a & 3b & 3c & 3d & 4a & 4b & 4c & 4d & 5b & 5c & 5d & 6d \\
\hline
\multirow{24}{*}{\begin{sideways}\textbf{Inclusive jet as function of \bQsq and \bptjet}~~~\end{sideways}}  & 1$\alpha$& 28& 7& 6& & -3& -2& -1& & 1& & & & 1& & & & & & & \\
 & 1$\beta$& 4& 21& 9& 3& & -2& -1& -1& 1& 1& & & & & & & & & & \\
 & 1$\gamma$& & 4& 13& 24& & -1& -2& -3& & & 1& 1& & & & & & & & \\
 & 1$\delta$& 6& & -3& 28& -1& & & -4& & & & 1& & & & & & & & \\
 & 2$\alpha$& -3& -1& -1& & 26& 10& 7& & -3& -1& -1& & & & & & & & & \\
 & 2$\beta$& & -3& -1& & 2& 19& 12& 2& -2& -2& -2& & & & & & & & & \\
 & 2$\gamma$& -1& & -2& -3& 2& 4& 12& 26& 1& -1& -2& -4& & & & & & & & \\
 & 2$\delta$& -2& & & -4& 5& 1& -2& 31& 1& & & -3& & & & & & & & \\
 & 3$\alpha$& 1& 1& & & -2& -1& -1& & 31& 14& 9& & -1& -1& -1& & & & & \\
 & 3$\beta$& & 1& & & -1& -2& -2& & 6& 22& 12& 1& & -2& -1& & & & & \\
 & 3$\gamma$& & & 1& 1& & & -2& -3& -1& 3& 15& 30& & & -1& -3& & & & \\
 & 3$\delta$& & & -1& 1& -1& & & -3& & 2& -4& 31& & & & -1& & & & \\
 & 4$\alpha$& & & & & & 1& & & -2& -1& -1& & 21& 20& 11& -1& -1& -1& -1& \\
 & 4$\beta$& & & & & & & & & & -1& -1& & -2& 16& 14& 6& -1& -1& -1& \\
 & 4$\gamma$& & & & & & & & & & & -1& -2& 1& 2& 16& 27& & & -1& \\
 & 4$\delta$& & & & & & & & & & & & -2& 2& 2& -2& 24& & & & \\
 & 5$\alpha$& & & & & & & & & & 1& & & -1& -1& & & 27& 19& 15& \\
 & 5$\beta$& & & 1& & & & & & & & & & & -1& -1& & 11& 20& 21& \\
 & 5$\gamma$& & & & & & & & -1& & & & & & & & -1& & 10& 24& -1\\
 & 5$\delta$& & & & & & & & & & & & & & & 1& -1& 2& -2& 18& \\
 & 6$\alpha$& & & & & & & & & & & & & & & & & & -1& 4& 34\\
 & 6$\beta$& & & & & & & & & & & & & & & & & & & & 27\\
 & 6$\gamma$& & & & & & & & & & & & & & & & & & & & 16\\
 & 6$\delta$& & & & & & & & & & & & & & & & & & & -1& 8\\
\end{tabular}
\caption{Correlation coefficients between data points of the inclusive jet measurement as a function of \Qsq\ and \ptjet\ and of the dijet measurement as a function of \Qsq\ and $\xidi$. 
 The bin labels are defined in \tab~\ref{tab:BinNumbering}. 
All values are multiplied by a factor of $100$.
\label{tab:CorrInclDijetXi}}
\end{table}

\begin{table}[htbp]
\footnotesize
\center
\def\arraystretch{1}
\def\tabcolsep{1pt}
\begin{tabular}{cc|cccccccccccccccccc}
 & & \multicolumn{15}{c}{\textbf{Trijet as function of \bQsq and \bxitri}} \\
 &  & 1A & 1B & 1C & 2A & 2B & 2C & 3A & 3B & 3C & 4A & 4B & 4C & 5B & 5C & 6C \\
\hline
\multirow{24}{*}{\begin{sideways}\textbf{Inclusive jet as function of \bQsq and \bptjet}~~~\end{sideways}}  & 1$\alpha$& 17& 6& 1& -3& -1& & -1& 1& & & & & & & \\
 & 1$\beta$& 8& 15& 8& -2& -3& -1& & -1& & & -1& & -1& & \\
 & 1$\gamma$& & 7& 21& & -2& -3& & & & & -1& -1& -1& -1& \\
 & 1$\delta$& 1& -1& 11& & & -2& & & & & & & & -1& -1\\
 & 2$\alpha$& -3& -1& & 14& 6& 2& -2& -1& & & & & & & \\
 & 2$\beta$& -2& -3& -1& 8& 14& 5& -1& -2& & & -1& & -1& & \\
 & 2$\gamma$& & -1& -4& -1& 8& 20& & -1& -2& & & & & -1& -1\\
 & 2$\delta$& & & -2& 1& 1& 13& & & -1& & & & & & \\
 & 3$\alpha$& & & & -1& -1& & 17& 8& 3& -1& -1& & & & \\
 & 3$\beta$& -1& & & -1& -2& -1& 5& 16& 5& -1& -2& -1& & -1& \\
 & 3$\gamma$& & & & & -1& -3& -1& 6& 16& & & -2& & -1& -1\\
 & 3$\delta$& & & & & & -1& & & 11& & & -1& & & \\
 & 4$\alpha$& & & & & & & -1& -1& & 14& 9& 4& -1& & \\
 & 4$\beta$& & -1& & & & & & -2& & 4& 15& 7& -1& -1& \\
 & 4$\gamma$& & & & & & & & & -2& -2& 6& 25& & -2& \\
 & 4$\delta$& & & & & & & & & -1& 2& 1& 8& & -1& \\
 & 5$\alpha$& & & & & & & & & & & -1& & 11& 9& -1\\
 & 5$\beta$& & & -1& & & -1& & & -1& & -1& -1& 13& 12& -1\\
 & 5$\gamma$& & & -1& & & & & & -1& & & -1& 3& 19& -1\\
 & 5$\delta$& & & & & & & & & & & & -1& -1& 7& \\
 & 6$\alpha$& & & & & & & & & & & & & & -1& 24\\
 & 6$\beta$& & & -1& & & -1& & & -1& & & -1& & -2& 18\\
 & 6$\gamma$& & & & & & & & & & & & & & & 21\\
 & 6$\delta$& & & & & & & & & & & & & & -1& 11\\
\end{tabular}
\caption{Correlation coefficients between data points of the inclusive jet measurement as a function of \Qsq\ and \ptjet\ and the data points of the trijet measurement as a function of \Qsq\ and $\xitri$. 
 The bin labels are defined in \tab~\ref{tab:BinNumbering}. 
All values are multiplied by a factor of $100$.
\label{tab:CorrInclTrijetXi}}
\end{table}

\begin{table}[htbp]
\footnotesize
\center
\def\arraystretch{1}
\def\tabcolsep{1pt}
\begin{tabular}{cc|cccccccccccccccccc}
 & & \multicolumn{15}{c}{\textbf{Trijet as function of \bQsq and \bxitri}} \\
 &  & 1A & 1B & 1C & 2A & 2B & 2C & 3A & 3B & 3C & 4A & 4B & 4C & 5B & 5C & 6C \\
\hline
\multirow{20}{*}{\begin{sideways}\textbf{Dijet as function of \bQsq and \bxidi}~~~\end{sideways}}  & 1a& 16& -2& & -3& & -1& -2& & & -1& & & & & \\
 & 1b& & 16& -2& & -3& 1& & & & & & & & & \\
 & 1c& -1& 5& 9& 1& -2& -2& 1& -1& & 1& -2& & -2& & \\
 & 1d& 4& -5& 17& -1& 1& -3& & & -1& & & -1& 1& -2& -1\\
 & 2a& -4& & & 12& -2& 1& -2& & & & & & & & \\
 & 2b& & -3& 1& 1& 15& -4& & -2& & & & & -1& 1& \\
 & 2c& 1& -2& -2& & 4& 11& 1& -2& -1& & -1& & -1& & \\
 & 2d& -1& 1& -3& 3& -2& 15& -1& 1& -2& & & & & -1& -1\\
 & 3a& -2& & & -2& 1& & 23& -7& 3& -3& & & & & \\
 & 3b& & & & & -2& 1& 2& 19& -4& & -2& & & & \\
 & 3c& & -1& & & -1& -2& -1& 5& 11& & -1& -1& -1& & \\
 & 3d& & & & & & -2& 3& -2& 13& & 1& -2& & -1& -1\\
 & 4a& -2& & & & & & -2& 1& & 17& -7& 2& & -1& \\
 & 4b& & -1& & & & & -1& -2& & 5& 15& -3& -1& & \\
 & 4c& 1& -1& & & -1& & & -2& -1& -3& 9& 12& -2& & \\
 & 4d& & 1& -1& & & & -1& 1& -1& 4& -3& 16& 1& -2& -1\\
 & 5b& -1& -1& & & & & & -1& & & -1& & 9& 1& -1\\
 & 5c& 1& -1& & 1& -1& -1& & -1& & & -2& -1& 15& 6& -1\\
 & 5d& -1& 1& -1& & & -1& & 1& -1& & & -2& 3& 16& -2\\
 & 6d& & & -1& & & -1& & & -1& & & -1& 1& -3& 29\\
\end{tabular}
\caption{Correlation coefficients between data points of the dijet measurement as a function of \Qsq\ and $\xidi$ and of the trijet measurement as a function of \Qsq\ and $\xitri$. 
 The bin labels are defined in \tab~\ref{tab:BinNumbering}. 
All values are multiplied by a factor of $100$.
\label{tab:CorrDijetXiTrijetXi}}
\end{table}

\clearpage

\begin{table}[tbhp]
\center
\footnotesize
\begin{tabular}[h!]{l|ccccc}
\multicolumn{6}{c}{\basmz \textbf{using different PDF sets} } \\[0.0cm] 
\hline\rule{0pt}{3ex}
 Measurement & $\as^{\rm MSTW2008}$
& $\as^{\rm CT10}$
& $\as^{\rm NNPDF 2.3}$
& $\as^{\rm HERAPDF1.5}$
& $\as^{\rm ABM11}$  \\
& \multicolumn{5}{c}{All PDF sets used were determined with $\asmz=0.1180$ }  \\[0.0cm] 
\hline
\sI	& 0.1174 & 0.1180 & 0.1167 & 0.1158 & 0.1136 \\
\sD	& 0.1137 & 0.1142 & 0.1127 & 0.1120 & 0.1101 \\
\sT	& 0.1178 & 0.1178 & 0.1169 & 0.1174 & 0.1176 \\
\hline
 & & \\[-2.8ex]
\sIN	& 0.1176 & 0.1185 & 0.1170 & 0.1183 & 0.1186 \\[1.2em]
\sDN	& 0.1135 & 0.1143 & 0.1127 & 0.1143 & 0.1150 \\[1.2em]
\sTN	& 0.1182 & 0.1185 & 0.1175 & 0.1191 & 0.1204 \\[0.6em]
\hline
$[\sI,\sD,\sT]$	& 0.1185 & 0.1187 & 0.1178 & 0.1180 & 0.1176 \\[0.4em]
$\left[\sIN,\sDN,\sTN\right]$   & 0.1165 & 0.1172 & 0.1158 & 0.1172 & 0.1177 \\[0.7em]
\hline
\end{tabular}
\caption{Values for \asmz\ obtained from fits to absolute and normalized cross sections using different PDF sets.
}
\label{tab:AsPDFGroups}
\end{table}

\begin{table}[htbp]
\center
\footnotesize
\resizebox{\columnwidth}{!}{
\begin{tabular}[htbp]{l @{~}ll|c@{\,$\,$}c@{\,$\,$}c@{\,$\,$}c@{\,$\,$}c@{\,$\,$}cc}
\multicolumn{10}{c}{\textbf{Summary of values of }$\basmz$\textbf{ and uncertainties}} \\
\hline 
Measurement & ${\asmz|_\kt}$ & ${\asmz|_\antikt}$ & \multicolumn{7}{|c}{\kd{PDF and theoretical uncertainties}}\\ 
 & & & \multicolumn{6}{|c}{\kd{Individual contributions}} & \kd{Total}\\
\hline 
\sI                   & $0.1174\,(22)_{\rm exp}$  & $0.1175\,(22)_{\rm exp}$ & $(7)_{\rm PDF}$ & $(7)_{\rm PDFset}$ & $(5)_{\rm PDF(\as)}$ & $(10)_{\rm had}$ & $(48)_{\mur}$ & $(6)_{\muf}$ & \kd{$(50)_{\rm pdf,theo}$}\\
\sD                   & $0.1137\,(23)_{\rm exp}$  & $0.1152\,(23)_{\rm exp}$ & $(7)_{\rm PDF}$ & $(7)_{\rm PDFset}$ & $(5)_{\rm PDF(\as)}$ & $(7)_{\rm had}$  & $(37)_{\mur}$ & $(6)_{\muf}$ & \kd{$(39)_{\rm pdf,theo}$}\\
\sT                   & $0.1178\,(17)_{\rm exp}$  & $0.1174\,(18)_{\rm exp}$ & $(3)_{\rm PDF}$ & $(5)_{\rm PDFset}$ & $(0)_{\rm PDF(\as)}$ & $(11)_{\rm had}$ & $(34)_{\mur}$ & $(3)_{\muf}$ & \kd{$(36)_{\rm pdf,theo}$}\\
\hline
 & & & & & & & & & \\[-2.8ex]
\sIN                   & $0.1176\,(9)_{\rm exp}$   & $0.1172\,(8)_{\rm exp}$  & $(6)_{\rm PDF}$ & $(7)_{\rm PDFset}$ & $(4)_{\rm PDF(\as)}$ & $(8)_{\rm had}$  & $(41)_{\mur}$ & $(6)_{\muf}$ & \kd{$(43)_{\rm pdf,theo}$}\\[1.2em]
\sDN                  & $0.1135\,(10)_{\rm exp}$  & $0.1147\,(9)_{\rm exp}$  & $(5)_{\rm PDF}$ & $(8)_{\rm PDFset}$ & $(3)_{\rm PDF(\as)}$ & $(6)_{\rm had}$  & $(32)_{\mur}$ & $(6)_{\muf}$ & \kd{$(34)_{\rm pdf,theo}$}\\[1.2em]
\sTN                  & $0.1182\,(11)_{\rm exp}$  & $0.1177\,(12)_{\rm exp}$ & $(3)_{\rm PDF}$ & $(5)_{\rm PDFset}$  & $(0)_{\rm PDF(\as)}$ & $(11)_{\rm had}$ & $(34)_{\mur}$ & $(3)_{\muf}$& \kd{$(36)_{\rm pdf,theo}$} \\[0.6em]
\hline
$[\sI,\sD,\sT]$      & $0.1185\,(16)_{\rm exp}$ &  $0.1181\,(17)_{\rm exp}$ & $(3)_{\rm PDF}$ & $(4)_{\rm PDFset}$ & $(2)_{\rm PDF(\as)}$ & $(13)_{\rm had}$ & $(38)_{\mur}$ & $(3)_{\muf}$& \kd{$(40)_{\rm pdf,theo}$}\\[0.4em]
$\left[\sIN,\sDN,\sTN\right]$  & $0.1165\,(8)_{\rm exp}$  &  $0.1165\,(7)_{\rm exp}$  & $(5)_{\rm PDF}$ & $(7)_{\rm PDFset}$ & $(3)_{\rm PDF(\as)}$ & $(8)_{\rm had}$ & $(36)_{\mur}$ & $(5)_{\muf}$& \kd{$(37)_{\rm pdf,theo}$}\\[0.7em]
\hline
\end{tabular}%
 } 
\caption{
  Values of \asmz\ obtained from fits to absolute and normalised single jet and multijet cross sections employing the \kt or the \antikt jet algorithm. Theoretical uncertainties are quoted for the fits to the \kt\ jet cross sections.
}
\label{tab:AsValuesAll}
\end{table}

\begin{table}[htbp]
\footnotesize
\center
\resizebox{\columnwidth}{!}{%
\begin{tabular}{c@{\,}r@{\,}cc|c@{\,$\,$}c@{\,$\,$}c@{\,$\,$}c@{\,$\,$}c@{\,$\,$}cc}
  \multicolumn{11}{c}{\basmz \textbf{ from data points with comparable } $\bmur$\textbf{-values}} \\
\hline \rule{0pt}{2ex} 
$\langle\mur\rangle$
        & \multicolumn{1}{c}{No. of}
        & $\asmz|_\kt$
        & $\asmz|_\antikt$
        & \multicolumn{7}{c}{{PDF and theoretical uncertainties}} \\[-0.15cm]
        $[\GeV]$
& \multicolumn{1}{c}{data points}
&
&
& \multicolumn{6}{c}{Individual contributions} & Total \\
\hline 
  11.9 &  9~~~~~~~ & $0.1168\,(10)_{\rm exp} $ & $0.1174\,(10)_{\rm exp}$ & $(6)_{\rm PDF}$ & $(10)_{\rm PDFset}$ & $(5)_{\rm PDF(\as)}$ & $(10)_{\rm had}$  & $(43)_{\mur}$ & $(6)_{\muf}$ & $(47)_{\rm pdf,theo}$ \\[0.0cm]
  14.1 &  6~~~~~~~ & $0.1155\,(16)_{\rm exp} $ & $0.1159\,(14)_{\rm exp}$ & $(6)_{\rm PDF}$ & $(11)_{\rm PDFset}$ & $(3)_{\rm PDF(\as)}$ & $(9)_{\rm had}$  & $(37)_{\mur}$ & $(5)_{\muf}$ & $(40)_{\rm pdf,theo}$ \\[0.0cm]
  17.4 & 18~~~~~~~ & $0.1174\,(13)_{\rm exp} $ & $0.1163\,(13)_{\rm exp}$ & $(5)_{\rm PDF}$ & $(12)_{\rm PDFset}$ & $(1)_{\rm PDF(\as)}$ & $(7)_{\rm had}$  & $(34)_{\mur}$ & $(5)_{\muf}$ & $(37)_{\rm pdf,theo}$ \\[0.0cm]
  25.6 & 22~~~~~~~ & $0.1153\,(14)_{\rm exp} $ & $0.1150\,(14)_{\rm exp}$ & $(4)_{\rm PDF}$ & $(11)_{\rm PDFset}$ & $(2)_{\rm PDF(\as)}$ & $(5)_{\rm had}$  & $(28)_{\mur}$ & $(5)_{\muf}$ & $(31)_{\rm pdf,theo}$ \\[0.0cm]
  59.6 &  9~~~~~~~ & $0.1169\,(66)_{\rm exp} $ & $0.1185\,(60)_{\rm exp}$ & $(10)_{\rm PDF}$ & $(9)_{\rm PDFset}$ & $(1)_{\rm PDF(\as)}$ & $(4)_{\rm had}$  & $(29)_{\mur}$ & $(8)_{\muf}$ & $(32)_{\rm pdf,theo}$ \\[0.0cm]
\hline
\multicolumn{11}{c}{} \\
\multicolumn{11}{c}{} \\
  \multicolumn{11}{c}{\basmur \textbf{ from data points with comparable } $\bmur$\textbf{-values}} \\
\hline
$\langle\mur\rangle$
        & \multicolumn{1}{c}{No. of}
        & $\asmur|_\kt$
        & $\asmur|_\antikt$
        & \multicolumn{7}{c}{{PDF and theoretical uncertainties at \mur}} \\[-0.15cm]
        $[\GeV]$
& \multicolumn{1}{c}{data points}
&
&
& \multicolumn{6}{c}{Individual contributions} & Total \\
\hline 
  11.9 &  9~~~~~~~ & $0.1684\,(22)_{\rm exp} $ & $0.1697\,(21)_{\rm exp}$ & $(13)_{\rm PDF}$ & $(21)_{\rm PDFset}$ & $(11)_{\rm PDF(\as)}$ & $(21)_{\rm had}$  & $(91)_{\mur}$ & $(13)_{\muf}$ & $(100)_{\rm pdf,theo}$ \\[0.0cm] 
  14.1 &  6~~~~~~~ & $0.1600\,(31)_{\rm exp} $ & $0.1605\,(28)_{\rm exp}$ & $(12)_{\rm PDF}$ & $(21)_{\rm PDFset}$ & $(6)_{\rm PDF(\as)}$  & $(18)_{\rm had}$  & $(72)_{\mur}$ & $(10)_{\muf}$ & $(79)_{\rm pdf,theo}$ \\[0.0cm] 
  17.4 & 18~~~~~~~ & $0.1567\,(24)_{\rm exp} $ & $0.1546\,(23)_{\rm exp}$ & $(9)_{\rm PDF}$ & $(22)_{\rm PDFset}$ & $(2)_{\rm PDF(\as)}$   & $(13)_{\rm had}$  & $(61)_{\mur}$ & $(9)_{\muf}$ & $(67)_{\rm pdf,theo}$ \\[0.0cm] 
  25.6 & 22~~~~~~~ & $0.1420\,(22)_{\rm exp} $ & $0.1415\,(21)_{\rm exp}$ & $(6)_{\rm PDF}$ & $(17)_{\rm PDFset}$ & $(3)_{\rm PDF(\as)}$   & $(8)_{\rm had}$  & $(43)_{\mur}$ & $(8)_{\muf}$ & $(47)_{\rm pdf,theo}$ \\[0.0cm] 
  59.6 &  9~~~~~~~ & $0.1248\,(76)_{\rm exp} $ & $0.1267\,(68)_{\rm exp}$ & $(10)_{\rm PDF}$ & $(10)_{\rm PDFset}$ & $(1)_{\rm PDF(\as)}$  & $(5)_{\rm had}$  & $(33)_{\mur}$ & $(9)_{\muf}$ & $(37)_{\rm pdf,theo}$ \\[0.0cm] 
\hline
\end{tabular}%
 } 
\caption{Values of \asmz\ \kd{and \asmur} from five fits to groups of data points with comparable value of the renormalisation scale from normalised multijet cross sections. The cross section weighted average value of the renormalisation scale is also given. Theoretical uncertainties are quoted for the fits to the normalised \kt\ jet cross sections.}
\label{tab:AsMu}
\end{table}

\clearpage

\clearpage

\clearpage


\clearpage

\clearpage

\clearpage

\clearpage

\clearpage

\clearpage

%

\begin{thebibliography}{10}

\bibitem{Fritzsch:72:135}
H.~Fritzsch and M.~Gell-Mann, Proceedings of "16th Int. Conference on
  High-Energy Physics", Batavia IL, 1972, 135,
  \href{http://arxiv.org/abs/hep-ph/0208010}{ [arXiv:hep-ph/0208010]}.

\bibitem{Gross:73:1343}
D.~Gross and F.~Wilczek,
  \href{http://dx.doi.org/10.1103/PhysRevLett.30.1343}{Phys. Rev.
  Lett.{\bfseries ~30} (1973) 1343}.

\bibitem{Politzer:73:1346}
H.~Politzer, \href{http://dx.doi.org/10.1103/PhysRevLett.30.1346}{Phys. Rev.
  Lett.{\bfseries ~30} (1973) 1346}.

\bibitem{Ali:2010tw}
A.~Ali and G.~Kramer, \href{http://dx.doi.org/10.1140/epjh/e2011-10047-1}{Eur.
  Phys. J.{\bfseries ~H 36} (2011) 245}, \href{http://arxiv.org/abs/1012.2288}{
  [arXiv:1012.2288]}, and references therein.

\bibitem{SchornerSadenius:2012de}
T.~Sc{h\"o}rner-Sadenius, \href{http://dx.doi.org/10.1140/epjc/s10052-012-2060-8}
  {Eur. Phys. J. C{\bfseries ~72} (2012) 2060}, Erratum ibid, 72 (2012) 2133.

\bibitem{Feynman:72}
R.~Feynman, {\em Photon -- Hadron Interactions}.
\newblock Benjamin, New York, 1972.

\bibitem{Streng:1979pv}
K.~Streng, T.~Walsh, and P.~Zerwas,
\href{http://dx.doi.org/10.1007/BF01474667}{Z. Phys.{\bfseries ~C 2} (1979)
  237}.

\bibitem{Abramowicz:10:965}
 ZEUS Collaboration, H.~Abramowicz {\em et~al.},
  \href{http://dx.doi.org/10.1140/epjc/s10052-010-1504-2}{Eur. Phys. J.
  C{\bfseries ~70} (2010) 965}, \href{http://arxiv.org/abs/1010.6167}{
  [arXiv:1010.6167]}.

\bibitem{Abramowicz:10:127}
 ZEUS Collaboration, H.~Abramowicz {\em et~al.},
  \href{http://dx.doi.org/10.1016/j.physletb.2010.06.015}{Phys. Lett.
  B{\bfseries ~691} (2010) 127}, \href{http://arxiv.org/abs/1003.2923}{
  [arXiv:1003.2923]}.

\bibitem{Aaron:10:1}
 H1 Collaboration, F.~D. Aaron {\em et~al.},
  \href{http://dx.doi.org/10.1140/epjc/s10052-010-1282-x}{Eur. Phys. J.
  C{\bfseries ~67} (2010) 1}, \href{http://arxiv.org/abs/0911.5678}{
  [arXiv:0911.5678]}.

\bibitem{Aaron:10:363}
 H1 Collaboration, F.~D. Aaron {\em et~al.},
  \href{http://dx.doi.org/10.1140/epjc/s10052-009-1208-7}{Eur. Phys. J.
  C{\bfseries ~65} (2010) 363}, \href{http://arxiv.org/abs/0904.3870}{
  [arXiv:0904.3870]}.

\bibitem{Ellis:93:3160}
S.~Ellis and D.~Soper, \href{http://dx.doi.org/10.1103/PhysRevD.48.3160}{Phys.
  Rev. D{\bfseries ~48} (1993) 3160},
  \href{http://arxiv.org/abs/hep-ph/9305266}{ [hep-ph/9305266]}.

\bibitem{Cacciari:08:063}
M.~Cacciari, G.~P. Salam, and G.~Soyez,
  \href{http://dx.doi.org/10.1088/1126-6708/2008/04/063}{JHEP{\bfseries ~0804}
  (2008) 063}, \href{http://arxiv.org/abs/0802.1189}{ [arXiv:0802.1189]}.

\bibitem{Kogler:10}
R.~Kogler, {\em Measurement of jet production in deep-inelastic $ep$ scattering
  at HERA}.
\newblock Dissertation, Universit{\"a}t Hamburg, DESY-THESIS-2011-003,
  MPP-2010-175, 2010.
\newblock Available at
  \url{http://www-h1.desy.de/publications/theses\_list.html}.

\bibitem{Abt:97:310}
 H1 Collaboration, I.~Abt {\em et~al.},
  \href{http://dx.doi.org/10.1016/S0168-9002(96)00893-5}{Nucl. Instr. and Meth.
  A{\bfseries ~386} (1997) 310}.

\bibitem{Abt:97:348}
 H1 Collaboration, I.~Abt {\em et~al.},
  \href{http://dx.doi.org/10.1016/S0168-9002(96)00894-7}{Nucl. Instr. and Meth.
  A{\bfseries ~386} (1997) 348}.

\bibitem{Andrieu:93:460}
 Calorimeter Group of H1, B.~Andrieu {\em et~al.},
  \href{http://dx.doi.org/10.1016/0168-9002(93)91257-N}{Nucl. Instr. and Meth.
  A{\bfseries ~336} (1993) 460}.

\bibitem{Andrieu:94:57}
 Calorimeter Group of H1, B.~Andrieu {\em et~al.},
  \href{http://dx.doi.org/10.1016/0168-9002(94)91155-X}{Nucl. Instr. and Meth.
  A{\bfseries ~350} (1994) 57}.

\bibitem{Andrieu:93:499}
 Calorimeter Group of H1, B.~Andrieu {\em et~al.},
  \href{http://dx.doi.org/10.1016/0168-9002(93)91258-O}{Nucl. Instr. and Meth.
  A{\bfseries ~336} (1993) 499}.

\bibitem{Appuhn:97:397}
 SPACAL Group of H1, R.~Appuhn {\em et~al.},
  \href{http://dx.doi.org/10.1016/S0168-9002(96)01171-0}{Nucl. Instr. and Meth.
  A{\bfseries ~386} (1997) 397}.

\bibitem{Appuhn:96:395}
 SPACAL Group of H1, R.~Appuhn {\em et~al.},
  \href{http://dx.doi.org/10.1016/S0168-9002(96)00769-3}{Nucl. Instr. and Meth.
  A{\bfseries ~382} (1996) 395}.

\bibitem{Pitzl:00:334}
D.~Pitzl {\em et~al.},
  \href{http://dx.doi.org/10.1016/S0168-9002(00)00488-5}{Nucl. Instr. and Meth.
  A{\bfseries ~454} (2000) 334}, \href{http://arxiv.org/abs/hep-ex/0002044}{
  [hep-ex/0002044]}.

\bibitem{Aaron:2012kn}
 H1 Collaboration, F.~D. Aaron {\em et~al.},
  \href{http://dx.doi.org/10.1140/epjc/s10052-012-2163-2}{Eur. Phys. J.
  C{\bfseries ~72} (2012) 2163}, \href{http://arxiv.org/abs/1205.2448}{
  [arXiv:1205.2448]}
Erratum ibid, 74 (2014) 2733.

\bibitem{Peez:03}
M.~Peez, {\em Search for deviations from the standard model in high transverse
  energy processes at the electron proton collider HERA (in French)}.
\newblock Dissertation, Univ. Claude Bernard, Lyon, DESY-THESIS-2003-023,
  CPPM-T-2003-04, 2003.
\newblock Available at
  \url{http://www-h1.desy.de/publications/theses\_list.html}.

\bibitem{Hellwig:04}
S.~Hellwig, {\em Investigation of the $D^{*} - \pi_{slow}$ double tagging
  method for the analysis of charm (in German}.
\newblock Diploma thesis, Hamburg University, 2004.
\newblock Available at
  \url{http://www-h1.desy.de/publications/theses\_list.html}.

\bibitem{Portheault:05}
B.~Portheault, {\em First measurement of charged and neutral current cross
  sections with the polarized positron beam at HERA II and QCD-electroweak
  analyses (in French)}.
\newblock Dissertation, Univ. Paris XI Orsay, LAL-05-05, 2005.
\newblock Available at
  \url{http://www-h1.desy.de/publications/theses\_list.html}.

\bibitem{Feindt:2004wla}
M.~Feindt,
\href{http://arxiv.org/abs/physics/0402093}{ physics/0402093}.

\bibitem{Feindt:06:190}
M.~Feindt and U.~Kerzel,
  \href{http://dx.doi.org/10.1016/j.nima.2005.11.166}{Nucl. Instr. and Meth.
  A{\bfseries ~559} (2006) 190}.

\bibitem{Bentvelsen:1992fu}
S.~Bentvelsen, J.~Engelen, and P.~Kooijman, Proceedings of "Physics at HERA",
  eds. Buchm{\"u}ller and G. Ingelman, DESY, Hamburg, vol. 1, 1992, 23
NIKHEF-H-92-02.

\bibitem{Hoeger:1992}
K.~C. Hoeger, Proceedings of "Physics at HERA", eds. Buchm{\"u}ller and G.
  Ingelman, DESY, Hamburg, vol. 1, 1992, 43.

\bibitem{Adloff:2003uh}
 H1 Collaboration, C.~Adloff {\em et~al.},
  \href{http://dx.doi.org/10.1140/epjc/s2003-01257-6}{Eur. Phys. J. C{\bfseries
  ~30} (2003) 1}, \href{http://arxiv.org/abs/hep-ex/0304003}{
  [hep-ex/0304003]}.

\bibitem{Aaron:2012qi}
 H1 Collaboration, F.~D. Aaron {\em et~al.},
  \href{http://dx.doi.org/10.1007/JHEP09(2012)061}{JHEP{\bfseries ~1209} (2012)
  061},
\href{http://arxiv.org/abs/1206.7007}{ [arXiv:1206.7007]}.

\bibitem{Bassler:95:197}
U.~Bassler and G.~Bernardi,
  \href{http://dx.doi.org/10.1016/0168-9002(95)00173-5}{Nucl. Instr. and Meth.
  A{\bfseries ~361} (1995) 197}, \href{http://arxiv.org/abs/hep-ex/9412004}{
  [hep-ex/9412004]}.

\bibitem{Bassler:99:583}
U.~Bassler and G.~Bernardi,
  \href{http://dx.doi.org/10.1016/S0168-9002(99)00044-3}{Nucl. Instr. and Meth.
  A{\bfseries ~426} (1999) 583}, \href{http://arxiv.org/abs/hep-ex/9801017}{
  [hep-ex/9801017]}.

\bibitem{Wobisch:00}
M.~Wobisch, {\em Measurement and QCD Analysis of Jet Cross Sections in
  Deep-Inelastic Positron-Proton Collisions at $\sqrt{s}=300$ GeV}.
\newblock Dissertation, RWTH Aachen, DESY-THESIS-2000-049, 2000.
\newblock Available at
  \url{http://www-h1.desy.de/publications/theses\_list.html}.

\bibitem{Cacciari:06:57}
M.~Cacciari and G.~P. Salam,
  \href{http://dx.doi.org/10.1016/j.physletb.2006.08.037}{Phys. Lett.
  B{\bfseries ~641} (2006) 57}, \href{http://arxiv.org/abs/hep-ph/0512210}{
  [hep-ph/0512210]}.

\bibitem{Chekanov:2006yc}
 ZEUS Collaboration, S.~Chekanov {\em et~al.},
 \href{http://dx.doi.org/10.1016/j.physletb.2007.03.039}{Phys. Lett. 
 B{\bfseries ~649} (2007) 12}, \href{http://arxiv.org/abs/hep-ex/0701039}{
  [hep-ex/0701039]}.

\bibitem{Frixione:97:315}
S.~Frixione and G.~Ridolfi,
  \href{http://dx.doi.org/10.1016/S0550-3213(97)00575-0}{Nucl. Phys.
  B{\bfseries ~507} (1997) 315}.

\bibitem{Gouzevitch:08}
M.~Gouzevitch, {\em Measurement of the strong coupling constant \as with jets
  in deep-inelastic scattering (in French)}.
\newblock Dissertation, Ecole Polytechnique Palaiseau, DESY-THESIS-2008-047,
  2008.
\newblock Available at
  \url{http://www-h1.desy.de/publications/theses\_list.html}.

\bibitem{Brun:87}
R.~Brun {\em et~al.}, CERN-DD/EE 84-1(1987) .

\bibitem{Pumplin:2002vw}
J.~Pumplin, D.~Stump, J.~Huston, H.~Lai, P.~M. Nadolsky, {\em et~al.},
  \href{http://dx.doi.org/10.1088/1126-6708/2002/07/012}{JHEP{\bfseries ~0207}
  (2002) 012},
\href{http://arxiv.org/abs/hep-ph/0201195}{ [hep-ph/0201195]}.

\bibitem{Charchula:94:381}
K.~Charchula, G.~A. Schuler, and H.~Spiesberger,
  \href{http://dx.doi.org/10.1016/0010-4655(94)90086-8}{Comput. Phys.
  Commun.{\bfseries ~81} (1994) 381}.

\bibitem{Lonnblad:92:15}
L.~L{\"o}nnblad, \href{http://dx.doi.org/10.1016/0010-4655(92)90068-A}{Comput.
  Phys. Commun.{\bfseries ~71} (1992) 15}.

\bibitem{Lonnblad:01}
L.~L{\"o}nnblad, Manual, Oct. 2001, available at
  \url{http://home.thep.lu.se/~leif/ariadne/ariadne.pdf}.

\bibitem{Jung:95:147}
H.~Jung, \href{http://dx.doi.org/10.1016/0010-4655(94)00150-Z}{Comput. Phys.
  Commun.{\bfseries ~86} (1995) 147}.

\bibitem{Jung:06}
H.~Jung, {Manual, Aug. 2006, available at
  \url{http://www.desy.de/~jung/rapgap/rapgap-desy.html} }.

\bibitem{Andersson:83:31}
B.~Andersson {\em et~al.},
  \href{http://dx.doi.org/10.1016/0370-1573(83)90080-7}{Physics
  Reports{\bfseries ~97} (1983) 31}.

\bibitem{Sjostrand:95iq}
T.~Sj{\"o}strand, \href{http://arxiv.org/abs/hep-ph/9508391}{
  arXiv:hep-ph/9508391}.

\bibitem{Schael:2004ux}
 ALEPH Collaboration, S.~Schael {\em et~al.},
\href{http://dx.doi.org/10.1016/j.physletb.2004.12.018}{Phys. Lett. B{\bfseries
  ~606} (2005) 265}.

\bibitem{Kwiatkowski:92:155}
A.~Kwiatkowski, H.~J. Spiesberger, and H.~J. M\"ohring,
  \href{http://dx.doi.org/10.1016/0010-4655(92)90136-M}{Comput. Phys.
  Commun.{\bfseries ~69} (1992) 155}.

\bibitem{Ingelman:97:108}
G.~Ingelman, A.~Edin, and J.~Rathsman,
  \href{http://dx.doi.org/10.1016/S0010-4655(96)00157-9}{Comput. Phys.
  Commun.{\bfseries ~101} (1997) 108},
  \href{http://arxiv.org/abs/hep-ph/9605286}{ [arXiv:hep-ph/9605286]}.

\bibitem{Schmitt:2012kp}
S.~Schmitt,
  \href{http://dx.doi.org/10.1088/1748-0221/7/10/T10003}{JINST{\bfseries ~7}
  (2012) T10003},
\href{http://arxiv.org/abs/1205.6201}{ [arXiv:1205.6201]}.

\bibitem{Britzger:13}
D.~Britzger, {\em Regularized unfolding of jet cross sections in deep-inelastic
  $ep$ scattering at HERA and determination of the strong coupling constant}.
\newblock Dissertation, Universit{\"a}t Hamburg, DESY-THESIS-2013-045, 2013.
\newblock Available at
  \url{http://www-h1.desy.de/publications/theses\_list.html}.

\bibitem{Kluge:2006xs}
T.~Kluge, K.~Rabbertz, and M.~Wobisch, Proceedings of "14th International
  Workshop on Deep Inelastic Scattering (DIS 2006)", eds. M. Kuze, K. Nagano
  and K. Tokushuku, Tsukuba, Japan, 2007, 483,
\href{http://arxiv.org/abs/hep-ph/0609285}{ [hep-ph/0609285]}.

\bibitem{Britzger2012}
D.~Britzger, K.~Rabbertz, F.~Stober, and M.~Wobisch, Proceedings of "20th
  International Workshop on Deep-Inelastic Scattering and Related Subjects (DIS
  2012)", ed. I. Brock, Bonn, Germany, 2013, 217,
\href{http://arxiv.org/abs/1208.3641}{ [arXiv:1208.3641]}.

\bibitem{Nagy:1998bb}
Z.~Nagy and Z.~Trocsanyi, \href{http://dx.doi.org/10.1103/PhysRevD.62.099902,
  10.1103/PhysRevD.59.014020}{Phys. Rev. D{\bfseries ~59} (1999) 014020},
  \href{http://arxiv.org/abs/hep-ph/9806317}{ [hep-ph/9806317]}
Erratum ibid, 62 (2000) 099902.

\bibitem{Nagy:2001xb}
Z.~Nagy and Z.~Trocsanyi,
  \href{http://dx.doi.org/10.1103/PhysRevLett.87.082001}{Phys. Rev.
  Lett.{\bfseries ~87} (2001) 082001},
\href{http://arxiv.org/abs/hep-ph/0104315}{ [hep-ph/0104315]}.

\bibitem{Whalley:2005nh}
M.~Whalley, D.~Bourilkov, and R.~Group, Proceedings for "HERA and the LHC",
  eds. A. De Roeck and H. Jung, Hamburg, Germany, 2005, 575,
\href{http://arxiv.org/abs/hep-ph/0508110}{ [hep-ph/0508110]}.

\bibitem{Martin:09:189}
A.~Martin, W.~Stirling, R.~Thorne, and G.~Watt,
  \href{http://dx.doi.org/10.1140/epjc/s10052-009-1072-5}{Eur. Phys. J.
  C{\bfseries ~63} (2009) 189},
\href{http://arxiv.org/abs/0901.0002}{ [arXiv:0901.0002]}.

\bibitem{Martin:2009bu}
A.~Martin, W.~Stirling, R.~Thorne, and G.~Watt,
  \href{http://dx.doi.org/10.1140/epjc/s10052-009-1164-2}{Eur. Phys. J.
  C{\bfseries ~64} (2009) 653},
\href{http://arxiv.org/abs/0905.3531}{ [arXiv:0905.3531]}.

\bibitem{Beringer:1900zz}
 Particle Data Group, J.~Beringer {\em et~al.},
  \href{http://dx.doi.org/10.1103/PhysRevD.86.010001}{Phys. Rev. D{\bfseries
  ~86} (2012) 010001}
and 2013 partial update for the 2014 edition.

\bibitem{Bodenstein:2012pw}
S.~Bodenstein, C.~A. Dominguez, K.~Schilcher, and H.~Spiesberger,
  \href{http://dx.doi.org/10.1103/PhysRevD.86.093013}{Phys. Rev. D{\bfseries
  ~86} (2012) 093013},
\href{http://arxiv.org/abs/1209.4802}{ [arXiv:1209.4802]}.

\bibitem{Botje:11:490}
M.~Botje, \href{http://dx.doi.org/10.1016/j.cpc.2010.10.020}{Comput. Phys.
  Commun.{\bfseries ~182} (2011) 490},
\href{http://arxiv.org/abs/1005.1481}{ [arXiv:1005.1481]}.

\bibitem{Gleisberg:2008ta}
T.~Gleisberg {\em et~al.},
  \href{http://dx.doi.org/10.1088/1126-6708/2009/02/007}{JHEP{\bfseries ~0902}
  (2009) 007},
\href{http://arxiv.org/abs/0811.4622}{ [arXiv:0811.4622]}.

\bibitem{Webber:84:492}
B.~Webber, \href{http://dx.doi.org/10.1016/0550-3213(84)90333-X}{Nucl. Phys.
  B{\bfseries ~238} (1984) 492}.

\bibitem{Collins:2002ey}
J.~C. Collins and X.-M. Zu,
  \href{http://dx.doi.org/10.1088/1126-6708/2002/06/018}{JHEP{\bfseries ~0206}
  (2002) 018},
\href{http://arxiv.org/abs/hep-ph/0204127}{ [hep-ph/0204127]}.

\bibitem{Campbell2007}
J.~M. Campbell, J.~W. Huston, and W.~J. Stirling,
  \href{http://dx.doi.org/10.1088/0034-4885/70/1/R02}{Rept. Prog.
  Phys.{\bfseries ~70} (2007) 89},
\href{http://arxiv.org/abs/hep-ph/0611148}{ [arXiv:hep-ph/0611148]}.

\bibitem{Lai:2010vv}
H.-L. Lai {\em et~al.},
  \href{http://dx.doi.org/10.1103/PhysRevD.82.074024}{Phys. Rev. D{\bfseries
  ~82} (2010) 074024},
\href{http://arxiv.org/abs/1007.2241}{ [arXiv:1007.2241]}.

\bibitem{Ball:2012cx}
R.~D. Ball {\em et~al.},
  \href{http://dx.doi.org/10.1016/j.nuclphysb.2012.10.003}{Nucl. Phys.
  B{\bfseries ~867} (2013) 244},
\href{http://arxiv.org/abs/1207.1303}{ [arXiv:1207.1303]}.

\bibitem{Aaron:2009aa}
 H1 and ZEUS Collaborations, F.~D. Aaron {\em et~al.},
  \href{http://dx.doi.org/10.1007/JHEP01(2010)109}{JHEP{\bfseries ~1001} (2010)
  109},
\href{http://arxiv.org/abs/0911.0884}{ [arXiv:0911.0884]}.

\bibitem{HERAPDF15}
 H1 and ZEUS Collaborations, {\em PDF fits including HERA-II high $Q^2$ data
  (HERAPDF1.5),} {Preliminary result, H1prelim-10-142, ZEUS-prel-10-018}, 2010.
\newblock See
  \url{http://www-h1.desy.de/publications/H1preliminary.short_list.html}.

\bibitem{Radescu:2010zz}
 H1 and ZEUS Collaborations, V.~Radescu, Proceedings of "35th International
  Conference on High Energy Physics (ICHEP 2010)", eds. B. Pire {\em et~al.},
  Paris, France, 2010, 168,
\href{http://arxiv.org/abs/1308.0374}{ [arXiv:1308.0374]}.

\bibitem{Alekhin:2012ig}
S.~Alekhin, J.~Bl\"umlein, and S.~Moch,
  \href{http://dx.doi.org/10.1103/PhysRevD.86.054009}{Phys. Rev. D{\bfseries
  ~86} (2012) 054009},
\href{http://arxiv.org/abs/1202.2281}{ [arXiv:1202.2281]}.

\bibitem{Barone:00:243}
V.~Barone, C.~Pascaud, and F.~Zomer,
  \href{http://dx.doi.org/10.1007/s100529900198}{Eur. Phys. J. C{\bfseries ~12}
  (2000) 243},
\href{http://arxiv.org/abs/hep-ph/9907512}{ [hep-ph/9907512]}.

\bibitem{Aktas:07:134}
 H1 Collaboration, A.~Aktas {\em et~al.},
  \href{http://dx.doi.org/10.1016/j.physletb.2007.07.050}{Phys. Lett.
  B{\bfseries ~653} (2007) 134},
\href{http://arxiv.org/abs/0706.3722}{ [arXiv:0706.3722]}.

\bibitem{Soper:97}
D.~E. Soper,
\href{http://arxiv.org/abs/hep-ph/9702203}{ hep-ph/9702203}.

\bibitem{Abramowicz:1900rp}
 H1 and ZEUS Collaborations, H.~Abramowicz {\em et~al.},
  \href{http://dx.doi.org/10.1140/epjc/s10052-013-2311-3}{Eur. Phys. J.
  C{\bfseries ~73} (2013) 2311},
\href{http://arxiv.org/abs/1211.1182}{ [arXiv:1211.1182]}.

\bibitem{Bethke:2012jm}
S.~Bethke, \href{http://dx.doi.org/10.1016/j.nuclphysbps.2012.12.020}{Nucl.
  Phys. Proc. Suppl.{\bfseries ~234} (2013) 229},
  \href{http://arxiv.org/abs/1210.0325}{ [arXiv:1210.0325]}.

\bibitem{Abramowicz:2012jz}
 ZEUS Collaboration, H.~Abramowicz {\em et~al.},
  \href{http://dx.doi.org/10.1016/j.nuclphysb.2012.06.006}{Nucl. Phys.
  B{\bfseries ~864} (2012) 1},
\href{http://arxiv.org/abs/1205.6153}{ [arXiv:1205.6153]}.



\bibitem{Dissertori:2007xa}
G.~Dissertori, A.~Gehrmann-De Ridder, T.~Gehrmann, E.~W.~N.~Glover, G.~Heinrich and H.~Stenzel,
\href{http://dx.doi.org/10.1088/1126-6708/2008/02/040} {JHEP {\bf 0802} (2008) 040},
\href{http://arxiv.org/abs/0712.0327}{ [arXiv:0712.0327]}.
  
\bibitem{Schieck:2006tc}
JADE Collaboration, J.~Schieck {\em et~al.},
\href{http://dx.doi.org/10.1140/epjc/s10052-007-0226-6}, \href{http://dx.doi.org/10.1140/epjc/s10052-007-0226-6}
{Eur.\ Phys.\ J.\ C {\bf 48} (2006) 3 [Erratum-ibid.\ C {\bf 50} (2007) 769]}
\href{http://arxiv.org/abs/0707.0392}{ [arXiv:0707.0392]}.  
  
\bibitem{OPAL:2011aa}
OPAL Collaboration, G.~Abbiendi {\em et~al.},
\href{http://dx.doi.org/10.1140/epjc/s10052-011-1733-z}  
{Eur.\ Phys.\ J.\ C {\bf 71} (2011) 1733}
\href{http://arxiv.org/abs/1101.1470}{ [arXiv:1101.1470]}.  
  
\bibitem{Chatrchyan:2013txa}
CMS Collaboration, S.~Chatrchyan {\em et~al.},
\href{http://dx.doi.org/10.1140/epjc/s10052-013-2604-6}{Eur. Phys. J. 
C{\bfseries ~73} (2013) 2604},
\href{http://arxiv.org/abs/1304.7498}{ [arXiv:1304.7498]}.

\bibitem{Abazov:2012lua}
D0 Collaboration, V.~Abazov {\em et~al.},
\href{http://dx.doi.org/10.1016/j.physletb.2012.10.003}{Phys. Lett.
B{\bfseries ~718} (2012) 56},
\href{http://arxiv.org/abs/1207.4957}{ [arXiv:1207.4957]}.


  


\end{thebibliography}
\end{document}